\documentclass{reviewarticle2}
\usepackage{graphicx,amssymb}

\def\be{\begin{equation}}
\def\ee{\end{equation}}
\def\bea{\begin{eqnarray}}
\def\eea{\end{eqnarray}}
\def\mbf{\mathbf}
\def\ovl{\overline}
\def\ovrw{\vec}

\makeindex

\begin{document}
\begin{frontmatter}

\title{Multiphoton Quantum Optics and \\ Quantum State Engineering}

\author{Fabio Dell'Anno, Silvio De Siena, and Fabrizio Illuminati}

\ead{dellanno@sa.infn.it, desiena@sa.infn.it, illuminati@sa.infn.it}

\address{Dipartimento di Fisica ``E. R. Caianiello'', Universit\`a
degli Studi di Salerno, CNISM and CNR-INFM, Coherentia, and INFN
Sezione di Napoli - Gruppo Collegato di Salerno, Via S. Allende,
I-84081 Baronissi (SA), Italy}

\begin{abstract}
We present a review of theoretical and experimental aspects
of multiphoton quantum optics. Multiphoton processes occur
and are important for many aspects of matter-radiation
interactions that include the efficient ionization of
atoms and molecules, and, more generally, atomic transition
mechanisms; system-environment couplings and
dissipative quantum dynamics; laser physics, optical
parametric processes, and interferometry.
A single review cannot account for all aspects of such an
enormously vast subject. Here we choose to concentrate our
attention on parametric processes in nonlinear media, with
special emphasis on the engineering of nonclassical states
of photons and atoms that are relevant for the
conceptual investigations as well as for the
practical applications of forefront aspects of
modern quantum mechanics.
We present a detailed analysis of the methods and techniques
for the production  of genuinely quantum multiphoton processes
in nonlinear media, and the corresponding models of
multiphoton effective interactions. We
review existing proposals for the classification,
engineering, and manipulation of nonclassical states,
including Fock states, macroscopic superposition states,
and multiphoton generalized coherent states.
We introduce and discuss the structure of canonical multiphoton quantum optics and
the associated one- and two-mode canonical multiphoton squeezed
states. This framework provides a consistent multiphoton
generalization of two-photon quantum optics and a consistent Hamiltonian
description of multiphoton processes associated to higher-order nonlinearities.
Finally, we discuss very recent advances that by combining
linear and nonlinear optical devices allow to realize multiphoton
entangled states of the electromnagnetic field, either in discrete or
in continuous variables, that are relevant for applications
to efficient quantum computation, quantum teleportation, and
related problems in quantum communication and information.
\end{abstract}

\end{frontmatter}

\newpage

\tableofcontents

\newpage

\section{Introduction}
\label{intro}

In this report we review and discuss recent developments
in the physics of multiphoton processes in nonlinear optical media and
optical cavities, and their manipulation in the presence of passive and active
optical elements. We review as well effective Hamiltonian models and Hamiltonian
dynamics of nonquadratic (anharmonic) multiphoton interactions,
and the associated engineering of nonclassical states of light
beyond the standard coherent and two-photon squeezed states of linear
quantum optics.\\
We present a detailed analysis of the methods and techniques
for the production of genuinely quantum multiphoton processes
in nonlinear media, and the corresponding models of
multiphoton effective nonlinear interactions.
Our main goal is to introduce the reader to the fascinating
field of quantum nonlinear optical effects (such as, e.g., quantized Kerr interactions,
quantized four-wave mixing, multiphoton down conversion, and electromagnetically induced
transparency) and their application to the engineering of (generally non Gaussian),
nonclassical states of the quantized electromagnetic field, optical Fock states,
macroscopic superposition states such as, e. g., optical Schr\"odinger cat states,
multiphoton squeezed states and generalized coherent states,
and multiphoton entangled states.\\
This review is mainly devoted to the theoretical aspects of
multiphoton quantum optics in nonlinear media and cavities, and
theoretical models of quantum state engineering. However,
whenever possible, we tried to keep contact with
experimental achievements and the more promising
experimental setups proposals. We tried to provide a self-contained
introduction to some of the most relevant and appropriate
theoretical tools in the physics of multiphoton quantum optics.
In particular, we have devoted a somewhat detailed
discussion to the recently introduced formalism of canonical
multiphoton quantum optics, a systematic and consistent
multiphoton generalization of standard one- and two-photon quantum
optics. We have included as well an introduction to
group-theoretical techniques and nonlinear
operatorial generalizations for the definition of some
types of nonclassical multiphoton states.
Our review is completed by a self-contained
discussion of very recent advances that by combining
linear and nonlinear optical devices have lead to the
realization of some multiphoton entangled states of the
electromnagnetic field. This multiphoton entanglement, that
has been realized either on discrete or on continuous variables
systems, is relevant for applications in efficient quantum
computation, quantum teleportation, and related problems in
quantum communication and information.\\
Multiphoton processes occur in a large variety of phenomena
in the physics of matter-radiation interactions. Clearly,
it is a task beyond our abilities and incompatible with
the requirements that a review article should be of
a reasonable length extension, and sufficiently self-contained.
We thus had to make a selection of topics, that was dictated
partly by our personal competences and tastes, and partly
because of the rapidly growing importance of research fields
including engineering and control of nonclassical states of
light, quantum entanglement, and quantum
information. Therefore, our review is concerned with that
part of multiphoton processes that leans towards the
``deep quantum'' side of quantum optics, and it does not
cover such topics as Rydberg states and atoms, intense fields,
multiple ionization, and molecular processes, that are all,
in some sense, on the ``semiclassical'' side of the discipline.
Moreover, we have not included sections or discussions specifically
devoted to quantum noise, quantum dissipative effects, and decoherence.
A very brief ``framing'' discussion with some essential bibliography on
these topics is included in the conclusions.

The plan of the paper is the following. In Section
\ref{section2} we give a short review of linear quantum
optics, introduce the formalism of quasi-probabilities in
phase space, and discuss the basics of homodyne and heterodyne
detections and of quantum state tomography.
In Section \ref{section3} we introduce the theory of
quantized macroscopic fields in nonlinear media, and
we discuss the basic properties of multiphoton parametric
processes, including the requirements of energy conservation
and phase matching, and the different experimental techniques
for the realization of these requirements and for the enhancement
of the parametric processes corresponding to higher-order
nonlinear susceptibilities.
In Section \ref{section4} we discuss in detail some
of the most important and used parametric processes
associated to second- and third-order optical nonlinearities,
the realization of concurring interactions, including
three- and four-wave mixing, Kerr and Kerr-like interactions,
three-photon down conversion, and a first introduction to
the engineering of mesoscopic quantum superpositions, and
multiphoton entangled states.
In Section \ref{section5} we describe group theoretical methods
for the definition of generalized (multiphoton) coherent states,
Hamiltonian models of higher-order nonlinear processes,
including degenerate $k$-photon down conversions with classical
and quantized pumps, Fock state generation in multiphoton
parametric processes, displaced-squeezed number states and
Kerr states, intermediate (binomial) states of the radiation
field, photon-added and photon-subtracted states, higher-power
coherent and squeezed states, and general $n$-photon schemes
for the engineering of arbitrary nonclassical states.
As already mentioned, in Section \ref{section6} we report on
a recently established general canonical formulation of multiphoton
quantum optics, that allows to introduce multiphoton squeezed
states associated to exact canonical structures and diagonalizable
Hamiltonians (multiphoton normal modes), we study their two-mode
extensions defining non Gaussian entangled states, and
we discuss some proposed setups for their experimental realization.
In Section \ref{section7} we give a bird-eye view on the
most relevant theoretical and experimental applications of
multiphoton quantum processes and multiphoton nonclassical states
in fields of quantum communication and information.
Finally, in Section \ref{section8} we present our conclusions and
discuss future perspectives.

\newpage

\section{A short review of linear quantum optics}
\label{section2}

In 1927, Dirac \cite{Dirac} was the first to carry out successfully
the (nonrelativistic) quantization of the free electromagnetic
field, by associating each mode of the radiation field with a
quantized harmonic oscillator. Progress then followed with the
inclusion of matter-radiation interaction \cite{Dirac2}, the
definition of the general theory of the interacting matter and
radiation fields \cite{HeisenbergPauli,Fermi}, and, after two
decades of strenuous efforts, the final construction of
divergence-free quantum electrodynamics in its modern form
\cite{Feynman,Schwinger,Tomonaga,Schweber,BjorkenDrell,ItzyksonZuber,Weinberg}.
However, despite these fundamental theoretical achievements and
the parallel experimental triumphs in the understanding of
electron-photon and atom-photon interactions, only in the sixties,
after the discovery of the laser \cite{Laser}, quantum optics
entered in its modern era when the theory of quantum optical
coherence was systematically developed for the first time by
Glauber, Klauder, and Sudarshan
\cite{Glauber1,Glauber2,Glauber3,Sudarshan,KlauderSpinor,KlauderSudar,KlauderSkager}.\\
Quantum electrodynamics predicts, and is necessary to describe
and understand such fundamental effects as spontaneous emission,
the Lamb shift, the laser linewidth, the Casimir effect, and the
photon statistics of the laser. The classical theory of radiation
fails to account for such effects which can only be explained in
terms of the perturbation of the atomic states due to the vacuum
fluctuations of the quantized electromagnetic field. Early
mathematical developments of quantum electrodynamics have relied
heavily on perturbation theory and have assigned a privileged role
to the orthonormal basis of Fock number states.
Such a formulation is however not very useful nor really appropriate
when dealing with coherent processes and structures, like laser beams
and nonlinear optical effects, that usually involve large numbers of photons,
large-scale space-time correlations, and different types of few- and
multi-photon effective interactions. For an exhaustive and comprehensive
phenomenological and mathematical introduction to the subjects of
quantum optics, see
\cite{Louiselltx,MandelWolftx,WallsMilburntx,BarnettRadmoretx,ScullyZubairytx,Loudontx,Sargenttx,Puritx,Haustx,Milonnitx,Yarivtx}.\\
Before we come to deal with multiphoton processes and multiphoton
quantum optics, we need to dedicate the remaining of this Section
to a brief review of the physical ``one-photon'' context in which
quantum optics was born. By shortly recalling the theory of
optical coherence and a self-contained tutorial on the formalism
of coherent states and their properties, this Section will serve
as an introduction to the notation and some of the mathematical
tools that will be needed in the course of this review.

\subsection{Generation of fully coherent radiation: an historical overview}

From a classical point of view, the coherence of light is
associated with the appearing of interference fringes. The
superposition of optical beams with equal frequency and steady
phase difference gives rise to an interference pattern, which can
be due to temporal or spatial coherence. Traditionally, an optical
field is defined to be coherent when showing first-order
coherence. The classic interference experiment of Young's double
slit can be described by means of the first-order correlation
function, using either classical or quantum theory. The
realization of experiments on intensity interferometry and
photoelectric counting statistics \cite{HanbTwiss,Rebka} led to
the introduction of higher-order correlation functions. In
particular, Hanbury Brown and Twiss \cite{HanbTwiss} verified the
bunching effect, showing that the photons of a light beam of
narrow spectral width have a tendency to arrive in correlated
pairs. A semiclassical approach was used by Purcell \cite{Purcell}
to explain the correlations observed in the photoionization
processes induced by a light beam. Mandel and Wolf
\cite{Mandelstoc} examined the correlations, retaining the
assumption that the electric field in a light beam can be
described as a classical Gaussian stochastic process. In 1963 came
the contributions by Glauber, Klauder and Sudarshan
\cite{Glauber1,Glauber2,Glauber3,Sudarshan,KlauderSpinor,KlauderSudar,KlauderSkager}
that were essential in opening a new and very fruitful path
of theoretical and experimental investigations.
Glauber re-introduced the coherent states,
first discovered by Schr\"{o}dinger \cite{Schrodcohst} in
the study of the quasi-classical properties of the harmonic oscillator,
to study the quantum coherence of optical fields as a cooperative
phenomenon in terms of many bosonic degrees of freedom.
Here we shortly summarize Glauber's procedure
for the construction of the electromagnetic field's coherent states
\cite{Glauber1,Glauber2,Glauber3}. The observable quantities of
the electromagnetic field are taken to be the electric and
magnetic fields $\ovrw{E}(\ovrw{r},t)$ and $\ovrw{B}(\ovrw{r},t)$,
which satisfy the nonrelativistic Maxwell equations in free space and in absence
of sources:
\be
\ovrw{\nabla}\cdot\ovrw{E}=0 \,, \quad\; \ovrw{\nabla}\cdot\ovrw{B}=0
\,, \quad\; \ovrw{\nabla}\times\ovrw{E}=-\frac{\partial \ovrw{B}}{\partial t}
\,, \quad\; \ovrw{\nabla}\times\ovrw{B}=\frac{\partial
\ovrw{E}}{\partial t} \; .
\label{freeMaxwell}
\ee
Here and throughout the whole report, we will adopt international
units with $c = \hbar = 1$.
The dynamics is governed by the
electromagnetic Hamiltonian \be H_{0}=\frac{1}{2} \int
(\ovrw{E}^{2}+\ovrw{B}^{2})d^{3}r \; , \label{freefieldHam}
\ee
and
the electric and magnetic fields can be expressed in terms of the
vector potential $\ovrw{A}$:
\be
\ovrw{E}=-\frac{\partial\ovrw{A}}{\partial t} \,\,\,, \quad
\ovrw{B}=\ovrw{\nabla}\times\ovrw{A} \,\,, \label{vectorpot} \ee
where the Coulomb gauge condition $\ovrw{\nabla}\cdot\ovrw{A}=0$
has been chosen. Quantization is obtained by replacing the
classical vector potential by the operator
\be
\ovrw{A}(\ovrw{r},t)=\sum_{\ovrw{k},\lambda}\frac{1}{\sqrt{2\omega_{\ovrw{k},\lambda}V}}
\{a_{\ovrw{k},\lambda}\hat{\epsilon}_{\ovrw{k},\lambda}
e^{i\ovrw{k}\cdot\ovrw{r}-i\omega_{\ovrw{k},\lambda}t}
+a_{\ovrw{k},\lambda}^{\dag}\hat{\epsilon}_{\ovrw{k},\lambda}^{*}
e^{-i\ovrw{k}\cdot\ovrw{r}+i\omega_{\ovrw{k},\lambda}t}\} \,\,,
\label{standquantiz} \ee with the transversality condition and the
bosonic canonical commutation relations given by
\be
\ovrw{k}\cdot\hat{\epsilon}_{\ovrw{k},\lambda}=0 \,\,, \quad\quad
[a_{\ovrw{k},\lambda},a_{\ovrw{k}',\lambda'}^{\dag}]
=\delta_{\ovrw{k},\ovrw{k}'}\delta_{\lambda,\lambda'} \,, \quad
(\hat{\epsilon}_{\ovrw{k},\lambda}^{*}\cdot\hat{\epsilon}_{\ovrw{k},\lambda'}
=\delta_{\lambda,\lambda'}) \; .
\label{QEDcommutat}
\ee
In Eq. (\ref{standquantiz}) $V$ is the
spatial volume, $\hat{\epsilon}_{\ovrw{k},\lambda}$ is the unit
polarization vector encoding the wave-vector $\ovrw{k}$ and the
polarization $\lambda$,
 $\omega_{\ovrw{k},\lambda}$ is the angular frequency, and
$a_{\ovrw{k},\lambda}$, $a_{\ovrw{k},\lambda}^{\dag}$ are the
corresponding bosonic annihilation and creation operators.
Denoting by $\mbf{k}$ the pair $(\ovrw{k},\lambda)$, the
Hamiltonian (\ref{freefieldHam}) reduces to
\be
H_{0} \, = \, \sum_{\mbf{k}} H_{0,\mbf{k}} \, = \,\frac{1}{2}\sum_{\mbf{k}}\omega_{\mbf{k}}
(a_{\mbf{k}}^{\dag}a_{\mbf{k}}+a_{\mbf{k}}a_{\mbf{k}}^{\dag}) \; .
\label{infharmosciH}
\ee
Equation (\ref{infharmosciH}) establishes
the correspondence between the mode operators of the
electromagnetic field and the coordinates of an infinite set of
harmonic oscillators. The number operator of the $k$-th mode
$n_{\mbf{k}}=a_{\mbf{k}}^{\dag}a_{\mbf{k}}$, when averaged over a
given quantum state, yields the number of photons present in mode
$\mbf{k}$, i.e. the number of photons possessing a given momentum
$\ovrw{k}$ and a given polarization $\lambda$. The operators
$n_{\mbf{k}}$, $a_{\mbf{k}}$, and $a_{\mbf{k}}^{\dag}$, together
with the identity operator $I_{\mbf{k}}$ form a closed algebra,
the Lie algebra $h_{4}$, also known as the Heisenberg-Weyl
algebra. The single-mode Hamiltonian $H_{0, \mbf{k}}$ has eigenvalues
$E_{\mbf{k}}=\omega_{\mbf{k}}\left(n_{\mbf{k}}+\frac{1}{2}\right)$,
$(n_{\mbf{k}}=0,1,2,...)$. The eigenstates of $H_{0}$ thus form a
complete orthonormal basis $\{|n_{\mbf{k}}\rangle\}$, and are
usually known as number or Fock states. The vacuum state
$|0_{\mbf{k}}\rangle$ is defined by the condition
\be
a_{\mbf{k}}|0_{\mbf{k}}\rangle=0 \,, \ee and the excited states
(excited number states) are given by successive applications of
the creation operators on the vacuum: \be |n_{\mbf{k}}\rangle=
\frac{(a_{\mbf{k}}^{\dag})^{n_{\mbf{k}}}}{\sqrt{n_{\mbf{k}}!}}|0_{\mbf{k}}\rangle
\,, \quad (n_{\mbf{k}}=0,1,2,...) \,.
\ee
From Eqs.
(\ref{vectorpot}) and (\ref{standquantiz}), the electric field
operator can be separated in the positive- and negative-frequency
parts
\bea
\ovrw{E}(\ovrw{r},t)&=&\ovrw{E}^{(+)}(\ovrw{r},t)+\ovrw{E}^{(-)}(\ovrw{r},t)
\,, \label{Electdecomp} \\ \ovrw{E}^{(+)}(\ovrw{r},t) &=&
i\sum_{\mbf{k}} \left[\frac{\omega_{\mbf{k}}} {2V}\right]^{1/2}
a_{\mbf{k}}\hat{\epsilon}_{\mbf{k}}
e^{i(\ovrw{k}\cdot\ovrw{r}-\omega_{\mbf{k}}t)} \,, \quad
\ovrw{E}^{(+)\dag}(\ovrw{r},t)=\ovrw{E}^{(-)}(\ovrw{r},t) \;.
\label{Eposnegfreq}
\eea The coherent states
$|\ovrw{\varepsilon}\rangle$ of the electromagnetic field are then
defined as the right eigenstates of $\ovrw{E}^{(+)}(\ovrw{r},t)$,
or equivalently as the left eigenstates of
$\ovrw{E}^{(-)}(\ovrw{r},t)$ :
\be
\ovrw{E}^{(+)}(\ovrw{r},t)|\ovrw{\varepsilon}\rangle =
\ovrw{\mathcal{E}}(\ovrw{r},t)|\ovrw{\varepsilon}\rangle \,, \quad
\langle\ovrw{\varepsilon}
|\ovrw{E}^{(-)}(\ovrw{r},t)=\ovrw{\mathcal{E}}^{*}(\ovrw{r},t)\langle\ovrw{\varepsilon}
| \,, \label{cohstdef1}
\ee
where the eigenvalue vector functions
$\ovrw{\mathcal{E}}(\ovrw{r},t)$ must satisfy the Maxwell
equations and may be expanded in a Fourier series with arbitrary
complex coefficients $\alpha_{\mbf{k}}$:
\be
\ovrw{\mathcal{E}}(\ovrw{r},t) = i\sum_{\mbf{k}}
\left[\frac{\omega_{\mbf{k}}} {2V}\right]^{1/2}
\alpha_{\mbf{k}}\hat{\epsilon}_{\mbf{k}}
e^{i(\ovrw{k}\cdot\ovrw{r}-\omega_{\mbf{k}}t)} \;.
\label{cohsteps}
\ee
From Eqs. (\ref{cohstdef1}) and
(\ref{cohsteps}), it follows that the coherent states are uniquely
identified by the complex coefficients $\alpha_{\mbf{k}}$:
$|\ovrw{\varepsilon}\rangle\equiv|\{\alpha_{\mbf{k}}\}\rangle$,
and, moreover, they are determined by the equations
\be
|\{\alpha_{\mbf{k}}\}\rangle
=\prod_{\mbf{k}}|\alpha_{\mbf{k}}\rangle , \quad
a_{\mbf{k}}|\alpha_{\mbf{k}}\rangle=
\alpha_{\mbf{k}}|\alpha_{\mbf{k}}\rangle , \quad
|\alpha_{\mbf{k}}\rangle=
e^{-\frac{|\alpha_{\mbf{k}}|^{2}}{2}}\sum_{n_{\mbf{k}}}
\frac{\alpha^{n_{\mbf{k}}}}{\sqrt{n_{\mbf{k}}!}}|n_{\mbf{k}}\rangle
. \label{cohstdef2}
\ee
Coherent states possess two fundamental
properties, non-orthogonality and over-completeness, expressed by
the following relations
\bea
&&\langle\alpha_{\mbf{k}}|\alpha_{\mbf{k}'}\rangle
=\exp\left[\alpha_{\mbf{k}}^{*}\alpha_{\mbf{k}'}
-\frac{1}{2}(|\alpha_{\mbf{k}}|^{2}+|\alpha_{\mbf{k}'}|^{2})\right]
\,, \nonumber \\
&& \nonumber \\
&&\int
|\{\alpha_{\mbf{k}}\}\rangle\langle\{\alpha_{\mbf{k}}\}|\prod_{\mbf{k}}\pi^{-1}d^{2}\alpha_{\mbf{k}}
= \hat{I} \; , \; \quad d^{2} \alpha_{\mbf{k}} = d Re[\alpha_{\mbf{k}}] d Im[\alpha_{\mbf{k}}] \; ,
\label{cohstprop}
\eea
where $\langle \cdot | \cdot
\rangle $ denotes the scalar product, and $\hat{I}$ is the
identity operator. Being indexed by a continuous complex
parameter, the set of coherent states is naturally over-complete,
but one can extract from it any complete orthonormal basis, and,
moreover, any arbitrary quantum state $|\psi\rangle$ can still be
expressed in terms of continuous superpositions of coherent
states:
\bea
&&|\psi\rangle=\int
f(\{\alpha_{\mbf{k}}^{*}\})|\{\alpha_{\mbf{k}}\}\rangle
\prod_{\mbf{k}}\pi^{-1} e^{-|\alpha|^{2}/2}d^{2}\alpha_{\mbf{k}}
\,, \nonumber \\
&& \nonumber \\
&&f(\{\alpha_{\mbf{k}}^{*}\})=\exp\{\frac{1}{2}\sum_{\mbf{k}}
|\alpha_{\mbf{k}}|^{2}\}\langle\{\alpha_{\mbf{k}}\}|\psi\rangle \,.
\eea
It is easy to show that the state (\ref{cohstdef2}) is
realized by the radiation emitted by a classical current. The
photon field radiated by an electric current distribution
$\ovrw{J}(\ovrw{r},t)$ is described by the interaction Hamiltonian
\be
H_{I}(t)=-\int \ovrw{J}(\ovrw{r},t)\cdot\ovrw{A}(\ovrw{r},t)
d^{3}r \; , \label{elecurrentH} \ee and thus the associated
time-dependent Schr\"{o}dinger equation is solved by the evolution
operator \bea U(t)&&=\exp\left\{-i\int_{-\infty}^{t}dt'\int
d^{3}r\ovrw{J}(\ovrw{r},t')\cdot\ovrw{A}(\ovrw{r},t')
+i\varphi(t')\right\} \nonumber \\
&&=\prod_{\mbf{k}}D_{\mbf{k}}(\alpha_{\mbf{k}})
=\prod_{\mbf{k}}\exp\{\alpha_{\mbf{k}}(t)a^{\dag}_{\mbf{k}}-\alpha^{*}_{\mbf{k}}(t)a_{\mbf{k}}\}
\; ,
\label{unitopchst}
\eea
where $\varphi(t)$ is an overall time-dependent
phase factor, the complex time-dependent amplitudes
$\alpha_{\mbf{k}}(t)$ read
\be
\alpha_{\mbf{k}}(t)=\frac{i}{\sqrt{2\omega_{\mbf{k}}V}}
\int_{-\infty}^{t}dt'\int d^{3}r
e^{-i\ovrw{k}\cdot\ovrw{r}
+i\omega_{\mbf{k}}t'}\hat{\epsilon}_{\mbf{k}}^{*}\cdot\ovrw{J}(\ovrw{r},t')
 \; ,
\ee
and, finally, the operator $D_{\mbf{k}}(\alpha_{\mbf{k}})
=\exp\{\alpha_{\mbf{k}}(t)a^{\dag}_{\mbf{k}}-\alpha^{*}_{\mbf{k}}(t)a_{\mbf{k}}\}$
is the one-photon Glauber displacement operator. Therefore the
coherent states of the electromagnetic field are generated by the
time evolution from an initial vacuum state under the action of
the unitary operator (\ref{unitopchst}). Looking at each of the
single-mode contributions in Eq. (\ref{unitopchst}), we see
clearly that the interaction corresponds to adding a linear
forcing part to the elastic force acting on each of the mode
oscillators, and that only one-photon processes are involved. Due
to this factorization, the single-mode coherent state
$|\alpha_{\mbf{k}}\rangle$, Eq. (\ref{cohstdef2}), associated to
the given mode $\mbf{k}$, is generated by the application of the
displacement operator $D_{\mbf{k}}(\alpha_{\mbf{k}})$ on the
single-mode vacuum $|0_{\mbf{k}}\rangle$. This can be easily
verified by resorting to the Baker-Campbell-Haussdorf relation
\cite{BarnettRadmoretx}, which in this specific case reads
$e^{\alpha_{\mbf{k}}a^{\dag}_{\mbf{k}}-\alpha^{*}_{\mbf{k}}a_{\mbf{k}}}=
e^{-\frac{1}{2}|\alpha_{\mbf{k}}|^{2}}
e^{\alpha_{\mbf{k}}a_{\mbf{k}}^{\dag}}e^{-\alpha_{\mbf{k}}^{*}a_{\mbf{k}}}$.
The decomposition of the electric field in the positive- and
negative-frequency parts not only allows to introduce the coherent
states in a very natural and direct way, but, moreover, leads, as
first observed by Glauber, to the definition of a sequence of
$n$-order correlation functions $G^{(n)}$
\cite{Glauber2,Glauber3,Glaubertit}. Let us consider the process
of absorption of $n$ photons, each one polarized in the
$\lambda_{j}$ direction $(j=1,...,n)$, and let the electromagnetic
field be in a generic quantum state (either pure or mixed)
described by some density operator $\rho$. The probability per
unit (time)$^{n}$ that $n$ ideal detectors will record $n$-fold
delayed coincidences of photons at points
$(\ovrw{r}_{1},t_{1}),...,(\ovrw{r}_{n},t_{n})$ is proportional to
\be
Tr[\rho
E_{\lambda_{1}}^{(-)}(\ovrw{r}_{1},t_{1})\cdot\cdot\cdot
E_{\lambda_{n}}^{(-)}(\ovrw{r}_{n},t_{n})E_{\lambda_{n}}^{(+)}(\ovrw{r}_{n},t_{n})
\cdot\cdot\cdot
E_{\lambda_{1}}^{(+)}(\ovrw{r}_{1},t_{1})] \,,
\label{corrfun}
\ee
where the polarization indices have been written explicitely.
Introducing the global variable
$x_{j}=(\ovrw{r}_{j},t_{j},\lambda_{j})$ that includes space, time
and polarization, the correlation function of order $n$ is easily
defined as a straightforward generalization of relation
(\ref{corrfun}):
\bea
&&G^{(n)}(x_{1},...,x_{n},x_{n+1},...,x_{2n})\equiv
G^{(n)}(x_{1},...,x_{2n})\nonumber \\
&& \nonumber \\
&&= Tr[\rho E^{(-)}(x_{1})\cdot\cdot\cdot
E^{(-)}(x_{n})E^{(+)}(x_{n+1})\cdot\cdot\cdot E^{(+)}(x_{2n})] \;
. \label{corrfundef}
\eea
These correlation functions are
invariant under permutations of the variables $(x_{1}...x_{n})$
and $(x_{n+1}...x_{2n})$ and their normalized forms are
conveniently defined as follows:
\be g^{(n)}(x_{1},...,x_{2n})=
\frac{G^{(n)}(x_{1},...,x_{2n})}{
\prod_{j=1}^{2n}\left\{G^{(1)}(x_{j},x_{j})\right\}^{1/2} } \; ,
\label{coherfunct} \ee so that, by definition, the necessary
condition for a field to have a degree of coherence equal to $n$
is \be |g^{(m)}(x_{1},...,x_{2m})|=1 \,,
\label{ncohercond1}
\ee
for every $m\leq n$. As $g^{(n)}(x_{1},...,x_{n};x_{n},...,x_{1})$
is a positive-defined function, other two alternative, but
equivalent, conditions for having coherence of order $n$ are:
\bea
&&g^{(m)}(x_{1},...,x_{m};x_{m},...,x_{1})=1 \; , \nonumber \\
&& \nonumber \\
&&G^{(m)}(x_{1},...,x_{m};x_{m},...,x_{1})=\prod_{i=1}^{m}G^{(1)}(x_{i},x_{i})
\; ,
\label{ncohercond2}
\eea
for every $m\leq n$. Relations
(\ref{ncohercond1}), or (\ref{ncohercond2}), are only necessary
conditions for $n$-order coherence, and mean that the detection
rate of $m$-fold delayed coincidences is equal to the products of
the detections rates of each photon counter. For a completely
coherent field, i.e., coherent at any arbitrary order, the
following equivalent conditions must be satisfied:
\bea
&&|g^{(n)}(x_{1},...,x_{2n})|=1 \; , \nonumber \\
&& \nonumber \\
&&|G^{(n)}(x_{1},...,x_{2n})|=\prod_{j=1}^{2n}\left\{G^{(1)}(x_{j},x_{j})\right\}^{1/2}
\; , \quad n=1,2,... \; .
\label{factorizcorrel}
\eea
Relations
(\ref{factorizcorrel}) indicate that an even stronger definition
of coherence can be adopted, by assuming the full factorization of
$G^{(n)}$ in terms of a complex function $\mathcal{E}(x)$ of the
global space-time-polarization variable:
\be
G^{(n)}(x_{1},...,x_{n},x_{n+1},...,x_{2n})=
\mathcal{E}^{*}(x_{1})\cdot\cdot\cdot\mathcal{E}^{*}(x_{n})
\mathcal{E}(x_{n+1})\cdot\cdot\cdot\mathcal{E}(x_{2n}) \,.
\ee
Coherent states defined by Eq. (\ref{cohstdef1}) imply the
complete factorization of the correlation functions, see e.g.
\cite{Glauber1,Glauber2,Glauber3}, and hence they describe a fully
coherent radiation field. Let us consider in more detail the
normalized second-order correlation function $g^{(2)}$. This
correlation is of particular importance when dealing with the
problem of discriminating the classical and the genuinely quantum,
or ``nonclassical'' in the quantum optics jargon, statistical
properties of a state. Let us consider only two modes $a_{j}$
$(j=1,2)$ of the field with frequencies $\omega_{j}$; then, the
normalized second order correlation function corresponding to the
probability of counting a photon in mode $i$ at time $t$ and a
photon in mode $j$ at time $t+\tau$ can be expressed in terms of the
creation and annihilation operators in the form:
\be
g_{ij}^{(2)}(t,t+\tau)=\frac{\langle
a_{i}^{\dag}(t)a_{j}^{\dag}(t+\tau)a_{i}(t)a_{j}(t+\tau)\rangle}{\langle
a_{i}^{\dag}(t)a_{i}(t)\rangle\langle
a_{j}^{\dag}(t+\tau)a_{j}(t+\tau)\rangle} \; , \quad (i,j=1,2) \;
,
\ee
where the notation is self-explanatory. For stationary
processes (processes invariant under time translations) $g^{(2)}$
is independent of $t$, i.e. $g^{(2)}(t,t+\tau) = g^{(2)}(0,\tau)
\equiv g^{(2)}(\tau)$. For zero time delay, we thus have
\be
g_{ij}^{(2)}(0)=\frac{\langle
a_{i}^{\dag}a_{j}^{\dag}a_{i}a_{j}\rangle}{\langle
a_{i}^{\dag}a_{i}\rangle\langle a_{j}^{\dag}a_{j}\rangle} \,.
\ee
In the case of a single-mode field, the equivalent relations are
obtained by eliminating the subscripts $i$ and $j$. For a coherent
state, all the correlation functions $g^{(n)}(\tau)=1$ at any
order $n$, and, in particular, $g^{(2)}(\tau)=1$. However, for a
generic state $g^{(2)}(\tau)\neq 1$ and the behavior of the
normalized second order correlation function is related to the
so-called bunching or antibunching effects \cite{antibPaul}. In
fact, for classical states photons exhibit a propensity to arrive
in pairs at a photodetector (bunching effect); deviations from
this tendency are then a possible signature of a genuinely
nonclassical behavior (antibunching effect: photons are revealed
each at a time at the photodetector). If one looks at second order
correlation functions, bunching is favored if
$g^{(2)}(\tau)<g^{(2)}(0)$, while antibunching is favored when
$g^{(2)}(\tau)>g^{(2)}(0)$ \cite{antibPaul}. Since
$\lim_{\tau\rightarrow\infty} g^{(2)}(\tau)= 1$, i.e. the
probability of joint detection coincides with the probability of
independent detection, a field for which $g^{(2)}(0)<1$ will
always exhibit photon antibunching on some time scale. As an
example, let us consider a single-mode, pure number state
described by the (projector) density operator
$\rho_{n}=|n\rangle\langle n|$, and a one-mode thermal state
described by the density operator
$\rho_{th}=\sum_{n}\frac{\bar{n}^{n}}{(1+\bar{n})^{n+1}}|n\rangle\langle
n|$, where $\bar{n}$ denotes the average number of thermal
photons. These two states both enjoy first-order coherence:
$|g_{n}^{(1)}(0)|=|g_{th}^{(1)}(0)|=1$, but it is easy to verify
that $g_{th}^{(2)}(0)=2$ while
$g_{n}^{(2)}(0)=\left\{\begin{array}{cc}1-\frac{1}{n} & (n\geq 2)
\, ,
\\ 0 & (n=0,1) \, .
\end{array}\right.$ \\ Then, the normalized second order correlation
function discriminates between the classical character of the
thermal state and the genuinely quantum nature of the number
states that exhibit antibunching. Similar results are obtained for
two-mode states by looking at the two-mode cross-correlation
functions. For a two-mode thermal state, described by the density
operator
$\rho_{2th}=\sum_{n_{1},n_{2}}\frac{\bar{n_{1}}^{n_{1}}}{(1+\bar{n_{1}})^{n_{1}+1}}
\frac{\bar{n_{2}}^{n_{2}}}{(1+\bar{n_{2}})^{n_{2}+1}}|n_{1},n_{2}\rangle\langle
n_{1},n_{2}|$, the second-order degree of coherence for the $i$-th
mode is given by $g_{ii}^{(2)}(0)=2$, while the intermode
cross-correlation is $g_{12}^{(2)}(0)=1$; therefore, for two-mode
thermal states, the direct- and cross-correlations satisfy the
classical Cauchy-Schwartz inequality \cite{repLoudon}
\be
g_{11}^{(2)}(0)g_{22}^{(2)}(0)>[g_{12}^{(2)}(0)]^{2} \; .
\ee
On
the contrary , for a two-mode number state
$|n_{1},n_{2}\rangle\langle n_{1},n_{2}|$ $(n_{i}\geq 2)$, we have
$g_{ii}^{(2)}(0)=1-\frac{1}{n_{i}}$ $(i=1,2)$ and
$g_{12}^{(2)}(0)=1$. Consequently, the nonclassicality of the
state emerges via the violation of the Cauchy-Schwartz inequality
for the direct- and the cross-correlations:
\be
g_{11}^{(2)}(0)g_{22}^{(2)}(0)=1-\frac{n_{1}+n_{2}-1}{n_{1}n_{2}}
<1=[g_{12}^{(2)}(0)]^{2} \; .
\ee
The most suitable framework for the description of the dynamical
and statistical properties of the quantum states of the radiation field
is established by introducing characteristic functions and appropriate quasi-probability
distributions in phase space \cite{Glauber1,Sudarshan,Cahillglaub},
that, among many other important properties, allow to handle and
compute expectation values of any kind of observable built from ordered products
of field operators. We will introduce and make use of some of the most important
quasi-probability distributions later on, but here we anticipate some remarks on the
phase-distribution function that Glauber and Sudarshan \cite{Glauber1,Sudarshan}
first introduced as the $P$-function representation of the density operator by
the equivalence
\be \rho=\int
P(\{\alpha_{\mbf{k}}\})|\{\alpha_{\mbf{k}}\}\rangle\langle\{\alpha_{\mbf{k}}\}|
\prod_{\mbf{k}}d^{2}\alpha_{\mbf{k}} \; . \label{Pfuncti} \ee In
particular, Sudarshan proved that, for any state $\rho$
of the quantized electromagnetic field, any expectation value
of any normally ordered operatorial function of the field operators
$\mathcal{O}_{N}(\{a^{\dag}_{\mbf{k}}\},\{a_{\mbf{k}}\})$
can be computed by means of a complex, classical distribution
functional:
\be
Tr[\mathcal{O}_{N}(\{a^{\dag}_{\mbf{k}}\},\{a_{\mbf{k}}\})\rho]=
\prod_{\mbf{k}}\int
d^{2}\alpha_{\mbf{k}}\mathcal{O}_{N}(\{\alpha^{*}_{\mbf{k}}\},\{\alpha_{\mbf{k}}\})P(\{\alpha_{\mbf{k}}\})
\; .
\ee
The correspondence between the quantum-mechanical and
classical descriptions, defined by the complex functional
$P(\{\alpha_{\mbf{k}}\})$, is at the heart of the optical
equivalence theorem \cite{Sudarshan,KlauderSudar}, stating that
the complete quantum mechanical description contained in the
density matrix can be recovered in terms of classical quasi-probability
distributions in phase space, as we will see in more detail in the following.

\subsection{More about coherence at any order, one-photon processes and coherent
           states}

The theory of coherent states marks the birth of modern quantum optics;
it provides a convenient mathematical formalism, and at the same time
it constitutes the standard of reference with respect to the degree
of nonclassicality of any generic quantum state of the electromagnetic
field. For this reason, we review here some
fundamental properties of the coherent states, restricting the
treatment to a single mode $a_{\mbf{k}}$ of the radiation field,
and dropping the subscript $\mbf{k}$. The coherent states can be
constructed using three different, but equivalent, definitions, each
of them shedding light on some of their most important physical
properties.
\begin{description}
\item[a)] The coherent states $|\alpha\rangle$ are eigenstates of
the annihilation operator $a$. The quadrature
representation $\psi_{\alpha}(x_{\lambda})$ of the coherent state
$|\alpha\rangle$, defined as the overlap between the quadrature
eigenstate $|x_{\lambda}\rangle$
and $|\alpha\rangle$: $\psi_{\alpha}(x_{\lambda}) =
\langle x_{\lambda}|\alpha \rangle$, can be
easily determined by solving the eigenvalue equation $\langle
x_{\lambda}|a|\alpha\rangle=\alpha\psi_{\alpha}(x_{\lambda})$.
Expressing the annihilation operator in terms of the quadrature
operators
\be
X_{\lambda}=\frac{e^{-i\lambda}a+e^{i\lambda}a^{\dag}}{\sqrt{2}}
\; , \quad P_{\lambda}=X_{\lambda+\frac{\pi}{2}} \; ,
\; \quad [X_{\lambda}, P_{\lambda}] \, = \, i \; ,
\label{QuadrComplamb}
\ee
the eigenvalue equation $\langle
x_{\lambda}|a|\alpha\rangle=\alpha\psi_{\alpha}(x_{\lambda})$ can
be expressed in the quadrature representation in the form
\be
\frac{e^{i\lambda}}{\sqrt{2}}
\left(x_{\lambda}+\frac{\partial}{\partial
x_{\lambda}}\right)\psi_{\alpha}(x_{\lambda})
=\alpha\psi_{\alpha}(x_{\lambda}) \; . \ee Its normalized solution
is \bea \psi_{\alpha}(x_{\lambda})& \, = \, &\pi^{-1/4}
e^{-\frac{1}{2}|\alpha|^{2}}e^{-\frac{1}{2}x_{\lambda}^{2}+\sqrt{2}e^{-i\lambda}\alpha
x_{\lambda}-\frac{1}{2}e^{-2i\lambda}\alpha^{2}} \nonumber
\\ &=&\pi^{-1/4}e^{-\frac{i}{2}\langle X_{\lambda}\rangle\langle
X_{\lambda+\pi/2}\rangle}e^{i\langle X_{\lambda+\pi/2}\rangle
x_{\lambda}}e^{-\frac{1}{2}(x_{\lambda}-\langle
X_{\lambda}\rangle)^{2}} \; ,
\label{cohstwvpack}
\eea
with
$\langle
X_{\lambda}\rangle=\langle\alpha|X_{\lambda}|\alpha\rangle=
\frac{\alpha e^{-i\lambda}+\alpha^{*}e^{i\lambda}}{\sqrt{2}}$.
Therefore coherent states are Gaussian in $x_{\lambda}$, in
the sense that they are characterized by a Gaussian probability distribution
$|\psi_{\alpha}(x_{\lambda})|^{2}$, and thus are completely specified
by the knowledge of the first and second statistical moments of the
quadrature operators. The free evolution of the wave
packet (\ref{cohstwvpack}) is given by
$|\alpha(t)\rangle=e^{-it\omega
a^{\dag}a}|\alpha\rangle=|e^{-i\omega t}\alpha\rangle$, and the
corresponding expectation values of the quadrature operators are
$\langle X\rangle(t)=\sqrt{2}Re[\alpha e^{i\omega t}]$, $\langle
P\rangle(t)=\sqrt{2}Im[\alpha e^{i\omega t}]$. Hence, the coherent
states $|\alpha(t)\rangle$ preserve the shape of the initial wave packet
at any later time, and the expectation values of the quadrature operators
evolve according to the classical dynamics of the pure harmonic
oscillator. \item[b)] The coherent
states can be obtained by applying the Glauber displacement
operator $D(\alpha)$ on the vacuum state of the quantum harmonic
oscillator, $|\alpha\rangle=D(\alpha)|0\rangle$. \item[c)] The
coherent states are quantum states of minimum Heisenberg
uncertainty,
\be
\langle\Delta
X_{\lambda}^{2}\rangle\langle\Delta
P_{\lambda}^{2}\rangle=\frac{1}{4} \; , \ee where \be \langle\Delta
F^{2}\rangle\equiv\langle F^{2}\rangle-\langle F \rangle^{2} \; , \ee
and, moreover,
\be
\langle \Delta X_{\lambda}^{2} \rangle =
\langle \Delta P_{\lambda}^{2} \rangle = \frac{1}{2} \; .
\ee
\end{description}
The three definitions $a)$, $b)$, $c)$ are equivalent, in the
sense that they define the same class of coherent states. Later on,
we will see that the equivalence between the
three definitions breaks down when generalized coherent
states will be defined for algebras more general than the
Heisenberg-Weyl algebra of the harmonic oscillator.

Let us consider now the photon number probability distribution
$P(n)$ for the coherent state $|\alpha\rangle$, i.e. the
probability that $n$ photons are detected in the coherent state
$|\alpha\rangle$. It is easy to see that it is a Poisson
distribution:
\be
P(n) \equiv |\langle
n|\alpha\rangle|^{2}=e^{-|\alpha|^{2}}\frac{|\alpha|^{2n}}{n!} \;
. \ee
The average number of photons $\langle n\rangle$ in the
state $|\alpha\rangle$ is therefore $\langle
n\rangle=|\alpha|^{2}$, with variance $\langle\Delta
n^{2}\rangle=\langle n\rangle$. Since the second order correlation
function, for zero time delay, can be easily expressed in terms of
the number operator as
\be
g^{(2)}(0)=1+\frac{\langle\Delta
n^{2}\rangle-\langle n\rangle}{\langle n\rangle^{2}} \; ,
\label{g2n}
\ee
we obtain for a coherent state, as expected,
$g^{(2)}(0)=1$. From the previous discussions it follows that
coherent states are, as anticipated, "classical" reference states,
in the sense that they share some statistical aspects together
with truly classical states of the radiation field, such as a
positive defined quasi-probability distribution, a Poissonian
photodistribution, and photon bunching. Thus, they can be used as
a standard reference for the characterization of the nonclassical
nature of other states, as, for instance, measured by deviations
from Poissonian statistics. In particular, a state with a photon
number distribution narrower than the Poissonian distribution
(which implies $g^{(2)}(0)<1$) is referred to as sub-Poissonian,
while if $g^{(2)}(0)>1$ (corresponding to a photon number
distribution broader than Poissonian distribution), it is referred
as super-Poissonian. It is clear that the phenomena of
sub-Poissonian statistics and photon antibunching are closely
related, because the first one implies the second for some time
scale. However, the reverse statement cannot be established in
general. In order to clarify this point, let us consider a generic
stationary field distribution, for which it can be shown that
\cite{ZouMandel}
\be
\langle\Delta n^{2}\rangle-\langle
n\rangle=\frac{\langle n\rangle^{2}}{T^{2}}\int_{-T}^{T}d\tau
(T-|\tau|)[g^{(2)}(\tau)-1] \; ,
\label{zoumand}
\ee where $T$ is
the counting interval. It is then possible for a state realizing
such a distribution, to be such that $g^{(2)}(\tau)>g^{(2)}(0)$,
but still with super-Poissonian statistics. A detailed review on
these subtleties and on sub-Poissonian processes in quantum optics
has been carried out by Davidovich \cite{Davidovich}. Another
important signature of nonclassicality has been introduced by
Mandel, who defined the $Q$ parameter \cite{MandelQpar}
\be
Q=\frac{\langle\Delta n^{2}\rangle-\langle n\rangle}{\langle
n\rangle} \,,
\label{Qmandel}
\ee
as a measure of the deviation of
the photon number statistics from the Poissonian distribution. The
interpretation of $Q$ is straightforward with respect to the field
statistics.

\subsection{Quasi-probability distributions, homodyne and heterodyne detection,
and quantum state tomography}

\begin{figure}[h]
\begin{center}
\includegraphics*[width=6cm]{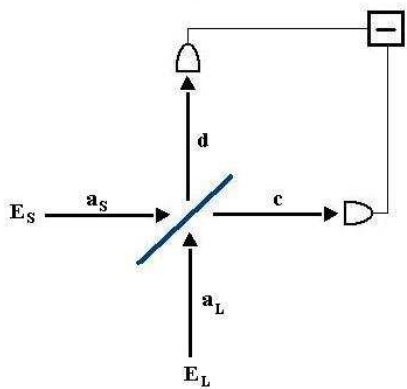}
\end{center}
\caption{Scheme of a homodyne detector.}
\label{Homodynescheme}
\end{figure}

The classical or nonclassical character of a
state can be tested on more general grounds by resorting to
quasi-probability distributions in phase space. As already
mentioned, the prototype of these distributions is the
$P$-function defined by Eq. (\ref{Pfuncti}), which provides the diagonal coherent
state representation \cite{Glauber1,Sudarshan}
\be
P(\alpha,\alpha^{*})=Tr[\rho\delta(\alpha^{*}-a^{\dag})\delta(\alpha-a)]
\; .
\label{Pfunc}
\ee
Here the Dirac $\delta$-function of an
operator is defined in the usual limiting sense in vector spaces.
For a coherent state $|\beta\rangle$ the $P$-representation is
then the two-dimensional delta function over complex numbers,
$\delta^{(2)}(\alpha-\beta)$, and this relation suggests a
possible definition of nonclassical state: \textit{``If the
singularities of $P(\alpha)$ are of types stronger than those of
delta functions, i.e. derivatives of delta function, the state
represented will have no classical analog''} \cite{Glauber3}.
Besides the $P$-function, other distribution functions associated
to different orderings of $a$ and $a^{\dag}$ can be defined. A
general quasi-probability distribution in phase space
$W(\alpha,p)$ is defined as the two-dimensional Fourier transform
of the corresponding $p-$ordered characteristic function
$\chi(\xi,p)$ \cite{Cahillglaub,BarnettRadmoretx}:
\be
W(\alpha,p)=\frac{1}{\pi^{2}}\int_{-\infty}^{\infty}d^{2}\xi
\chi(\xi,p)e^{\alpha\xi^{*}-\alpha^{*}\xi} , \quad
\chi(\xi,p)=Tr[\rho e^{\xi a^{\dag}-\xi^{*}a}]e^{p|\xi|^{2}/2} ,
\label{pOrderedDistributions}
\ee
where $\alpha$ and $\xi$ are complex variables, and $p=1,0,-1$
correspond, respectively, to normal, symmetric and antinormal
ordering \cite{Cahillglaub,BarnettRadmoretx}
in the product of bosonic operators. Moreover, it can
been shown that $W(\alpha,p)$ is normalized and real for all
complex $\alpha$ and real $p$. Statistical moments of any
$p-$ordered product of annihilation and creation operators $a$ and
$a^{\dag}$ can be obtained exploiting the relation
\be
\langle
a^{\dag m}a^{n}\rangle_{p}=\int_{-\infty}^{\infty}d^{2}\alpha
W(\alpha,p)\alpha^{* m}\alpha^{n} \; .
\ee
For $p=1$,
$W(\alpha,1)$ reduces to the Glauber-Sudarshan $P$ distribution;
for $p=-1$, $W(\alpha,-1)$ defines the Husimi distribution
$Q(\alpha)=\frac{1}{\pi}\langle\alpha|\rho|\alpha\rangle$
\cite{Husimi}. Finally, for $p=0$, $W(\alpha,0)=W(\alpha)$ defines the
Wigner distribution \cite{Wigner,Wigner2}. The Wigner distribution
can be viewed as a joint distribution in phase space for the two
quadrature operators $X_{\lambda}$ and $P_{\lambda}$ and can be
written in the form
\be
W(x_{\lambda},p_{\lambda})=\frac{1}{\pi}\int dy
e^{-2ip_{\lambda}y}\langle x_{\lambda}+y|\rho|x_{\lambda}-y\rangle
\; .
\label{Wigxplambda}
\ee
The probability distribution for the
quadrature component $X_{\lambda}$ is given by $\frac{1}{2}\int
dp_{\lambda}W(x_{\lambda},p_{\lambda})$, and an equivalent,
corresponding definition holds as well for the quadrature
$P_{\lambda}$. In order to obtain a classical-like description and
equip them with the meaning of true joint probability
distributions in classical phase space, the Wigner functions
$W(\alpha)$ should be nonnegative defined. In fact, in general the
Wigner function can take negative values, in agreement with the
basic quantum mechanical postulate on the complementarity of
canonically conjugated observables. However, it is easy to see
that $W(\alpha)$ is non negative for all Gaussian states, and
thus, in particular, for coherent states. Therefore, another
important measure of nonclassicality can be taken to be the
negativity of $W(x_{\lambda},p_{\lambda})$ \cite{negativeWig}.
This criterion turns out to be of practical importance, after
Vogel and Risken succeeded to show that the Wigner function can be
reconstructed from a set of measurable quadrature-amplitude
distributions, achieved by homodyne detection \cite{homotomogr}.\\
In quantum optics, homodyne detection is a fundamental technique
for the measurement of quadrature operators $X_{\lambda}$ of the
electromagnetic field \cite{homodynedet,homodynedet2}. The scheme
of a balanced homodyne detection is depicted in Fig.
(\ref{Homodynescheme}).
Two electromagnetic field inputs of the same frequency $\omega$, a
signal field $E^{(+)}_{S}\propto a_{S} \; e^{-i\omega t} \;$ and a
strong coherent laser beam $E^{(+)}_{L}\propto a_{L} \;
e^{-i\omega t} \;$ (local oscillator), enter the two input ports
of a beam splitter. The input modes $a_{S}$ and $a_{L}$ are
converted in the output modes $c$ and $d$ by the unitary
transformation $U_{BS}$:
\be
\left(\begin{array}{c}
  c \\
  d \\
\end{array} \right) \, = \, U_{BS}\left(\begin{array}{c}
  a_{S} \\
  a_{L} \\
\end{array} \right) \; , \quad
U_{BS}=\left( \begin{array}{cc}
 \sqrt{\eta}  & \sqrt{1-\eta}e^{i\delta}  \\
  -\sqrt{1-\eta}e^{-i\delta}  & \sqrt{\eta} \\
\end{array} \right)
\; ,
\label{beamsplit}
\ee
where $\eta$ is the transmittance of
the beam splitter, and $\delta$ is the phase shift between
transmitted and reflected waves. The detected difference of the
output intensities $\langle I\rangle=\langle
c^{\dag}c\rangle-\langle d^{\dag}d\rangle$ is
\be
\langle
I\rangle= (1-2\eta)(\langle a^{\dag}_{L}a_{L}\rangle -
\langle a^{\dag}_{S}a_{S}\rangle)+2\sqrt{\eta(1-\eta)}\langle
e^{-i\delta}a_{S}^{\dag}a_{L}+e^{i\delta}a_{S}a_{L}^{\dag}\rangle
\; .
\label{homodphcurr}
\ee
If the local oscillator mode $a_{L}$
can be approximated by an intense coherent field of complex
amplitude $\alpha_{L}$ ($a_{L}\rightarrow\alpha_{L}$),
exploiting $|\alpha_{L}|^2 \gg \langle a^{\dag}_{S}a_{S}\rangle$, the
difference current can be written in the form
\be
\langle I\rangle
\, \approx \,
(1-2\eta)|\alpha_{L}|^{2}+\sqrt{2\eta(1-\eta)}|\alpha_{L}|\langle
X_{\xi-\delta}\rangle \; ,
\label{homodetX}
\ee
where $\xi=\arg
\alpha_{L}$, and $X_{\xi-\delta}$ is the quadrature operator
$X_{\lambda}$, associated to the signal mode $a_{S}$, with
$\lambda \equiv \xi -\delta$. Exploiting the freedom in tuning the
angles $\xi$ and $\delta$, the mean amplitude of any quadrature
phase operator can be measured. Thus, the homodyne detector allows
the direct experimental measurement of the field quadratures. Now,
as shown in Ref. \cite{homotomogr}, the Wigner function,
corresponding to a state $\rho$ of the signal mode $a_{S}$, can be
reconstructed via an inverse Radon transform from the quadrature
probability distribution $p(x,\lambda )\doteq \langle
x_{\lambda}|\rho|x_{\lambda}\rangle$, which in turn is determined
by the homodyne measurements. This procedure for the recontruction
of a quantum state is the core of the so-called \textit{quantum
homodyne tomography}. It has been widely studied, refined, and
generalized \cite{tomography,tomography2,tomography3}, and
experimentally implemented in several different instances
\cite{Wigfexp1,Wigfexp2,Wigfexp3,Wigfexp4}. One can show that the
Wigner function can be written as the inverse Radon transform of
$p(x,\lambda )$ in the form
\be
W(x,y) \, = \,
\int_{0}^{\pi}\frac{d\lambda}{\pi}\int_{-\infty}^{\infty}dx'
p(x',\lambda)\int_{-\infty}^{\infty}\frac{dk}{4}e^{ik(x'-x
\cos\lambda-y\sin\lambda)} \; .
\ee
A further aim of quantum
tomography is to estimate, for arbitrary quantum states, the
average value $\langle\mathcal{O}\rangle$ of a generic operator
$\mathcal{O}$. This expectation can be computed as
\be
\langle\mathcal{O}\rangle \, = \,
\int_{0}^{\pi}\frac{d\lambda}{\pi}\int_{-\infty}^{\infty}dx
R[\mathcal{O}](x,\lambda) \; ,
\ee
 where the estimator
$R[\mathcal{O}](x,\lambda)$ is given by
\be
R[\mathcal{O}](x,\lambda) \, = \, Tr[\mathcal{O}K(X_{\lambda}-x)] \; ,
\ee
and
$$
K(x) \, \equiv \, \int_{-\infty}^{\infty} (dk /4) \exp (ikx)
\, = \, - \mathcal{P}[1/(2x^{2})] \; ,
$$
where $\mathcal{P}$ denotes the Cauchy principal value.

\begin{figure}
\begin{center}
\includegraphics*[width=8cm]{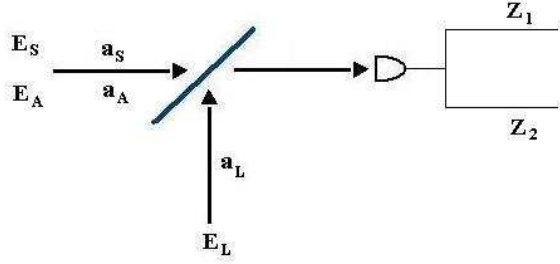}
\end{center}
\caption{Scheme of heterodyne detector.} \label{Heteroscheme}
\end{figure}

For the sake of completeness, we briefly outline the description
of another detection method, the so-called heterodyne detection
\cite{heterodet1,heterodet2}. It allows simultaneous measurements
of two orthogonal quadrature components, whose statistics is
described by the Husimi $Q$-function. Heterodyne detection can be
realized by the following device. The signal field and another
field, the auxiliary field, feed the same port of a beam
splitter, as depicted in Fig.(\ref{Heteroscheme}). Moreover, as in
the case of homodyne detection, a local field oscillator enters
the other port of the beam splitter.
In this configuration, at variance with the homodyne instance, the
frequencies of the signal, auxiliary and local oscillator fields
are different. The signal field $E_{S}$ is associated to the mode
$a_{S}$ at the frequency $\omega_{S}$, the auxiliary field $E_{A}$
is associated to the mode $a_{A}$ at the frequency $\omega_{A}$,
and the local oscillator field $E_{L}$ is associated to the mode
$a_{L}$ at the frequency $\Omega_{L}$, where
\be
\omega_{S} \, +
\, \omega_{A} \, = \, 2\Omega_{L} \; , \quad \omega_{S} \, - \,
\omega_{A} \, = \, 2\Omega \; , \quad \Omega_{L} \, \gg \, \Omega
\; .
\ee
A broadband detector is placed in one output port of the
beam splitter to detect beats at the frequency $\Omega \ll
\Omega_{L}$. After demodulation, the time dependent components,
proportional to $\cos(\Omega t)$ and $\sin(\Omega t)$, can be
detected simultaneously, yielding the measured variable
\be
 Z \, \equiv \, Z_{1} + iZ_{2} \, = \,
a_{S}\otimes I_{a_{A}} \, + \, I_{a_{S}}\otimes a_{A}^{\dag} \; .
\ee
It is to be remarked that, formally, the same quantum
measurement can be obtained by a double homodyne detection.

\newpage

\section{Parametric processes in nonlinear media}
\label{section3}

The advent of laser technology allowed to begin the study of nonlinear
optical phenomena related to the interaction of matter with intense coherent
light, and extended the field of conventional linear
optics (classical and quantum) to nonlinear optics (classical
and quantum). Historically, the fundamental events,
which marked such a passage, were the realization of the first laser
device (a pulsed ruby laser) in 1960 \cite{Laser} and the
production of the second harmonic, through a pulsed laser incident
on a piezoelectric crystal, in 1961 \cite{SHGFranken}. \\
As the main body of this report will be concerned with quantum
optical phenomena in nonlinear media, relevant for the generation
of nonclassical multiphoton states, this Section is dedicated to a
a self-contained, yet somewhat detailed discussion of the basic
aspects of nonlinear optics that are of importance in the quantum
domain. For a much more thorough and far more complete examination
of the classical aspects of nonlinear optics the reader is
referred to Refs.
\cite{txtHeLiu,txtBloemb,txtShen,txtButchCott,txtBoyd}.
\\ All linear and nonlinear optical effects arise in the processes
of interaction of electromagnetic fields with matter fields. The
physical characteristics of the material system determine its
reaction to the radiation; therefore, the effect on the field can
provide information about the system. On the other hand, the
medium can be used to generate a new radiation field with
particular features. Harmonic generation, wave mixing,
self-focusing, optical phase-conjugation, optical bistability, and
in particular optical parametric amplification and oscillation,
can all be described by studying the properties of a fundamental
object, the nonlinear polarization. For this reason, we will
dedicate this Section to review the essential notions of nonlinear
quantum optics, including the general description of optical
field-induced electric polarization, the standard phenomenological
quantization procedures, the effective Hamiltonians associated to
two- and three-wave mixing processes, and the theory of the
nonlinear susceptibilities. We will pay much attention to the
lowest order processes, i.e. second and third order processes,
which will be widely used in the next Section, and we will finally
discuss the properties and the relative orders of magnitude of the
different nonlinear susceptibilities.

Let us begin by considering a system constituted by an atomic medium and the applied
optical field; the Hamiltonian of the whole system can be written as
\be
H=H_{0}+H_{I}(t) \; ,
\label{Hintmediumradiation}
\ee
where $H_{0}$
is the unperturbed Hamiltonian of the medium without an applied
field and $H_{I}(t)$ is the matter-radiation interaction part.
Considering, for simplicity, only the interaction between an
outer-shell electron and the applied optical field, the
interaction operator takes the form \cite{txtHeLiu}
\be
H_{I}(t)=-\frac{e}{2m_{e}}(\ovrw{p}_{e}\cdot\ovrw{A}+\ovrw{A}\cdot\ovrw{p}_{e})+
\frac{e^{2}}{2m_{e}}\ovrw{A}\cdot\ovrw{A}+e V \; ,
\ee
where $e$
and $m_{e}$ are the charge and the mass of the electron,
$\ovrw{p}_{e}$ is the momentum operator of the electron, and
$\ovrw{A}$ and $V$ are the vector and the scalar potential of the
optical field. Choosing the Coulomb gauge so that
$\ovrw{\nabla}\cdot\ovrw{A}=0$ and $V=0$, and neglecting the
smaller diamagnetic quadratic term $\ovrw{A}\cdot\ovrw{A}$, we get
\be
H_{I}(t)=-\frac{e}{m_{e}}\ovrw{A}\cdot\ovrw{p}_{e} \; .
\label{HintAp}
\ee
For optical fields the wavelength is generally much
larger than the molecular radius; as a consequence, the electric-quadrupole and
the magnetic-dipole contributions can be neglected and the
interaction Hamiltonian (\ref{HintAp}) reduces to the electric-dipole
interaction:
\be
H_{I}(t)=-\ovrw{p}\cdot\ovrw{E}(t) \; ,
\label{electricdipole}
\ee
where $\ovrw{p}=e\,\ovrw{r}$ is the microscopic
electric dipole vector of a single atom. The
meaning, the applicability and the different properties of the
interactions (\ref{HintAp}) and (\ref{electricdipole}) have been
discussed at length in the literature
\cite{ediplamb,ediphaken,edipyang,edipkobe,edipgiacobino,edipMandel,edipbecker,edipdrummond},
and it turns out that the vast majority of nonlinear optical processes can be
adequately described by applying the electric-dipole approximation to the
matter-radiation interaction Hamiltonian. The interaction is considered within a polarizable
unit of the material system, that is a volume in which the
electromagnetic field can be assumed to be uniform at any given
time. In a solid this volume is large in comparison with the
atomic dimensions, but small with respect to the wavelength of the
optical field. As the field is uniform within a polarizable unit,
the radiation interaction looks like that of an electric dipole in
a constant field, and the electric-dipole approximation is fully justified
in the spectral region going from the far-infrared to the ultraviolet; for
shorter wavelengths $\ovrw{E}$ cannot be assumed to be uniform
over atomic dimensions. When applicable, the electric-dipole
approximation allows to identify the nonlocal macroscopic polarization
with the local, macroscopic electric-dipole polarization
$\ovrw{P}(t) \equiv \sum_{i=1}^{N}\ovrw{p}_{i}(t)$, given by the
sum of all the $N$ induced atomic dipole moments that constitute
the dielectric medium. The macroscopic polarization is in general
a complicated nonlinear function of the electric fields; however,
in the electric-dipole approximation, it is possible to describe
the polarization and the dynamics of radiation in material media with
the help of the dielectric susceptibilities, to be defined in the
following. In particular, it is possible to expand the macroscopic dielectric
polarization in a power series of the electric field amplitudes
\cite{txtHeLiu,txtBloemb,txtShen,txtButchCott,txtBoyd}:
\be
\ovrw{P}(t)=\ovrw{P}^{(1)}(t)+
\ovrw{P}^{(2)}(t)+...+\ovrw{P}^{(n)}(t)+... \; , \label{Polarizt}
\ee
where the the generic term $\ovrw{P}^{(n)}(t)$ in the power
series expansion reads, for each spatial component $P_{j}^{(n)}(t)$
($j=1,2,3$),
\be
P_{j}^{(n)}(t)=\int_{0}^{\infty}...\int_{0}^{\infty}
d\tau_1
 ... d\tau_n R_{j\alpha_1 ...\alpha_n}^{(n)}(\tau_1
,...,\tau_n) E_{\alpha_1}(t-\tau_1)...E_{\alpha_n}(t-\tau_n)
 .\label{nresponsefun}
\ee
Here the subscript $\alpha_i$ is a short-hand notation for the
spatial and polarization components of the electric field vector
$\ovrw{E}$ at time $t-\tau_i$, the object $R_{j\alpha_1
\alpha_2...\alpha_n}^{(n)}(\tau_1 ,\tau_2 ,...,\tau_n)$ is the
$j$-th spatial component of the $n$-th order response function, a
tensor of rank $(n+1)$, that takes into account the reaction of
the medium to the applied electromagnetic field. In Eqs.
(\ref{Polarizt}) and (\ref{nresponsefun}) only the temporal
dependence is retained, while the spatial dependence is not
explicitated for ease of notation. Moreover, it is assumed that
the medium reacts only by a local response, that is the
polarization at a point is completely determined by the electric
field at that same point. Moving from the time domain to the
frequency domain, Eq. (\ref{nresponsefun}) becomes
\be
P_{j}^{(n)}(\omega)=\chi_{j\alpha_1
\alpha_2...\alpha_n}^{(n)}(\omega_{\sigma};\omega_1 ,...,\omega_n)
E_{\alpha_1}(\omega_1)...E_{\alpha_n}(\omega_n) \,,
\label{nchifun}
\ee
where
$\omega_{\sigma}=\sum_{i=1}^{n}\omega_{i}$, and the $n$-th order
nonlinear susceptibility tensor is defined by
\bea
&&\chi_{j\alpha_1 ...\alpha_n}^{(n)}(\omega_{\sigma};\omega_1
,...,\omega_n) \nonumber \\
&& \nonumber \\
&&= \int ...\int d\tau_1  ... d\tau_n \Theta(\tau_1
)...\Theta(\tau_n ) R^{(n)}_{j\alpha_1 ...\alpha_n}(\tau_1
,...,\tau_n)e^{i(\omega_1 \tau_1 +...+\omega_n \tau_n)} \; ,
\label{phenomdefchi}
\eea
where $\Theta$ is the Heaviside step
function. For a lossless, nondispersive and uniform medium, the
susceptibilities $\underline{\chi}^{(n)}$ are symmetric tensors of
rank $(n+1)$, while the polarization vector $\ovrw{P}$ provides
the macroscopic description of the interaction of the
electromagnetic field with matter \cite{txtBloemb}. There are
various physical mechanisms which are responsible for nonlinear
polarization responses in the medium \cite{txtHeLiu}: the
distortion of the electronic clouds, the intramolecular motion,
the molecular reorientation, the induced acoustic motion, and the
induced population changes. For our purposes, here and in the
following only the first two mechanisms will have to be taken
into account.

\subsection{Quantized macroscopic fields in nonlinear dielectric
            media}

Several approaches have been proposed for the quantization of the
electromagnetic field in nonlinear, inhomogeneous, or dispersive
media
\cite{QEDBornInf,QEDJauchWats,QEDShen,QEDTucker,QEDNienhuis,QEDHillery,QEDAbram,QEDKnoll,QEDDrummond,QEDGlauberLew,QEDAbramCoh,QEDHuttenr,QEDSantos,QEDGruner,QEDMatloob,QEDDuan}.
The standard phenomenological macroscopic quantum theory, widely used in nonlinear optics,
was formulated by Shen \cite{QEDShen} and was later elaborated by Tucker
and Walls \cite{QEDTucker} for the description of parametric
frequency conversion. Classical electrodynamics in a dielectric medium is
described by the macroscopic Maxwell equations
\bea
&&\ovrw{\nabla} \cdot \ovrw{D}=\rho_{ext} \,, \label{Max3} \\
&& \nonumber \\
&&\ovrw{\nabla} \times \ovrw{E}=-\frac{\partial\ovrw{B}}{\partial t}
\,, \label{Max2} \\
&& \nonumber \\
&&\ovrw{\nabla} \cdot \ovrw{B}=0 \,, \label{Max1} \\
&& \nonumber \\
&&\ovrw{\nabla} \times \ovrw{B}=\frac{\partial\ovrw{D}}{\partial
t}+\ovrw{J}_{ext} ,
\label{Max4}
\eea
where
$\ovrw{D}=\ovrw{E}+\ovrw{P}$ is the displacement field,
$\rho_{ext}$ and $\ovrw{J}_{ext}$ represent charge and current
sources external to
the dielectric medium, $\ovrw{P}$ is the polarization of the
medium, and Heaviside-Lorentz units have been used throughout. We can also
write the polarization, whose $j$-th component is given by Eq.
(\ref{nchifun}), in the more synthetic form
\be
\ovrw{P}=\underline{\chi}^{(1)}:\ovrw{E}
+\underline{\chi}^{(2)}:\ovrw{E}\ovrw{E}
+\underline{\chi}^{(3)}:\ovrw{E}\ovrw{E}\ovrw{E} + ...
\; ,
\label{polarization}
\ee
where the first term denotes the contraction of the electric field
vector with the first-order susceptibility tensor $\underline{\chi}^{(1)}$
(which is a tensor of rank $2$), the second term denotes
the contraction of two electric field vectors with the second-order
susceptibility tensor $\underline{\chi}^{(2)}$ (a
tensor of rank $3$), the third term denotes the contraction of three
electric field vectors with the third-order susceptibility tensor
$\underline{\chi}^{(3)}$ (a tensor of rank $4$), and so on.
Equations (\ref{Max1}), (\ref{Max2}), (\ref{Max3}),
(\ref{Max4}) and (\ref{polarization}) constitute the basis of the
theory of nonlinear optical effects in matter. The standard method
to derive a macroscopic quantum theory is to quantize the
macroscopic classical theory. The Hamiltonian is
\be
H=H_{0}+H_{I}=\frac{1}{2}\int
d^{3}r[\ovrw{E}^{2}+\ovrw{B}^{2}]-\int d^{3}r
\ovrw{E}\cdot\ovrw{P} \; .
\ee
The first term is the free quadratic Hamiltonian $H_{0}$, while
the second term $H_{I}$ represents the interaction in the medium.
Recalling the expression of the polarization vector $\ovrw{P}$
in terms of the electric fields and of the susceptibilities, one
sees that the interaction Hamiltonian contains, in principle,
nonlinear, anharmonic terms of arbitrary order. The lowest-order,
cubic power of the electric field in the interaction Hamiltonian,
is associated to the second-order susceptibility
$\underline{\chi}^{(2)}$.
The standard phenomenological quantization is
achieved by introducing the vector potential operator
(\ref{standquantiz}), where
 $\omega_{\ovrw{k},\lambda}= c|\ovrw{k}|
/n_{\ovrw{k},\lambda}$ is the angular frequency in the medium
($n_{\ovrw{k},\lambda}$ being the index of refraction). Effective
Hamiltonians associated to nonlinear quasi-steady-state processes
of different orders are widely used in quantum optics, and are
based on this simple quantization procedure. However there are
some difficulties with this theory; the most serious one being
inconsistency with Maxwell equations. For instance, it is easy to
verify that Eqs. (\ref{vectorpot}) and the Coulomb condition
$\ovrw{\nabla}\cdot\ovrw{A}=0$ imply that, in absence of external
charges, $\ovrw{\nabla}\cdot\ovrw{E}=0$ rather than
$\ovrw{\nabla}\cdot\ovrw{D}=0$, and, moreover, Eq. (\ref{Max2}) is
not satisfied. In the following we review the main progresses
achieved to overcome such inconsistencies and to provide an
exhaustive and consistent formulation of quantum electrodynamics
in nonlinear media. A first successful solution to the
shortcomings of the standard quantization scheme has been
introduced by Hillery and Mlodinow \cite{QEDHillery}, who assume
the displacement field $\ovrw{D}$ as the canonical variable for
quantization in a homogeneous and nondispersive medium. Starting
from an appropriate Lagrangian density, they introduce the
interaction Hamiltonian
\be
H_{I}=\int d^{3}r
\left(\frac{1}{2}\chi_{ij}^{(1)}E_{i}E_{j}
+\frac{2}{3}\chi_{ijk}^{(2)}E_{i}E_{j}E_{k}
+\frac{3}{4}\chi_{ijkl}^{(3)}E_{i}E_{j}E_{k}E_{l}+...\right) ,
\label{HamHillery}
\ee
where $\ovrw{E}$ and $\ovrw{B}$ are to be
considered as functions of $\ovrw{A}$ and $\ovrw{D}$. Performing a
mode expansion, annihilation (and creation) operators can be
defined by
\be
a_{\ovrw{k},\lambda}(t)=\int d^{3}r
e^{-i\ovrw{k}\cdot\ovrw{r}}\hat{\epsilon}_{\ovrw{k},\lambda}
\left(\sqrt{\frac{\omega_{\ovrw{k}}}{2V}}\ovrw{A}(\ovrw{r},t)
-\frac{i}{\sqrt{2\omega_{\ovrw{k}}V}}\ovrw{D}(\ovrw{r},t)\right)
\; .
\ee
It is easy to check that these newly defined annihilation
and creation operators obey the bosonic canonical commutation
relations (\ref{QEDcommutat}). It is also important to notice
that, as $a_{\ovrw{k},\lambda}(t)$ depends on $\ovrw{D}$, it
contains both field and matter degree of freedom. One can
introduce a further, alternative quantization procedure
\cite{QEDHillery} by redefining the four-vector potential, the
so-called "dual potential" $\Lambda=(\Lambda_{0},\ovrw{\Lambda})$:
\be
\ovrw{D}=\ovrw{\nabla}\times\ovrw{\Lambda} \,, \quad
\ovrw{B}=\frac{\partial \ovrw{\Lambda}}{\partial
t}+\ovrw{\nabla}\Lambda_{0} \; .
\ee
Expressing the polarization
in a more convenient form:
\be
\ovrw{P}=\underline{\eta}^{(1)}:\ovrw{D}
+\underline{\eta}^{(2)}:\ovrw{D}\ovrw{D}
+\underline{\eta}^{(3)}:\ovrw{D}\ovrw{D}\ovrw{D}+ ... \; ,
\label{polarizD}
\ee
where the quantities $\eta$ can be expressed
uniquely in terms of the susceptibilities $\chi$, one can derive
the following canonical Hamiltonian density
\bea
H=&&\frac{1}{2}(\ovrw{B}^{2}+\ovrw{D}^{2})-
\frac{1}{2}\ovrw{D}\cdot\underline{\eta}^{(1)}:\ovrw{D}
-\frac{1}{3}\ovrw{D}\cdot\underline{\eta}^{(2)}:\ovrw{D}\ovrw{D}
\nonumber \\
&& \nonumber \\
&&-\frac{1}{4}\ovrw{D}\cdot\underline{\eta}^{(3)}:\ovrw{D}\ovrw{D}\ovrw{D}-
...-\ovrw{B}\cdot\ovrw{\nabla}\Lambda_{0} \; .
\eea
The gauge
condition can be chosen so that $\Lambda_{0}=0$ and
$\ovrw{\nabla}\cdot\ovrw{\Lambda}=0$. The theory can be quantized
in the same way as the free quantum electrodynamics, and the
commutation relations are
\be
[\dot{\Lambda}_{i}(\ovrw{r},t),\Lambda_{j}(\ovrw{r}',t)]=i\delta_{ij}^{tr}(\ovrw{r}-\ovrw{r}')
\; ,
\ee
where
$\delta_{ij}^{tr}(\ovrw{r})=\frac{1}{(2\pi)^{3}}\int
d^{3}k\left(\delta_{ij}-\frac{k_{i}k_{j}}{|\ovrw{k}|^{2}}\right)e^{i\ovrw{k}\cdot\ovrw{r}}$
is the transverse delta function. In this theory, however, some
problems arise with operator ordering, and it is difficult to
include dispersion. However, as shown by Drummond
\cite{QEDDrummond}, this quantization method can be generalized to
include dispersion, by resorting to a field expansion in a slowly
varying envelope approximation, including an arbitrary number of
envelopes, and assuming lossless propagation in the relevant
frequency bands. The final quantum Hamiltonian is written in terms
of creation and annihilation operators corresponding to
group-velocity
photon-polariton excitations in the dielectric. \\
Concerning nonlinear and inhomogeneous media, we briefly discuss a
general procedure for quantization \cite{QEDDuan} that extends the
approach of Glauber and Lewenstein \cite{QEDGlauberLew}. This
method is based on the assumption of medium-independent
commutation relations for the fields $\ovrw{D}$ and $\ovrw{B}$,
which from Eqs. (\ref{Max1}) and (\ref{Max2}) (for the source-free
case) can be expanded in terms of a complete set of transverse
spatial functions $\{\ovrw{f}_{\ovrw{k},\lambda}(\ovrw{r})\}$ and
$\{\ovrw{\nabla}\times\ovrw{f}_{\ovrw{k},\lambda}(\ovrw{r})\}$:
\be
\ovrw{D}(\ovrw{r},t)=-\sum_{\ovrw{k},\lambda}
\Pi_{\ovrw{k},\lambda}(t)\ovrw{f}^{*}_{\ovrw{k},\lambda}(\ovrw{r})
\; , \quad \ovrw{B}(\ovrw{r},t)=\sum_{\ovrw{k},\lambda}
Q_{\ovrw{k},\lambda}(t)\ovrw{\nabla}\times\ovrw{f}_{\ovrw{k},\lambda}(\ovrw{r})
\; .
\label{duanexpans}
\ee
Both the functions $\ovrw{f}_{\ovrw{k},\lambda}$ and the operatorial
coefficients $Q$ and $\Pi$ satisfy Hermiticity conditions:
$\ovrw{f}_{\ovrw{k},\lambda}^{*}=\ovrw{f}_{-\ovrw{k},\lambda}$,
$Q_{\ovrw{k},\lambda}^{\dag}=Q_{-\ovrw{k},\lambda}$, and
$\Pi_{\ovrw{k},\lambda}^{\dag}=\Pi_{-\ovrw{k},\lambda}$. Moreover,
the spatial functions satisfy transversality, orthonormality, and
completeness conditions, and the commutation relations read
\be
[Q_{\ovrw{k},\lambda}(t),\Pi_{\ovrw{k}',\lambda'}(t)]=
i\delta_{\ovrw{k}\ovrw{k}'}\delta_{\lambda\lambda'} \; .
\label{duancomm}
\ee
The energy density and the Hamiltonian of the
electromagnetic field in the medium are given by
\be d
U(\ovrw{r},t)=\ovrw{E}(\ovrw{r},t)\cdot
d\ovrw{D}(\ovrw{r},t)+\ovrw{H}(\ovrw{r},t)\cdot
d\ovrw{B}(\ovrw{r},t), \quad H=\int d^{3}r U(\ovrw{r},t)  .
\ee
Finally, being $H$ a functional of $\ovrw{D}$ and $\ovrw{B}$, the
full field quantization is obtained from Eqs. (\ref{duanexpans})
and (\ref{duancomm}).

\subsection{Effective Hamiltonians and multiphoton processes}

In this Subsection, by applying the standard quantization
procedure to Hillery and Mlodinow's Hamiltonian (\ref{HamHillery})
treated in the rotating wave approximation, we will show how to
obtain various phenomenological Hamiltonian models describing
effective multiphoton processes in nonlinear media. The electric
contribution to the electromagnetic energy in the nonlinear
medium, Eq. (\ref{HamHillery}), can be written in the form
\bea
&&H=\int_{V}d^{3}r\left[\frac{1}{2}E^{2}(\ovrw{r},t)+
\sum_{n}X_{n}(\ovrw{r},t)\right]  ,  \nonumber \\
&& \nonumber \\
&&X_{n}(\ovrw{r},t)=\frac{n}{n+1}\underline{\chi}^{(n)}:
\ovrw{E}\ovrw{E}...\ovrw{E}  .
\label{nonlinearpurpe}
\eea
In
terms of the Fourier components of the electric fields in the
frequency domain, the scalar field $X_{n}(\ovrw{r},t)$ becomes
\bea
X_{n}(\ovrw{r},t)=\frac{n}{n+1}\int\int ...\int && d\omega
d\omega_{1}... d \omega_{n} e^{- i(\omega+\omega_{\sigma})t}\,
\nonumber \\
&& \nonumber \\
\times
&&\underline{\chi}^{(n)}(\omega_{\sigma};\omega_{1},...,\omega_{n}):
\ovrw{E}(\omega)\ovrw{E}(\omega_{1})...\ovrw{E}(\omega_{n}) \; ,
\label{Xn}
\eea
where the spatial dependence of the fields has
been omitted. The canonical quantization of the macroscopic field
in a nonlinear medium is obtained by replacing the classical field
$\ovrw{E}(\ovrw{r},t)$ with the corresponding free-field Hilbert
space operator
\be
\ovrw{E}(\ovrw{r},t) = i\sum_{\ovrw{k},\lambda}
\left[\frac{\omega_{\ovrw{k},\lambda}} {2V}\right]^{1/2}
\{a_{\ovrw{k},\lambda}\hat{\epsilon}_{\ovrw{k},\lambda}
e^{i(\ovrw{k}\cdot\ovrw{r}-\omega_{\ovrw{k},\lambda}t)}
-a_{\ovrw{k},\lambda}^{\dag}\hat{\epsilon}_{\ovrw{k},\lambda}^{*}
e^{-i(\ovrw{k}\cdot\ovrw{r}-\omega_{\ovrw{k},\lambda}t)}\} \; .
\label{Elfieldop}
\ee
Denoting by $\mbf{k}$ the pair ($\ovrw{k},
\lambda$), and introducing $\Lambda_{\mbf{k}} \equiv
\sqrt{\frac{\omega_{\mbf{k}}}{2V}}$, the Fourier components of the
quantum field are given by
\be
\ovrw{E}(\ovrw{r},\omega) =
i\sum_{\mbf{k}} \Lambda_{\mbf{k}}
\{a_{\mbf{k}}\hat{\epsilon}_{\mbf{k}} e^{i {\ovrw{k}}\cdot
{\ovrw{r}}}\delta(\omega-\omega_{\mbf{k}})-
a_{\mbf{k}}^{\dag}\hat{\epsilon}_{\mbf{k}}^{*} e^{-i
{\ovrw{k}}\cdot {\ovrw{r}}} \delta(\omega+\omega_{\mbf{k}})\} \; .
\label{Four_comp}
\ee
The contribution of the $n$-th order
nonlinearity to the quantum Hamiltonian can thus be obtained by
replacing the Fourier components of the quantum field
Eq.~(\ref{Four_comp}) in Eq.~(\ref{Xn}). Because of the phase
factors $e^{i\omega t}$, many of the resulting terms in
Eq.~(\ref{Xn}) can be safely neglected (rotating wave
approximation), as they are rapidly oscillating and average to
zero. The effective processes involve the annihilation of $s$
photons $(1\leq s\leq n)$ and the creation of $(n-s+1)$ photons,
as imposed by the constraint of total energy conservation. Thus
the nonvanishing contributions correspond to sets of frequencies
satisfying the relation
\be
\sum_{i=1}^{s}\omega_{\mbf{k}_{i}}=\sum_{i=s+1}^{n+1}\omega_{\mbf{k}_{i}}
\; , \label{freq_con} \ee and involve products of boson operators
of the form \be a_{\mbf{k}_{1}}a_{\mbf{k}_{2}}\cdot\cdot\cdot
a_{\mbf{k}_{s}} a_{\mbf{k}_{s+1}}^{\dag}\cdot\cdot\cdot
a_{\mbf{k}_{n+1}}^{\dag} \; ,
\ee
and their hermitian conjugates.
The occurrence of a particular multiphoton process is selected by
imposing the conservation of total momentum. This is the so-called
phase matching condition and, classically, corresponds to the
synchronism of the phase velocities of the electric field and of
the polarization waves. These conditions can be realized by
exploiting the birefringent and dispersion properties of
anisotropic crystals. The relevant modes of the radiation involved
in a nonlinear parametric process can be determined by the
condition (\ref{freq_con}) and the corresponding phase-matching
condition
\be
\sum_{i=1}^{s}{{\ovrw{k}}}_{i}=
\sum_{i=s+1}^{n+1}{{\ovrw{k}}}_{i} \; .
\label{phas_match}
\ee
In
principle, the highest order $n$ of the processes involved can be
arbitrary. However, in practice, due to the fast decrease in order
of magnitude of the nonlinear susceptibilities $\chi^{(n)}$ with
growing $n$, among the nonlinear contributions the second- and
third-order processes (three- and four-wave mixing) usually play
the most relevant roles. In fact, the largest part of both
theoretical and experimental efforts in nonlinear quantum optics
has been concentrated on these processes. We will then now move to
calculate explicitely the contributions $X_{n}$ associated to the
first two nonlinearities $(n=2,3)$, and, by using the expansion
(\ref{Four_comp}) and exploiting the matching conditions
(\ref{freq_con}) and (\ref{phas_match}), we will determine the
effective Hamiltonians associated to three- and four-wave mixing.

\vspace{0.3cm}

\begin{description}

\item[a)] {\it Second order processes}

\vspace{3mm}

Let us consider an optical field composed of three
quasi-monochromatic frequencies $\omega_{\mbf{k}_{1}}$,
$\omega_{\mbf{k}_{2}}$, $\omega_{\mbf{k}_{3}}$, such that
$\omega_{\mbf{k}_{1}}+\omega_{\mbf{k}_{2}}=\omega_{\mbf{k}_{3}}$.
Ignoring the oscillating terms, and apart from inessential
numerical factors, the second-order effective interaction
Hamiltonian reads
\bea
\int_{V}d^{3}r &&X_{2}(\ovrw{r},t)\simeq 2i
\sum_{\mbf{k}_{1},\mbf{k}_{2},\mbf{k}_{3}}\Lambda_{\mbf{k}_{1}}
\Lambda_{\mbf{k}_{2}}\Lambda_{\mbf{k}_{3}} \nonumber \\
&& \nonumber \\
 \times &&\,
\underline{\chi}^{(2)}(\omega_{\mbf{k}_{3}};\omega_{\mbf{k}_{1}},
\omega_{\mbf{k}_{2}}):
\hat{\epsilon}_{\mbf{k}_{1}}\hat{\epsilon}_{\mbf{k}_{2}}
\hat{\epsilon}_{\mbf{k}_{3}}^{*}
a_{\mbf{k}_{1}}a_{\mbf{k}_{2}}a_{\mbf{k}_{3}}^{\dag}
\int_{V}d^{3}r e^{i\Delta\ovrw{k}\cdot\ovrw{r}}+H.c.\; ,
\label{X2}
\eea
where
\be
\Delta\ovrw{k}=\ovrw{k}_{1}+\ovrw{k}_{2}-\ovrw{k}_{3}
\label{phasmismatch}
\ee
is the phase mismatch, which vanishes
under the phase matching condition (\ref{phas_match}). This is in
fact a fundamental requirement for the effective realization of
nonlinear interactions in material media: if the phase matching
condition does not hold, then the integral appearing in
Eq.(\ref{X2}) is vanishingly low on average, and the interaction
process is effectively suppressed. Following Eq. (\ref{X2}), we
see that the resulting nonlinear parametric processes (in a
three-wave interaction) are described by generic trilinear
Hamiltonians of the form
\be
H_{mix}^{3wv}\propto\kappa^{(2)}a^{\dag}bc+H.c. \; ,\quad
\omega_{a}=\omega_{b}+\omega_{c} \; ,
\label{H2}
\ee
where $a$,
$b$, and $c$ are three different modes with frequency $\omega_{i}$
and momentum-polarization $\mbf{k}_{i}$ $(i=a,b,c)$, and
$\kappa^{(2)}\propto\chi^{(2)}$. Here the connection with the
previous notation is straightforward. The Hamiltonian (\ref{H2})
can describe the following three-wave mixing processes:
sum-frequency mixing for input $b$ and $c$ and
$\omega_{b}+\omega_{c}=\omega_{a}$; non-degenerate parametric
amplification for input $a$, and
$\omega_{a}=\omega_{b}+\omega_{c}$; difference-frequency mixing
for input $a$ and $c$ and $\omega_{b}=\omega_{a}-\omega_{c}$. If
some of the modes in Hamiltonian (\ref{H2}) degenerate in the same
mode (i.e. at the same frequency, wave vector and polarization),
one obtains degenerate parametric processes as : second harmonic
generation for input $b=c$ and $2\omega_{b}=\omega_{a}$;
degenerate parametric amplification for input $a$, and
$\omega_{a}=2\omega_{b}$, with $b=c$; and other effects as optical
rectification and Pockels effect involving d.c. fields
\cite{txtHeLiu,txtBloemb,txtShen,txtButchCott,txtBoyd}.

\vspace{0.25cm}

\item[b)] {\it Third order processes}

\vspace{3mm}

In the case of four-wave mixing, the relation (\ref{freq_con})
can give rise to two distinct conditions
\bea
\omega_{\mbf{k}_{1}}
+\omega_{\mbf{k}_{2}}&+&\omega_{\mbf{k}_{3}}=
\omega_{\mbf{k}_{4}}\; , \label{Freq_con31} \;
\\ && \nonumber\\
\omega_{\mbf{q}_{1}}+\omega_{\mbf{q}_{2}}
&=&\omega_{\mbf{q}_{3}}+\omega_{\mbf{q}_{4}} \; ,
\label{Freq_con32} \;
\eea
Following the same procedure
illustrated in the case of three-wave mixing, the third-order
effective interaction Hamiltonian reads
\bea
&& H_{mix}^{4wv}
\propto \Gamma_{\mbf{k}_{1},\mbf{k}_{2},\mbf{k}_{3},\mbf{k}_{4}}
\underline{\chi}^{(3)}(\omega_{\mbf{k}_{4}};\omega_{\mbf{k}_{1}},
\omega_{\mbf{k}_{2}},\omega_{\mbf{k}_{3}}):
\hat{\epsilon}_{\mbf{k}_{1}}\hat{\epsilon}_{\mbf{k}_{2}}
\hat{\epsilon}_{\mbf{k}_{3}}\hat{\epsilon}_{\mbf{k}_{4}}^{*}
a_{\mbf{k}_{1}}a_{\mbf{k}_{2}}a_{\mbf{k}_{3}}a_{\mbf{k}_{4}}^{\dag}
\nonumber \\
&& \nonumber \\
&& + \Gamma'_{\mbf{q}_{1},\mbf{q}_{2},\mbf{q}_{3},\mbf{q}_{4}}
\underline{\chi}^{(3)}(\omega_{\mbf{q}_{4}};\omega_{\mbf{q}_{1}},
\omega_{\mbf{q}_{2}},-\omega_{\mbf{q}_{3}}):
\hat{\epsilon}_{\mbf{q}_{1}}
\hat{\epsilon}_{\mbf{q}_{2}}\hat{\epsilon}_{\mbf{q}_{3}}^{*}
\hat{\epsilon}_{\mbf{q}_{4}}^{*}
a_{\mbf{q}_{1}}a_{\mbf{q}_{2}}a_{\mbf{q}_{3}}^{\dag}a_{\mbf{q}_{4}}^{\dag}
+H.c., \nonumber \\
\label{H3}
\eea
where $a_{\mbf{k}_{i}}$
$(a_{\mbf{q}_{i}})$ ($i=1,..,4$) are different modes at
frequencies $\omega_{\mbf{k}_{i}}$ $(\omega_{\mbf{q}_{i}})$, and
$\Gamma$ and $\Gamma'$ are proportional to products of
$\Lambda_{\mbf{k}}$. The four-wave mixing can generate a great
variety of multiphoton processes, including third harmonic
generation, Kerr effect, and coherent Stokes and anti-Stokes Raman
spectroscopy
\cite{txtHeLiu,txtBloemb,txtShen,txtButchCott,txtBoyd}.

\end{description}

In principle, by considering higher order nonlinearities, the
variety of possible multiphoton interaction becomes enormous. On the
other hand, as already mentioned, and as we will see soon in more
detail, the possibility of multiphoton processes of very high
order is strongly limited by the very rapidly decreasing magnitude of
the susceptibilities $\chi^{(n)}$ with growing $n$.

\subsection{Basic properties of the nonlinear susceptibility
tensors}

Nonlinear susceptibilities of optical media play a fundamental
role in the description of nonlinear optical phenomena. For this
reason, in this Section we briefly summarize the essential points
of the theory of nonlinear susceptibilities and discuss some of
their basic properties like their symmetries and their resonant
enhancements. The following results are valid under the
assumptions that the electronic-cloud distortion and the
intramolecular motion are the main sources of the polarization of
the medium, and that the nonlinear polarization response of the
medium is instantaneous and localized with respect to an applied
optical field. A complete quantum formulation of the theory can be
obtained using the density matrix approach
\cite{txtHeLiu,txtBloemb,txtShen,txtButchCott,txtBoyd}. The
density-matrix operator $\rho(t)$, describing the system composed
by the medium interacting with the applied optical field, evolves
according to the equation:
\be
i\frac{\partial\rho(t)}{\partial
t}=[H,\rho(t)]+i\left(\frac{\partial\rho}{\partial
t}\right)_{relax}\equiv[H_{0}+H_{I}(t),\rho(t)]+
i\left(\frac{\partial\rho}{\partial
t}\right)_{relax} \; ,
\label{rhomxeq}
\ee
where $H$ is the
Hamiltonian (\ref{Hintmediumradiation}), the first term is the
unitary, Liouvillian part of the dynamics, and the last term
represents the damping effects. The interaction Hamiltonian
$H_{I}(t)$ can be viewed as a time-dependent perturbation and the
density matrix can then be expanded in a power series as
$\rho(t)=\rho^{(0)}+\rho^{(1)}(t)+...+\rho^{(r)}(t)+...$, where
the generic term $\rho^{(r)}(t)$ is proportional to the $r$-th
power of $H_{I}(t)$, and usually the series is taken up to a
certain maximum, finite order $n$, which is determined by the
highest-order susceptibility that one needs in practice to
compute. The first term $\rho^{(0)}$ is the initial value of the
density matrix in absence of the external field, and at thermal
equilibrium we have
\be
\rho^{(0)}=Z e^{-H_{0}/k_{B}T} \; .
\ee
Inserting the series expansion of $\rho(t)$ into Eq.
(\ref{rhomxeq}) and collecting terms of the same order with
$H_{I}(t)$ treated as a first-order perturbation, one obtains the
following equation for $\rho^{(r)}$:
\be
i\frac{\partial\rho^{(r)}}{\partial
t}=[H_{0},\rho^{(r)}]+[H_{I},\rho^{(r-1)}]+i\Gamma\rho^{(r)} \; ,
\quad r=1,...,n \; ,
\label{densitymatrixeqs}
\ee
where the last
term again represents the explicit form of the damping effect, and
$\Gamma$ is a phenomenological constant. Clearly, given
$\rho^{(0)}$ and $H_{I}(t)$, one can in principle reconstruct the
complete density-matrix $\rho$. Let us next consider a volume $V$
of the medium, large with respect to the molecular dimensions and
small with respect to the wavelength of the field; such a volume
contains a number $\ovl{n}$ of identical and independent
molecules, each with electric-dipole momentum $\ovrw{p}$, so that
the $n$-th order dielectric polarization vector is given by
\be
\ovrw{P}^{(n)}(t)=Tr[N\,\ovrw{p}\,\rho^{(n)}(t)]=
N\sum_{a}[\ovrw{p}\,\rho^{(n)}(t)]_{aa}
=N\sum_{a,b}(\ovrw{p})_{ab}\,(\rho^{(n)}(t))_{ba} \,,
\label{Polariznav}
\ee
where $N=\ovl{n}/V$. In Eq.
(\ref{Polariznav}) the expressions $(\cdot)_{ab}$ indicates the
matrix elements $\langle a|\cdot|b \rangle$, where $|a\rangle$ and
$|b\rangle$ belong to a set of basis vectors, and a completeness
relation has been inserted in the last equality in Eq.
(\ref{Polariznav}). We are interested in the response to a field
that can be decomposed into Fourier components
\be
\ovrw{E}(t)=\sum_{q}\mathcal{E}_{q}(\omega_{q})e^{-i\omega_{q}t}
\,.
\ee
Since $H_{I}(t)$ can also be expressed as a Fourier series
$\sum_{q} H_{I}(\omega_{q}) e^{-i\omega_{q}t}$, analogously we can
write $\rho^{(r)}(t)=\sum_{q} \rho^{(r)}(\omega_{q})
e^{-i\omega_{q}t}$. Eq. (\ref{densitymatrixeqs}) can be solved for
$\rho^{(r)}(\omega_{q})$ in successive orders. The full
microscopic formulas for the nonlinear polarizations and
susceptibilities are then derived directly from the expressions of
$\rho^{(r)}(\omega_{q})$. Here we give the final expressions for
the second and third order susceptibilities \cite{txtHeLiu}:
\bea
&&\chi_{ijk}^{(2)}(\omega_{1},\omega_{2})=\frac{N}{2}\mathbf{S}
\sum_{a,b,c}\rho_{aa}^{(0)}
\left[\frac{(p_{i})_{ab}(p_{j})_{bc}(p_{k})_{ca}}{(\omega_{ba}-
\omega_{1}-\omega_{2}-i\Gamma_{ba})
(\omega_{ca}-\omega_{2}-i\Gamma_{ca})}\right. \nonumber \\
&& \nonumber \\
&&+
\frac{(p_{j})_{ab}(p_{i})_{bc}(p_{k})_{ca}}{(\omega_{cb}-
\omega_{1}-\omega_{2}-i\Gamma_{bc})}
\left(\frac{1}{\omega_{ac}+\omega_{2}+i\Gamma_{ac}}+
\frac{1}{\omega_{ba}+\omega_{1}+i\Gamma_{ba}}\right)
\nonumber \\
&& \nonumber \\
&&+ \left.
\frac{(p_{k})_{ab}(p_{j})_{bc}(p_{i})_{ca}}{(\omega_{ca}+\omega_{1}+
\omega_{2}+i\Gamma_{ca})
(\omega_{ba}+\omega_{2}+i\Gamma_{ba})}\right] \; .
\label{chi2formula}
\eea
\bea
&&\chi_{ijkl}^{(3)}(\omega_{1},\omega_{2},\omega_{3})= \;
\frac{N}{6}\mathbf{S} \sum_{a,b,c,d}\rho_{aa}^{(0)} \nonumber \\
&& \nonumber \\
&& \times \;
\left[\frac{(p_{i})_{ab}(p_{j})_{bc}(p_{k})_{cd}(p_{l})_{da}}{(\omega_{ba}
-\omega_{1}-\omega_{2}-\omega_{3}-i\Gamma_{ba})
(\omega_{ca}-\omega_{2}-\omega_{3}-i\Gamma_{ca})(\omega_{da}-\omega_{3}-
i\Gamma_{da})}\right. \nonumber \\
&& \nonumber \\
&&+
\frac{(p_{j})_{ab}(p_{i})_{bc}(p_{k})_{cd}(p_{l})_{da}}{(\omega_{ba}+
\omega_{1}+i\Gamma_{ba})
(\omega_{ca}-\omega_{2}-\omega_{3}-i\Gamma_{ca})(\omega_{da}-
\omega_{3}-i\Gamma_{da})}
\nonumber \\
&& \nonumber \\
&&+
\frac{(p_{j})_{ab}(p_{k})_{bc}(p_{i})_{cd}(p_{l})_{da}}{(\omega_{ba}+
\omega_{1}+i\Gamma_{ba})
(\omega_{ca}+\omega_{1}+\omega_{2}+i\Gamma_{ca})(\omega_{da}-
\omega_{3}-i\Gamma_{da})}
\nonumber \\
&& \nonumber \\ &&+ \left.
\frac{(p_{j})_{ab}(p_{k})_{bc}(p_{l})_{cd}(p_{i})_{da}}{(\omega_{ba}+
\omega_{1}+i\Gamma_{ba})
(\omega_{ca}+\omega_{1}+\omega_{2}+i\Gamma_{ca})(\omega_{da}+
\omega_{1}+\omega_{2}+\omega_{3}+i\Gamma_{da})}\right]
\; . \nonumber  \\
\label{chi3formula}
\eea
Here $\omega_{ab}$ is
the transition frequency from the state $a$ to the state $b$,
$\rho^{(0)}_{aa}$ denotes a diagonal element of the zero-order
density matrix, and $\Gamma_{ab}$ is the damping factor
corresponding to the off-diagonal element of the density matrix.
The symmetrizing operator $\mathbf{S}$ indicates that the
expressions which follow it must be summed over all the possible
permutations of the pairs $(j,\,\omega_{1})$, $(k,\,\omega_{2})$,
and $(l,\,\omega_{3})$. For nonresonant interactions, the
frequencies $\omega_{1}$, $\omega_{2}$, $\omega_{3}$, and their
linear sums, are far from the molecular transition frequencies;
hence, the damping factors $\Gamma$ can be neglected in
expressions (\ref{chi2formula}) and (\ref{chi3formula}). An
alternative method to perform perturbative calculations and to
determine the density matrices $\rho^{(n)}$ and thus the
susceptibilities $\chi^{(n)}$ is through a diagrammatic technique,
devised by Yee and Gustafson \cite{txtShen,YeeGust}. The results
reported so far have been obtained in the framework of
perturbation theory, and are correct only for dilute media. In
fact, in dense matter, induced dipole-dipole interactions arise
and cannot be neglected: as a consequence, the local field for a
single molecule may differ from the macroscopic averaged field in
the medium. In this case, local-field corrections have to be
introduced \cite{txtShen}. A simple analytical treatment can be
obtained for isotropic or cubic media, for which the so-called
Lorentz model can be applied \cite{txtShen}. In this framework,
the local field at a spatial point $\ovrw{E}_{loc}(\ovrw{r})$ is
determined by the applied field $\ovrw{E}(\ovrw{r})$ and the field
generated by the neighboring dipoles $\ovrw{E}_{dip}(\ovrw{r})$
\be
\ovrw{E}_{loc}(\ovrw{r})=\ovrw{E}(\ovrw{r})+\ovrw{E}_{dip}(\ovrw{r})
\; .
\ee
Exploiting the Lorentz model, one can write
$\ovrw{E}_{dip}(\ovrw{r})=\frac{1}{3}\ovrw{P}(\ovrw{r})$, where
the local polarization $\ovrw{P}(\ovrw{r})$ can be again expanded
in a power series. The $n$-th order susceptibility is then given
by
\bea
&&\chi^{(n)}_{loc}(\omega_{\sigma};\omega_{1},...,\omega_{n})=
\nonumber \\
&& \nonumber \\
&&NL^{(n)}[\varepsilon^{(1)}(\omega_{\sigma}),
\varepsilon^{(1)}(\omega_{1}),...,\varepsilon^{(1)}(\omega_{n})]
\chi^{(n)}(\omega_{\sigma};\omega_{1},...,\omega_{n}) \,,
\label{susceptlfc}
\eea
where
$L^{(n)}[\varepsilon^{(1)}(\omega_{\sigma}),
\varepsilon^{(1)}(\omega_{1}),...,\varepsilon^{(1)}(\omega_{n})]=
\left[\frac{\varepsilon^{(1)}(\omega_{\sigma})}{3}\right]\left[
\frac{\varepsilon^{(1)}(\omega_{1})}{3}\right]\cdots
\left[\frac{\varepsilon^{(1)}(\omega_{n})}{3}\right]$
and the linear dielectric constant is $\varepsilon^{(1)}(\omega)
=\left[1+\frac{2N}{3}\chi^{(1)}(\omega)\right]
\left[1-\frac{N}{3}\chi^{(1)}(\omega)\right]^{-1}$.
Relation (\ref{susceptlfc}) is valid also in more general cases,
but then $L^{(n)}$ will be a tensorial function depending on the
symmetry of the system.

Concerning the main symmetry properties of the $n$-th order
susceptibilities
$\chi_{j\alpha_{1}...\alpha_{n}}^{(n)}(\omega;\omega_{1},...,\omega_{n})$,
the obvious starting point is that they have to remain unchanged
under the symmetry operations allowed by the medium. We begin by
discussing the influence of the spatial symmetry of the material
system on
$\chi_{j\alpha_{1}...\alpha_{n}}^{(n)}(\omega;\omega_{1},...,\omega_{n})$.
Relation (\ref{nchifun}) implies that $\chi^{(n)}$ is a polar
tensor of $(n+1)-$th rank since $\ovrw{P}$ and $\ovrw{E}$
transform as polar tensors of the first rank (vectors) under
linear orthogonal transformations of the coordinate system. If $T$
denotes such transformation represented by the orthogonal matrix
$T_{\kappa\lambda}$, we then have
\be
\tilde{E}_{\kappa}=T_{\kappa\lambda}E_{\lambda} \,, \quad
\tilde{P}_{\kappa}^{(n)}=T_{\kappa\lambda}P_{\lambda}^{(n)} \; ,
\ee
where the tilde denotes that the quantity is expressed in the
new coordinate system. Upon direct substitution, one finds that
\be
\chi_{j' \alpha_1 ' ... \alpha_n '}^{(n)}=T_{j' j}T_{\alpha_1
' \alpha_1}...T_{\alpha_n ' \alpha_n}\chi_{j \alpha_1 ... \alpha_n
}^{(n)} \; ,
\ee
which shows explicitly, as already anticipated,
that $\chi^{(n)}$ transforms like a tensor of $(n+1)-$th rank.
Furthermore, the susceptibilities $\chi^{(n)}$ must be invariant
under the symmetry operations which transform the medium into
itself. This implies a number of relations between the components
of $\chi^{(n)}$ from which the nonvanishing independent components
can be extracted. The simplest example is the case of a medium
invariant under mirror inversion. This transformation corresponds
to an orthogonal matrix $T_{\lambda' \lambda}=(-1)\delta_{\lambda'
\lambda}$ and thus leads to
\be
\chi_{j \alpha_1 ... \alpha_n
}^{(n)}=(-1)^{n+1}\chi_{j \alpha_1 ... \alpha_n }^{(n)} \; ,
\ee
which implies $\chi^{(n)}=0$ for even $n$. A symmetry of more
general nature, that applies to any kind of medium, is the
intrinsic index/frequency permutation symmetry of the
susceptibility tensors; in the case of non degenerate frequencies,
the $n$ fields that enter in the product defining the $n$-th order
susceptibility can be arranged in $n!$ ways, with a corresponding
rearrangement of the indices and frequency arguments of the
susceptibility tensor. This consideration leads to the permutation
symmetry of the $\chi$ tensors: the interchange of any pair of the
last $n$ frequencies and of the corresponding cartesian
coordinates leaves $\chi$ invariant:
\bea
&&\chi_{j\alpha_1 ...
\alpha_p ...\alpha_q ...\alpha_n}^{(n)}(\omega;\omega_1 ,...,
\omega_p ,...,\omega_q ,...,\omega_n) \nonumber \\
&& \nonumber \\
&&= \chi_{j\alpha_1 ... \alpha_q ... \alpha_p
...\alpha_n}^{(n)}(\omega;\omega_1 ,..., \omega_q ,...,\omega_p
,...,\omega_n) \; .
\label{intpermsymm}
\eea
For nonresonant
interactions, property (\ref{intpermsymm}) can be further extended
by including also the pair $(j,\,\omega)$ in the possible
index/frequency interchanges. Again for nonresonant interactions,
it can also be proven that the nonlinear susceptibility tensor is
real
\be
\chi_{j\alpha_1 \alpha_2
...\alpha_n}^{(n)}(\omega;\omega_1 , \omega_2 ,...,\omega_n)=
\chi_{j\alpha_1 \alpha_2 ...\alpha_n}^{(n)}(\omega;\omega_1 ,
\omega_2 ,...,\omega_n)^{*} \; .
\label{nonreschireal}
\ee
Moreover, from the general phenomenological definition of the
$n$-th order susceptibility given in Eq. (\ref{phenomdefchi}), the
so-called complex conjugation symmetry implies that
\be
\chi_{j\alpha_1 \alpha_2 ...\alpha_n}^{(n)}(\omega;\omega_1 ,
\omega_2 ,...,\omega_n)^{*}=\chi_{j\alpha_1 \alpha_2
...\alpha_n}^{(n)}(-\omega;-\omega_1 , -\omega_2 ,...,-\omega_n)
\; .
\label{complexconjusymchi}
\ee
Relations
(\ref{nonreschireal}) and (\ref{complexconjusymchi}) lead to the
fundamental symmetry under time reversal:
\be
\chi_{j\alpha_1
\alpha_2 ...\alpha_n}^{(n)}(\omega;\omega_1 , \omega_2
,...,\omega_n)= \chi_{j\alpha_1 \alpha_2
...\alpha_n}^{(n)}(-\omega;-\omega_1 , -\omega_2 ,...,-\omega_n)
\; .
\label{timerevsymchi}
\ee
Finally, assuming that the
frequencies $\omega$, $\omega_{1}$,...,$\omega_{n}$ are small with
respect to the molecular resonance frequencies, the susceptibility
tensor is invariant also under interchange of cartesian
coordinate:
\bea
&&\chi_{j\alpha_1 ... \alpha_p ...\alpha_q
...\alpha_n}^{(n)}(\omega;\omega_1 ,..., \omega_p ,...,\omega_q
,...,\omega_n) \nonumber \\
&& \nonumber \\
&&= \chi_{j\alpha_1 ... \alpha_q ... \alpha_p
...\alpha_n}^{(n)}(\omega;\omega_1 ,..., \omega_p ,...,\omega_q
,...,\omega_n) \; .
\label{kleinman}
\eea
Such a property is again
valid for nonresonant interactions, and is known as the Kleinman
symmetry \cite{Kleinman}.

\subsubsection{Magnitude of nonlinear susceptibilities}

\begin{figure}[h]
\begin{center}
\includegraphics*[width=5.5cm]{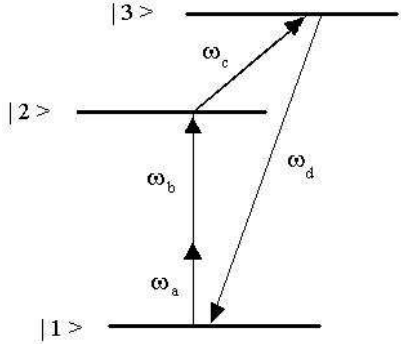}
\end{center}
\caption{Energy-level diagram of a prototype atomic system for the
sum frequency process
$\omega_{d}=\omega_{a}+\omega_{b}+\omega_{c}$.} \label{EIT}
\end{figure}

In order to give an idea of the experimental feasibility of
$n$-order interactions, let us now consider the orders of
magnitude of nonlinear susceptibilities. Recently, Boyd
\cite{suscestimBoyd} has developed simple mathematical models to
estimate the size of the electronic, nuclear, and electrostrictive
contributions to the optical nonlinearities. Typical values for
the susceptibilities in the Gaussian system of units or
electrostatic units $(esu)$ are $\chi^{(1)}\simeq 1$,
$\chi^{(2)}\simeq 10^{-8}esu$, and $\chi^{(3)}\simeq 10^{-15}esu$
\cite{txtHeLiu,txtBoyd}. We remind that the electrostatic units
corresponding to $\chi^{(r)}$ are $(cm/statvolt)^{r-1}$ and that
$1\,statvolt/cm= 3 \times 10^{3}\,V/m$. In general, the following
approximate relation holds
\be
|\chi^{(n-1)}|\approx|\chi^{(n)}|
|E_{atom}| \; ,
\label{chiestimat}
\ee
where $|E_{atom}|\approx
10^{7}esu\;(10^{11}V/m)$ is the magnitude of the average electric
field inside an atom. The ratio between two polarizations of
successive orders is
\be
|P^{(n)}|/|P^{(n-1)}|\approx
|E|/|E_{atom}| \,,
\label{polarizestimat}
\ee
where $|E|$ is the
magnitude of an applied optical field. Many optical effects are
generated through the action on the nonlinear medium of intense
coherent fields; commonly used laser pumps have magnitudes of the
order of $10^{8}V/m$. In such cases the Hamiltonian contributions
due to second and third order susceptibilities may become
relevant. \\ From the order-of-magnitude relations
(\ref{chiestimat}) and (\ref{polarizestimat}), it is clear that,
to analyze nonlinear phenomena involving phase-matched processes
like the third harmonic generation or the third order
sum-frequency generation, an enhancement of the magnitude of
$\chi^{(3)}$ is needed. This goal can be reached by exploiting
resonant interactions: when the frequencies of the applied optical
fields, as well as of their linear combinations, are close to the
molecular resonant frequencies of the medium, the susceptibilities
are complex - See e.g. Eqs. (\ref{chi2formula}) and
(\ref{chi3formula}) - and their effective value can grow very
sharply. For instance, let us denote with $\omega_{eg}$ a
molecular transition, where the subscripts $g$ and $e$ denote the
ground and the excited states involved in the transition. Let us
next consider the two-photon sum-frequency resonance effect
$\omega_{eg}=\omega_{I}+\omega_{II}$. In this case, in equations
(\ref{chi2formula}) and (\ref{chi3formula}), taking for instance
$\omega_{ba} = \omega_{eg}$ in formula (\ref{chi2formula}) and
$\omega_{ca} = \omega_{eg}$ in formula (\ref{chi3formula}), only
the resonant terms proportional to
$1/(\omega_{eg}-\omega_{I}-\omega_{II}-i\Gamma_{eg})$ can be
retained. In the specific instance of third harmonic generation
(THG), in the case of two-photon absorptive transition
$\omega_{eg}=2\omega$, the enhanced magnitude is given by
$|\chi^{(3)}_{THG}|\propto
1/\sqrt{(\omega_{eg}-2\omega)^{2}+\Gamma^{2}}$. This leads to
resonant values of $\chi^{(3)}$ which can attain $10^{-10} esu$.
Obviously, the structure of the susceptibilities shows that,
beyond the two-photon sum-frequency resonance, the enhancement can
be obtained as well by one-photon resonance, two-photon
difference-frequency resonance, Raman resonance, Brillouin
resonance, and so on. Although the resonant interaction causes a
remarkable increase of the magnitude of the susceptibility, an
exact resonance can also lead to a depletion both of the input
optical pump and of the output signal wave. In practice, a
near-resonance condition is experimentally preferred. Recent
efforts in the fabrication of composite materials as layered
dielectric-dielectric composite structures \cite{chi3ena1},
metal-dielectric photonic crystals \cite{chi3ena2}, and
metal-dielectric nanocomposite films \cite{chi3ena3}, have
succeeded in obtaining fast response, strongly enhanced
$\chi^{(3)}\simeq 10^{-7}esu$. Large and extremely fast response
third order optical nonlinearity has been obtained also in
$Au:TiO_{2}$ composite films with varying $Au$ concentration
\cite{chi3enafemto}; by measurements on a femtosecond time scale,
it has been found a maximum value for $\chi^{(3)}$ of $6\times
10^{-7} esu$. Besides composite structures, also photonic
crystals, that are systems with spatially periodic dielectric
constant, seem to be very promising materials to realize nonlinear
optical devices. They may have photonic band gaps, and, by the
introduction of defects, it is possible to engineer waveguides and
cavities with them. Moreover, they can be useful for the
enhancement of $\chi^{(3)}$ nonlinearities. Nonlinear interactions
of femtosecond laser pulses have been demonstrated in photonic
crystal fibers \cite{chi3enafemto2}, and a considerable
enhancement of $\chi^{(3)}$ has been observed in fully
three-dimensional photonic crystals \cite{chi3ena5}. Another
interesting technique to produce an effective third-order
nonlinearity is by means of cascading second order processes
$\chi^{(2)}:\chi^{(2)}$, see for instance Refs.
\cite{cascadechi2,cascadechi22,cascadechi23,cascadechi24}, that
have been extended even to third-order cascaded processes
exploited to enlarge the range of possible frequency generations
\cite{cascadechi3,cascadechi32}. \\ Finally, we want to emphasize
the importance of coherent atomic effects such as coherent
population trapping (CPT) \cite{Gozzini,CPTArimondo} (first
discovered by Gozzini and coworkers in the context of optical
pumping experiments on $Na$), and the related effect of
electromagnetically induced transparency (EIT)
\cite{EITImamoglu,EITtoday,EITHarris,EITnature,EITBoller} for
nonlinear optics. In a resonant regime, light propagation in a
nonlinear medium suffers strong absorption and dispersion, due to
the growing importance of the linear dissipative effects
associated to the linear susceptibility $\chi^{(1)}$. Fortunately,
thanks to EIT, it is possible to realize processes with resonantly
enhanced susceptibilities while at the same time inducing
transparency of the medium \cite{EITHarris,EITnature}. Such
remarkable result can be in the end traced back to quantum
mechanical interference. Let us briefly outline the basic
mechanism at the basis of EIT. The scheme in Fig. (\ref{EIT})
represents an energy-level diagram for an atomic system
\cite{EITHarris}; a strong electromagnetic coupling field of
frequency $\omega_{c}$ is applied between a metastable state
$|2\rangle$ and a lifetime-broadened state $|3\rangle$, and the
sum frequency $\omega_{d}=\omega_{a}+\omega_{b}+\omega_{c}$ is
generated.
It can be shown \cite{EITHarris} that when the field at
$\omega_{c}$ is applied, the $\chi^{(3)}$ medium becomes
transparent to the resonant transition $|1\rangle \rightarrow
|3\rangle$, while in the absence of $\omega_{c}$ the radiation at
$\omega_{d}$ is strongly absorbed. The transparency is due to the
destructive interference of the two possible absorption
transitions $|1\rangle \rightarrow |3\rangle$ and $|2\rangle
\rightarrow |3\rangle$. Exploiting EIT can provide large third- or
higher-order nonlinear susceptibilities and a minimization of
absorption losses. Several proposals, based on the three-level
$\Lambda$ configuration as in Fig. (\ref{EIT}), or on generalized
multi-level schemes, have been made for the enhancement of the
Kerr nonlinearity, that shall be discussed in the next Section
\cite{EITkerr1,EITkerr2,EITkerr3}. Several successful experimental
realizations based on the transparency effect have been achieved;
for instance, high conversion efficiencies in second harmonic
\cite{EITshg} and sum-frequency generation \cite{EITsumfreq} have
been obtained in atomic hydrogen, and the experimental observation
of large Kerr nonlinearity with vanishing linear susceptibilities
has been observed in four-level rubidium atoms \cite{EITkerrexp}.
In order to give an idea of the efforts in the direction of
producing higher-order nonlinearities of appreciable magnitude, we
also mention the proposal of resonant enhancement of $\chi^{(5)}$,
based on the effect of coherent population trapping
\cite{chi5ena}.

\subsection{Phase matching techniques and experimental
            implementations}

\begin{figure}[h]
\begin{center}
\includegraphics*[width=12cm]{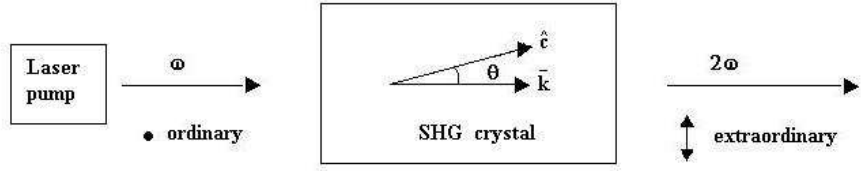}
\end{center}
\caption{Type I second harmonic generation in negative uniaxial
crystal. $\hat{c}$ is the optic axis. The process is phase matched
if $n^{ext}(2\omega,\theta)=n^{ord}(\omega)$.}
\label{angletunphasmatch}
\end{figure}

The phase matching condition (\ref{phas_match}), that is the
vanishing of the phase mismatch $\Delta \ovrw{k}$, is an essential
ingredient for the realization of effective, Hamiltonian nonlinear parametric
processes. For this reason, we briefly discuss here the most used
techniques and some experimental realizations of phase matching. In the
three-wave interaction the phase matching condition writes
\be
\omega_{a}n_{a}\hat{\imath}_{a}=\omega_{b}n_{b}\hat{\imath}_{b}+\omega_{c}n_{c}\hat{\imath}_{c}
\; ,
\label{Glande}
\ee
where $n_{q}=n(\omega_{q})$ and
$\hat{\imath}_{q}=\ovrw{k}_{q}/|\ovrw{k}_{q}|$, $(q=a,b,c)$. For
normal dispersion, i.e. $n_{a}>\{n_{b}\,,n_{c}\}$, relation
(\ref{Glande})
can never be fulfilled. In the process of collinear sum-frequency
generation $(\hat{\imath}_{a}=\hat{\imath}_{b}=\hat{\imath}_{c})$,
described quantum-mechanically by the Hamiltonian (\ref{H2}), the
intensity of the wave $a$ can be computed in the slowly-varying
amplitude approximation and is of the form \cite{txtShen}
\be
I_{a} \propto \frac{\sin^{2}(\Delta k L)}{(\Delta k L)^{2}}
\,,
\label{sumfreqeff}
\ee
where $L$ is the effective path length of the light propagating
through the crystal. The phase mismatch defines a coherence length
$L_{coh}=1/\Delta k$, which must be sufficiently long in order to
allow the sum-frequency process. Commonly, in optically
anisotropic crystals, phase matching is achieved by exploiting the
birefringence, i.e. the dependence of the refractive index on the
direction of polarization of the optical field. In 1962, by
exploiting the birefringence and the dispersive properties of the
crystal, Giordmaine \cite{SHGGiordmaine} and Maker
 \textit{et al.} \cite{SHGMaker} independently observed second
harmonic generation in potassium dihydrogen phosphate. The
relation (\ref{sumfreqeff}) was experimentally verified by Maker
\textit{et al} \cite{SHGMaker}. In order to illustrate the
phenomenon of birefringence in brief, let us consider the class of
uniaxial crystals (trigonal, tetragonal, hexagonal). Ordinary
polarized light (with polarization orthogonal to the plane
containing $\ovrw{k}$ and the optical axis) undergoes ordinary
refraction with index $n^{ord}$; extraordinary polarized light
(with polarization parallel to the plane containing $\ovrw{k}$ and
the optical axis) experiences refraction with the extraordinary
refractive index $n^{ext}$; the latter depends on the angle
$\theta$ between $\ovrw{k}$ and the optical axis; if they are
orthogonal, the ordinary and extraordinary refraction indices
coincide to the same value $n_{o}$, and when they are parallel,
the extraordinary refraction index assumes its maximum value
$n_{e}$ (obviously both $n_{o}$ and $n_{e}$ depend on the
material). For generic angles, we have
\be
n^{ord}\equiv n_{o} \,,
\quad
n^{ext}(\theta)=\frac{n_{o}n_{e}}{(n_{e}^{2}\sin^{2}\theta+
n_{o}^{2}\cos^{2}\theta)^{1/2}}
\; ,
\label{phasmatchbiref}
\ee
where the values of $n_{o}$ and
$n_{e}$ are known at each frequency. In Fig.
(\ref{angletunphasmatch}), the scheme represents the experimental
setup for second harmonic generation by angle-tuned phase
matching.
In a series of papers, Midwinter and Warner
\cite{midwinterI,midwinterII} analyzed and classified the phase
matching techniques for three- and four-wave interactions in
uniaxial crystals. Tables \ref{tabphasmatchI} and
\ref{tabphasmatchII} summarize the possible phase
matching methods for positive and negative uniaxial crystals,
i.e., respectively, uniaxial crystals with $(n_{e}>n_{o})$ and
$(n_{e} < n_{o})$.
\begin{table}[h]
  \begin{tabular}{ccc}
  \hline
     & Positive uniaxial & Negative uniaxial \\
   & $(n_{e}>n_{o})$ & $(n_{e}<n_{o})$ \\
   \hline
  Type I & $n_{a}^{ord}\omega_{a}=
n_{b}^{ext}\omega_{b}+ n_{c}^{ext}\omega_{c}$ &
$n_{a}^{ext}\omega_{a}= n_{b}^{ord}\omega_{b}+ n_{c}^{ord}\omega_{c}$ \\
  Type II & $n_{a}^{ord}\omega_{a}=
n_{b}^{ord}\omega_{b}+ n_{c}^{ext}\omega_{c}$ &
$n_{a}^{ext}\omega_{a}= n_{b}^{ext}\omega_{b}+ n_{c}^{ord}\omega_{c}$ \\
  \hline \\
  \end{tabular}
  \caption{Phase matching methods for three-wave interaction in
uniaxial crystals.}
\label{tabphasmatchI}
\end{table}

\begin{table}[h]
\begin{tabular}{ccc}
  \hline
     & Positive uniaxial & Negative uniaxial \\
      \hline
  Type I & $n_{4}^{ord}\omega_{4}= n_{3}^{ext}\omega_{3}+ n_{2}^{ext}\omega_{2}+n_{1}^{ext}\omega_{1}$ &
  $n_{4}^{ext}\omega_{4}= n_{3}^{ord}\omega_{3}+ n_{2}^{ord}\omega_{2}+n_{1}^{ord}\omega_{1}$ \\
  Type II & $n_{4}^{ord}\omega_{4}= n_{3}^{ext}\omega_{3}+ n_{2}^{ext}\omega_{2}+n_{1}^{ord}\omega_{1}$ &
  $n_{4}^{ext}\omega_{4}= n_{3}^{ord}\omega_{3}+ n_{2}^{ord}\omega_{2}+n_{1}^{ext}\omega_{1}$ \\
  Type III & $n_{4}^{ord}\omega_{4}= n_{3}^{ext}\omega_{3}+ n_{2}^{ord}\omega_{2}+n_{1}^{ord}\omega_{1}$ &
  $n_{4}^{ext}\omega_{4}= n_{3}^{ord}\omega_{3}+ n_{2}^{ext}\omega_{2}+n_{1}^{ext}\omega_{1}$ \\
  \hline \\
\end{tabular}
\caption{Phase matching methods for four-wave interaction in
uniaxial crystals.}
\label{tabphasmatchII}
\end{table}
Concerning more complex configurations, we should mention
phase-matched three-wave interactions in biaxial crystals
discussed in Refs. \cite{biaxphmatch,biaxphmatch2} and phase
matching via optical activity, first proposed by Rabin and Bey
\cite{OAphmatch}, and successively further investigated by Murray
\textit{et al.} \cite{OAphmatch2}.

\begin{figure}[h]
\begin{center}
\includegraphics*[width=11cm]{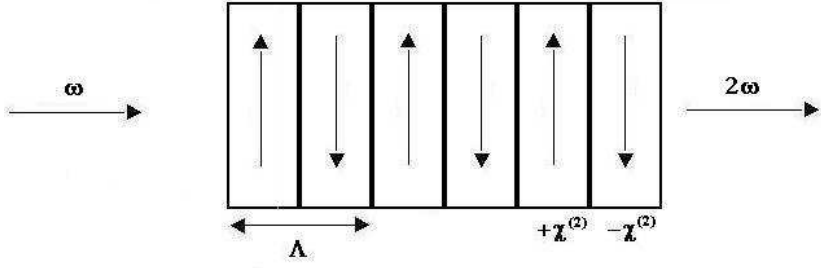}
\end{center}
\caption{Quasi phase matching for the second harmonic generation
in a periodically poled nonlinear crystal. $\Lambda$ is the period
of modulation of the susceptibility $\chi^{(2)}$.} \label{PPNC}
\end{figure}

In order to circumvent the difficulties in realizing exact
phase matching, for instance when trying to realize
concurrent interactions or in those frequency ranges
where it does not hold, one can resort to the technique
of so-called quasi phase matching. This idea was introduced in a
seminal work on media with periodic
modulation of the nonlinearity by
Armstrong \textit{et al.} \cite{ArmstrongBloemb}, already in 1962;
the same approach was proposed by Franken and Ward \cite{FrankWard} one
year later. Media with periodic modulation of the nonlinearity
consist of a repeated chain of elementary blocks, where in each block
of charachteristic linear dimension $\Lambda$ the susceptibility takes
opposite signs in each of the two halves of the block. This kind of structure allows
for a quasi phase matching condition in the sense that
the destructive interference
caused by dispersive propagation is compensated by
the inversion of the sign of the nonlinear susceptibility. As an
example, Fig. (\ref{PPNC}) represents a scheme for second
harmonic generation exploiting collinear quasi phase matching in a
periodically poled nonlinear crystal.
When the quasi phase matching condition
\be
k(2\omega)-2k(\omega)=\frac{2\pi}{\Lambda}
\ee
is satisfied, then the oscillating factor appearing in the
integral in Eq.(\ref{X2}) is of order one, guaranteeing the
non vanishing of the effective interaction much in the same
way as exact phase matching.
This technique
has been applied to several materials, as $LiNbO_{3}$, $KTP$,
$LiTaO_{3}$, fibres, polymers, and semiconductors (see, e.g.
Refs. \cite{QPM}). Being applicable to a very large class of
material media, the methods of quasi phase matching help
to use very high nonlinear coefficients, otherwise not
accessible with the standard techniques based on birefringence.
Finally, we wish to mention that quasi phase-matching conditions
have been recently realized in quasi-periodic optical superlattices
in order to generate second \cite{2harmsuperlatt} and third harmonic
\cite{3harmsuperlatt}.

\newpage

\section{Second and third order optical parametric processes}
\label{section4}

Moving on from the basic aspects introduced in Section
\ref{section3}, we now begin to discuss in some detail the most
important multiphoton processes occurring in nonlinear media. In
this Section we restrict the analysis to processes generated by
the strongest optical nonlinearities, i.e. those associated to the
second- and third-order susceptibilities as described by the
trilinear Hamiltonian (\ref{H2}) and the quadrilinear Hamiltonian
(\ref{H3}). Historically, after the experimental generation of the
second harmonic of the laser light \cite{SHGFranken}, the first
proposal for a quantum-mechanical model of the frequency amplifier
and frequency converter was presented by Louisell, Yariv, and
Siegman \cite{louisell}. Several papers dedicated to the analysis
of the statistical properties of these models rapidly followed
\cite{louisell2,mollowglauber,tuckerwalls2}. Frequency down
conversion was observed for the first time in 1970, in photon
coincidence counting experiments \cite{expDCburnham}, and
successively it was observed in time-resolved correlation
measurements \cite{fribergmandel}. Models based on four-wave
interactions were considered by Yuen and Shapiro
\cite{4wmYuenShap} and Yurke \cite{4wmYurke}, and in the 1980's
several experiments using four-wave mixing in nonlinear media were
reported (See e.g. \cite{4wmexpSS1,4wmexpSS2,4wmexpSS3}. Finally,
we need to recall that detailed analysis of the three- and
four-wave interactions was given by Armstrong \textit{et al.}
\cite{ArmstrongBloemb}, who discussed and found the exact
solutions for the classical coupled equations. In the following we
give an overview on the quantum models and quantum states
associated with three- and four-wave interactions, as well as the
exact and approximate mathematical methods to study the
corresponding dynamics. We will also recall some important
experimental realizations and proposals.

The lowest order nonlinearity $\chi^{(2)}$ is responsible of
three-photon processes whose dynamics is governed by Hamiltonians
of the form (\ref{H2}), which, in the pure
quantum case, can be exactly solved only by numerics (See below
for more details). Most
theoretical analyses have thus been concerned with physical
situations such that one mode, the pump mode, is highly excited
and can be considered in a high-amplitude coherent state.
In such a case, we can resort to the so-called
parametric approximation: the pump mode is treated
classically as a $c$-number, thus neglecting the depletion
mechanism and the quantum fluctuations. Consequently, for instance,
bilinear and trilinear models are greatly simplified and reduce,
respectively, to linear and bilinear ones. This fact allows
exact solvability by the application of standard methods like the
disentangling formulas for Lie algebras.
The range of validity of the parametric
approximation has been investigated in Ref.
\cite{degtrizubairy,scharf,degtricrouch,parappParis,padarianoparis}.
In particular, D'Ariano \textit{et al.} have shown that the usual
requirements, i.e. short interaction time and strong classical
undepleted pump are too restrictive, and that the main requirement is
only that the pump remains coherent after interaction with the medium
has occurred
\cite{padarianoparis}.

\subsection{Three-wave mixing and the trilinear Hamiltonian}

The fully quantized, lowest order multiphoton process is described
by the trilinear Hamiltonian
\be
H^{trl} \; = \; H_{0} +
H_{mix}^{3wv} \; = \; \omega_{a}a^{\dag}a+\omega_{b}
b^{\dag}b+\omega_{c}c^{\dag}c+\kappa^{(2)}(a^{\dag}bc+ab^{\dag}c^{\dag})
\; ,
\label{trilinearH}
\ee
where
$\omega_{a}=\omega_{b}+\omega_{c}$, and the coupling constant
$\kappa^{(2)}$ is assumed to be real. The Hamiltonian
(\ref{trilinearH}) describes the two-photon down-conversion
process in the crystal (one photon of frequency $\omega_{a}$ is
absorbed, and two photons of frequencies $\omega_{b}, \omega_{c}$
are emitted), and the sum-frequency generation (two photons of
frequencies $\omega_{b}, \omega_{c}$ are absorbed and one photon
of frequency $\omega_{a}$ is emitted). Several aspects of model
(\ref{trilinearH}) have been thoroughly studied in the literature
\cite{scharf,wallsbarak,agarwmehta,gambini,kumarmehta,34wmKatriel,mcnielgard,jurco,carusotto,tridrobnyjex1,tridrobnyjex2,kinsler,trihilleryyu,kinsler2,tridrobnyjex3,tribrif}.
The first description of the parametric amplifier and frequency
converter, without the classical approximation for the pumping
field, was performed by Walls and Barakat \cite{wallsbarak}, who
solved exactly the quantum-mechanical problem by the technique of
the integrals of motion. In the following we briefly outline this
method \cite{wallsbarak}, considered also by other authors
\cite{tridrobnyjex1,tridrobnyjex2,tridrobnyjex3,tribrif}, by
applying it to the case of parametric amplification, with $a$
denoting the laser mode, $b$ the idler mode, and $c$ the signal
mode. The system (\ref{trilinearH}) possesses five integrals of
motion, with three of them being independent. The five invariants
are the operators $H_{0}$, $H_{mix}^{3wv}$,
$N_{ab}=a^{\dag}a+b^{\dag}b$, $N_{ac}=a^{\dag}a+c^{\dag}c$, and
$D_{bc}\equiv N_{ac}-N_{ab}=c^{\dag}c-b^{\dag}b$, the last three
being known as Manley-Rowe invariants \cite{ManlRowe}. Having
three independent integrals of motion, the system can be
characterized by three independent quantum numbers. Exploiting
such numbers, the dynamics of the system can be studied by
decomposing the Hilbert space $\mathcal{H}$ associated with
Hamiltonian (\ref{trilinearH}) in a direct sum of
finite-dimensional subspaces. Let us then choose as independent
conserved quantities the operators $N_{ab}$, $D_{bc}$, and
$H_{mix}^{3wv}$, let us fix the (integer) eigenvalues $N$ of
$N_{ab}$, and $D$ of $D_{bc}$, and let us finally write down the
eigenvalue equations in this subspace $\mathcal{H}_{N,D}$ of fixed
$N$ and $D$:
\bea
&&
N_{ab}|\lambda_{j}(N,D)\rangle=N|\lambda_{j}(N,D)\rangle \; ,
\quad
D_{bc}|\lambda_{j}(N,D)\rangle=D|\lambda_{j}(N,D)\rangle \; , \\
&& \nonumber \\
&& H_{mix}^{3wv}|\lambda_{j}(N,D)\rangle=
\lambda_{j}(N,D)|\lambda_{j}(N,D)\rangle
\; .
\eea
For fixed values of $N$ and $D$, the eigenvalues
$\lambda_{j}(N,D)$ of $H_{mix}^{3wv}$ and the
common eigenvectors $|\lambda_{j}(N,D)\rangle$
are obtained by diagonalizing $H_{mix}^{3wv}$ on the
subspace $\mathcal{H}_{N,D}$. The latter is
spanned by the set of vectors of
the form $\{|n\rangle_{N,D}=|N -n \rangle_{a}|n\rangle_{b}|D + n \rangle_{c}\,,
0 \leq n \leq N\}$, where $|n\rangle_{a}|m\rangle_{b}|l\rangle_{c}$
are the three-mode number states. In this basis, the matrix
representation of $H_{mix}^{3wv}$ is \be
H_{mix}^{3wv} \; = \; \kappa^{(2)}
\left(%
\begin{array}{ccccccc}
  0 & h_{0} & 0 & 0 & 0 & \cdots & 0 \\
  h_{0} & 0 & h_{1} & 0 & 0 & \cdots & 0 \\
  0 & h_{1} & 0 & h_{2} & 0 & \cdots & 0 \\
  0 & 0 & h_{2} & 0 & h_{3} & \cdots & 0 \\
  \vdots & \vdots & \ddots & \ddots & \ddots & \ddots & h_{N-1} \\
  0 & \cdots & \cdots & \cdots & 0 & h_{N-1} & 0 \\
\end{array}%
\right) \, ;
\label{matrixH3wv}
\ee
where
$h_{r}=\sqrt{(N-r)(D+r+1)(r+1)}$, $(0 \leq r \leq N-1)$. The final form of
the eigenvalue equation for $H^{trl}$ can be written as:
\bea
H^{trl}|\lambda_{j}(N,D)\rangle
&=&[(\omega_{a}N_{ab}+\omega_{c}D_{bc})+H_{mix}^{3wv}]|\lambda_{j}(N,D)\rangle
\nonumber \\
&& \nonumber \\
&=&[\omega_{a}N+\omega_{c}D+\lambda_{j}(N,D)]|\lambda_{j}(N,D)\rangle
\; .
\eea
The eigenvectors $|\lambda_{j}(N,D)\rangle$ in the subspace,
can be expressed as superpositions of the number-state basis vectors,
$|\lambda_{j}(N,D)\rangle=\sum_{i=0}^{n}u_{ji}(N,D)|n\rangle_{N,D}$,
where the values of $u_{ji}(N,D)$ have to be determined numerically.
Therefore, fixed finite values of $N$ and $D$, the dynamics can
be solved exactly, albeit numerically, by determining the eigenvalues
$\lambda_{j}(N,D)$ and the orthonormal eigenvectors
$|\lambda_{j}(N,D)\rangle$ of the $(N+1)\times(N+1)$ matrix
(\ref{matrixH3wv}). Clearly, the dynamics depends crucially on the
nature of the initial state. Typical choices are
three-mode number states (for their simplicity),
and states of the coherent form
$|\alpha\rangle|0\rangle|0\rangle$ that represent the
initial condition for spontaneous parametric down conversion
(here $|\alpha\rangle$ denotes the coherent
state for the pump mode). This technique has been exploited to
study the statistics and the squeezing properties of the signal
mode \cite{tridrobnyjex1}, and the time evolution of the
entanglement between the modes \cite{tridrobnyjex2}. The system
can exhibit sub-Poissonian statistics and anticorrelation (roughly
speaking, two-mode antibunching as measured by the cross-correlation
function), as well as strong entanglement between the pump and signal
or idler modes. Moreover, the model is characterized by the appearance
of collapses and revivals in the mean photon numbers
\cite{wallsbarak,tridrobnyjex1}. \\
Analytic solutions of the
dynamics generated by the Hamiltonian (\ref{trilinearH})
can be accomplished under short-time approximations, which
are however realistic due to the usually very short interaction
times \cite{agarwmehta,gambini,tridrobnyjex3}. The nonlinear, coupled
dynamical equations
\bea
&&i\frac{d
a(t)}{dt}=\omega_{a}a(t)+\kappa^{(2)}b(t)c(t) \; , \nonumber \\
&& \nonumber \\
&& i\frac{d b(t)}{dt}=\omega_{b}b(t)+\kappa^{(2)}a(t)c^{\dag}(t)
\; , \nonumber \\
&& \nonumber \\
&&i\frac{d c(t)}{dt}=\omega_{c}c(t)+\kappa^{(2)}a(t)b^{\dag}(t) \; ,
\eea
can be solved by expanding each mode in a Taylor series up
to quadratic terms and by exploiting the equations of motion;
for instance, the time evolution of the signal mode operator $c$, up to
second order in the dimensionless reduced time
$\kappa^{(2)}t$, is given by
\be
c(t) \; = \; c \, - \, i\kappa^{(2)}t
b^{\dag}a \, + \,
\frac{(\kappa^{(2)}t)^{2}}{2!}c(a^{\dag}a \, - \,
b^{\dag}b) \, + \, \mathcal{O}[(\kappa^{(2)}t)^{3}]
\; , \label{trishortct}
\ee
where $a$, $b$, $c$ in the right hand side
of Eq. (\ref{trishortct}) denote the initial values of the
operators at time $t=0$. Given
Eq. (\ref{trishortct}), the dynamics of the signal mode $c$
is determined, for example by resorting to its diagonal
coherent state representation \cite{agarwmehta}. \\ Another
interesting method \cite{34wmKatriel} to obtain analytic solutions
for three-wave mixing is based on the use of generalized Bose
operators introduced by Brandt and Greenberg \cite{brandtgreenb},
and put in a closed form by Rasetti \cite{gBoRasetti}. Let us
define the pair of operators \cite{34wmKatriel}
\be
A=F(n_{b},n_{c})bc \; , \quad \quad
A^{\dag}=c^{\dag}b^{\dag}F^{*}(n_{b},n_{c}) \; ,
\ee
where
$F(n_{b},n_{c})$ is an operatorial function of the number
operators for modes $b$ and $c$, that is fixed by imposing
the bosonic canonical commutation relation $[A,A^{\dag}]=1$.
It can then be shown that
$|F(n_{b},n_{c})|^{2}(n_{b}+1)(n_{c}+1)-|F(n_{b}-1,n_{c}-1)|^{2}
n_{b}n_{c}=1$, and by induction that
$|F(n_{b},n_{c})|^{2}=(n_{>}+1)^{-1}$ with
$n_{>}=\max\{n_{b},n_{c}\}$. As $D_{bc}\equiv n_{c}-n_{b}$ is a
constant of the motion, under the hypothesis that the mode $c$ is
much more intense than the mode $b$, we can write $b^{\dag}b\simeq
A^{\dag}A$ and $c^{\dag}c\simeq n_{0}+A^{\dag}A$, where $n_{0}$,
the difference in the number of photons in the two modes,
satisfies $n_{0}\gg 1$. The Hamiltonian (\ref{trilinearH}) can
then be expressed in the form
\bea
H^{trl}  \,& = & \, \omega_{c}n_{0}+\omega_{a}(A^{\dag}A+a^{\dag}a)
\nonumber \\
&& \nonumber \\
&&+\kappa^{(2)}\left[a^{\dag}
\frac{1}{F(n_{b},n_{c})}A+A^{\dag}\frac{1}{F^{*}(n_{b},n_{c})}a
\right].
\label{triHkatriel}
\eea
Under the imposed condition
$n_{0}\gg 1$, $F$ can be treated as a constant and $H^{trl}$
reduces to
\be \tilde{H}^{trl}=\omega_{a}(A^{\dag}A+a^{\dag}a)
+\tilde{\kappa}^{(2)}(a^{\dag}A+A^{\dag}a) \; ,
\ee
where
$\tilde{\kappa}^{(2)}=\kappa^{(2)}/F\simeq\kappa^{(2)}/\sqrt{n_{0}}$,
and the constant term $\omega_{c}n_{0}$ has been dropped. The
solution of the Heisenberg equation of motion for mode $a$ is
\be
a(t)=e^{-i\omega_{a}t}[a\cos(\tilde{\kappa}^{(2)}t)-
iA\sin(\tilde{\kappa}^{(2)}t)]
\; ,
\ee
where in the r.h.s $a$ and $A$ are the initial-time
operators. Therefore, the mean number of photons in mode $a$,
taking an initial two-mode Fock number state
$|n_{b},n_{c}\rangle$, is $\langle
n_{a}(t)\rangle=\sin^{2}(\tilde{\kappa}^{(2)}t)n_{<}$, where
$n_{<}$ denotes the mean number of photons in the less intense
between the modes $b$ and $c$. The number of photons in mode $a$
hence exhibits oscillations with period
$\pi/\tilde{\kappa}^{(2)}$, in close agreement with the result of
Walls and Barakat \cite{wallsbarak}.
\\ Other approaches have been used to analyze the dynamics of the
system (\ref{trilinearH}): Jur\v{c}o found exact solutions using
the Bethe ansatz \cite{jurco}; Gambini and Carusotto solved the
coupled nonlinear equations of motion by using iteration methods
\cite{gambini,carusotto}; McNeil and Gardiner have studied the
process in a cavity by finding the solution of a Fokker-Planck
equation \cite{mcnielgard}; Hillery \textit{et al.} determined
exact relations between the number fluctuations of the three modes
\cite{trihilleryyu}. \\ With the particular choice
$\omega_{b}=\omega_{c}=\omega_{a}/2$ \cite{tridrobnyjex1,tribrif},
the Hamiltonian (\ref{trilinearH}) can be expressed in the form
\be
H^{trl}_{su11}=\omega_{a}(a^{\dag}a+K_{0})+\kappa^{(2)}(aK_{+}+a^{\dag}K_{-})
\; , \label{trilinearHsu11}
\ee
where the operators
\be
K_{+}=b^{\dag}c^{\dag} \; , K_{-}=bc \; ,
K_{0}=\frac{1}{2}(b^{\dag}b+c^{\dag}c+1) \; , \label{su11algebra}
\ee
span the $SU(1,1)$ Lie algebra:
\be
[K_{0},K_{\pm}]=\pm
K_{\pm} \; , \quad [K_{-},K_{+}]=2K_{0} \; . \label{su11algebracom}
\ee
Clearly, the form (\ref{trilinearHsu11}) is suitable for the
description of interactions with underlying $SU(1,1)$ symmetry,
such as parametric amplification. The Casimir operator
\be
C=K_{0}^{2}-\frac{1}{2}(K_{+}K_{-}+K_{-}K_{+})=\frac{1}{4}(D_{bc}^{2}-1)=k(k-1)
\; ,
\ee
is an invariant of the motion. The Bargmann index
$k=\frac{1}{2}(|D_{bc}|+1)$ takes the discrete
values $k=\frac{1}{2},1,\frac{3}{2},2,...$ \cite{bargmann}.
Given the quantum number $L=N_{ac}$, the diagonalization
method proceeds along lines similar to those of Walls
and Barakat that we have already illustrated:
the whole Hilbert space can be decomposed into
a direct sum of finite-dimensional subspaces $\mathcal{H}_{L}$.
Each subspace is spanned by the complete orthonormal set
$\{|n\rangle_{a}|k,L-n\rangle,\;n=0,...,L\}$, where
$|k,m\rangle$ denotes the two-mode Fock number state
of the following form $|m+2k-1\rangle_{b}|m\rangle_{c}$. A set of
generalized coherent
states associated with the Hamiltonian (\ref{trilinearHsu11}) has
also been introduced \cite{tribrif}:
\be
|z;k,L\rangle=\mathcal{N}^{-1/2}e^{zaK_{+}}|L\rangle_{a}|k,0\rangle
\; , \label{cohstbrif}
\ee
where $z=-i\kappa^{(2)}t$ and
$|L\rangle_{a}|k,0\rangle$ is the initial reference state. It can
be shown that the system evolves from the reference state into the
coherent state (\ref{cohstbrif}) during the very early stage of the
dynamics with respect to $1/\kappa^{(2)}$. \\
For $b=c$, the $SU(1,1)$ operators read:
\be
K_{+}=\frac{1}{2}b^{\dag 2} \; , K_{-}=\frac{1}{2}b^{2} \; ,
K_{0}=\frac{1}{2}\left(b^{\dag}b+\frac{1}{2}\right) \; ,
\label{su11degalgebra}
\ee
and $k=\frac{1}{4}$ or $\frac{3}{4}$.
The dynamics induced by the corresponding, fully
quantized Hamiltonian, gives rise to degenerate parametric
amplification or second harmonic generation, and has been
studied in detail in Refs.
\cite{degtrizubairy,degtricrouch,degtriKatriel,degtritindle,degtrimostow,degtriyubergou,degtricohen}.

\subsection{Four-wave mixing and the quadrilinear Hamiltonians}

\begin{figure}[h]
\begin{center}
\includegraphics*[width=7cm]{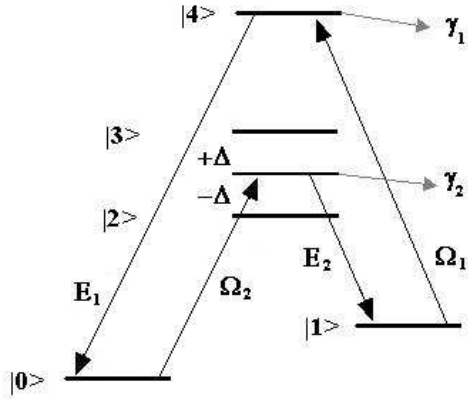}
\end{center}
\caption{Double-$\Lambda$ transition scheme.
$\Omega_{j}$ $(j=1,2)$ represent
the pump fields, $E_{j}$ $(j=1,2)$
represent the signal and the idler
fields, and $|0\rangle$ and $|1\rangle$ are
two metastable ground states.
The scheme allows the cancellation of destructive phase
shifts.}
\label{4wm2Lamb}
\end{figure}

We now move on to discuss four-photon processes and four-wave
mixing, that is the parametric interaction between four photons in
third-order nonlinear media. After introducing the fundamental aspects,
we will focus on the most recent theoretical and experimental
progresses. The naming ``Four-wave mixing'' stands for
many different types of interactions that can all be
described by Hamiltonian terms of the general form (\ref{H3}).
Taking into account also the free Hamiltonian part, we can
rewrite the total Hamiltonians, comprising the free and
the interaction parts, for the two different types
of four-wave mixing as follows, with obvious meaning of
the notations:
\bea
&&H_{Tj}^{4wm} \; = \; H_{0}^{4wm} \, + \, H_{Ij}^{4wm} \; , \;
\quad (j=1,2) \; , \label{HTj4wm} \\
&& \nonumber \\
&&H_{0}^{4wm} \; = \; \sum_{q}\omega_{q}q^{\dag}q \; , \;
\quad (q=a,b,c,d) \; , \label{H04wm} \\
&& \nonumber \\
&&H_{I1}^{4wm} \; = \; \kappa_{1}^{(3)}a^{\dag} b^{\dag} c^{\dag} d+
\kappa_{1}^{(3)*}abcd^{\dag} \; \; , \quad
\omega_{a}+\omega_{b}+\omega_{c}=\omega_{d}
\label{H14wmx} \\
&& \nonumber \\
&&H_{I2}^{4wm} \; = \; \kappa_{2}^{(3)}a^{\dag} b^{\dag} c d+
\kappa_{2}^{(3)*}abc^{\dag} d^{\dag} \; \; , \quad
\omega_{a}+\omega_{b}=\omega_{c}+\omega_{d} \; ,
\label{H24wmx}
\eea
where $a$, $b$, $c$, $d$ are four distinct quantum modes,
and the complex
couplings $\kappa_{j}^{(3)}$ $(j=1,2)$ are proportional to the third order
susceptibilities. The sum-rule conditions on the frequencies are due to
energy conservation, and the corresponding phase matching
conditions have been tacitly assumed.
\begin{figure}[h]
\begin{center}
\includegraphics*[width=9cm]{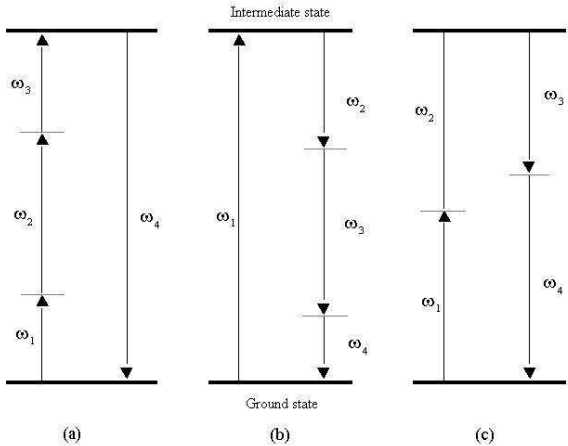}
\end{center}
\caption{Schematic description of quantum transitions for general
nondegenerate four-wave mixing processes: (a) sum-frequency
generation, (b) three-photon down conversion, (c) two-photon down
conversion from two pumping waves.} \label{4wvmixschem}
\end{figure}
In Fig. (\ref{4wvmixschem}) we report some typical schemes for
quantum parametric transitions associated to the
most important four-wave mixing interactions. In a parametric
process, the medium, initially in its ground state, reaches,
due to the interaction with the radiation field, an intermediate,
excited state, and then finally decays back to the ground state.
Each process is characterized by different input and output
frequency combinations, consistent with the physical
matching conditions.
Fig. (\ref{4wvmixschem})
(a) and (b), corresponding to the Hamiltonian (\ref{H14wmx}),
represent, respectively, the sum-frequency process and the
three-photon down conversion process. Fig. (\ref{4wvmixschem}) (c),
corresponding to the Hamiltonian (\ref{H24wmx}), refers to the
interaction event in which two input photons are annihilated with
the consequent creation of the signal and idler photons in the
nonlinear medium. It is evident that the degeneracy of two or more
modes in Eqs. (\ref{H14wmx}) and (\ref{H24wmx}) leads to a large
subclass of models, such as Kerr or cross-Kerr interaction and
degenerate three-photon down conversion process. Moreover, the
single or repeated application of the parametric approximation
(high-intensity coherent regime) for the pump modes can
simplify the Hamiltonians, reducing them to trilinear or bilinear
forms. Four-wave mixing was first proposed by Yuen and Shapiro
for the generation of squeezed light \cite{4wmYuenShap}; they
considered the so-called backward configuration in which both the
two pump fields and the two signal fields are counterpropagating.
Successively, Yurke proposed the generation of squeezed states by
means of four-wave mixing in an optical cavity \cite{4wmYurke}. A
fully quantum-mechanical theory of nondegenerate mixing in an
optical cavity containing a nonlinear medium of two-level atoms
was introduced by Reid and Walls \cite{4wmReidWal}. The first
experimental observations of squeezed light by means of four-wave
mixing are due to Bondurant \textit{et al.} \cite{4wmexpSS1},
Levenson \textit{et al.} \cite{4wmexpSS2}, Slusher \textit{et al.}
\cite{4wmexpSS3}. Since these successful experimental
realizations, great attention has been dedicated to the
study of the quantum statistics of this process: a review and a
bibliographic guide can be found in Ref. \cite{txtPerina}.
One should remark the importance of four-wave mixing as
a tool for nonlinear spectroscopy, due to the enhancement
of the process occurring at the resonances characteristic of
the medium; see for instance \cite{4wmspectrosc,nobelBloemb}. \\
As already seen for three-wave mixing, exact treatments of the
dynamics of the interaction models (\ref{H14wmx}) and
(\ref{H24wmx}) are possible only in a numerical framework.
Analytical results can be obtained in some approximation schemes.
For instance, in the short-time approximation \cite{girigupta},
one can investigate the squeezing power of the four-wave
interaction $H_{I2a}^{4wm}=\kappa (a^{\dag
2}cd+a^{2}c^{\dag}d^{\dag})$, that is, the model (\ref{H24wmx})
with a single pump $(a=b)$ and
$2\omega_{a}=\omega_{c}+\omega_{d}$. Alternatively, the four-wave
interactions (\ref{H14wmx}) and (\ref{H24wmx}) can be linearized
by using a specific form of the generalized Bose operators,
already introduced for trilinear Hamiltonians in the previous
Subsection \cite{34wmKatriel}. In this approach, one can derive
integral equations for the time evolution of the photon number
operators, that are solvable in terms of Jacobian elliptic
functions.  \\
It is important to discuss in some detail recent theoretical and
experimental results related to the question of possible large
enhancement of four-wave mixing. This can be obtained by an
ingenious application of electromagnetically induced transparency
(EIT, see also the previous Section) to suppress the single- and
multi-photon absorption that limits the efficiency of third-order
processes. In the last years, different schemes of four-wave
mixing enhancement have been proposed, based on $\Lambda$ and
double-$\Lambda$ transitions.
\cite{4wmEIT,4wmEIT2,4wmEIT3,4wmEIT4,4wmEIT5}. Here we
review a method due to Johnsson and Fleischauer for the
realization of resonant, forward four-wave mixing
\cite{4wmEIT2,4wmEIT3}. These authors have considered the
double-$\Lambda$ configuration depicted in Fig. (\ref{4wm2Lamb});
it represents a symmetric five-level setup
$(|0\rangle\, ,|1\rangle\, ,|2\rangle\, ,|3\rangle\, ,|4\rangle)$,
which is particularly convenient because in this configuration the
ac-Stark shifts, that reduce the conversion efficiency, are
cancelled.
Here $\Omega_{j}$ $(j=1,2)$ denote the driving fields, with
$\Omega_{1}$ assumed to be in resonance with the $|1\rangle\rightarrow
|4\rangle$ transition, i.e. $\omega_{\Omega_{1}} = \omega_{41}$,
while $\Omega_{2}$ has a detuning $\Delta$, i.e.
$\omega_{\Omega_{2}}=\omega_{30}-\Delta=\omega_{20}+\Delta$.
The $E_{j}$s are the signal and idler fields with $E_{1}$
assumed to be in resonance with the $|4\rangle\rightarrow
|0\rangle$ transition, i.e. $\omega_{E_{1}}=\omega_{40}$, while
$E_{2}$ has a detuning $\Delta$, i.e.
$\omega_{E_{2}}=\omega_{31}-\Delta=\omega_{21}+\Delta$. The finite
detuning $\Delta$ is assumed to be large compared to the Rabi
frequencies in order to ensure the minimization of linear losses
due to single-photon absorption. Moreover, it is assumed that the
pump and the generated fields, propagating in the same direction
$z$, are pairwise in two-photon resonance, and thus, globally,
in four-photon resonance, i.e.
$\omega_{\Omega_{1}}+\omega_{\Omega_{2}}=\omega_{E_{1}}+\omega_{E_{2}}$.
One can thus proceed first to derive an effective
classical interaction Hamiltonian in the adiabatic limit,
and then to quantize it
\cite{4wmEIT2}. The effective non Hermitian interaction
Hamiltonian, in a rotating wave approximation corresponding
to slowly varying amplitudes of the basis
$(|0\rangle\,,|1\rangle\,,|2\rangle\,,|3\rangle\,,|4\rangle)^{T}$,
can be written in the following form:
\be
H_{I}=
-\left(\begin{array}{ccccc}
  0 & 0 & \Omega_{2}^{*} & \Omega_{2}^{*} & E_{1}^{*} \\
  0 & 0 & E_{2}^{*} & -E_{2}^{*} & \Omega_{1}^{*} \\
  \Omega_{2} & E_{2} & -\Delta+i\gamma_{2} & 0 & 0 \\
  \Omega_{2} & -E_{2} & 0 & \Delta+i\gamma_{2} & 0 \\
  E_{1} & \Omega_{1} & 0 & 0 & i\gamma_{1}
\end{array}\right) \; .
\ee
At the input the signal and the idler
fields are assumed to have zero amplitudes, and all atoms are in
the ground state $|0\rangle$. This state results to be an approximate
adiabatic eigenstate of $H_{I}$, which can then be replaced
by the corresponding eigenvalue. Solving the characteristic
equations for the eigenvalues, and expanding the ground-state
one in a power series to the lowest order in $\Delta^{-1}$,
one obtains
\be
H_{I}=\frac{1}{\Delta}\left(\frac{\Omega_{1}^{*}
\Omega_{2}^{*}E_{1}E_{2}+\Omega_{1}
\Omega_{2}E_{1}^{*}E_{2}^{*}}{|\Omega_{1}|^{2}+|E_{1}|^{2}}\right)
\; .
\ee
Here we see EIT at work: as a consequence of
the quantum interference leading to the induced
transparency, the resonant interaction
has no imaginary component left, and therefore
there is no linear loss. The
quantization of $H_{I}$ is achieved by replacing the complex
amplitudes by the corresponding positive and negative operators in
normal ordered form, and by inserting the density of atoms $N$,
the effective cross section of the beams $A$, and by
integrating over the interaction volume. The effective interaction
Hamiltonian then reads \cite{4wmEIT2}
\be
H_{I}=\frac{NA}{\Delta}\int dz
\left(\frac{\Omega_{1}^{\dag}\Omega_{2}^{\dag}E_{1}E_{2}+
\Omega_{1}\Omega_{2}E_{1}^{\dag}E_{2}^{\dag}}{\Omega_{1}^{\dag}\Omega_{1}+
E_{1}^{\dag}E_{1}}\right)
\; .
\label{HFleisch}
\ee
It is possible to verify that there are
four independent constants of motion
$(\Omega_{1}^{\dag}\Omega_{1}+E_{1}^{\dag}E_{1})$,
$(\Omega_{2}^{\dag}\Omega_{2}+E_{2}^{\dag}E_{2})$,
$(\Omega_{1}^{\dag}\Omega_{1}-\Omega_{2}^{\dag}\Omega_{2})$, and
$(\Omega_{1}^{\dag}\Omega_{2}^{\dag}E_{1}E_{2}+
\Omega_{1}\Omega_{2}E_{1}^{\dag}E_{2}^{\dag})$.
By exploiting them under stationary
conditions, the dynamics of the system has been analyzed
numerically up to $10^{3}$ input photons \cite{4wmEIT2}. An
oscillatory exchange between the modes has been observed both for
initial number states and for initial coherent states; moreover,
the statistics turned out to be superPoissonian in most of
the time regimes. The possible
application of this setup for the realization of
a photonic phase gate has also been
studied \cite{4wmEIT2,4wmEIT3}, and will be discussed
in more detail in Section (\ref{section7}) together
with other proposed applications of four-wave mixing
interactions to entanglement generation,
quantum information protocols, and quantum gates
for quantum computation.
Remarkably, experimental
demonstrations of four-wave mixing using the EIT effect have been recently realized in
systems of ultracold atoms \cite{4wmEITexp,4wmEITexp2}. In the experiment described
in Ref. \cite{4wmEITexp} backward four-wave mixing with
EIT is obtained in a double-$\Lambda$
system engineered by four levels of increasing energy,
$|0\rangle, |1\rangle, |2\rangle, |3\rangle$, of ultracold atoms of
$\,^{87}Rb$. A coupling laser of frequency $\omega_c$, tuned to the transition
$|1\rangle \rightarrow  |2\rangle$ between
the intermediate energy states, sets up a quantum interference, and provides EIT for an
anti-Stokes laser of frequency $\omega_{AS}$. The anti-Stokes laser is tuned
to the transition $|2\rangle \rightarrow  |0\rangle$,
and a pump laser of frequency $\omega_p$ is detuned from the $|0\rangle \rightarrow  |3\rangle$
resonance. If the pump and coupling lasers are strong, and the anti-Stokes laser is weak,
the anti-Stokes beam  generates a counterpropagating Stokes beam of frequency $\omega_S$, that satisfies
phase matching, and energy conservation: $\omega_S = \omega_p + \omega_c - \omega_{AS}$.
The four-wave mixing process is thus realized, with the anti-Stokes laser producing
photons with a nearly $100\%$ transmission at line center of the resonant transition
$|2\rangle \rightarrow  |0\rangle$; this fact can be of relevance
for transmission of quantum information. A similar scheme is realized in Ref. \cite{4wmEITexp2}.

The time-independent quantum mechanics of four-wave
mixing Hamiltonians is of particular interest, because
it is possible in certain cases to obtain analytical results
for their spectra. In the case of various, fully
quantized multiwave-mixing models, the so-called Bethe
ansatz is commonly used \cite{AndreevJPA}.
Such an approach is based on the assumption that the
form of the energy eigenstates can be expressed
in terms of several Bethe parameters.
However, the equations for these parameters are very complicated
even for numerical solution. An alternative, algebraic method
has been recently proposed and applied to obtain explicit
analytical expressions both for the eigenenergies and the eigenstates
of the total Hamiltonians $H_{Tj}^{4wm}$ Eq. (\ref{HTj4wm})
\cite{WuYangOL}. Here we briefly describe
this method for the case of the interaction Hamiltonian
$H_{I1}^{4wm}$, with $\kappa^{(3)}_{1}$ taken to be
real for simplicity. The same procedure can be applied
along similar lines to the model with interaction $H_{I2}^{4wm}$.
Concerning the total Hamiltonian $H_{T1}^{4wm}$,
the following operators are integrals of
motion: $N_{ad}=n_{a}+n_{d}$, $N_{bd}=n_{b}+n_{d}$,
$N_{cd}=n_{c}+n_{d}$,
$H_{0}^{4wm}=\omega_{a}N_{ad}+\omega_{b}N_{bd}+\omega_{c}N_{cd}$,
and $H_{I1}^{4wm}$. Therefore, these operators share a complete
set of eigenstates
$|\Psi_{N_{ad},N_{bd},N_{cd},\lambda}\rangle$ satisfying the
relations:
\bea
&&H_{T1}^{4wm}|\Psi_{N_{ad},N_{bd},N_{cd},\lambda}\rangle \; = \;
E_{N_{ad},N_{bd},N_{cd},\lambda}|\Psi_{N_{ad},N_{bd},N_{cd},\lambda}\rangle
\; , \nonumber \\
&& \nonumber \\
&&H^{4wm}_{I1}|\Psi_{N_{ad},N_{bd},N_{cd},\lambda}\rangle \; = \;
\lambda|\Psi_{N_{ad},N_{bd},N_{cd},\lambda}\rangle \; , \nonumber \\
&& \nonumber \\
&&E_{N_{ad},N_{bd},N_{cd},\lambda} \; = \;
\omega_{a}N_{ad}+\omega_{b}N_{bd}+\omega_{c}N_{cd}+\kappa_{1}^{(3)}\lambda
\; , \nonumber \\
&& \nonumber \\
&&N_{ad},N_{bd},N_{cd}=0,1,2,... \; ,
\label{SpectrumI14wm}
\eea
where, without danger of confusion, we
have used the same symbols for the operators and for the
associated quantum numbers. The eigenstates
$|\Psi_{N_{ad},N_{bd},N_{cd},\lambda}\rangle$ can be expressed in
terms of an operatorial function of the mode creation operators
applied to the vacuum, in the form \cite{WuYangXiao}
\bea
&&|\Psi_{N_{ad},N_{bd},N_{cd},\lambda}\rangle \; = \;
S(a^{\dag},b^{\dag},c^{\dag},d^{\dag})|0,0,0,0\rangle
\; , \nonumber \\
&& \nonumber \\
&&S(a^{\dag},b^{\dag},c^{\dag},d^{\dag}) \; = \;
\sum_{j=0}^{M}\frac{\alpha_{j}}{j!}(a^{\dag})^{N_{ad}-j}
(b^{\dag})^{N_{bd}-j}(c^{\dag})^{N_{cd}-j}(d^{\dag})^{j} \; , \;
\nonumber \\ && \nonumber \\
&&M=\min\{N_{ad},N_{bd},N_{cd}\} \; .
\label{WuYangeig}
\eea
As
$H_{I1}^{4wm}|\Psi_{N_{ad},N_{bd},N_{cd},\lambda}\rangle
=[H_{I1}^{4wm},S]|0,0,0,0\rangle$ it can be shown, using
expression (\ref{H14wmx}), that $S$ satisfies the operatorial
differential equation
\be
\left(d^{\dag}\frac{\partial^{3}}{\partial
a^{\dag}b^{\dag}c^{\dag}}+a^{\dag}b^{\dag}c^{\dag}\frac{\partial}{\partial
d^{\dag}}\right)S \; = \; \lambda S \; .
\label{WuYangS}
\ee
Relations (\ref{WuYangeig}) and (\ref{WuYangS}) lead to the
following recursive equations
\be \alpha_{j+1} \; = \;
\lambda\alpha_{j}-p_{j-1}\alpha_{j-1} \; , \; \quad 0\leq j\leq M
\; ,
\label{WuYangiter}
\ee
where $\alpha_{-1}=\alpha_{M+1}=0$ and
$p_{j}=-(j+1)j[(j-1)(j+1-N_{ad}-N_{bd}-N_{cd})
+(N_{ad}N_{bd}+N_{bd}N_{cd}+N_{ad}N_{cd}-N_{ad}-N_{bd}
-N_{cd}+1)]+(j+1)N_{ad}N_{bd}N_{cd}$. Equation (\ref{WuYangiter})
implies
\be
\alpha_{j} \; = \; \alpha_{0}\det A^{(j)}(\lambda) \;
, \; \quad j=0,1,2,...,M \; ,
\ee
where $\det A^{(j)}(\lambda)$ is
defined as follows: $\det A^{(0)}(\lambda)=1$, $\det
A^{(1)}(\lambda)=\lambda$, while for $k\geq 2$ it is the
determinant of the $k\times k$ matrix $A^{(k)}(\lambda)$ with
elements
$A_{ij}^{(k)}(\lambda)=\lambda\delta_{i,j}-\delta_{i+1,j}-p_{j}\delta_{i,j+1}$
$(i,j=0,1,...,k-1)$. The energy eigenvalue $\lambda$ of the
interaction part of the Hamiltonian is determined by finding the
roots of the polynomial equation
\be
\det A^{(M+1)}(\lambda)\; =
\; 0 \; , \; \quad M = \min\{N_{ad},N_{bd},N_{cd}\} \; .
\ee
In
conclusion, one obtains an analytical expression for the energy
spectrum (\ref{SpectrumI14wm}) and for the energy eigenstates in
terms of the parameter $\lambda$ \cite{WuYangOL}
\bea
&&|\Psi_{N_{ad},N_{bd},N_{cd},\lambda}\rangle \; = \;
\sum_{j=0}^{\min\{N_{ad},N_{bd},N_{cd}\}}
c_{j}|N_{ad}-j,N_{bd}-j,N_{cd}-j,j\rangle \; , \\
&& \nonumber \\
&&c_{j} \; = \;
c_{0}\left[\frac{(N_{ad}
-j)!(N_{bd}-j)!(N_{cd}-j)!}{j!N_{ad}!N_{bd}!N_{cd}!}\right]^{1/2}
\det A^{(j)}(\lambda) \; ,
\eea
where $c_{0}$ is a normalization
factor and $N_{ad},N_{bd},N_{cd}=M,M+1,...$
$(M=\min\{N_{ad},N_{bd},N_{cd}\})$.

\subsection{Two-photon squeezed states by three- and four-wave
mixing}

Although squeezed states are the simplest
example of nonclassical multiphoton states of
light, and have thus been extensively
investigated both theoretically and experimentally in the
literature, here, for completeness, we shall rapidly
recall their main properties, and cite some applications
and experimental realizations. Extensive treatments on the
subject can be found in works
\cite{Yuen,naturewalls,cavschum,YamHaus,WallsMilburntx} and references
therein.

Squeezed states can be generated either by considering trilinear
interactions with one of the modes in a classical configuration
(e.g. intense pump laser $E_{p}$), or by considering quadrilinear
(four-wave mixing) interactions with two modes in a classical
configuration leading to a classical amplitude $E_{p}^{2}$.
In both cases, the Hamiltonian reduces to the
quadratic form
\be
H_{I}^{2ph} \; = \; \eta^{*}(t) a_{1}a_{2} +
\eta(t) a_{1}^{\dag}a_{2}^{\dag} \; ,
\label{nondegPA}
\ee
where, in the trilinear instance the complex
parameter $\eta$ is proportional to the $\chi^{(2)}$ nonlinearity
and to the coherent pump: $\eta\propto\chi^{(2)}E_{p}$, while,
in the quadrilinear case, $\eta\propto\chi^{(3)}E_{p}^{2}$.
This approximated Hamiltonian, thoroughly studied in the literature
\cite{QEDTucker,louisell,louisell2,mollowglauber,tuckerwalls2,bonifacio,vonforester,lu,agarmetha,persico},
describes the nondegenerate
parametric amplifier, whose temporal dynamics
generates two-mode squeezed states, which reduce to
single-mode squeezed states in the degenerate case $a_{1}=a_{2}\equiv a$.
Many experimental schemes for generating
squeezed states of light have been proposed, such as the resonance
fluorescence \cite{SSfluoresc}, the use of free-electron laser
\cite{SSfrelaser}, the harmonic generation \cite{SSharmgen},
two-photon and multiphoton absorption
\cite{SSphabsoramp,SSphabsorption}, four-wave mixing
\cite{4wmYuenShap} and parametric amplification
\cite{SSphabsoramp,SSparampl}.

The first successful experiments on the generation
and detection of squeezed states were realized in the 1980s.
In 1985, Slusher
\textit{et al.} reported the observation of quadrature squeezing
of an optical field by degenerate four-wave mixing in an optical
cavity filled with $Na$ atoms \cite{slusher}. In 1986, Shelby
\textit{et al.} obtained squeezing in an optical fiber via the Kerr
effect \cite{shelby}. In the same year, in a crucial experiment,
Wu \textit{et al.} succeeded in realizing a very large squeezing in parametric
down conversion, up to more than $50$ \% squeezing in a below-threshold
optical parametric oscillator (OPO) \cite{wu}.
Successively, a number of experiments with analogous
results were performed by using both second order \cite{kurz,kim}
and third order nonlinearities \cite{rosenbluh,bergman}. A
squeezing ratio of about $70$ \% can be currently obtained
in silica fibers \cite{bergman2}, and in OPOs \cite{polzik}.
It should be noted that noise reduction with respect
to classical light can be routinely observed not only in
OPOs, i.e. in a continuous-wave
oscillator configuration \cite{heidmann}, but also in
optical parametric amplifiers (OPAs), i.e. in a pulsed amplifier
configuration \cite{aytur}. Number-phase squeezing has been
generated as well in diode-laser based devices \cite{atomizzailcinesino}.
Finally, for a comprehensive review of nonlinear quantum optics
applied to the control and reduction of quantum noise and the production
of squeezed states in artificially phase-matched and quasi phase-matched
materials, see Ref. \cite{squeezingexperiments}. \\
Historically, squeezed states were originated from the analysis,
in the degenerate instance, of unitary and linear Bogoliubov
transformations and from the investigation of the possible ways to
generate minimum uncertainty states of the radiation field more
general than coherent states. The so-called two-photon coherent
states of Yuen \cite{Yuen} are based on the Bogoliubov linear
transformation $b=\mu a+\nu a^{\dag}$, with $\mu$ and $\nu$
complex parameters that must satisfy the relation $
|\mu|^{2}-|\nu|^{2}=1$ for the transformation to be canonical. The
transformation $b(a,a^{\dag})$ can also be obtained by action of a
unitary operator $S$ on the mode operators:
$b(a,a^{\dag})=SaS^{\dag}$. The transformed operators $b$ and
$b^{\dag}$ can be interpreted as "quasi-photon" annihilation and
creation operators, and determine a quasi-photon number operator
$n_{g}=b^{\dag}b$ with integer positive eigenvalues and a
quasi-photon ground state (or squeezed vacuum) $|0_{g}\rangle$
defined by the relation $b|0_{g}\rangle = 0$. The quasi-photon
number states $|m_{g}\rangle \doteq \frac{b^{\dag
m}}{\sqrt{m!}}|0_{g} \rangle \equiv S|m\rangle$ form an
orthonormal basis. The two-photon coherent states (TCS)
$|\beta\rangle_{g}=|\beta;\mu,\nu\rangle$ are defined as the
eigenstates of $b$ with complex eigenvalue $\beta$, such that
\be
b|\beta\rangle_{g} \; = \; \beta|\beta\rangle_{g} \; ,
\label{TCS}
\ee
and, of course, they reduce to the ordinary coherent states
for $\nu=0$. In analogy with the case of the standard one-photon
coherent states, the TCS can be expressed as well in terms of the
action of a displacement operator depending on the transformed
mode variables $(b,b^{\dag})$, i.e.
\be
|\beta\rangle_{g} =
e^{\beta b^{\dag}-\beta^{*} b}|0_{g}\rangle \equiv
S(S^{\dag}e^{\beta b^{\dag}-\beta^{*} b}S)|0 \rangle \equiv
SD(\beta)|0 \rangle \; ,
\ee
where $D(\beta) = e^{\beta
a^{\dag}-\beta^{*} a}$, is the Glauber displacement operator in
terms of the original mode variables $(a,a^{\dag})$. The TCS obey
non-orthogonality and over-completeness relations similar to those
holding for coherent states (\ref{cohstprop}). The condition of
canonicity leads to the standard parametrization $\mu=\cosh r \; ,
\; \nu=e^{i\phi}\sinh r$, and the unitary squeezing operator $S$
can be written in terms of $a$ and $a^{\dag}$ as
$S(\varepsilon)=e^{-\frac{1}{2}\varepsilon a^{\dag
2}+\frac{1}{2}\varepsilon^{*} a^{2}}$ with $\varepsilon=r
e^{i\phi}$. Thus, finally, the TCS can be written as
$|\beta\rangle_{g}=S(\varepsilon)D(\beta)|0\rangle$. Clearly, it
is legitimate to consider switching the order of application of
the two unitary operators $D$ and $S$. Doing so, provides an
alternative definition that yields in principle a different class
of states, the so-called two-photon squeezed states, thus defined
as $|\alpha , \varepsilon \rangle = D(\alpha) S(\varepsilon) |0
\rangle$, where $D(\alpha)$ is the Glauber displacement operator
with {\it generic} complex coherent amplitude $\alpha$. However,
the two classes of states coincide as soon as one simply lets
$\alpha=\mu^{*}\beta-\nu\beta^{*}$. With this identification, in
the following, without ambiguity, we will always refer to both
classes of states as two-photon squeezed states. According to
whether $\alpha$ is null or finite, we have, respectively,
squeezed vacuum or squeezed coherent states.

The uncertainty properties of two-photon squeezed states are of
particular interest because such states allow noise reduction
below the standard quantum limit. In fact, in a squeezed state,
the uncertainty on the generalized quadrature $X_{\lambda}$ reads
\bea
\langle \Delta X_{\lambda}^{2}\rangle
&=&\frac{1}{2}\{|\mu|^{2}+|\nu|^{2}
-2Re[e^{2i\lambda}\mu\nu^{*}]\} \nonumber \\
&& \nonumber \\
&=& \frac{1}{2}\{\cosh 2r-\sinh 2r \cos(\phi-2\lambda)\} \; ,
\eea
leading, for a pair of canonically conjugated quadrature variables
$(X_{\lambda},X_{\lambda+\frac{\pi}{2}})$, to the following form
of the Heisenberg uncertainty relation:
\be
\langle\Delta
X_{\lambda}^{2}\rangle\langle\Delta
X_{\lambda+\frac{\pi}{2}}^{2}\rangle=
\frac{1}{4}\{\cosh^{2}2r-\sinh^{2}2r\cos^{2}(\phi-2\lambda)\} \; .
\label{SSuncert}
\ee
Note that, for $\phi = 2\lambda + k\pi$, one
has $\langle\Delta X_{\lambda}^{2}\rangle=\frac{1}{2}e^{\mp 2r}$
and $\langle\Delta
X_{\lambda+\frac{\pi}{2}}^{2}\rangle=\frac{1}{2}e^{\pm 2r}$.
Therefore, depending on the sign in front of the squeezing
parameter $\; r$ in the exponentials, the quantum noise on one of
the quadrature is lowered below the standard quantum limit $1/2$,
while increasing of the same amount the uncertainty on the other
quadrature, in such a way that the uncertainty product stays fixed
at its minimum Heisenberg value. This is the essence of the
``Quadrature Squeezing'' phenomenon. Thus, the states $\{
|\beta\rangle_{g} \} $ define a class of  minimum uncertainty
states \cite{StolerMUS} more general than the coherent states.
Note that, remarkably, the squeezed states associated to the
degenerate parametric amplifier (both with classical and with
quantum optical pump) exhibit also generalized
\textit{higher-order squeezing} \cite{HongMandelSqueez}. In the
most common meaning, a state shows $2N$-th order squeezing if
\be
\langle [\Delta X_{\lambda}]^{2N}\rangle <
\left(\frac{1}{2}\right)^{N}(2N-1)!! \; .
\label{highordsqueez}
\ee
This definition follows from a comparison with the Gaussian
coherent state, with the odd squeezing degrees all vanishing due
to the normal ordering.

\begin{figure}
\begin{center}
\includegraphics*[width=7.5cm]{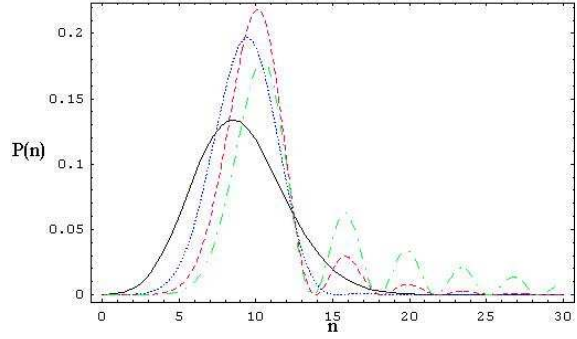}
\end{center}
\caption{Photon number distribution for a squeezed state
$|\alpha,r\rangle$, with $\alpha=3$ for several values of $r$: (a)
$r=0$, i.e. coherent state (full line); (b) $r=0.5$ (dotted line);
(c) $r=1$ (dashed line); (d) $r=1.5$ (dash-dotted line).}
\label{PndTCS}
\end{figure}

Photon statistics characterizes squeezed states, again at variance with
the coherent states, as the simplest instance of nonclassical states.
Namely, the photon number distribution of a squeezed state is
\bea
&& P(n) = |\langle n|\alpha,\varepsilon\rangle|^{2}  \nonumber \\
&& \nonumber \\
&&
=\left|\frac{e^{-\frac{1}{2}(|\alpha|^{2}+\alpha^{*2}e^{i\phi}\tanh
r)}}{\sqrt{n!\cosh r}}\left(\frac{1}{2}e^{i\phi}\tanh
r\right)^{n/2} H_{n}\left[\frac{\alpha+\alpha^{*}e^{i\phi}\tanh
r}{\sqrt{2e^{i\phi}\tanh r}}\right]\right|^{2} ,
\eea
where
$H_{n}[x]$ denote the Hermite polynomial of order $n$. In Fig.
(\ref{PndTCS}) $P(n)$ is plotted, with $\alpha=3$ and $\phi=0$,
for several values of the squeezing parameter $r$. Firstly, we see
that, contrary to the case of coherent states, now $P(n)$ is not a
Poisson distribution any more. Depending on the choice of the
parameters, the photon number distribution can be super- or
sub-Poissonian. In particular, the squeezed vacuum always exhibits
super-Poissonian statistics, while the squeezed coherent states at
fixed squeezing $r$ and sufficiently large coherent amplitude
$\alpha$ can exhibit sub-Poissonian statistics. Furthermore, for
increasing $r$, the distribution exhibits oscillations, see Fig.
(\ref{PndTCS}), which can be interpreted as interference in phase
space \cite{schleichnature,schleich}.
The mean photon number in the state $|\alpha,\varepsilon\rangle$
is made of two contributions: $\langle
n\rangle=|\alpha|^{2}+|\nu|^{2}$, that is the sum of the average
numbers of coherent and squeezed photons. The variance reads
\be
\langle\Delta n^{2}\rangle=|\alpha \cosh
r-\alpha^{*}e^{i\phi}\sinh r|^{2}+2\cosh^{2}r\sinh^{2}r \; .
\label{nvarSS}
\ee
By making use of the Hermitian phase operator
$\Phi$ introduced by Pegg and Barnett \cite{PeggBarn}, relation
(\ref{nvarSS}) allows to define number-phase squeezed
states. These are states of minimum Heisenberg number-phase
uncertainty
$\langle\Delta
n^{2}\rangle\langle\Delta\Phi^{2}\rangle=\frac{1}{4}$, just like
coherent states \cite{BarnPegg}, but one of the dispersions can
get smaller than that of a coherent state. For particular choices of the
phase $\theta$ of the complex field amplitude
$\alpha=|\alpha|e^{i\theta}$ in Eq. (\ref{nvarSS}), the
corresponding state is called phase-squeezed state if the
squeezing is $\pi/2$ out of phase with the complex field amplitude
$(\phi=2\theta+\pi)$, or amplitude-squeezed state if the squeezing
is in phase $(\phi=2\theta)$. The relation between quadrature and
number squeezing is thoroughly explored in Ref. \cite{Davidovich}.

A further signature of nonclassicality of squeezed states is that,
for them, the Glauber $P$-representation cannot be defined.
However, other quasi-probability distribution functions exist
for squeezed states, in particular the Husimi $Q$-function and
the Wigner function. The latter, being the squeezed states
Gaussian, is positive-defined.

\begin{figure}[h]
\begin{center}
\includegraphics*[width=15cm]{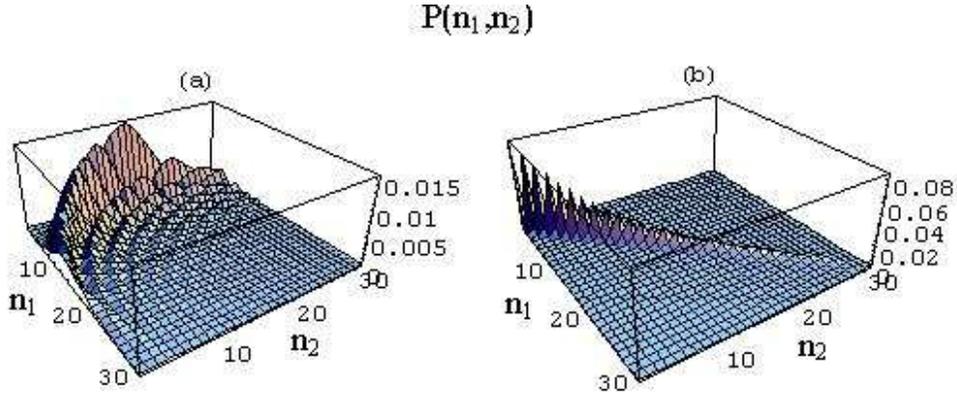}
\end{center}
\caption{Photon number distributions for the two-mode squeezed
state $|\alpha_{1},\alpha_{2};r\rangle$ with $r=1.5$ and (a)
$\alpha_{1}=1$, $\alpha_{2}=3$, (b) $\alpha_{1}=-\alpha_{2}=2$.}
\label{Pnd2mSS}
\end{figure}

The general quadratic Hamiltonian associated to the general
Bogoliubov transformation, including a $c$-number term $\xi$: $b =
\mu a + \nu a^{\dag} + \xi$ takes the form \cite{Yuen}
\be
H_{q}=f_{1}\left(a^{\dag}a+\frac{1}{2}\right)+f_{2}^{*}a^{2}+f_{2}a^{\dag
2}+f_{3}^{*}a+f_{3}a^{\dag} \; ,
\label{quadH}
\ee
where $f_{i}$
are $c$-numbers, possibly time-dependent. The real coefficient
$f_{1}$ is the free radiation energy of the mode $a$; the complex
factor $f_{2}$ is the two-photon interaction energy; and, finally,
$f_{3}$ is a linear forcing field (pump) associated to one-photon
processes. The condition $f_{1}>2|f_{2}|$ assures that the
Hamiltonian (\ref{quadH}) is physical (positive-definite and
bounded from below). It is moreover diagonalized to the form
$H_{q}=\Omega(f_{1},f_{2}) b^{\dag}b+C(f_{1},f_{2},f_{3})$ by the
transformation $b=\mu(f_{1},f_{2}) a+\nu(f_{1},f_{2})
a^{\dag}+c(f_{1},f_{2},f_{3})$. A typical physical system
associated to the quadratic form (\ref{quadH}) is the degenerate
parametric amplifier, which can be conveniently described by the
Hamiltonian
\be
H^{2ph}_{deg}=\omega a^{\dag}a-[\eta^{*}
e^{2i\omega t}a^{2}+\eta e^{-2i\omega t}a^{\dag 2}] \; .
\label{H2DCdeg} \ee The general solution of the Heisenberg
equation of motion is then \be a(t)=\cosh(|\eta|t)e^{-i\omega
t}a+i\frac{\eta}{|\eta|}\sinh(|\eta|t)e^{-i\omega t}a^{\dag} \; .
\label{degPAHeis}
\ee
Exploiting relation (\ref{degPAHeis}), it
can be easily shown that squeezed light generated from an initial
vacuum state exhibits photon bunching, while for an initial
coherent state the output light may be antibunched. This last
effect marks a further characterization of the nonclassicality of
squeezed states \cite{Stolerantib}. The methods introduced above
can be generalized to define nondegenerate, multimode squeezed
states. These states are obtained by successive applications on
the vacuum state of the $n$-mode squeezing operator
$S(\underline{\varepsilon})=
e^{-\frac{1}{2}\varepsilon_{ij}a^{\dag}_{i}a^{\dag}_{j}
+\frac{1}{2}\varepsilon_{ij}^{*}a_{i}a_{j}}$
and of the generalized displacement operator
$D(\underline{\alpha})=e^{\alpha_{i}a^{\dag}_{i}-\alpha_{i}^{*}a_{i}}$,
where the Einstein summation convention on the repeated indices
has been adopted, $\underline{\alpha} \equiv \{ \alpha_{i} \}$,
and $\underline{\varepsilon} \equiv \{ \varepsilon_{ij} \}$ ($i,j
= 1, ..., n$). We can then conveniently denote them as
$|\underline{\alpha},\underline{\varepsilon}\rangle=
D(\underline{\alpha})S(\underline{\varepsilon})|0\rangle$. We
approach in some detail the important two-mode case, by
considering two correlated modes $a_{1}$ and $a_{2}$ interacting
according to bilinear Hamiltonians of the general form
(\ref{nondegPA}). A comprehensive framework for two-photon quantum
optics was introduced by Caves and Schumaker in a beautiful series
of papers \cite{cavschum}. They defined two-mode, linear canonical
transformations as
\be
b_{1}=\mu a_{1} + \nu a_{2}^{\dagger} \; ,
\quad \quad b_{2} =\mu a_{2} + \nu a_{1}^{\dagger} \; ,
\ee
with
the same parametrization of the complex coefficients as in the
single-mode case. The transformed quasi-photon modes
$(b_{1},b_{2})$ can also be obtained from the original mode
variables by acting with the unitary operator
\bea
&&S_{12}(\zeta) = e^{{\zeta}^{*} a_{1} a_{2} - \zeta a_{1}^{\dagger}
a_{2}^{\dagger}} \, ,
\quad \zeta = r e^{i\phi} \; , \nonumber \\
&& \nonumber \\
&& S_{12}\,a_{i}\,S_{12}^{\dag} =a_{i}\cosh r +
a_{j}^{\dag}e^{i\phi}\sinh r \; , \quad i,j=1,2 \; , i \neq j \; ,
\eea
and two-mode squeezed states are defined as
$|\alpha_{1},\alpha_{2};\zeta\rangle=
D(\alpha_{1})D(\alpha_{2})S_{12}(\zeta)|0\rangle$.
Equivalently, these states can be labelled by the complex
eigenvalues of $b_{i}$: $b_{i}|\alpha_{1},\alpha_{2};\zeta\rangle
=\beta_{i}|\alpha_{1},\alpha_{2};\zeta\rangle$ $(i=1,2)$, where
$\beta_{i}=\alpha_{i}\cosh r+\alpha_{j}^{*}e^{i\phi}\sinh r$,
$(i,j=1,2 \; , \; i \neq j)$. The existence of nontrivial
correlations between the modes is elucidated by computing the
expectation values:
\bea
&&\langle a_{i}\rangle=\alpha_{i}\; , \nonumber \\
&& \nonumber \\
&&\langle
a_{i}^{\dag}a_{j}\rangle=\alpha_{i}^{*}\alpha_{j}+\delta_{ij}\sinh^{2}r \; , \nonumber \\
&& \nonumber \\
&&\langle
a_{i}a_{j}\rangle=\alpha_{i}\alpha_{j}-(1-\delta_{ij})e^{i\phi}\sinh
r\cosh r \; .
\eea
Among the interesting statistical properties of
these states, an important role is played by the two-mode
photon-number distribution, defined as the joint probability to
find $n_{1}$ photons in the mode $a_{1}$ and $n_{2}$ photons in
the mode $a_{2}$ \cite{schleich}:
\bea
&&P(n_{1},n_{2})=|\langle
n_{1},n_{2}|\alpha_{1},\alpha_{2};r\rangle|^{2} \nonumber \\
&& \nonumber \\
&&=\left|\frac{(-\tanh r)^{p}}{\cosh
r}\left(\frac{p!}{q!}\right)^{1/2}\mu_{1}^{n_{1}-p}\mu_{2}^{n_{2}-p}
L_{p}^{(q-p)}\left(\frac{\mu_{1}\mu_{2}}{\tanh r}\right)
e^{-\frac{1}{2}(\alpha_{1}^{*}\mu_{1}+\alpha_{2}^{*}\mu_{2})}\right|^{2}
,
\eea
where $\mu_{i}=\alpha_{i}+\alpha_{j}^{*}\tanh r$ with
$(i,j=1,2 \; , \; i \neq j)$, $p=min(n_{1},n_{2})$,
$q=max(n_{1},n_{2})$ and $L_{l}^{(k)}(x)$ are generalized Laguerre
polynomials. For a two-mode squeezed vacuum
($\alpha_{1}=\alpha_{2}=0$), the joint probability has only
diagonal elements: $P(n,n)= (\tanh r)^{2n}/\cosh^{2}r$. More in
general, for $\alpha_{1}=\alpha_{2}\neq 0$ the joint probability
is obviously symmetric. In Figs. \ref{Pnd2mSS} (a)-(b)
$P(n_{1},n_{2})$ is plotted at fixed squeezing $r=1.5$ and for two
choices of the coherent amplitudes. Fig. \ref{Pnd2mSS} (a) shows
asymmetric oscillations with the peaks being shifted due to the
different coherent amplitudes. Fig. \ref{Pnd2mSS} (b) is instead
characterized by the absence of oscillations.
Two-mode squeezed states are generated in the dynamical evolution
of a nondegenerate parametric amplifier, described by the total
Hamiltonian
\be
H^{2ph}_{T} = H^{2ph}_{0} + H^{2ph}_{I} = \omega_1
a_{1}^{\dag}a_{1}+\omega_2
a_{2}^{\dag}a_{2}+i\chi(a_{1}^{\dag}a_{2}^{\dag}e^{-2i\omega t}
-a_{1}a_{2}e^{2i\omega t}) \; ,
\ee
where the coupling constant
$\chi$ is assumed real without loss of generality. The solutions
of the Heisenberg equations of motion in the interaction picture
are
\be
a_{1}(t)=a_{1}\cosh \chi t +a_{2}^{\dag}\sinh \chi t \; ,
\quad a_{2}(t)=a_{2}\cosh \chi t +a_{1}^{\dag}\sinh \chi t \; ,
\ee
so that, as expected, the conservation law
$n_{1}(t)-n_{2}(t)=n_{1}(0)-n_{2}(0)$ holds. Choosing an initial
two-mode coherent state $|\alpha_{1}\rangle |\alpha_{2}\rangle$,
the mean photon number in each mode evolves as
\be \langle
n_{i}(t)\rangle \, = \, |\alpha_{i}\cosh \chi t
+\alpha_{j}^{*}\sinh \chi t |^{2} \, + \, \sinh^{2}\chi t \; ,
\quad i,j=1,2 \; , i \neq j \; ,
\ee
with the last term
representing the amplification of vacuum fluctuations. A further,
important signature of nonclassicality stems from the fact that,
if the system is initially in the vacuum state, the intensity
cross-correlation, defined by $\langle n_{1}(t)n_{2}(t)\rangle$,
maximally violates the classical Cauchy-Schwartz inequality
$(\langle n_{1} n_{2} \rangle )^{2} \leq \langle a^{\dag 2}_{1}
a_{1}^{2} \rangle \langle a^{\dag 2}_{2} a_{2}^{2} \rangle$
\cite{WallsMilburntx}.

\subsection{An interesting case of
four-wave mixing: degenerate three-photon down conversion}

\begin{figure}[h]
\begin{center}
\includegraphics*[width=7cm]{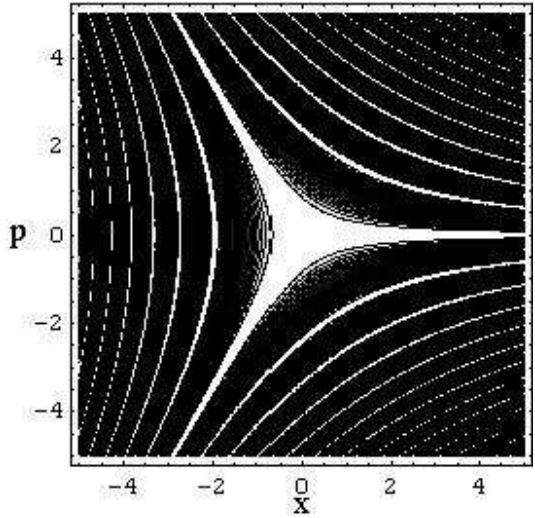}
\end{center}
\caption{Contour plot of the interference pattern $W_{int}(x,p)$.}
\label{WigBanaKnig}
\end{figure}

The degenerate three-photon down conversion
$3\omega\rightarrow\omega+\omega+\omega$ is the next natural step
to be considered beyond the two-photon down conversion processes.
The interaction Hamiltonian of degenerate three-photon down
conversion reads
\be
H_{I}^{3dc} = \kappa a^{\dag 3}b+\kappa^{*}
a^{3}b^{\dag} \; ,
\label{HI3phDC}
\ee where the modes $a$ and $b$
are the down-converted signal and the quantized pump,
respectively. The interaction can be realized in a $\chi^{(3)}$
medium and the Hamiltonian (\ref{HI3phDC}) can be derived by Eq.
(\ref{H3}). The process has been studied theoretically for running
waves (OPA)
\cite{FisherNieto,BraunMcLac,BraunCav,Elyutin,DrobnyJex1,DrobnyJex2,TanasGantsog,BanaKnig}
and for optical cavities (OPO) \cite{Felbinger}, and in general
the time evolution generated by Hamiltonian (\ref{HI3phDC}) cannot
be expressed in closed, analytical form. Analogously to the
two-photon case, a physical simplification is obtained by applying
the parametric approximation to the pump mode (considered coherent
and intense) $b\rightarrow\beta$ ($\beta$ $c$-number). In this
approximation $H_{I}^{3dc}$ reduces to
\be
H_{I}^{3dcp} \, = \,
\xi a^{\dag 3} \, + \, \xi^{*}a^{3} \; ,
\label{H3phDCparapp}
\ee
where the time-independent complex parameter $\xi$ is proportional
to the pump amplitude $\beta$. Since in both cases disentangling
formulas for the evolution operators do not exist, it is then
necessary to resort to numerical methods in order to study the
dynamics of the process. Moreover, the evolution operator for the
Hamiltonian (\ref{H3phDCparapp}) suffers of divergences
\cite{FisherNieto}, that need to be treated by specific summation
techniques \cite{BraunMcLac}. In Ref. \cite{Elyutin} Elyutin and
Klyshko showed that the average output energy $\langle n\rangle$
of the parametric amplifier $H_{I}^{3dcp}$ diverges infinity after
a finite time lapse $t_{0}$. The dynamics for $n$ can be cast in
the form:
\be
\ddot{n} \, = \, 6n^{2}+6n+4 \; ,
\label{elyuheisn}
\ee
and, after some manipulations, an exact expression for the
average value $N(\tau) \equiv \langle n\rangle(\tau)$ (with
$\tau=\kappa t$ rescaled, dimensionless time) can be expressed in
the form of a Taylor series. As an example, if we choose the
vacuum as the initial state, then
\be
N(\tau)=2\tau^{2}+4\tau^{4}+11.2\tau^{6}+34.8\tau^{8} + ... \; .
\label{elyuexplst}
\ee
We see that the explosion time $\tau_{0}$
is clearly finite; it can be defined as the convergence radius of
the series (\ref{elyuexplst}) obtained by an extrapolation of the
coefficients in Eq. (\ref{elyuexplst}) that yields $\tau_{0}\simeq
0.53$ \cite{Elyutin}. This is at striking variance with the case
of two-photon down conversion, where such a divergence occurs only
in the limit of infinite time. The existence of a finite explosion
time shows a pathology of the perturbative techniques, that
consequently need to be treated with some care.  This and related
problems will be discussed further in the next Section, but here
we first move on to illustrate some interesting properties of the
Wigner function for the states generated by the fully quantized
Hamiltonian (\ref{HI3phDC}). Concerning OPO dynamics, Felbinger
\textit{et al.} \cite{Felbinger} studied, using quantum trajectory
simulations, three-photon down conversion in an optical cavity
resonant at the frequencies $\omega$ and $3\omega$. The authors
included homodyne detection and quantum state reconstruction
setups, and considered, besides the three-photon down conversion
interactions, the simultaneous presence of Kerr and cross-Kerr
terms of the type $a^{\dag 2}a^{2} \; , b^{\dag 2}b^{2} \; ,
a^{\dag}ab^{\dag}b$. They were able to determine the Wigner
function for the intracavity and extracavity fields, and, in the
case of the intracavity mode, showed that it is non Gaussian,
non-negative, and exhibits a threefold symmetry with respect to
three directions in phase space (``star states''). Concerning OPA
dynamics, Banaszek and Knight \cite{BanaKnig} studied the
evolution of the signal mode under the action of the Hamiltonian
(\ref{HI3phDC}) for the initial state
$|0\rangle_{a}|\beta\rangle_{b}$ and calculated numerically the
Wigner function for the reduced density operator
$\rho_{b}(t)=
Tr_{b}[e^{-itH_{I}^{3dc}}|0\rangle_{a}|\beta\rangle_{b}\,_{b}
\langle\beta|_{a}\langle
0|e^{itH_{I}^{3dc}}]$. The Wigner function exhibits a deep
nonclassical behavior, again possessing three arms with a
star-symmetry in phase space, and a typical interference pattern
in the regions delimited by the three arms. The authors developed
a very interesting approximate analytical description of such a
pattern (based on a coherent superposition of distinct phase-space
components), that here we briefly summarize. Let us assume that
the wave function $\psi(x)$ can be approximately written as
superposition of a finite number of components:
\be
\psi(x) \simeq
\sum_{i}\mathcal{A}_{i}(x)e^{i\Phi_{i}(x)} \; ,
\ee
where
$\mathcal{A}_{i}(x)$ are slowly varying positive envelopes and
$\Phi_{i}(x)$ are real functions defining the phases. The
corresponding Wigner function reads
\bea
&&W(x,p)= \nonumber \\
&& \nonumber \\
&&\frac{1}{2\pi}\sum_{i,j}\int dy
\mathcal{A}_{i}\left(x-\frac{y}{2}\right)
\mathcal{A}_{j}\left(x+\frac{y}{2}\right)
e^{-ipy
-i\Phi_{i}\left(x-\frac{y}{2}\right)+i\Phi_{j}\left(x+\frac{y}{2}\right)}
\; .
\eea
Assuming the stationary-phase approximation
$\Phi'_{i}\left(x-\frac{y}{2}\right)+\Phi'_{j}\left(x+\frac{y}{2}\right)=2p$
(where the prime denotes the first derivative), the contribution
to the Wigner function at the point $(x,p)$ comes from the points
of the trajectories $[x_{i};p_{i}=\Phi'_{i}(x)]$ and
$[x_{j};p_{j}=\Phi'_{j}(x)]$, satisfying the relations
$x_{i}+x_{j}=2x$ and $p_{i}+p_{j}=2p$. By fixing a pair of values
$(x,p)$, the corresponding pair $x_{i},x_{j}$ is determined by the
two previous constraints and by the expression of the phases.
Taking the value of the envelope at the point $x_{i}$, expanding
the phases up to quadratic terms, and performing the associated
Gaussian integrals, the approximate form of the Wigner function is
\bea
W(x,p)&&\approx\sum_{i}\mathcal{A}_{i}^{2}(x)\delta(p-\Phi'_{i}(x))
\nonumber \\
&& \nonumber \\
&&+ \sum_{i \neq j}
\sum_{x_{i},x_{j}}\frac{\sqrt{2}\mathcal{A}_{i}(x_{i})
\mathcal{A}_{j}(x_{j})}{\sqrt{\pi
i[\Phi''_{i}(x_{i})-\Phi''_{j}(x_{j})]}}e^{ip(x_{i}-x_{j})-
i\Phi_{i}(x_{i})+i\Phi_{j}(x_{j})}
\; .
\label{BanKniWig}
\eea
Specializing to the case of
three-photon down conversion processes, it can be shown that the
arms of the Wigner function can be modelled by the three
components of the wave function $\psi(x)$
\bea
\psi(x)  &=&
\sum_{i=1}^{3}\psi_{\theta_{i}}(x) \nonumber \\
&& \nonumber \\
&=& \mathcal{A}(x) +\sqrt{2}\mathcal{A}(-2x)e^{\frac{\sqrt{3}}{2}i
x^{2}-i\frac{\pi}{6}}+\sqrt{2}\mathcal{A}(-2x)e^{-\frac{\sqrt{3}}{2}i
x^{2}+i\frac{\pi}{6}} \; ,
\label{psiKnight}
\eea
where, as we
will see soon, the angles $\theta_{i}$ are connected to rotations
in phase space. The three components $\psi_{\theta_{i}}$ are
obtained in the following way. By considering a slowly varying
positive function $\mathcal{A}(x)$ one defines
$\psi_{\theta_{1}}(x)\equiv\psi_{0}(x)=\mathcal{A}(x)$. Then one
obtains the other components performing rotations around the
origin of the phase space: $U(\theta)=e^{-i\theta a^{\dag}a}$.
Using the configuration representation of the rotation operator,
and choosing in a suitable way the rotation angles $\theta_{2}$
and $\theta_{3}$, the other two components in Eq.
(\ref{psiKnight}) are obtained. Applying Eq. (\ref{BanKniWig}),
the Wigner function is approximated by the expression
\be
W_{\psi}(x,p) \, = \,
W_{\psi_{\theta_{1}}}(x,p)+W_{\psi_{\theta_{2}}}(x,p)+
W_{\psi_{\theta_{3}}}(x,p)+W_{int}(x,p)
\; ,
\ee
where $W_{int}(x,p)$ is the interference term responsible
for the nonclassical nature of the state: due to $W_{int}(x,p)$,
plotted in Fig. (\ref{WigBanaKnig}), the total Wigner function
exhibits strong interference patterns and oscillations, and
becomes negative in several regions of phase space.
Banaszek and Knight have shown that there is excellent
agreement between the Wigner function computed by
exact numerical methods and the one obtained by the approximate
analytical method that we have just described. This fact suggests
that such a technique might be suitable
for application to other instances of (degenerate) multiphoton processes.

\subsection{Kerr nonlinearities as a particularly interesting
case of four-wave mixing. A first discussion on the
engineering of nonclassical states and
macroscopic quantum superpositions}

\begin{figure}[h]
\begin{center}
\includegraphics*[width=11cm]{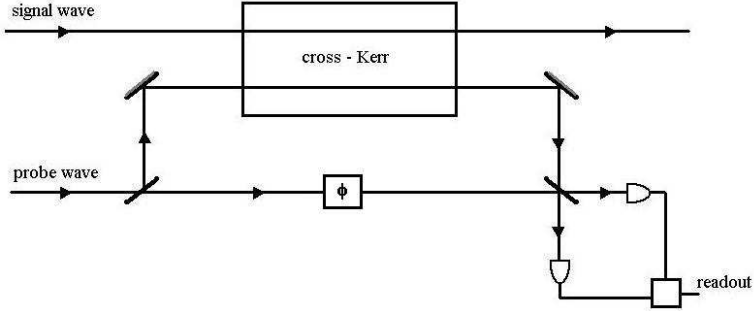}
\end{center}
\caption{Experimental setup (Mach-Zehnder interferometric
detector) for a quantum nondemolition measurement of the photon
number through the cross-Kerr effect.} \label{QND}
\end{figure}

A particular and important subclass of third-order multiphoton processes
described by the four-wave mixing Hamiltonian (\ref{H24wmx}) is
selected in the cases of total degeneracy, yielding the
Kerr interaction $a^{\dag 2}a^{2}$ (self-phase modulation effect),
and in the partially degenerate case, yielding the cross-Kerr
interaction $a^{\dag}b^{\dag}ab$ (cross-phase modulation effect).
In the presence of two optical frequencies, $\omega_{s}$ (signal)
and $\omega_{p}$ (probe), and by using the expression
(\ref{Elfieldop}) for the electric field operator, the interacting part
of Hamiltonian (\ref{nonlinearpurpe}) with $n=3$ reads
\bea
H_{Kerr}& = & \lambda_{s}[\chi^{(3)}(\omega_{s};\omega_{s},-
\omega_{s},\omega_{s})a^{\dag}_{s}a_{s}a^{\dag}_{s}a_{s}
\, + \, 5\mathcal{F}(a^{\dag}_{s},a_{s})] \nonumber \\
&& \nonumber \\
& + & \lambda_{p}[\chi^{(3)}(\omega_{p};\omega_{p},-
\omega_{p},\omega_{p})a^{\dag}_{p}a_{p}a^{\dag}_{p}a_{p} \, +
\, 5\mathcal{F}(a^{\dag}_{p},a_{p})] \nonumber \\
&& \nonumber \\
& + & \lambda_{sp}[\chi^{(3)}(\omega_{p};\omega_{p},-
\omega_{s},\omega_{s})a^{\dag}_{p}a_{p}a^{\dag}_{s}a_{s}+
23\mathcal{F}(a^{\dag}_{p},a_{p},a^{\dag}_{s},a_{s})]
\; ,
\label{HKerrtot}
\eea
where $\lambda_{s}$, $\lambda_{p}$, and
$\lambda_{sp}$ are real constants, $\mathcal{F}$ contains
terms in which the order of the arguments is interchanged with respect
to the Kerr terms written down explicitely, and the
associated nonlinear processes are automatically phase
matched. The final form of the normal ordered Hamiltonian is
\be
H_{Kerr} \, = \, \chi_{s} a^{\dag
2}_{s}a^{2}_{s} \, + \, \chi_{p}a^{\dag 2}_{p}a^{2}_{p}
\, + \, \chi_{sp}a^{\dag}_{s}a^{\dag}_{p}a_{s}a_{p}
\; ,
\label{HKerrtotf}
\ee
where the nonlinear couplings
$\chi_{s}$, $\chi_{p}$, and $\chi_{sp}$
are simple combinations of the original couplings.
It is worth noting that Kerr interactions are important
in quantum optics because of their crucial role
in the realization of nonlinear devices, such as quantum
nondemolition meters, nonlinear couplers, and interferometers.
Moreover, they are important in quantum state engineering,
e.g. in the realization of macroscopic superpositions
and entangled states.
In some applications, as quantum nondemolition
measurements of the photon number, it could be necessary to remove
one or both of the self-phase modulation effects. This task can be
accomplished by resorting to different strategies; for example, the
cross-Kerr process can be favored by resonance conditions under
which the other terms become negligible. Another possibility
consists in the cancellation of the undesired self-phase
modulation term by means of an auxiliary medium with
negative $\chi^{(3)}$ (two-crystal configuration), in the sense
that the undesired term
is cancelled out if the corresponding field passes through another
negative $\chi^{(3)}$ medium. \\
Here we begin by discussing the possibility of using a Kerr
medium to perform quantum nondemolition (QND) measurements.
The latter are important for many applications in modern
quantum physics, and essentially consist in measuring an
observable with exact precision at the expense of an increasing
uncertainty of its canonically conjugated observable.
An excellent discussion of QND measurements can be found in the review
by Braginsky and Khalili \cite{QNDrev}.
Here we illustrate the Kerr-based scheme
proposed by Imoto, Haus, and Yamamoto \cite{QNDImoto,mesoscopQO}
and depicted in Fig. (\ref{QND}). It represents a nonlinear
interferometer, that is a Mach-Zehnder interferometer with a Kerr
medium placed in one of the arms \cite{YamHaus,McZend1,McZend2}.
 A signal and a probe wave, respectively
at frequencies $\omega_{s}$ and $\omega_{p}$, propagate in the
optical Kerr medium. Because of the cross-Kerr effect $\chi_{sp}
n_{s}n_{p}$ ($n_{s}=a^{\dag}_{s}a_{s}\;,\;
n_{p}=a^{\dag}_{p}a_{p}$), the photon number $n_{s}$ modulates the
phase of the probe wave, and a measurement of this modulation
yields the information about the $n_{s}$ itself. Let $a_{s}^{in}$
and $a_{p}^{in}$ denote the input operators for the signal and the
probe waves; the corresponding output operators can be determined
explicitly, by solving the Heisenberg equations, and read:
\be
a_{s}^{out}(\tau) \, = \, e^{i\tau n_{p}}a_{s}^{in} \; , \quad
a_{p}^{out}(\tau) \, = \, e^{i\tau n_{s}}a_{p}^{in} \; ,
\ee
with the dimensionless time $\tau=\chi_{sp} t$. Moreover, a phase
difference $\phi=\pi/2$ is imposed between the two arms of the
detector, and the two photodiode currents are subtracted as shown
in the scheme Fig. (\ref{QND}). Introducing the quadrature
operators $X_{i}^{l}\equiv(a_{i}^{l}+a_{i}^{\dag \, l})/\sqrt{2}$
and $P_{i}^{l}\equiv i(a_{i}^{\dag \, l}-a_{i}^{l})/\sqrt{2}$,
with $i=s,p$ and $l=in\,,out$, the readout observable
corresponding to the difference current is the output probe
quadrature phase amplitude
\be
P_{p}^{out} \, = \,
X_{p}^{in}\sin(\tau n_{s}) \, + \, P_{p}^{in}\cos(\tau n_{s}) \; .
\label{QNDPout}
\ee
Fixing the input probe phase such that
$\langle P_{p}^{in}\rangle=0$, relation (\ref{QNDPout}) can be
approximated (for small $\chi_{sp}$) as
\be
P_{p}^{out}\simeq\langle X_{p}^{in}\rangle\tau n_{s}+\Delta
P_{p}^{in} \; ,
\label{QNDns}
\ee
 where $\Delta
P_{p}^{in}=P_{p}^{in}-\langle P_{p}^{in}\rangle\equiv P_{p}^{in}$.
A normalized readout observable $n_{s}^{(obs)}$ corresponding to
the signal photon number can be defined as
\be
n_{s}^{(obs)}\equiv
\frac{P_{p}^{out}}{\langle X_{p}^{in}\rangle
\tau}=n_{s}+\frac{\Delta P_{p}^{in}}{\langle X_{p}^{in}\rangle
\tau} \,.
\ee
Thus the measured observable is $\langle
n_{s}^{(obs)}\rangle=\langle n_{s}\rangle$, and its variance is
given by
\be
\langle(\Delta n_{s}^{(obs)})^{2}\rangle=\langle
\Delta n_{s}^{2}\rangle+\frac{\langle(\Delta
P_{p}^{in})^{2}\rangle}{\langle n_{p}\rangle \tau^{2}} \; ,
\ee
which holds, for instance, if the probe wave is in a coherent
state, with $\langle(\Delta P_{p}^{in})^{2}\rangle=1/4$.
Therefore, for increasing $\langle n_{p}\rangle \tau^{2}$, the
measured uncertainty tends to the ideal value $\langle \Delta
n_{s}^{2}\rangle$, that is the intrinsic uncertainty of the
observable $n_{s}$. At the same time, as a result of the
measurement, an increase of the quantum uncertainty of the probe
wave photon number is obtained. The scheme exposed above well
represents the main features of a quantum nondemolition
measurement. Of course, many other proposals have been used based
on the extension or modification of this scheme, and for different
purposes \cite{QNDKitag,QNDloss,QNDcav1,QNDcav2,QNDinterf}.
Kitagawa, Imoto, and Yamamoto have exploited the quantum
nondemolition measurement via the cross-Kerr effect to produce
number-phase minimum uncertainty states and near-number states
\cite{QNDKitag}. Losses have been taken into account as well, for
example in Ref. \cite{QNDloss}, the scheme has been extended to
optical cavities \cite{QNDcav1,QNDcav2}, and the effect of quantum
nondemolition measurements on quantum interference has been also
investigated \cite{QNDinterf}. Finally, photon-number measurements
have been realized in optical fibers \cite{QNDexpfib1,QNDexpfib2},
and by exploiting atoms in a cavity
\cite{QNDexpatom1,QNDexpatom2}.

Many other Kerr-based nonlinear devices, beyond the interferometric
setup described in  Fig. (\ref{QND}), have been proposed in the
literature; here we limit
ourselves to the discussion of two significant examples.
The nonlinear directional
coupler \cite{dircoupler,dircoupler2} consists of two parallel
waveguides, exhibiting third-order nonlinearity, that exchange
energy by means of evanescent waves. The general
Hamiltonian reads
\be
H_{coup} \, = \, H_{Kerr} \, + \,
\lambda a_{s}a^{\dag}_{p} \, + \, \lambda^{*} a^{\dag}_{s} a_{p}
 \; ,
\label{Hdircoupl}
\ee
where $H_{Kerr}$ is given by Eq.
(\ref{HKerrtotf}), and the last two terms
are the evanescent-waves contribution. This system presents
interesting properties such as self-trapping, self-modulation, and
self-switching of the energy of the coupled modes
\cite{dircoupler3}, and it can be useful for the generation and
transmission of nonclassical light. The Kerr couplers can produce
sub-Poissonian squeezed light
\cite{dircoupler4,dircoupler5,dircoupler6}, and entangled states
\cite{dircoupler7}. We want also to mention another interesting
device, named nonlinear quantum scissor, that has been proposed
to perform optical state truncation
\cite{qscissors,qscissors2}. It can be realized by means of a fully
degenerate Kerr
medium in an optical cavity, pumped by external ultra-short pulses of
laser light \cite{qscissors2}. The optical state truncation that
can be achieved
allows, under suitable conditions \cite{qscissors2}, the
reduction of an initial coherent state even up to a single-photon
Fock state. It is worth noting that such a device can
be realized by means of linear optical components and
photo-detections as well \cite{LQSciss}.

Concerning Kerr-based quantum state engineering, nonlinear
interferometric devices have been largely considered for the
generation of nonclassical states of the radiation field
\cite{nonlintSanders,nonlintSanders2,Fredkingerry,nlintgerry1,nlintgerry2},
such as, specifically, the entangled coherent states of the form
$(e^{-i\pi/4}|i\beta\rangle_{1}|i\alpha\rangle_{2}
+ e^{i\pi/4}|-\alpha\rangle_{1}|\beta\rangle_{2})/\sqrt{2}$
discussed in Ref. \cite{nonlintSanders}. Nonlinear Mach-Zehnder-type
interferometers have been proposed for the production of maximally
entangled number states of the form
\be
|\Psi\rangle_{MES} \,
= \,
\frac{1}{\sqrt{2}}(|N\rangle_{a}|0\rangle_{b}
\, + \, e^{i\Phi_{N}}|0\rangle_{a}|N\rangle_{b})
\; .
\label{MES}
\ee
A first proposal to produce them \cite{Fredkingerry} is based on a
setup employing the Fredkin gate \cite{Fredkingate} (based on a
cross-Kerr interaction) in a double-interferometer configuration.
Another proposal \cite{nlintgerry1} relies on the use of a
four-wave mixer, described by the interaction Hamiltonian
$H_{int}=\chi^{(3)}(a^{\dag}b+ab^{\dag})^{2}$, in one arm
of the interferometer, that contains both cross-Kerr and nonlinear
birefringence terms. States (\ref{MES}) are potentially
important for applications in metrology, as atomic frequency
measurements \cite{Bollinger}, and interferometry \cite{Holland}.
Moreover, these states show phase super-sensitivity, in the sense
that they reduce the phase uncertainty to the Heisenberg limit
$\Delta \phi_{HL}=1/N$. These states have not yet been produced
in nonlinear devices; however, in recent successful experiments,
their three-photon ($N=3$) \cite{3MESnature} and four-photon ($N=4$)
\cite{4MESnature} versions have been realized by means of linear
optical elements and photodetections.

\begin{figure}[h]
\begin{center}
\includegraphics*[width=13cm]{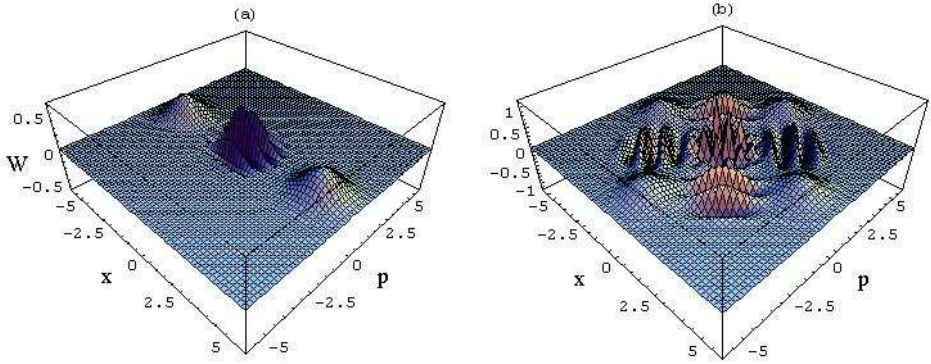}
\end{center}
\caption{Figure (a): Wigner function for the even Schr\"{o}dinger
cat states Eq. (\ref{kerrstateYS}). Figure (b): Wigner function
for the odd Schr\"odinger cat states Eq. (\ref{kerrstateYS2}). In
both cases, the coherent amplitude $\alpha=3$.} \label{kerrCatsW}
\end{figure}

\begin{figure}[h]
\begin{center}
\includegraphics*[width=14cm]{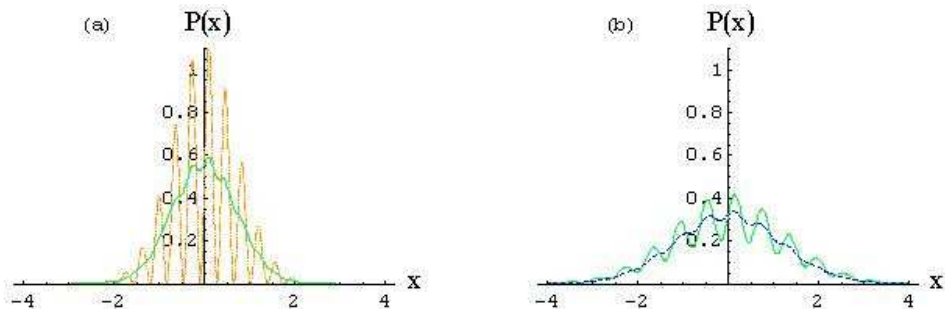}
\end{center}
\caption{In these figures we compare the behavior of the
interference fringes associated to the cat states of Yurke and
Stoler (Fig. (a)), and those associated to the superposition
states generated through the modified quasi-particle Kerr
Hamiltonian $H_{gK}$ (Fig. (b)). In both cases, the superposition
states are sent in one input port of a beam splitter of
transmittance $\eta$, while the other input port is taken empty.
The beam splitter can be used to model medium or detector losses,
which are quantified by the transmittance $\eta$. The $x$-axes
corresponds to the homodyne variable measured at the first output
port of the beam splitter. The $y$-axes corresponds to the
probability distributions $P(x)$ for the homodyne variable $x$
detected at the first output port independently of what has left
the second output port. In Figure (a) the dotted line represents
the interference fringes obtained with the Yurke-Stoler scheme in
the case of perfect transmittance $\eta=1$, and the full line
represents the fringes in the same scheme with transmittance
$\eta=0.96$. In Figure (b), the full line represents the fringes
associated to the scheme based on the modified quasi-particle Kerr
Hamiltonian $H_{gK}$ for $\eta=0.96$, and the dashed line
represents the fringes obtained in the same scheme for $\eta=0.9$.
In either case the squeezing parameter is fixed at the value
$r=0.4$. The mean number of photons in the initial states, in all
cases here considered, is equal to $36$. From Figure (a), we see
that the interference fringes, which are evident in the ideal
$\eta=1$ case, practically disappear even for a slight amount of
loss in the transmittance. Figure (b) shows that the action of the
squeezing, although lowering the primary peak of the fringes,
moves down the threshold of complete decoherence from $\eta=0.96$
to $\eta=0.9$.} \label{fringesdeco}
\end{figure}

The engineering of quantum states by use of Kerr media includes
the generation of optical ``macroscopic'' superpositions of
coherent states. Macroscopic (or mesoscopic) superpositions take
the form $|\psi\rangle=c_{1}|\alpha e^{i\theta}\rangle +
c_{2}|\alpha e^{-i\theta}\rangle$, where $|\alpha e^{\pm i \theta}
\rangle $ are coherent states with a sufficiently high average
number of photons $|\alpha|^{2}$. In the course of the years, many
proposals have been put forward for the generation of such
"Schr\"{o}dinger cat" superpositions. Among them, we should
mention schemes based on state reduction methods
\cite{cohsuperposbystatreduct}, and on conditional measurements
performed on entangled states \cite{cohsuperposbycondmeas}.
However, the simplest way (at least from a theoretical point of
view) to obtain macroscopic superpostions is the propagation of
the radiation field through an optical fiber associated with nonlinear
Kerr effects and interactions
\cite{McZend2,kerrMilburnHolmes,kerrYurkeStoler,kerrTombesiMecozzi,YurkeStoler4wmx,cohsuperposgen}.
Schr\"odinger cat states possess important nonclassical properties
\cite{catsnonclasprop,catsnonclasprop2} like squeezing and sub-Poissonian
statistics, moreover they should provide the crucial playground
for the testing of the quantum-classical transition and the theory
of decoherence.
\\ In 1986, generalizing the model studied by Milburn and Holmes \cite{kerrMilburnHolmes},
Yurke and Stoler \cite{kerrYurkeStoler} considered the time
evolution of an initial coherent state under the influence of the
anharmonic-oscillator Hamiltonian:
\be
H_{anh} \, = \, \omega n \,
+ \, \Omega n^{k} \; , \quad (k \geq 2) \; ,
\label{HanharmYurkeStoler}
\ee
where $\omega$ is the energy-level
splitting for the harmonic-oscillator part of the Hamiltonian,
$\Omega$ is the strength of the anharmonic term, and $n$ is the
number operator. In a nonlinear medium $\Omega$ is proportional to
the $(k+1)$-th order nonlinearity. Note that, for $k=2$, the
anharmonic part of the Hamiltonian is of the degenerate,
single-mode Kerr form, apart an additive term linear in $n$ that
simply amounts to an irrelevant constant rotation in phase space.
Yurke and Stoler evaluated the response of a homodyne detector to
the evolved state
\be
|\alpha,t\rangle \, = \, e^{-it\Omega
n^{k}}|\alpha \rangle \, = \, e^{-\frac{|\alpha|^{2}}{2}}
\sum_{n=0}^{\infty}\frac{\alpha^{n}
e^{-i\phi_{n}(t)}}{\sqrt{n!}}|n\rangle \; ,
\label{YSkerrstat}
\ee
where $\phi_{n}(t) = \Omega t n^{k}$. The state vector is periodic
with period $2\pi/\Omega$, and coherent superpositions of
distinguishable states of the radiation field emerge for special
values of $t$; for instance:
\bea
|\alpha,\pi/2\Omega\rangle
&=&\frac{1}{\sqrt{2}}
[e^{-i\pi/4}|\alpha\rangle+e^{i\pi/4}|-\alpha\rangle] \; , \quad k
\; \;
even \; , \label{kerrstateYS} \\
&& \nonumber \\
|\alpha , \pi /2 \Omega \rangle &=&\frac{1}{2} [|\alpha \rangle
-|i \alpha \rangle + |-\alpha\rangle + |-i\alpha \rangle ] \; ,
\quad k \; \; odd \; .
\label{kerrstateYS2}
\eea A very useful
insight on the properties of states (\ref{kerrstateYS}) and
(\ref{kerrstateYS2}) is gained by looking at their Wigner
functions, which are plotted, respectively for even and odd $k$ in
Fig. (\ref{kerrCatsW})-(a) and (\ref{kerrCatsW})-(b). They show a
typical two- and four-lobe structure, representing the coherent
state components, and evident fringes due to quantum interference.
The coherence properties are very sensitive to losses as shown,
using coherent superpositions of the form (\ref{kerrstateYS}), in
Ref. \cite{kerrYurkeStoler} by Yurke and Stoler: for macroscopic
coherent inputs, corresponding to large mean photon numbers
$|\alpha|^{2}$, even a very small amount of loss, corresponding to
nonideal detection efficiency, destroys the interference fringes
and leads to a probability distribution for the
homodyne-detector's output current indistinguishable from that
of a classical statistical mixture of
coherent states, see Fig. (\ref{fringesdeco}) (a). \\
In Ref. \cite{kerrTombesiMecozzi} Mecozzi and Tombesi have
suggested to extend the method of Yurke and Stoler by considering
more general Kerr-like Hamiltonians $H_{gK}$ of the form
$H_{gK}=H_{0}+\lambda (H_{0})^{2}$, with
$H_{0}$ any diagonalizable Hamiltonian. Although the original aim
of the authors was to discuss the decoherence effects due to vacuum
fuctuations, as we will see in the following, we can take a special
realization of their general Hamiltonian $H_{gK}$ in order to compare
directly with the original scheme of Yurke and Stoler and highlight the role
of squeezing (and in general of nonclassicality)
in improving the preservation of coherence against losses.
To this aim, let us in fact choose the particular
$H_{0}=b^{\dag}b$, with $b$ denoting the canonically
transformed Bogoliubov quasi-photon mode: $b=\mu a+\nu a^{\dag}$.
With this choice, following the line of thought of Yurke and Stoler, we
consider the evolution generated by the Hamiltonian $H_{gK}$ applied to
an initial squeezed state (the eigenstate of mode $b$). Comparing Figures
(\ref{fringesdeco}) (a) and (b) shows that, for the same amount of losses that
destroy the coherence in the Yurke-Stoler scheme, a finite amount of
squeezing is more effective in fighting decoherence, as
witnessed by the persistent visibility of the interference fringes.
For a degenerate Kerr Hamiltonian $\Omega\; a^{\dag 2}a^{2}$, a
more general class of superposition states is generated at times
$t=\pi/m\Omega$ with $m$ an arbitrary, nonzero integer
\cite{superposKerrTaraAgar}:
\bea
&&|\psi(t=\pi/m\Omega)\rangle=e^{-\frac{i\pi}{m}n(n-1)}|\alpha\rangle
\nonumber \\ && \nonumber \\
&&=\left\{
\begin{array}{rrr}
\sum_{q=0}^{m-1}f_{q}^{(e)}|\alpha e^{-\frac{2\pi i
q}{m}}e^{i\frac{\pi}{m}}\rangle \; ,
& f_{q}^{(e)}=
\frac{1}{m}\sum_{n=0}^{m-1}e^{\frac{2\pi i q}{m}n}e^{-\frac{i \pi}{m}n^{2}}
\; ,
& \quad m \; even\; , \\
&& \nonumber \\
 \sum_{q=0}^{m-1}f_{q}^{(o)}|\alpha e^{-\frac{2\pi i
q}{m}}\rangle \; , &
f_{q}^{(o)}=\frac{1}{m}\sum_{n=0}^{m-1}e^{\frac{2\pi i
q}{m}n}e^{-\frac{i \pi}{m}n(n-1)}\; , & \quad m \; odd \; .
\end{array}
\right. \\
\label{gattiagarwal}
\eea
Interestingly, these
superpositions of arbitrary finite length are eigenstates of
$a^{m}$, and thus have a suggestive interpretation in terms of
``higher-order'' coherent states. In fact, the nonlinear
refractive index of the Kerr medium acts on the input field by
modifying the phase-sensitive quantum noise. For an input coherent
state this leads to "self-squeezed" light \cite{Tanasselfsqueez},
in the sense that the initial coherent state acts itself as a
``self-pump'', playing the role of a linear driving field.
However, the Kerr effect does not change the photon statistics,
which, if the input is coherent,
remains Poissonian. \\
The production of two-mode (entangled) optical Schr\"{o}dinger
cats via the two-mode Kerr interaction (\ref{HKerrtotf}) has been
investigated as well \cite{2modesuperposkerr}. To insure the
time-periodicity of the state vectors, the parameters
$\Delta=\omega_{s}-\omega_{p}$, $\chi_{s}$, $\chi_{p}$, and
$\chi_{sp}$ must be chosen mutually commensurate. In the resonant
case $\Delta=0$, with the choices $\chi_{s}=\chi_{p}$,
$\chi_{sp}/\chi_{p}=1.2$, and initial two-mode coherent state
$|\alpha , \beta \rangle$, the following coherent superposition is
generated at the reduced, adimensional
time $\tau \doteq \chi_{s} t = 5\pi/2$:
\be
|\alpha,\beta,\,\tau=5\pi/2 \rangle \, = \, \frac{1}{\sqrt{2}}
\{e^{-i\pi/4}|i\alpha,i\beta\rangle \, + \,
e^{i\pi/4}|-i\alpha,-i\beta\rangle\} \; .
\ee
Macroscopically
distinguishable quantum states can also be produced exploiting
nonlinear birefringence, as shown by Mecozzi and Tombesi
\cite{kerrTombesiMecozzi} and by Yurke and Stoler
\cite{YurkeStoler4wmx}. Mecozzi and Tombesi considered the
effective Hamiltonian
\be
H \, = \, \omega(a^{\dag}a + b^{\dag}b)
\, + \, \kappa (a b^{\dag}-a^{\dag}b) \, + \, \frac{\Omega}{2}(a
b^{\dag}+a^{\dag}b)^{2} \; ,
\label{Hnonlinbiref}
\ee
where the
modes $a$ and $b$ represent the two orthogonal polarizations of a
coherent light beam entering the birefringent medium. Moreover,
Mecozzi and Tombesi showed that the vacuum fluctuations can be
suitably removed by injecting in the
unused port a squeezed vacuum. \\
A similar Hamiltonian was considered
by Agarwal and Puri to study the one-dimensional propagation of
elliptically polarized light in a Kerr medium
\cite{ellitpolarkerr}. The study of this
kind of processes leads to an interesting result, essentially due
to the quantum nature of the field; it consists in the possibility
that the degree of polarization may change with time,
leading, for some times, to a partially polarized field, in
contrast to the one coming from semiclassical theory
which predicts a constant polarization. \\
For completeness, we point out that multicomponent entangled
Schr\"odinger cat states can be in principle generated by the
dynamics associated to a fully quantized nondegenerate four-wave
mixing process, described by the effective interaction Hamiltonian
$H_{ndg}^{4wm}=\kappa (a^{\dag}b+b^{\dag}a)(c^{\dag}d+d^{\dag}c)$,
with $a$ and $c$ taken as the pump modes, and $b$ and $d$ taken as
the signal modes \cite{4wmcatgen}.

Regarding the possibility of experimental realizations of
macroscopic superpositions in a Kerr medium, the quite small
values of the available Kerr nonlinearities would require, for the
generation of a cat state, a long interaction time, or
equivalently a large interaction length. For an optical frequency
of $\omega\approx 5\times 10^{14} \; rad\; sec^{-1}$, one would need
an optical fiber of about $1500$ km, and, consequently, losses
and decoherence would destroy completely the quantum superpositions.
However, in recent years there have been some interesting
theoretical proposals aimed at overcoming these
difficulties and obtaining cat states in the presence of
small Kerr effects. One of the most recent and feasible proposals
\cite{superpossmallkerr} is based on the following scheme:
a Kerr-evolved state (\ref{gattiagarwal}) and the vacuum feed
the two input ports of a $50-50$ beam splitter;
after passing the beam splitter, the real part of
the coherent amplitude
of the resulting two-mode output state is measured by
homodyne detection, reducing to a new finite superposition
of coherent states. Suitably engineering the coefficients
of this superposition leads to a Schr\"odinger cat.
This scheme allows to produce cat states of high quality without
requiring a strong nonlinearity. In fact, in this scheme, based on
the successive application of Kerr interaction, beam splitter,
and, finally, homodyne detection, the time needed to produce the
Schr\"odinger cat state practically coincides with the time needed
to produce the Kerr-evolved state of the form
(\ref{gattiagarwal}). Therefore, choosing for instance in Eq.
(\ref{gattiagarwal}) $m \simeq
10^{2}$, the equivalent length of the optical fiber needed to
produce the cat is reduced to few tens of kilometers, and this in
turn greatly reduces the effect of losses and dispersion. Further
improvements along this line may be thus expected in the near
future.
However, as we will see in the next Section, methods based on
high-$Q$ cavity fields have proved more effective than fibers
in the experimental production of mesoscopic coherent superpositions.

\subsection{Simultaneous and cascaded multiphoton processes
by combined three- and four-wave mixing}

\begin{figure}[h]
\begin{center}
\includegraphics*[width=13cm]{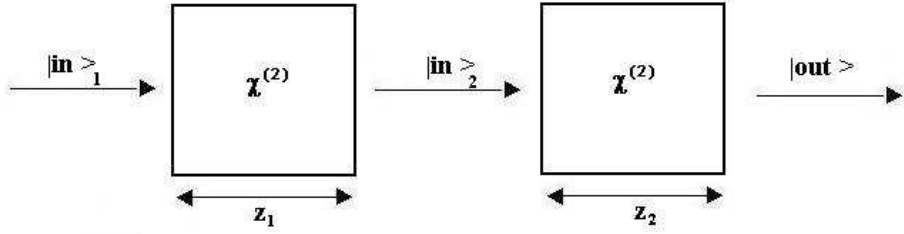}
\end{center}
\caption{Scheme for a cascade of two $\chi^{(2)}$ nonlinear
crystal. The input of the second medium is the output of the first
one.} \label{cascade2nonlin}
\end{figure}

Most of the interaction Hamiltonians used in multiphoton quantum
optics deal with a single parametric process, but in the two last
Subsections we have already met some examples of coexisting,
concurring interactions. At present the possibility to produce
quantum states with enhanced nonclassical properties, the need to
take into account more than one nonlinear contribution, and the
complexity in experimental engineering, have promoted the idea to
exploit composite, concurrent multiphoton interactions. These can
be realized in several ways through multi-step, cascaded schemes
either in the same nonlinear medium, or by using different media
in a suitable experimental configuration. In this Subsection we
describe some models and schemes based on multiphoton multiple
processes and interactions.\\
Let us first consider some examples of combinations of two
different nonlinear processes occurring inside the same crystal. A
simple multiphoton interaction Hamiltonian involving two different
nonlinear terms was studied by Tombesi and Yuen
\cite{TombYuenOPOkerr} in order to improve the maximum available
squeezing of the standard optical parametric amplifier (OPA). They considered
the short-time interaction of a single-mode coherent light with an
optically bistable two-photon medium that combines Kerr-type effects
and degenerate down conversion, and gives rise to the interaction
Hamiltonian
\be
H_{I}^{SK} \, = \, i \kappa
(a^{2}-a^{\dag 2}) \, + \, \Omega a^{\dag 2}a^{2} \; ,
\label{OPOKerr}
\ee
with $\kappa$ and $\Omega$ real. Later, the time
evolution of the Hamiltonian (\ref{OPOKerr}) was studied by Gerry
and Rodrigues \cite{GerryRodOPOKerr} for longer times by means of
a numerical method.
Assuming a coherent state as the initial
input state, Gerry and Rodrigues verified the presence of squeezing and
antibunching effects recurring on a longer time scale. Hamiltonian
models of the form (\ref{OPOKerr}) have also been studied in cavity,
both for the one-mode \cite{KryuchOPOKerr} and the two-mode case
\cite{2modecavOPOKerr}. \\
Concurrent, two-step optical $\chi^{(2)}$-interactions were
observed for the first time in 1970, using ammonium dihydrogen phosphate
as a medium \cite{simUpDwnCexp}. In this experiment the crystal
was illuminated by a classical laser pump at frequency
$\omega_{p}$ with equal ordinary and extraordinary polarization
components $\omega_{p}^{o}\equiv\omega_{p}^{e}$; the simultaneous
collinear phase matching was obtained for the spontaneous down
conversion process
$\omega_{p}^{e}\rightarrow\omega_{i}^{o}+\omega_{1}^{e}$, and the
successive up conversion process
$\omega_{p}^{o}+\omega_{i}^{o}\rightarrow\omega_{2}^{e}$ where
$\omega_{i}^{o}$, $\omega_{1}^{e}$, $\omega_{2}^{e}$ represent,
respectively, the one idler, and the two signal modes.
Few years later, the quantum
theory of coupled parametric down-conversion and up-conversion
with simultaneous phase matching was formulated
\cite{SmithersLu}.
The interaction Hamiltonian describing this two-step process is
\be
H_{I}^{2step} \, = \, [\kappa^{e}a_{1}^{\dag}a_{i}^{\dag} \, + \,
\kappa^{o}a_{i}a_{2}^{\dag} \; + \; H.c.] \; ,
\label{smitherHI}
\ee
where the real parameters $\kappa^{e}$ and
$\kappa^{o}$ are proportional to the extraordinary and ordinary
components of the pump amplitude.
The dynamics of the system can be solved by means of group-theoretical
methods \cite{SmithersLu}.
Assuming the three-mode vacuum
$|0\rangle_{1}|0\rangle_{2}|0\rangle_{i}\equiv |0,0,0\rangle$ as
initial state, the photon statistics of the system in each mode is
super-Poissonian because $\langle\Delta
n_{j}^{2}\rangle(t)/\langle n_{j}\rangle(t)=\langle
n_{j}\rangle(t)+1\geq 1$, $n_{j}=a^{\dag}_{j}a_{j}$, $(j = 1,2,i)$.
Moreover, the temporal behavior of the output signals depends on which
of the two processes is predominant. In fact,
the intensities $\langle n_{j} \rangle$, for $\kappa^{e}>\kappa^{o}$,
are exponentially increasing functions at long times,
while they are oscillating functions for $\kappa^{o}>\kappa^{e}$.
Recently, the interaction (\ref{smitherHI}) has been exploited
to generate three-mode entangled states \cite{FerraroBondani1}, as
we will see in more detail in Section \ref{section7}.\\
Another interesting
theoretical proposal, later experimentally implemented,
is due to Marte \cite{Marte} who investigated
the possibility to generate sub-Poissonian light via competing
$\chi^{(2)}$-nonlinearities, in particular through the interaction
of doubly resonant second harmonic generation and nondegenerate
down-conversion in a cavity. Experimental observation of some
quantum (and classical) effects arising in a cascaded $\chi^{2}:\chi^{2}$
system are reported in Ref. \cite{Marteexp}. \\
The generation of sub-Poissonian light has been studied by other
authors in consecutive quasi-phase matched wave interactions
\cite{Chirkin1,Chirkin2,Chirkin3}. They considered the interaction
of waves with multiple frequencies $\omega$, $2\omega$ and
$3\omega$, described by the Hamiltonian:
\be
H_{I}^{m\omega} \, =
\, \kappa_{1}a_{1}^{\dag 2} \, + \, \kappa_{3}a_{3}^{\dag}a_{1} \,
+ \, H.c. \; ,
\label{Hmultipfreq}
\ee
where $a_{1}$ denotes the
mode at frequency $\omega$, and $a_{3}$ corresponds to the mode at
frequency $3\omega$. The coupling constants $\kappa_{1}$ and
$\kappa_{3}$ are proportional, respectively, to the strength of
the parametric processes $2\omega\rightarrow\omega+\omega$ and
$\omega+2\omega\rightarrow 3\omega$, and to the amplitude of the
classical pump at frequency $2\omega$. With a scheme similar to
that depicted in Fig. (\ref{PPNC}), the model (\ref{Hmultipfreq})
can be implemented in a periodically poled nonlinear crystal, with
second order nonlinearity, illuminated by an intense pump at
frequency $2\omega$ \cite{Chirkin2}. The conditions of quasi-phase
matching are
\bea
&&\Delta k(2\omega)=k(2\omega)-2k(\omega)=2\pi
m_{2}/\Lambda \; , \nonumber \\
&& \nonumber  \\
&&\Delta k(3\omega)=k(3\omega)-k(2\omega)-k(\omega)=2\pi
m_{3}/\Lambda \; ,
\eea
where $\Lambda$ is the period of
modulation of the nonlinear susceptibility, and $m_{j}=\pm 1,\pm
3,... \; $. These conditions are satisfied in a periodically poled
$LiNbO_{3}$ crystal, where fields with sub-Poissonian photon
statistics are formed at frequencies $\omega$ and $3\omega$, and,
furthermore, the degree of entanglement for the output fields can
be studied and determined
\cite{Chirkin3}. \\
Besides the single-crystal configuration, also
multicrystal configurations have been considered. Some experiments
\cite{2crystexp,2crystexp1,2crystexp2,2crystexp3,2crystexp4,2crystexp5},
employing a setup with two nonlinear crystals, have been realized
to study entanglement properties and interference effects of down
converted light beams. For instance, Zou \textit{et al.}
\cite{2crystexp1}, using an experimental device
whose core is based on the exploitation of a cascaded
process of the kind depicted in Fig. (\ref{cascade2nonlin}),
reported the experimental observation of second-order
interference in the superpositions of signal photons from two
coherently pumped parametric down converters, under the condition
of alignment of the idler photons. Later, it was shown that in this
experimental arrangement the emerging states can be related to
$SU(2)$ and $SU(1,1)$ coherent and minimum-uncertainty states
\cite{luisPerina}. The simplified scheme depicted in Fig.
(\ref{cascade2nonlin}) represents a two-crystal experimental setup
based on cascaded second order nonlinearities.
Given a state $|in\rangle_{1}$ at the input of the first crystal,
the evolved state $|out\rangle$ at the output of the second
crystal is of the form
\be
|out\rangle \, = \, e^{-iH_{I}^{(2)}z_{2}}e^{-iH_{I}^{(1)}z_{1}}|in\rangle_{1}
\; ,
\ee
where $H_{I}^{(j)}$ are the interaction Hamiltonians
corresponding to the $j$-th crystal, and $z_{j}$ $(j=1,2)$ are the
propagation lengths, proportional to the travelling times.
This scheme has been proposed by D'Ariano \textit{et al.}
\cite{darianoPCS} to synthesize the phase-coherent states
$|\gamma\rangle=\sqrt{1-|\gamma|^{2}}\sum_{n}\gamma^{n}|n\rangle$,
introduced by Shapiro \textit{et al.} \cite{shapiroPCS}. Each of
the processes occurring in the two crystals are described in this
case by the three-wave Hamiltonian (\ref{H2}). In the first step,
the parametric down-conversion from the vacuum produces a two-mode
squeezed vacuum (a so-called twin beam).
In the second step, the up-conversion of the twin beam gives
origin to the phase coherent states. The two-crystal configuration
has also been exploited to engineer macroscopic superpositions
\cite{DeMartini,DeMartini2}. In particular, De Martini has suggested an
experimental setup, using two coupled down converter, to generate
a field state showing Schr\"odinger-cat behavior and enhanced
entanglement \cite{DeMartini}. \\ A suitable multicrystal
configuration, including both parametric amplifier and symmetric
directional couplers, can realize the multiphoton interaction
model
\be
H_{I}^{3m} \, = \, \frac{i\dot{r}}{2}
\big[ a_{1}a_{2}+a_{2}a_{3}+a_{3}a_{1}
+a_{1}^{\dag}a_{2}+a_{2}^{\dag}a_{3}+a_{3}^{\dag}a_{1}-H.c. \big] \; ,
\label{3modesqueezH}
\ee
where the dot denotes the time derivative, and $\dot{r}/2$ is
the coupling. As shown in Ref. \cite{3modeSvac}, the time evolution
of an initial vacuum state driven by this Hamiltonian generates the
three-mode squeezed vacuum defined as
\be
S_{3}|000\rangle \, \equiv \,
e^{ir(X_{1}P_{2}+X_{2}P_{3}+X_{3}P_{1})}|000\rangle \; ,
\label{3mSvac}
\ee
where $X_{i}$ and $P_{i}$ are the quadrature
operators corresponding to the $i$-th mode. It can be easily
proven that $S_{3}$ is a proper squeezing operator for the
three-mode quadratures $\tilde{X}=\frac{1}{\sqrt{6}}\sum_{i}X_{i}$
and $\tilde{P}=\frac{1}{\sqrt{6}}\sum_{i}P_{i}$
$([\tilde{X},\tilde{P}]=\frac{i}{2}$). It is evident that the
state (\ref{3mSvac}) is a simple generalization of the two-mode
squeezed vacuum state $S_{2}|00\rangle =
e^{ir(X_{1}P_{2}+X_{2}P_{1})}|00\rangle $, with
$r$ real squeezing parameter. \\
Before ending this Section, we want to mention another recent
theoretical development in the realization of multiphoton
parametric oscillators \cite{Kryuch1}, again based on the scheme
depicted in Fig. (\ref{cascade2nonlin}), but, this time, inserted
in a cavity. The model consists of two sequential parametric
processes of photon down conversion in $\chi^{(2)}$-nonlinear
media placed inside the same cavity, and realizes in principle,
by suitably tuning the experimental conditions,
the generation either of three-photon, or of four-photon (entangled) states.
The quantum statistical properties of these models
have been studied by computing the Wigner function resorting to
methods of quantum-jump simulations
\cite{Kryuch1,MolmerJump,KnightPlenioJump}. This study has
revealed complete phase-space symmetry for the different arms of
the Wigner functions, and multiple stability zones of the
subharmonic components. Extensions are obviously possible; in
particular, the cascaded frequency doubler generating the second
and the fourth harmonics has been recently
studied \cite{Kryuch2}.

\newpage

\section{Models of multiphoton interactions and
engineering of multiphoton nonclassical states}
\label{section5}

The many fascinating properties of nonclassical states of light
allow, at least in principle, innovative and far-reaching
applications in quantum control, metrology, interferometry, quantum information
and communication. In a way, although the very concept of
nonclassicality presents many subtleties and its quantification
is somehow still controversial, one might go so far to say that
it should be considered a physical resource to
be exploited, much in the same sense as energy and entropy.
In Section \ref{section2} we have listed some
quantities that can be taken as indices of nonclassicality of
a state. Among them, we recall the
deviation from Poissonian statistics, the Mandel's $Q$ parameter,
the nonexistence of a positive Glauber-Sudarshan $P$ function, and
the negativity of the Wigner function. A rich collection of references on
nonclassical states in quantum optics has been recently published
by Dodonov \cite{revDodonov,Dodonovtx}, and a review of optical correlation experiments
distinguishing between the predictions of classical and quantum
theory can be found in Ref. \cite{ReidWallsviol}. In the latter paper,
the authors consider an optical field produced by an intracavity four-wave
mixing, and discuss the Cauchy-Schwartz and Bell's inequalities,
as well as nonclassical properties as squeezing and antibunching.
All these probes of nonclassical behaviors share a
common idea: the reference states that are taken as a standard
of comparison are the Glauber coherent states, which provide the
boundary line of separation between the classical and the quantum
domains. \\
In the previous Section we have reviewed the simplest multiphoton
processes arising in nonlinear media, and their ability to
generate some standard nonclassical states of light. We have
limited our analysis to consolidated experimental settings
associated to second- and third-order nonlinearities, and up to
four-wave interactions. In this Section we review various methods,
models, experimental proposals and realizations, introduced in
order to engineer arbitrary nonclassical states of light
associated to general multiphoton processes and interactions. A
large part of these generalized classes of multiphoton
Hamiltonians and multiphoton nonclassical states will be
associated to higher-order nonlinearities and will exhibit a great
variety of nonclassical properties.

\subsection{Nonclassical states by group-theoretical
methods}

The basic idea of group-theoretical methods is to generalize the
concept of coherent state to physical systems with more general
symmetries and group-theoretical structures.
In this framework, the construction of multiphoton coherent states
associated with Lie groups has been carried out by several authors,
most notably Klauder \cite{Klauder},
Perelomov \cite{Perelomov}, and Gilmore \cite{Gilmore}.
The most direct way to construct generalized coherent states is
to extend the three Glauber's original
definitions. It is to be remarked, however,
that, at variance with the case of the harmonic-oscillator
$h_{4}$ algebra, in more general algebraic contexts the three
Glauber definitions can define inequivalent classes of states.
The simplest procedure is to obtain coherent
states as eigenstates of a lowering operator.
This approach was used by Barut and
Girardello \cite{BarutGirard} to define a class of generalized
coherent states of the group $SU(1,1)$, whose generators
$\{K_{+}\,,K_{-}\,,K_{0}\}$ obey the commutation
relations (\ref{su11algebracom}). As is well known,
the group $SU(1,1)$ is associated with the dynamics of the
parametric amplifier, in which two photons are created or annihilated.
Barut and Girardello thus defined the eigenstates $\{ |\varsigma \rangle \}$
of the lowering (or deexcitation) operator $K_{-}$ as the $SU(1,1)$
coherent states:
\be
K_{-}|\varsigma \rangle \, = \, \varsigma|\varsigma \rangle \; .
\label{barutgirst}
\ee
If the group $SU(1,1)$ is in the degenerate realization Eq. (\ref{su11degalgebra}),
for which $K_{-} = \frac{1}{2}a^{2}$, the states
$|\varsigma\rangle$ are the cat-like states \cite{evenoddcohstat}, while if
the group is in the nondegenerate realization Eq. (\ref{su11algebra}),
for which $K_{-}=ab$, the states $|\varsigma\rangle$ are the pair coherent
states \cite{paircohst}. We will discuss these two classes of states in more
detail later on. Here we limit ourselves to mention that
pair coherent states can be produced by a dissipative dynamics
engineered by exploiting the strong
competition between four-wave mixing and amplified spontaneous emission
in resonant two-photon excitations. These states are deeply entangled and
show strong squeezing and antibunching. Extensions of pair coherent states
to multimode systems are in principle possible, and have been carried out
for three modes (``trio coherent states'') \cite{triovietnam}.
Generalizations of this approach include the eigenstates
of linear combinations of the $SU(1,1)$
operators \cite{SU11eigenst1,SU11eigenst2}, and the general concept
of algebra eigenstates introduced by Brif
\cite{h6Brif}. \\
A different approach is based on the generalization of
minimum uncertainty relations. Given two operators $A$ and $B$
associated to a Lie algebra, minimum uncertainty coherent states
are defined as the states that saturate the relation
$\Delta A^{2}\Delta B^{2}\geq\frac{1}{4}|\langle [A,B]\rangle|^{2}$
\cite{aragone,NTgensyst,MUSpuri,SU11intellig}; these states are
also known as intelligent states.
Both approaches to generalized coherent states, as eigenstates of
a lowering operator or as minimum uncertainty states, can be limited
in some cases by several constraints, including constraints on the
Lie algebra, nonuniqueness of the generalized coherent wave packets,
possible nonexistence of a standard resolution of unity. \\
In fact, the most convenient starting point to construct
generalized coherent states is based on the action of a
suitable displacement operator on a reference state.
Given an algebra with generators
$T_{i}$ $(i=1,...,n)$, the generic group coherent state can be defined as
\be
|\xi_{1},...,\xi_{n}\rangle\, = \, e^{\sum_{i}\xi_{i}T_{i}}|\Phi_{0}\rangle
\; ,
\label{groupcohst}
\ee
where $\xi_{i}$ are complex parameters
and $|\Phi_{0}\rangle$ is a fiducial state.
To be more specific, we follow Ref.
\cite{zhangfeng}, where Zhang, Feng, and Gilmore present a general
algorithm for constructing coherent states of dynamical groups for
a given quantum physical system. To this aim, let us consider a
quantum system described by a certain Hamiltonian
expressed in terms of a set of generators
$\mbf{g}\equiv\{T_{i}\}$, spanning a closed algebra, i.e.
$[T_{i},T_{j}]=\sum_{k}C_{ij}^{k}T_{k}$, where $C_{ij}^{k}$ are
the structure constants of $\mathbf{g}$. The
Hamiltonian of the system is typically
expressed in linear or quadratic form
\be
H \, = \, H(T_{i}) \, = \, \sum_{i}c_{i}T_{i} \, + \,
\sum_{i,j}c_{ij}T_{i}T_{j} \; .
\label{Hgroup}
\ee
Let $G$ the dynamical group associated with
$\mathbf{g}$. If $\mathbf{g}$ is a semisimple Lie algebra, the
operators $\{T_{i}\}$ can be expressed in the standard Cartan
basis
$\{H_{i}\,,E_{\alpha}\,,E_{-\alpha}\}$ with commutation relations
\bea
&&[H_{i},H_{j}]=0 \; , \\
&&[H_{i},E_{\alpha}]=\alpha_{i}E_{\alpha} \; , \\
&&[E_{\alpha},E_{-\alpha}]=\alpha^{i}H_{i} \; , \label{cartanE+E-} \\
&&[E_{\alpha},E_{\beta}]=N_{\alpha,\beta}E_{\alpha+\beta} \; .
\eea The operators $H_{i}$ are diagonal in any unitary irreducible
representation $\Gamma^{\Lambda}$ of $G$, and $E_{\alpha}$ are the
shift operators. If the representation is Hermitian, we have
$H_{i}^{\dag}=H_{i}$ and $E_{\alpha}^{\dag}=E_{-\alpha}$. The
first step in the construction of generalized coherent states
requires choosing an arbitrary normalized reference state
$|\Phi_{0}\rangle$ within the Hilbert space
$\mathcal{H}^{\Lambda}$. This reference state will determine the
structure of the generalized coherent states and of the phase
space of the dynamical system. The maximum stability subgroup $D$
of $G$ is defined as the set of group elements $d$ that leave
invariant the reference state $|\Phi_{0}\rangle$ (and will thus
depend on the choice of $|\Phi_{0}\rangle$) up to a phase factor:
\be
d|\Phi_{0}\rangle \, = \, e^{i\phi(d)}|\Phi_{0}\rangle \; , \;
\quad d\in D \; . \ee Therefore the group $G$ is uniquely
decomposed in the two subgroups $D$ and $G/D$ such that \be
\mbf{g}=\Omega d \; , \; \quad \mbf{g}\in G \; , \; \quad d\in D
\; , \; \quad \Omega\in G/D \; .
\ee
Having introduced the
complement of $G$ with respect to $D$, the group definition of the
generalized coherent states $|\Lambda,\Omega\rangle$ is
\be
|\Lambda,\Omega\rangle \, \doteq \, \Omega|\Phi_{0}\rangle \; .
\label{cohstLie}
\ee
The most useful choice for the reference
state $|\Phi_{0}\rangle$ is an unperturbed physical ground state.
For this choice the Hamiltonian consists of linear terms in terms
of the generators $T_{i}$, and the ground state is an extremal
state, defined, at least for Hamiltonians with a discrete
spectrum, as the highest-weight state $|\Lambda,\Lambda\rangle$ of
the irreducible representation $\Gamma^{\Lambda}$ of the Lie group
$G$. Consequently, $ E_{\alpha}|\Lambda,\Lambda\rangle =0$, where
$\alpha$ belongs to the positive root set of $G$. With this
choice, the coset space as well as the generalized coherent states
(\ref{cohstLie}) are constructed by the exponential map of shift
operators that do not annihilate the extremal state, together with
their Hermitian-conjugates. Following this procedure, the
generalized coherent states $|\Lambda,\Omega\rangle$ can be
written in terms of a displacement operator in one-to-one
correspondence to the coset representatives $\Omega$ of $G/D$,
acting on the extremal state $|\Lambda,\Lambda\rangle$:
\be
\Omega
\, = \, \exp \left\{
\sum_{\beta}\eta_{\beta}E_{\beta}-\eta_{\beta}^{*}E_{-\beta}
\right\} \; ,
\ee
where $\eta_{\beta}$ are complex parameters, and
the sum is restricted to those shift operators which do not
annihilate the extremal state. We review the procedure for the
trivial case of the harmonic-oscillator coherent states. We know
that a single-mode coherent state is realized during the time
evolution generated by the one-photon Hamiltonian applied to the
initial vacuum:
\be
H(t) \, = \, \omega a^{\dag}a \, + \,
\kappa(t)a^{\dag} \, + \, \kappa^{*}(t)a \; .
\label{cohstgenH}
\ee
The evolved state is the standard Glauber coherent state (see
Section \ref{section2}). The Hamiltonian (\ref{cohstgenH}) is a
linear combination of the harmonic oscillator operators $a$,
$a^{\dag}$, and $n$, which, together with the identity operator
$I$, span the Lie algebra denoted as $h_{4}$; the corresponding
Lie group, obtained from the algebra through the operation of
exponentiation, is the Heisenberg-Weyl group $H_{4}$. We consider
as extremal state the vacuum $|0\rangle$. The stability subgroup
which leaves $|0\rangle$ invariant is $U(1)\otimes U(1)$, whose
elements take the form $d=e^{i\xi n+i\phi I}$. A representative of
the coset space $H(4)/(U(1)\otimes U(1))$ is the Glauber
displacement operator $D(\alpha)=e^{\alpha a^{\dag}-\alpha^{*}a}$.
Therefore the Glauber coherent states can be obtained by the
action on the extremal state $|0\rangle$ of the element
$\mbf{g}\in H_{4}$, $\mbf{g}=D(\alpha)d$:
\be
\mbf{g}|0\rangle \,
= \, D(\alpha)d|0\rangle \, = \, e^{i\phi}|\alpha\rangle \; .
\ee
Another simple case of generalized coherent state is the
single-mode squeezed state that is generated during the time
evolution driven by the positive definite quadratic Hamiltonian
(\ref{quadH}) which is a linear combination of the operators
spanning the algebra $h_{6}=\left\{n,a,a^{\dag},a^{2},a^{\dag
2},I\right\}$, a subalgebra of the symplectic algebra $sp(4)$. The
evolution operator can be disentangled by using the equivalent
matrix representations of the group $H_{6}$. In analogy with the
case of the Heisenberg-Weyl group, the coherent states of $H_{6}$
are obtained by choosing the vacuum state as extremal state, and
by considering a representative in the coset space
$H_{6}/(U(1)\otimes U(1))$, that is the product
$D(\alpha)S(\varepsilon)$ of the displacement and squeezing
operators. The single-mode algebra $h_{6}$ can be extended to the
$n$-mode case \cite{zhangfeng}, by considering the subgroup of
$sp(2n+2)$, generated by the operators
$\left\{a_{i}^{\dag}a_{j}^{\dag},a_{i}^{\dag}a_{j}+
\frac{1}{2}\delta_{ij},a_{i}a_{j},a_{i}^{\dag},a_{j},I\right\}$,
which are the independent elements needed to construct the
multimode generalization of Hamiltonian (\ref{quadH}):
\bea H \,
&&= \,
\sum_{i}\omega_{i}\left(a^{\dag}_{i}a_{i}+
\frac{1}{2}\right)+\sum_{ij}(p_{ij}\,a_{i}^{\dag}a_{j}^{\dag}+h.c.)
\nonumber \\
&& \nonumber \\
&&+\sum_{ij}r_{ij}\left(a_{i}^{\dag}a_{j}+
\frac{1}{2}\delta_{ij}\right)+\sum_{i}(s_{i}a^{\dag}_{i}+h.c.)
\; .
\eea
More recently, the formalism of {\it nonlinear}
algebras has been applied to multiphoton optical
Hamiltonians associated to symmetries which can be described
by polynomially deformed $SU(1,1)$ and $SU(2)$ algebras \cite{karassiov}.
A general approach to construct the multiphoton coherent states of such
algebras has been introduced in Ref. \cite{nonlinalgeb}. Here we
want to briefly recall the basic concepts of this method. In the Cartan
basis, a deformation of a Lie algebra is obtained by replacing Eq.
(\ref{cartanE+E-}) with $[E_{\alpha},E_{-\alpha}]=f(H_{i})$, where
$f(H_{i})$ is a polynomial function of $H_{i}$. In particular, the
polynomial deformation of the Jordan-Schwinger realizations of the
$SU(2)$ and $SU(1,1)$ algebras takes the form
\be
[J_{0},J_{\pm}] \, = \, \pm J_{\pm} \; , \; \quad
[J_{+},J_{-}] \, = \, F(J_{0}) \, = \, \sum_{i=0}^{n}C_{i}J_{0}^{i} \; ,
\ee
where $J_{0}$ is the diagonal operator, $J_{\pm}$ are the scaling
operators, and the coefficients $C_{i}$ are real constants.
Interesting cases are the quadratic algebra ($F(J_{0})$ quadratic
in $J_{0}$) and  the cubic, or Higgs, algebra ($F(J_{0})$ cubic in
$J_{0}$). For example, the coherent states associated with the
trilinear Hamiltonian (\ref{trilinearHsu11}) have been defined by
using the generators of the polynomial quadratic algebra,
$J_{0}=\frac{1}{2}(a^{\dag}a-K_{0})$,
$J_{+}=a^{\dag}K_{-}=J_{-}^{\dag}$ \cite{nonlinalgeb}. The
spectrum and the eigenfunctions of the general two-mode
multiphoton Hamiltonian
\be
H_{Hig} \, = \,
\omega_{1}a_{1}^{\dag}a_{1}+\omega_{2}a_{2}^{\dag}a_{2}+\kappa
a_{1}^{\dag 2}a_{2}^{2}+\kappa^{*}a_{1}^{2}a_{2}^{\dag 2} \; ,
\ee
can be found \cite{Debergh} by exploiting the Higgs
algebra, generated by the operators $J_{+}=a_{1}^{\dag
2}a_{2}^{2}=J_{-}^{\dag}$, and
$J_{0}=\frac{1}{4}(a_{1}^{\dag}a_{1}-a_{2}^{\dag}a_{2})$. \\
Another class of nonlinear coherent and squeezed states associated
with a simple nonlinear extension of the harmonic-oscillator
algebra can be obtained as follows. Given a well defined,
Hermitian function $f(n)$ of the number operator $n$, let us
define the pair of lowering and raising operators
\be
A \, = \, a
f(n) \; , \; \quad A^{\dag} \, = \, f(n)a^{\dag} \; , \ee which
satisfy the following commutation relations \bea &&[n,A]=-A \; ,
\; \quad [n,A^{\dag}]=A^{\dag} \; ,  \nonumber \\
&& \nonumber \\
&&[A,A^{\dag}] \, = \, (n+1)f(n+1)^{2}-nf(n)^{2} \; .
\eea
Nonlinear coherent and squeezed states can be constructed either
as eigenstates of the generalized annihilation operator $A$
\cite{nlchst1,nlchst2}, namely
\be
A|\gamma,f\rangle \, = \, a
f(n)|\gamma,f\rangle \, = \, \gamma|\gamma,f\rangle \; ,
\ee
or by
the action of a displacement-type operator on the vacuum
\cite{nlchst3,nlchst4}. In this case, we first need the definition
of a dual algebra by defining the canonical conjugate $B^{\dag}$
of $A$:
\be
B^{\dag} \, \doteq \, \frac{1}{f(n)}a^{\dag} \; , \;
\; \quad [A,B^{\dag}]=1 \; .
\ee
Nonlinear coherent and squeezed
states are then defined as
\bea
&&|\zeta,f\rangle \, = \, e^{\zeta
A^{\dag}-\zeta^{*}B}|0\rangle \; , \; \; \quad
|\tilde{\zeta},f\rangle \, = \, e^{\tilde{\zeta}
B^{\dag}-\tilde{\zeta}^{*}A}|0\rangle \; , \\
&& \nonumber \\
&&|\xi,f\rangle \, = \, e^{\frac{1}{2}(\xi A^{\dag
2}-\xi^{*}B^{2})}|0\rangle \; , \; \; \quad
|\tilde{\xi},f\rangle \, = \, e^{\frac{1}{2}(\tilde{\xi} B^{\dag
2}-\tilde{\xi}^{*}A^{2})}|0\rangle \; .
\eea
An extension of this approach that yields a larger class of
multiphoton coherent state was formulated by Shanta, Chaturvedi,
Srinivasan, and Agarwal \cite{srinivasanMCS}. They developed a
method to derive the eigenstates of a generalized (multiphoton)
annihilation operator $F$ consisting of products of annihilation
operators and of functions of the number operators. The method is
based on constructing a family of operators $\{ G_{i}^{\dag} \} $
such that $[F,G_{i}^{\dag}]=1$. Let $|v\rangle_{i}$, $i=0,1,...$,
denote the states annihilated by $F$, and let $\{ S_{i} \}$ the
collection of subspaces of states realized by repeated application
of the operator $F^{\dag}$ on $|v\rangle_{i}$. The Fock space is
then decomposed into the mutually orthogonal sectors $S_{i}$. Let
$G_{i}^{\dag}$ be an operator such that $[F,G_{i}^{\dag}]=1$ holds
in the sector $S_{i}$. The conjugate $G_{i}$ annihilates the same
states as $F$: $G_{i}|v\rangle_{i}=0$ $\forall$ $i$. The following
states
\be
|f\rangle_{i} \, = \, e^{f G_{i}^{\dag}}|v\rangle_{i}
\; , \; \; \quad |g\rangle_{i} \, = \, e^{gF^{\dag}}|v\rangle_{i}
\ee
are eigenstates of $F$ and $G_{i}$, respectively, i.e.
$F|f\rangle_{i}=f|f\rangle_{i}$ and
$G_{i}|g\rangle_{i}=g|g\rangle_{i}$, and they satisfy the relation
$_{i}\langle f|g\rangle_{k}=e^{f^{*}g}\delta_{ik}$. For
simplicity, we consider the single-mode case with $F$ of the form
$f(a^{\dag}a)a^{p}$ (with $f(x)\neq 0$ at $x=0$ and at positive
integer values of $x$). The vacua $|v\rangle_{i}$ coincide with
the Fock states $|i\rangle$, $i=0,1,...,(p-1)$, and the sectors
$S_{i}$ are built out of these vacua by repeated applications of
$F^{\dag}$: $S_{i}=\{|pn+i\rangle\}$, ($n=0,1,...$). The operators
$G_{i}^{\dag}$ can be explicitly constructed
\bea
G_{i}^{\dag} \,
&=& \, \frac{1}{p}[a^{\dag}a-i]F^{\dag}\frac{1}{FF^{\dag}}
\nonumber \\
&& \nonumber \\
&=& \, \frac{1}{p}F^{\dag}\frac{1}{FF^{\dag}}[a^{\dag}a+p-i] \, =
\, G_{0}^{\dag}\frac{a^{\dag}a+p-i}{a^{\dag}a+p} \; .
\eea
A
clarifying example is obtained in the particular case $F=a^{2}$
$(f=1\,,\;\;p=2)$. The two vacua of $F$ are the states $|0\rangle$
and $|1\rangle$ and the corresponding sectors of the Fock space
are $S_{0}=\{|2n\rangle\}_{n=0}^{\infty}$ and
$S_{1}=\{|2n+1\rangle\}_{n=0}^{\infty}$, respectively, the sets of
even and odd number states. The canonical conjugates of $F$ in the
two sectors are
\be
G_{0}^{\dag} \, = \, \frac{1}{2}a^{\dag
2}I_{a} \; , \; \; \quad G_{1}^{\dag} \, = \,
\frac{1}{2}a^{\dag}I_{a}a^{\dag} \; ,
\ee
where
$I_{a}=\frac{1}{1+a^{\dag}a}$. Consequently, the eigenstates of
$F$ are
\be
|f\rangle_{0} \, = \, e^{\frac{f}{2}a^{\dag
2}I_{a}}|0\rangle \; , \; \; \quad |f\rangle_{1} \, = \,
e^{\frac{f}{2}a^{\dag}I_{a}a^{\dag}}|1\rangle \; ,
\ee
namely,
linear combinations of the even and odd Fock states, respectively.
The corresponding eigenstates of $G_{0}$ and $G_{1}$ are \be
|g\rangle_{0} \, = \, e^{g a^{\dag 2}}|0\rangle \; , \; \; \quad
|g\rangle_{1} \, = \, e^{g a^{\dag 2}}|1\rangle \; . \ee The state
$|g\rangle_{0}$ with $|g|\leq 1$ is the squeezed vacuum, while
$|g\rangle_{1}$ is a squeezed number state. This scheme allows for
a systematic construction of the eigenstates of products of
annihilation operators.

\subsection{Hamiltonian models of higher-order nonlinear processes}

We will now introduce the study of phenomenological theories of
$k$-photon parametric amplification, for arbitrary $k$, based on
the quantization of expression (\ref{Xn})
\cite{FisherNieto,BraunMcLac,BraunCav,Elyutin,DrobnyJex1,DrobnyJex2,TanasGantsog,multiphrasetti1,multiphrasetti2,kphTombMec,kphHillery,BuzekDrobny,DrobnyBuzek}.
\subsubsection{Nondegenerate, fully quantized $k$-photon down conversion}

Already in 1984, a dissipative multimode parametric amplifier  was
studied by Graham \cite{Graham} to derive a general relation
between the intensity cross-correlation function and the intensity
autocorrelations for any pair of simultaneously excited modes. The
nondegenerate, $k$-photon parametric amplifier is described by the
interaction Hamiltonian of the general form
\be
H_{I}^{ndg,kph} \,
= \, \kappa^{(k)}a_{1}^{\dag}a_{2}^{\dag}\cdots a_{k}^{\dag} b \,
+ \, H.c. \; ,
\label{qndegkparamp}
\ee
with real $\kappa^{(k)}$
and under condition of approximate resonance
$|\omega_{b}-\sum_{i=1}^{k}\omega_{i}|\ll\omega_{j}$,
$j=1,\ldots,k$. In the absence of losses, by using the conserved
quantities $D_{ij}=a_{i}^{\dag}a_{i}-a_{j}^{\dag}a_{j}$, it can be
shown that the intensities of pairs of excited modes are maximally
correlated at all times, namely $\langle
a_{i}^{\dag}a_{i}\rangle=\langle a_{j}^{\dag}a_{j}\rangle$ and
$\langle a_{i}^{\dag}a_{j}^{\dag}a_{i}a_{j}\rangle=\langle
a_{i}^{\dag}a_{i}^{\dag}a_{i}a_{i}\rangle+\langle
a_{i}^{\dag}a_{i}\rangle$ $(i\neq j)$. If finite losses are
considered, solution of the master equation in the steady state
yields
\bea
\eta_{i}\langle a_{i}^{\dag}a_{i}\rangle \, &=& \,
\eta_{j}\langle a_{j}^{\dag}a_{j}\rangle
\label{graham1} \\
&& \nonumber \\
\langle a_{i}^{\dag}a_{j}^{\dag}a_{i}a_{j}\rangle \, &=& \,
\frac{\eta_{i}}{\eta_{i} +\eta_{j}}\langle
a_{i}^{\dag}a_{i}^{\dag}a_{i}a_{i}\rangle
+\frac{\eta_{j}}{\eta_{i}+\eta_{j}}\langle
a_{j}^{\dag}a_{j}^{\dag}a_{j}a_{j}\rangle\nonumber \\ && \nonumber
\\ &+&\frac{\eta_{i}}{\eta_{i}+\eta_{j}}\langle
a_{i}^{\dag}a_{i}\rangle \,, \quad\quad i\neq j \,,
\label{graham2}
\eea
where $\eta_{i}$ is a phenomenological
constant relative to the mode $a_{i}$. Since the total power
$P_{i}$ extracted from the mode $a_{i}$ is given by
$P_{i}=2\eta_{i}\omega_{i}\langle a_{i}^{\dag}a_{i}\rangle$, Eq.
(\ref{graham1}) implies the Manley-Rowe's relation \cite{ManlRowe}
\be
\frac{P_{1}}{\omega_{1}} \, = \, \frac{P_{2}}{\omega_{2}}\, =
\, \ldots \, \frac{P_{k}}{\omega_{k}} \; .
\ee
Eqs. (\ref{graham2}) show the presence of strong correlations between
excited modes due their joint, simultaneous generation by the same
quantum process. It is quite evident that, for reasons of
convenience and practicality, most of the studies are devoted to
the simplest case of complete degeneracy, described by the
Hamiltonian
\be
H_{I}^{dg, kph} \, = \, \kappa^{(k)}a^{\dag k} b \,
+ \, H.c. \; ,
\label{qdegkparamp}
\ee
where $\kappa^{(k)}$ is
proportional to the $k$-th order nonlinearity. After these
preliminary considerations on the most general, nondegenerate,
fully quantized $k$-photon interaction, in the following we will review more
in detail the approaches and the methods used to study the
dynamics of the degenerate $k$-photon amplifier, either in the
fully quantized version or in the parametric approximation for the
pump mode.

\subsubsection{Degenerate $k$-photon down conversion with classical pump}

In 1984, Fisher, Nieto, and Sandberg \cite{FisherNieto} claimed
the impossibility of generalizing the squeezed coherent states by
means of generalized $k$-photon ($k>2$), unitary ``squeezing''
operators of the form:
\be
U_{k} \, = \, \exp \{ i A_{k} \} \, =
\, \exp \{ z_{k}a^{\dag k} \, - \, z_{k}^{*}a^{k} + \, i h_{k-1}
\} \; , \; \; \; k > 2 \; ,
\label{UkFisher}
\ee
where $h_{q}$ is
a Hermitian operator polynomial in $a$ and $a^{\dag}$ with degree
$q$. In particular, they considered the vacuum expectation value
of $U_{k}$ with $h_{q}=0$ and showed that this operator is
unbounded. In fact, expanding the exponential of Eq.
(\ref{UkFisher}) in a power series and taking the expectation
value in the vacuum state, it results
\bea
\langle 0|
U_{k}|0\rangle\, &=& \,
1-|z_{k}|^{2}\frac{k!}{2!}+|z_{k}|^{4}\frac{1}{4!}[(k!)^{2}+(2k)!]
\nonumber \\
&& \nonumber \\
&+&...+(-1)^{n}|z_{k}|^{2n}\frac{1}{(2n)!}C_{n}+... \; ,
\label{0U0series}
\eea
where $C_{n}$ is formed of positive terms.\\
The largest one
is of order $(kn)!$, so that
$\lim_{n\rightarrow\infty}|z_{k}|^{2n}\frac{1}{(2n)!}C_{n}\neq 0$
for all $k>2$ and $z_{k}\neq 0$. The terms of large order $2n$ in
the series with alternating signs (\ref{0U0series}) are bounded from
below by $(kn)!/(2n)!$,
which has no convergent limit for $n\rightarrow\infty$ when $k>2$.
In conclusion, the series is divergent, and $|0\rangle$ is not an
analytic vector of the generator $A_{k}$ in the Fock space.
Therefore, a "naive" generalization of one-photon coherent states and
two-photon squeezed states to many-photon generalized squeezed and coherent
states appears to be impossible.

\begin{figure}[h]
\begin{center}
\includegraphics*[width=13cm]{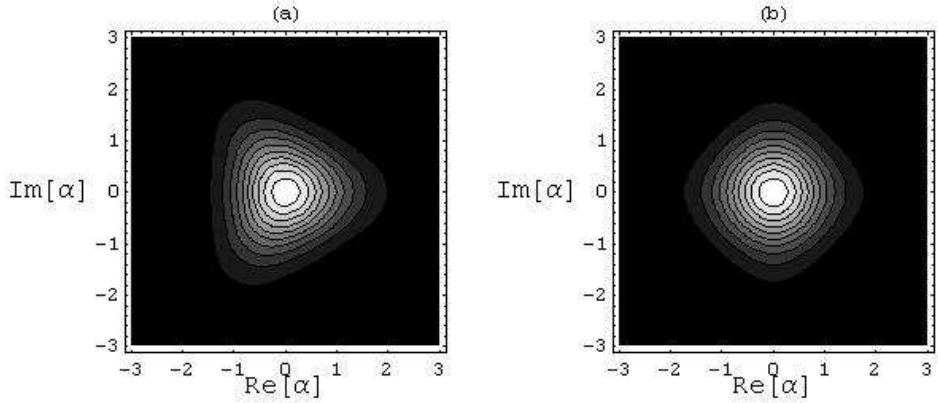}
\end{center}
\caption{Contour plots of the $Q$-function for the initial vacuum
state and for $(a)$ $k=3$, $r=0.1$; $(b)$ $k=4$, $r=0.025$.}
\label{ContourQBraunCav}
\end{figure}

The problem raised in Ref. \cite{FisherNieto} inspired the
introduction of new types of generalized $k$-photon
squeezed coherent states \cite{multiphrasetti1,multiphrasetti2}, whose
definition is based on the generalized multiphoton operators of
Brandt and Greenberg \cite{brandtgreenb}, including those defined
for Holstein-Primakoff realizations \cite{holstprim} of
$SU(2)$, $SU(1,1)$ and $SU(n)$ \cite{multiphrasetti2}.\\
However, the problem of the divergence of the vacuum to vacuum matrix
element (\ref{0U0series}) can be approached by more standard
methods, and has in fact been solved by Braunstein and McLachlan
\cite{BraunMcLac}. They studied the matrix element by numerical
techniques and, using Pad\'{e} approximants, obtained good
convergence of the asymptotic series.
They considered the degenerate multiphoton parametric amplifier
(\ref{qdegkparamp}) that, in the Schr\"{o}dinger
picture, for $k$-photon processes, and within the parametric approximation
takes the form
\be
H_{P}^{dg,kph} \, = \, \omega a^{\dag}a \, + \, i[z_{k}(t)a^{\dag k}-z_{k}^{*}(t)a^{k}]
\; ,
\label{Hkdegampl}
\ee
where $z_{k}(t)=|z|e^{i(\phi-k\omega
t)}$. In the interaction picture, the time evolution operator
corresponding to Eq.~(\ref{Hkdegampl}) reads
\be
U_{k}(z,z^{*};t) \, = \, \exp \{(za^{\dag k} - z^{*}a^{k})t \} \; ,
\label{Hktevol}
\ee
where $z=|z|e^{i\phi}$. The properties of the states generated
by the time evolution Eq.~(\ref{Hktevol}) can be studied using the
Husimi function
$Q(\alpha)=Tr[\rho(t)\delta(a-\alpha)\delta(a^{\dag}-\alpha^{*})]$,
where $\rho(t)=U_{k}(t)\rho(0)U_{k}^{\dag}(t)$. If the initial state
is the vacuum $\rho(0)=|0\rangle\langle 0|$, $Q$ is invariant under
a rotation of $2\pi/k$, leading to a  $k$-fold symmetry in phase
space. Assuming a real $z$, defining a scaled time $r=|z|t$, and exploiting
the equations of motion for the density operator in the interaction picture
and standard operatorial relations,
the evolution equation for the $Q-$function reads
\be
\frac{\partial
Q}{\partial r} \, = \, L Q \, \equiv \,
\left\{\alpha^{k}-\left(\alpha+\frac{\partial}{\partial
\alpha^{*}}\right)^{k}+\alpha^{*k}-\left(\alpha^{*}+\frac{\partial}{\partial
\alpha}\right)^{k}\right\}Q \; ,
\label{liouvill}
\ee
where $L$ is
the Liouvillian operator. The general solution of
Eq.~(\ref{liouvill}) is of the form
\be
Q(r) \, = \, \exp \{rL\}
Q(0) \, = \, Q(0) \, + \, L Q(0)r \, + \,
\frac{1}{2}L^{2}Q(0)r^{2} \, + \, ... \; ,
\label{asympser}
\ee
where $Q(0)$ is the initial value of the $Q-$function. Being that
the series (\ref{asympser}) is in fact asymptotic, it can be
truncated in order to determine the small-time behavior. For the
initial vacuum state $\rho(0)=|0\rangle \langle 0|$, this
truncation yields
\bea
Q(r) \, &\sim& \,
e^{-|\alpha|^{2}}[1+r(\alpha^{k}+\alpha^{*k})] \, + \,
\mathcal{O}(r^{2}) \nonumber \\
&& \nonumber \\
&=& \, e^{-|\alpha|^{2}}[1+2r|\alpha|^{k}\cos(k\theta)] \, + \,
\mathcal{O}(r^{2}) \; ,
\eea
where $\alpha=|\alpha|e^{i\theta}$.
For $r\ll 1$ the $Q-$function exhibits $k$ lobes along the
directions $\theta=0,\frac{2\pi}{k},...,\frac{2\pi(k-1)}{k}$,
showing a strong nonclassical behavior even for short times. In
Fig. (\ref{ContourQBraunCav}) we show contour-plots of the
$Q$-function corresponding to $U_{3}$ and $U_{4}$, with $r=0.1$
and $r=0.025$ respectively; the presence of the arms, revealing
the symmetry of the system, can be observed already for short
times.
There is no analytic technique to study the full dynamics at all
times; one possible method is to expand the $Q$-function in a
power series in terms of Fock space matrix elements of the
evolution operator:
\be
Q(r) \, = \, \exp \{-|\alpha|^{2}\} \left|
\sum_{n=0}^{\infty} \frac{\alpha^{* n}}{\sqrt{n!}}\langle n|
U_{k}(r)|0\rangle  \right|^{2} \; .
\label{urka}
\ee
The problem
is then further reduced because the matrix elements are
nonvanishing only for $n$ multiple of $k$, $n=mk$. In this case
one has
\bea
&&\langle
k|U_{k}(r)|0\rangle=\frac{1}{\sqrt{k!}}\frac{d}{dr}\langle
0|U_{k}(r)|0\rangle \,, \nonumber \\
&& \nonumber \\
&&\langle
2k|U_{k}(r)|0\rangle=\frac{1}{\sqrt{(2k)!}}\left(k!+
\frac{d^{2}}{dr^{2}}\right)\langle
0|U_{k}(r)|0\rangle \,, \;.... \,.
\eea
The divergence of the
vacuum-to-vacuum matrix element is due to the singular behavior on
the imaginary time axis, since the Taylor series converges only up
to the nearest pole. An analytic continuation can be obtained
using the Pad\'{e} approximants \cite{Padè} which reproduce the
pole structure that limits the convergence of the Taylor series.
Therefore, one can conclude, following Braunstein and McLachlan,
that the generalization of one-photon coherent states and
two-photon squeezed states to the many-photon case is possible,
and that the
resulting non Gaussian states show evident nonclassical features.\\
Successively, Braunstein and Caves \cite{BraunCav} have studied in
detail the statistics of direct, heterodyne and homodyne
detection for the generalized squeezed states associated with the
cubic and quartic interaction Hamiltonians
\bea
H_{P}^{dg,3ph} &=&i\kappa(a^{\dag 3}-a^{3}) \; , \label{BrauCavH3} \\
&& \nonumber \\
H_{P}^{dg,4ph} &=&i\kappa(a^{\dag 4}-a^{4}) \; , \label{BrauCavH4}
\eea
($\kappa$ real), and to the interaction Hamiltonian introduced by
Tombesi and Mecozzi \cite{kphTombMec}
\be
H_{P}^{dg,(2-2)ph} \, = \, \kappa[i(a^{\dag 2}-a^{2})]^{2} \,
= \, -\kappa(a^{\dag 4} + a^{4}) \, +
\, \kappa(a^{\dag 2}a^{2}+a^{2}a^{\dag 2})
\; .
\label{TombMecH22}
\ee
The Hamiltonians $H_{P}^{dg,3ph}$ and $H_{P}^{dg,4ph}$ can be studied
by numerical methods, while the Hamiltonian $H_{P}^{dg,(2-2)ph}$
describes an exactly solvable model
\cite{kphTombMec} based on both four-photon and Kerr processes in
nonlinear crystals without inversion center. It can be obtained by
the model Hamiltonian
\be
H_{M} \, = \, \omega a^{\dag}a+\gamma^{(3)}(a^{\dag
2}a^{2}+a^{2}a^{\dag 2})+\gamma^{(4)}[E^{*}(t)a^{4}+E(t)a^{\dag 4}]
\; ,
\label{TombMecH22model}
\ee
where $\gamma^{(3)}$ and
$\gamma^{(4)}$ are proportional, respectively, to the third- and
fourth-order susceptibility tensors, and $E(t)=E e^{i\phi-4\omega
t}$ is a classical external pump field. In the interaction
picture, setting the phase of the pump $\phi=\pi$ and choosing the
amplitude $E$ such that $E=\gamma^{(3)}/\gamma^{(4)}$, the
Hamiltonian (\ref{TombMecH22model}) leads to
Eq.~(\ref{TombMecH22}), defining $\kappa=\gamma^{(4)}E$. The time
evolution operators corresponding to the Hamiltonians of
Eqs.~(\ref{BrauCavH3}), (\ref{BrauCavH4}) and (\ref{TombMecH22})
are respectively
\bea
&&U_{3}(r) \, = \, e^{r(a^{\dag 3}-a^{3})} \; , \label{BCU3} \\
&& \nonumber \\
&&U_{4}(r) \, = \, e^{r(a^{\dag 4}-a^{4})} \; , \label{BCU4} \\
&& \nonumber \\
&&U_{2,2}(r) \, = \, e^{ir(a^{\dag 2}-a^{2})^{2}} \; ,
\label{BCU22}
\eea
where $r=\kappa t$ is a scaled time. The
evolution operator $U_{2,2}(r)$ can be written as a Gaussian
average over the analytic continuation of the quadratic evolution
operator \cite{kphTombMec}
\be
U_{2,2}(r) \, = \,
\int_{-\infty}^{\infty}d\xi\frac{1}{\sqrt{\pi}}
e^{-\xi^{2}}e^{2\sqrt{ir}\xi(a^{\dag 2}-a^{2})} \; .
\ee
In this
case, the vacuum-to-number-state matrix elements are given by
\bea
&&\langle n|U_{2,2}(r)|0\rangle =\nonumber \\
&& \nonumber \\
&& \pi^{-1/2}\sqrt{n!}[(n/2)!]^{-1}2^{-n/2}\int_{-\infty}^{\infty}
d\xi e^{-\xi^{2}} \frac{\left( \tanh \tau \right)^{n/2}}{\left(
\sqrt{\cosh\tau} \right)} \left|_{\tau=4\sqrt{ir\xi}} \right.
\quad n \; even ,
\eea
and $0$ for $n$ odd. The statistics of
direct, heterodyne, and homodyne detection correspond,
respectively, to determine the photon number distribution, the
$Q$-function, and the projections of the Wigner function
\cite{BraunCav}. For the Hamiltonian $H_{P}^{dg,3ph}$, the photon
number distribution shows photon triplets, while for
$H_{P}^{dg,4ph}$ and $H_{P}^{dg,(2-2)ph}$ it shows photon
quadruplets. In the same way, the $Q$-function exhibits the
characteristic shape with three and four arms (lobes). Finally,
the Wigner function shows interference fringes due to coherent
superposition effects in phase space.

\subsubsection{Degenerate $k-$photon down-conversion with quantized pump}

\begin{figure}[h]
\begin{center}
\includegraphics*[width=9cm]{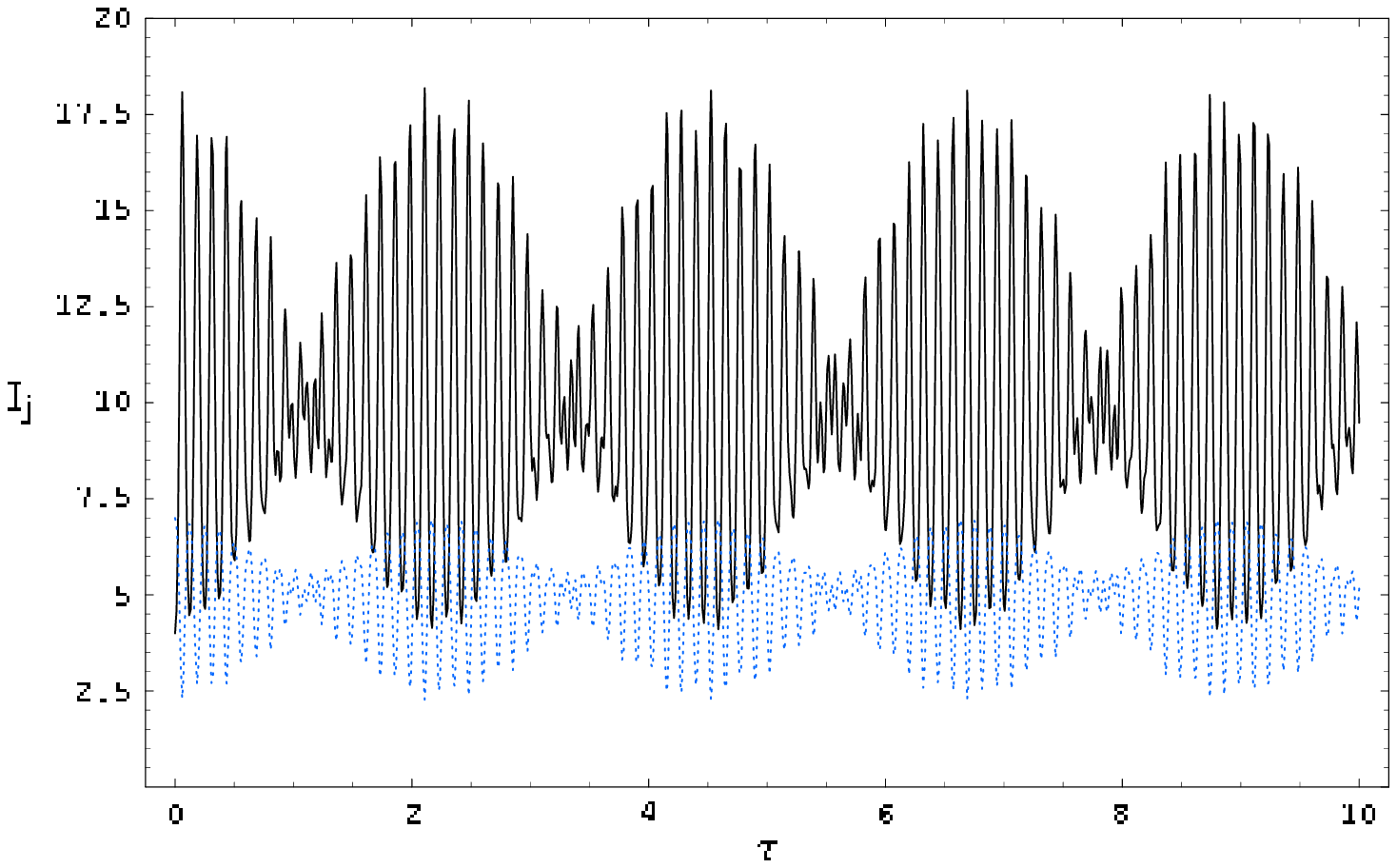}
\end{center}
\caption{$I_{j}(\tau)$ as a function of the dimensionless time $\tau$ for
an initial Fock state $|4,7\rangle$, nonlinearity of order $k=3$,
and constant of the motion $N_{i}=25$. The
upper curve (full line) corresponds to $I_{a}(\tau)$ in the degenerate
mode $a$; the lower curve
(dotted line) corresponds to $I_{b}(\tau)$ in the quantized pump mode $b$.
The beats' envelope and the sequence of collapses and revivals
are clearly observable.} \label{collrev}
\end{figure}

\begin{figure}[h]
\begin{center}
\includegraphics*[width=15cm]{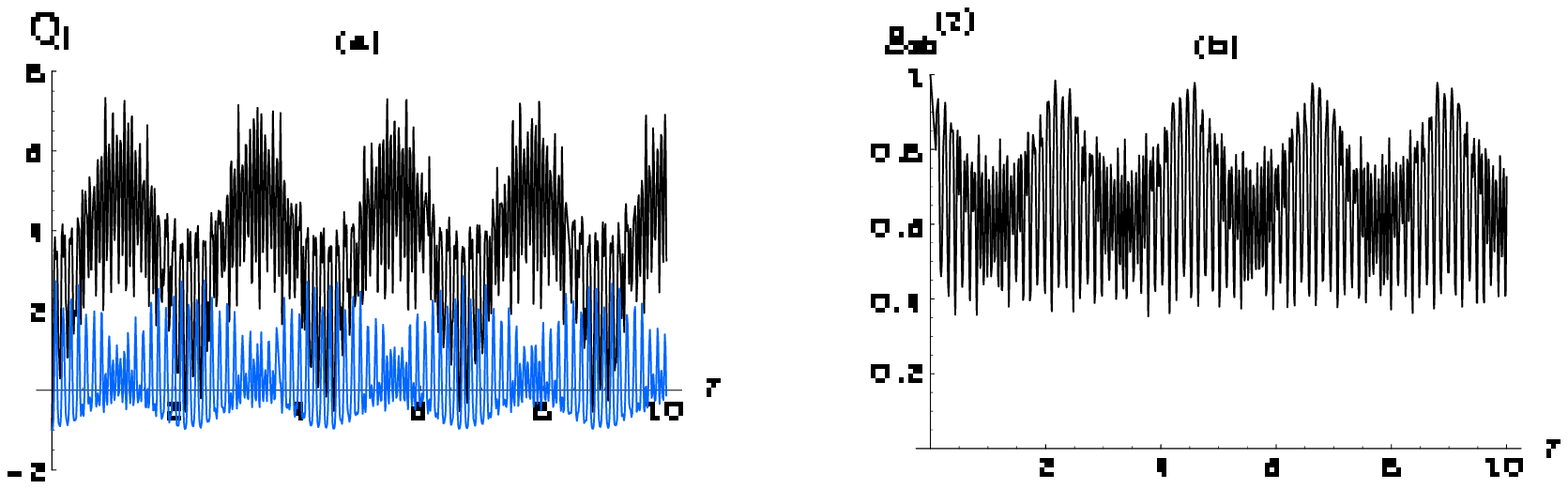}
\end{center}
\caption{Figure (a): $Q$ parameters of mode $a$ and mode $b$ as functions of
the rescaled, dimensionless time $\tau$
for the initial Fock state $|4,7\rangle$, $k=3$, and $N_{i}=25$.
The upper curve (full line) corresponds to $Q_{a}(\tau)$, the
lower (dotted line) to $Q_{b}(\tau)$. Strong oscillations between
super- and sub-Poissonian statistics
in mode $b$ are clearly observable.
Figure (b): Cross-correlation
$g^{(2)}_{ab}$ as a function of the rescaled, dimensionless
time $\tau$, for the same initial state.
Anticorrelation of the modes is observed at all times.}
\label{Qg2ab}
\end{figure}

\begin{figure}[h]
\begin{center}
\includegraphics*[width=8cm]{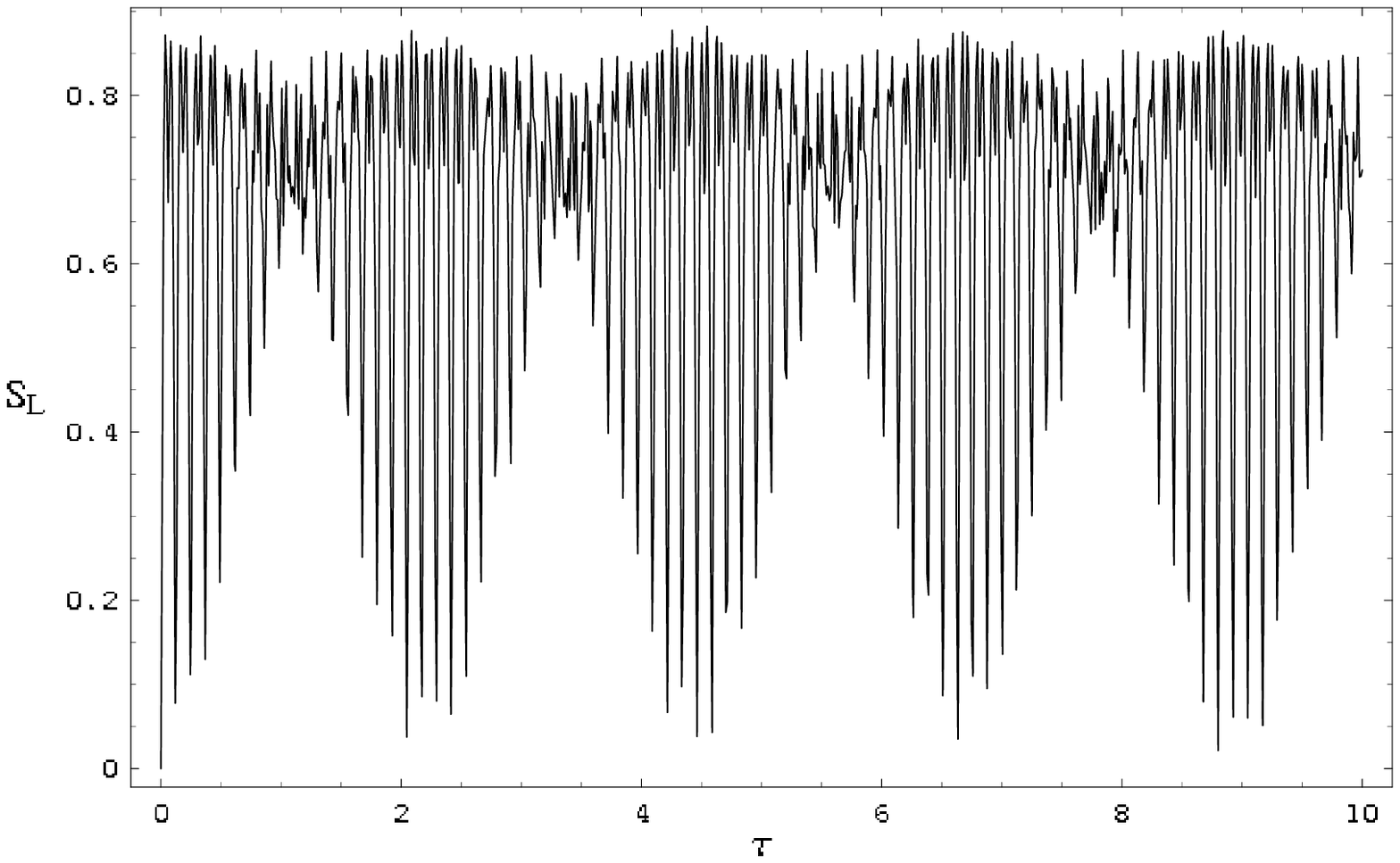}
\end{center}
\caption{Linear entropy $S_{L}$ as a function of the scaled, dimensionless
time $\tau$ for the initial factorized two-mode Fock state $|4,7\rangle$, and
nonlinearity of order $k=3$. The entanglement is
highly oscillatory, but remains finite at all times $\tau >0$.}
\label{3DCentFock}
\end{figure}

We now turn to the properties of $k-$photon down-conversion with
quantized pump in a high-\textit{Q} cavity in the presence of a
Kerr-like medium. Among others, the collapses and revival
phenomenon in the energy exchange of two field modes
\cite{DrobnyJex1}, the statistics of the process
\cite{DrobnyJex2}, the phase properties \cite{TanasGantsog}, the
signal-pump entanglement \cite{DrobnyBuzek}, and the limit on the
energy transfer between the modes \cite{DrobnyBuzek} have been
studied. In the rotating wave approximation, the process of
$k-$photon down conversion, with $k\geq3$, can be described,
including the free parts and the Kerr terms, by the effective
total Hamiltonian
\bea
H_{Q}^{dg,kph} \, &=& \, \omega_{a}
a^{\dag}a+\omega_{b} b^{\dag}b+\lambda_{k}(a^{k}b^{\dag}+a^{\dag
k}b) \nonumber \\
&& \nonumber \\
&+&g_{a}a^{\dag 2}a^{2}+g_{b}b^{\dag
2}b^{2}+g_{ab}a^{\dag}ab^{\dag}b \; ,
\label{Hk-phDC}
\eea
where
exact resonance $\omega_{b}=k\omega_{a}$ is assumed, and
$\lambda_{k}$, $g_{a}$, and $g_{b}$ are real constants, with
$\lambda_{k}\propto \chi^{(k)}$, and the $g_{i}s \propto
\chi^{(3)}$. The time evolution generated by Hamiltonian
(\ref{Hk-phDC}) is not plagued by the divergences that appear in
the parametric approximation studied in the previous paragraph.
The integral of motion
\be
N^{(k)} \, = \, a^{\dag}a \, + \, k
b^{\dag}b
\ee
decomposes the Hilbert space of the whole system
$\mathcal{H}=\mathcal{H}_{a}\otimes\mathcal{H}_{b}$ in the direct
sum $\mathcal{H}=\oplus_{i}\mathcal{H}_{i}$, where each subspace
$\mathcal{H}_{i}$, associated to a fixed, positive integer value
$N_{i}$ ($N_{i}=0,...,\infty$) of $N^{(k)}$, is defined as the set
of all state vectors that can be expressed as linear combinations
of the basis vectors $\{|N_{i}-nk,n\rangle\;,n=0,...,[N_{i}/k]\}$,
where $|N_{i}-nk,n\rangle$ is a compact notation for the two-mode
number state $|N_{i}-nk\rangle_{a}|n\rangle_{b}$, and $[x]$ stands
for the integer part of $x$. The dimension of the subspace
$\mathcal{H}_{i}$ is $[N_{i}/k]+1$. This decomposition of the
Hilbert space enables to diagonalize the Hamiltonian on each
subspace $\mathcal{H}_{i}$, with eigenvalue equation
\be
H_{Q}^{dg,kph}|\Psi_{\gamma,N_{i}}\rangle \, = \,
E_{\gamma,N_{i}}|\Psi_{\gamma,N_{i}}\rangle \; ,
\ee
where
$\gamma=0,...,[N_{i}/k]$. In terms of the eigenvalues and
eigenvectors, the evolution operator of the system can be written
in the form
\be
U(t) \, = \, e^{-itH_{Q}^{dg,kph}} \, = \,
\sum_{N_{i}=0}^{\infty}\sum_{\gamma=0}^{[N_{i}/k]}
e^{-itE_{\gamma,N_{i}}}|\Psi_{\gamma,N_{i}}\rangle\langle\Psi_{\gamma,N_{i}}|
\; ,
\ee
and the eigenstates $|\Psi_{\gamma,N_{i}}\rangle$ can be
expressed in the basis $\{|N_{i}-n k,n\rangle\}$ as
\be
|\Psi_{\gamma,N_{i}}\rangle \, = \, \sum_{m=0}^{[N_{i}/k]}\langle
N_{i}-m k,n|\phi_{\gamma,N_{i}}\rangle |N_{i}-mk,n\rangle  \; ,
\ee
so that, finally, given any initial state $|\Phi\rangle_{0}$
of the system, it is possible to reconstruct the density matrix
$\rho(t)$ and to analyze the statistical properties of the field
at all times. In the following analysis, all the Kerr terms in the
Hamiltonian (\ref{Hk-phDC}) will be put to zero for simplicity. On
the other hand, this choice will be physically justified {\it a
posteriori} in the case $k=3$ (possibly resonant). Moreover,
concerning the mathematical analysis of the problem with generic
$k$, such terms are harmless because they always commute with the
integral of motion $N^{(k)}$. Clearly, a convenient choice for
$|\Phi\rangle_{0}$ is the pure two-mode Fock state
$|N_{i}-kl,l\rangle$; however, a more realistic choice is the
state $|0\rangle_{a} |\beta\rangle_{b}$. Every initial state
$|\Phi\rangle_{0}$ can be expressed as
$|\Phi\rangle_{0}=\sum_{N_{i}=0}^{\infty}c_{N_{i}}|\Phi_{N_{i}}\rangle_{0}$,
where $|\Phi_{N_{i}}\rangle_{0}$ is the normalized projection of
the initial state on the subspace associated to the value $N_{i}$
of the integral of motion, and with real weight factor $c_{N_{i}}$
($\sum_{N_{i}=0}^{\infty}c_{N_{i}}^{2}=1$). It is worth noting
that the form of the initial state will strongly influence the
dynamics of the system. As a simple example, we will study the
time evolution driven by the three-photon down conversion  ($k=3$)
Hamiltonian (\ref{Hk-phDC}), without Kerr-like terms, of the
initial Fock state $|\Phi\rangle_{0}=|4,7\rangle$, and thus with a
constant of motion $N_{i}=25$. In Fig. (\ref{collrev}) we show the
time evolution of the mean intensity $I_{j}(\tau)$ $(j=a,b)$ in
the two modes as a function of the scaled, dimensionless time
$\tau=\lambda_{3}t$.
A very sharp sequence of collapses and revivals in the energy exchange between the field
and the pump modes is clearly observable. The introduction of the Kerr terms can lead to an
enhancement of the phenomenon; on the other hand, for too high
values of the Kerr couplings $g_{i}$, the collapse-revival behavior is lost
and replaced by a modulated oscillatory behavior. To establish the
statistical properties of the
fields, we can evaluate the Mandel's $Q$ parameter.
Fig. (\ref{Qg2ab}) (a) shows the evolution of the $Q$ parameter for
the same initial Fock state. We see that the pump mode $b$ exhibits
strong oscillations between the super-Poissonian (positive $Q$) and
the sub-Poissonian regime (negative $Q$), while the statistics
of mode $a$, although presenting fast oscillations, remains almost
always super-Poissonian.
However, it can be shown that for small values of $N_{i}$ the statistics in
both modes remains almost always sub-Poissonian, while for increasing $N_{i}$ the
statistics tends to lose its sub-Poissonian character in either mode.
Another interesting
physical feature can be investigated through the time evolution
of the cross-correlation
function $g_{ab}^{(2)}=\langle a^{\dag}a b^{\dag}b\rangle/\langle
a^{\dag}a\rangle\langle b^{\dag}b\rangle$. As shown in Fig.
(\ref{Qg2ab}) (b), $g_{ab}^{(2)}<1$ at all times, indicating a
permanent anticorrelation of the field and pump modes, so that
there is a strong tendency for photons in the two different modes
{\it not} to be created/detected simultaneously. When
$|\Phi\rangle_{0}=|0\rangle_{a}|\beta\rangle_{b}$ and the intensity
$|\beta|$ of the initial coherent state for mode $b$ is high enough,
the anticorrelation can be violated \cite{DrobnyJex2}.
The quantum dynamics realized by the
Hamiltonian (\ref{Hk-phDC}) is also able to produce
strong entangled states of
the signal and pump modes. To quantify quantum correlations in a two-mode
pure state (and, in general, in any bipartite pure state) we can resort
to the von Neumann entropy of the reduced single-mode mixed state of
any of the two modes:
$S^{(\alpha)} = -Tr_{\alpha}[\rho_{\alpha}\ln\rho_{\alpha}]$,
where $\rho_{\alpha}=Tr_{\alpha'\neq\alpha}[\rho]$
is the reduced density operator for mode $\alpha$
($\alpha = a,b$). If the two modes are assumed to be
initially in a pure product state, then $S^{(a)}(t=0)=S^{(b)}(t=0)=0$.
As the evolution is unitary (Hamiltonian), the total entropy $S$ of the system is
a conserved quantity. From the Araki-Lieb theorem \cite{Araki},
which can be expressed in the form $|S^{(a)}-S^{(b)}|\leq S\leq
S^{(a)}+S^{(b)}$, it follows that $S(t)=0$ and $S^{(a)}=S^{(b)}$
for any $t>0$. An equivalent measure of entanglement for pure
bipartite states is the reduced linear entropy
$S_{L}$, that, for simplicity, we will use in the following,
defined as:
\be
S_{L} \, = \, 1 -Tr_{a}[\rho_{a}^{2}] \, = \, 1-Tr_{b}[\rho_{b}^{2}] \; ,
\ee
where $Tr[\rho^{2}]$ measures the purity of
a state: $Tr[\rho^{2}] \leq 1$, and saturation
is reached only for pure states. The linear entropy is in fact
a lower bound to the von Neumann entropy: $S_{L} \leq S^{(\alpha)}$,
so that the higher $S_{L}$, the higher is the entanglement between
the pump and the signal mode. In Fig. (\ref{3DCentFock}) we plot
the time evolution of the linear entropy $S_{L}$ during the process of
three-photon  down conversion, starting from the initial, factorized Fock state $|4,7\rangle$.
We see that the Hamiltonian (\ref{Hk-phDC}) has a strong entangling
power, and, as long as the interaction stays on, the entanglement
remains nonvanishing at all times, although exhibiting a fast oscillatory
behavior.
More generally, for an initial coherent state of the pump mode,
the entanglement is an
increasing function both of the order of the process and of the
intensity of the pump \cite{BuzekDrobny}. \\
We conclude this short discussion of multiphoton down conversion
processes with fully quantized pump mode, by discussing some
general spectral properties of the interaction Hamiltonians.
Following a method similar to that developed in Ref.
\cite{WuYangOL} to determine eigenvalues and eigenvectors of
four-wave mixing Hamiltonians, the same authors have determined
the energy spectra and the eigenstates for a general class of
two-mode multiphoton models \cite{WuYangmph}. This procedure
successfully applies to the Hamiltonian
\be
H_{Qr}^{dg,kph} \, =
\, \omega_{a}a^{\dag}a+\omega_{b}b^{\dag}b+g(a^{\dag
k}b^{r}+b^{\dag r}a^{k})\, = \, \omega N+gH_{int} \; ,
\ee
where
$r$, $k$ are positive integers, $\omega_{a}=r\omega$,
$\omega_{b}=k\omega$, and $N=r a^{\dag}a+kb^{\dag}b$ is an
integral of motion. Clearly, for $r=1$, $H_{Qr}^{dg,kph}$ reduces
to the $k$-photon down converter (\ref{Hk-phDC}), without the Kerr
terms. This setting describes the effective interaction of a
$k$-degenerate signal field with a $r$-degenerate pump field. We
can impose the following simultaneous eigenvalue equations
\cite{WuYangmph}
\bea
&&n_{a}|\Psi_{\lambda,M}\rangle \, = \,
(nk+n_{a0})|\Psi_{\lambda,M}\rangle \; , \nonumber \\
&& \nonumber \\
&&n_{b}|\Psi_{\lambda,M}\rangle \, = \, [(M-n)r+n_{b0}]|\Psi_{\lambda,M}\rangle \nonumber \\
&& \nonumber \\
&&H_{int}|\Psi_{\lambda,M}\rangle \, = \,
\lambda|\Psi_{\lambda,M}\rangle \; , \nonumber \\
&& \nonumber \\
&&M=0,1,..., \quad n_{a0}=0,1,...,(k-1), \quad
n_{b0}=0,1,...,(r-1).
\eea
Writing the eigenstates
$|\Psi_{\lambda,M}\rangle$ in the form
\be
|\Psi_{\lambda,M}\rangle \, = \, S(a^{\dag},b^{\dag})|0,0\rangle
\, = \, \sum_{n=0}^{M}\alpha_{n}(\lambda)\frac{a^{\dag
(nk+n_{a0})}b^{\dag [(M-n)r+n_{b0}]}}{(nk+n_{a0})!}|0,0\rangle \; ,
\ee
with $\alpha_{n}(\lambda)$ real parameters, and exploiting
the relation
$[H_{int},S]|0,0\rangle=\lambda|\Psi_{\lambda,M}\rangle$, all the
energy eigenvalues and eigenstates are determined in terms of the
real parameter $\lambda$, which in turn can be determined as the
root of a simple polynomial. The same approach can be adopted to
study the three-mode Hamiltonian generalization of
$H_{Qr}^{dg,kph}$ involving a $k$-degenerate signal field $a$ and
two pump fields $b$ and $c$, respectively $r$-fold and $s$-fold
degenerate (the opposite interpretation obviously holds as well,
taking $a$ as the pump field and $b$, $c$ as the signal fields)
\cite{WuYangmph2}:
\be
H_{Qrs}^{dg,kph} \, = \,
\omega_{a}a^{\dag}a+\omega_{b}b^{\dag}b+\omega_{c}c^{\dag}c+g(a^{\dag
k}b^{r}c^{s}+c^{\dag s}b^{\dag r}a^{k}) \; ,
\ee
with
$k\omega_{a}=r\omega_{b}+s\omega_{c}$.

\subsection{Fock state generation in multiphoton parametric processes}

Number states are the natural basis in separable Hilbert spaces,
constitute the most fundamental instance of multiphoton
nonclassical states, and are in principle crucial in many concrete
applications; therefore, they deserve a particularly detailed
study. On the other hand, the experimental production of number
states presents extremely challenging difficulties. In this
Subsection we will discuss the problem of Fock state engineering,
reviewing some of the most important theoretical and experimental
proposals. Most of the methods are based either on nonlinear
interactions, or conditional measurements, or state preparation in
cavity.
\cite{fockgeneration,fockgen0,fockgen1,fockgen2,fockgen3,fockgen4,fockgen5,fockgen6,fockgen7}.
Concerning state preparation in a high-$Q$ cavity, here we limit
ourselves to mention that number states $|n\rangle$ with $n$ up to
2 photons have been unambiguously observed by the Garching group
lead by H. Walther \cite{fockgeneration}. Rather, following the main
line of this review, we will briefly describe those theoretical
proposals that involve parametric processes in nonlinear media and
optical devices \cite{fockgen5,fockgen6,fockgen7}. Intuitively,
the simplest method to prepare a generic quantum state
$|\psi\rangle$ is to realize a Hamiltonian operator $H$, which
governs the time evolution starting from an initial vacuum state:
$|\psi\rangle=e^{itH}|0\rangle$. This idea has been applied by
Kilin and Horosko to devise a scheme for Fock state production
\cite{fockgen5}. These authors introduce the following Hamiltonian
operator
\be
H_{n} \, = \, \lambda a^{\dag}a \, - \, \lambda
\frac{(a^{\dag}a)^{2}}{n} \, + \,
\left[\frac{a^{\dag n}}{\sqrt{n!}}\left(1-\frac{a^{\dag}a}{n}\right) \,
+ \, H. c. \right] \; ,
\label{kilinH}
\ee
with $\lambda$ real and
$n \geq 1$ positive integer. It is easy to see that
\be
H_{n}|0\rangle \, = \, |n\rangle \; , \quad H_{n}|n\rangle \, = \,
|0\rangle \; , \quad \exp\{i\tilde{t}H_{n}\}|0\rangle \, = \,
|n\rangle \; ,
\ee
with $\tilde{t}=\pi/2+2\pi m$, $m$ integer. For
$\lambda=0$, the Hamiltonian (\ref{kilinH}) can be realized in a
nonlinear medium via two phase-matched processes in which a
classical pump at frequency $\Omega$ is simultaneously converted
in $n$ photons at frequency $\omega=\Omega/n$. This conversion can
be realized by either one of the two processes $\Omega\rightarrow
n\omega$ and $\Omega+\omega\rightarrow (n+1)\omega$. In this case
$H_{n}$ takes the form \be H_{n} \, = \, \chi^{(n)}a^{\dag n}E \,
+ \, \chi^{(n+2)}a^{\dag n+1}aE + H. c. \; , \ee where the
nonlinear susceptibilities are taken to be real (without loss of
generality), the classical pump field impinging on the crystal is
$E^{(+)}=E e^{-i\Omega t}$, and the medium must be manipulated in
such a way that the nonlinear susceptibilities satisfy the
constraint $\chi^{(n+2)}=-\chi^{(n)}/n$. At times
$$
\tilde{t} \, = \,
\frac{\pi}{\chi^{(n)}|E|\sqrt{n!}}\left(m+\frac{1}{2}\right) \; ,
$$
($m$ integer) an initial vacuum state evolves in a $n$-photon Fock
state \cite{fockgen5}. \\
Another dynamical model \cite{fockgen6} for $k$-photon Fock state
generation is related to the following interaction Hamiltonian
\cite{fockgen6}: \be H_{I} \, = \, \varepsilon (a^{k} \, + \,
a^{\dag k}) \, + \, \frac{\delta}{2} a^{\dag}a(a^{\dag}a - k) \; ,
\label{leonK} \ee where the real, classical pump $\varepsilon$ and
the nonlinear Kerr coupling $\delta$ must satisfy the condition
$\varepsilon \ll \delta$, so that the model consists of a Kerr
term, perturbed by a weak parametric process of order $k$. The
time evolution of the system, starting from an initial vacuum
state, can then be studied in perturbation theory. The unperturbed
Kerr Hamiltonian for the Kerr process couples the states
$|0\rangle$ and $|k\rangle$, providing a kind of resonance between
these states. It can be shown that at some time, the system is in
the state $|k\rangle$ with probability one \cite{fockgen6}. \\ We
mention also an optical device \cite{fockgen7}, whose scheme is
based on an avalanche-triggering photodetector and a ring cavity
coupled to an external wave though a cross-Kerr medium. Such
device can synthesize Fock states and their superpositions. \\
Finally, it is worth to remark that, following a different
line of thought, a proposal of generation of Fock states by linear
optics and quantum nondemolition measurement has been presented in
Ref. \cite{fockgen8}.

\subsection{Displaced--squeezed number states}

\begin{figure}[h]
\begin{center}
\includegraphics*[width=15cm]{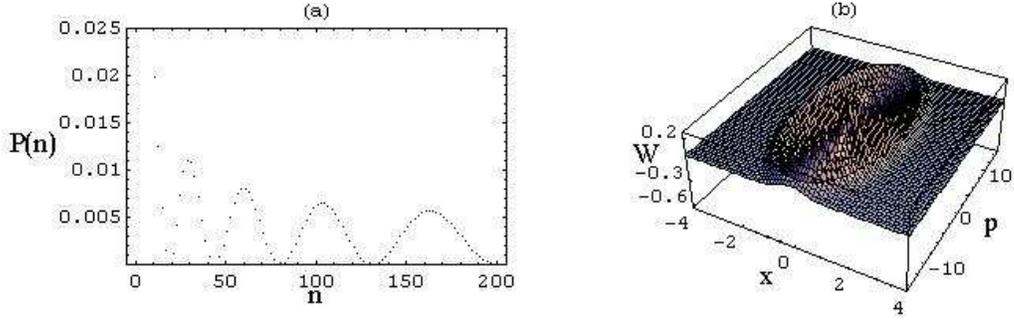}
\end{center}
\caption{(a) Photon number distribution in the state
$|\alpha,\varepsilon,m\rangle=|0,1.7,15\rangle$. A high value of
the squeezing parameter $\varepsilon$ is chosen
in order to magnify the oscillatory
behavior. (b) Wigner function for the state $|0,1,15\rangle$.}
\label{DSFockPW}
\end{figure}

After the introduction of squeezed number states $|m_{g}\rangle=S(\varepsilon)|m\rangle$
originally defined by Yuen \cite{Yuen}, displaced-squeezed
number states have received growing attention, due
to their marked nonclassical properties
\cite{DSFock1,DSFock2,DSFock22,DSFock3,DSFock32,NietoDSFock}.
These states are defined as
\be
|\alpha,\varepsilon,m\rangle \, = \, D(\alpha)S(\varepsilon)|m\rangle \; .
\label{DSFockdef}
\ee
Obvious subclasses of these states are
the displaced number states and the squeezed number states, which
are obtained by simply letting $\varepsilon=0$ or $\alpha=0$,
respectively. Relation (\ref{DSFockdef}) immediately suggests a
possible generation scheme by successive evolutions of the
Fock state $|m\rangle$ in quadratic and linear media. Starting
from an initial number state, the generation of displaced--squeezed
number states by the action of a forced time-dependent
oscillator has been proposed in Ref. \cite{DSFockgenLo}. The
coefficients of the expansion of states
$|\alpha,\varepsilon,m\rangle$ in the number-state basis have been
explicitly computed in Ref. \cite{DSFock1}, while the kernel of
their coherent state representation has been obtained in Ref.
\cite{DSFock22}. Here we review the main aspects of the particular
realization $|0,\varepsilon,m\rangle$ (i.e. squeezed number
states) with $\varepsilon$ real. Fig.
(\ref{DSFockPW}) (a) shows the continuous-variable approximation
of the photon number distribution $P(n)=|\langle
n|0,\varepsilon,m\rangle|^{2}$ for the state $|0,1.7,15\rangle$.
It is characterized by regular oscillations for high
$n$ and it is nonvanishing only for even
values of $|n-m|$; hence, for the initial number state
$|15\rangle$, $P(n)\neq 0$ for odd $n$.
As usual, we can compute the average number of photons $\langle
n\rangle = m\cosh(2r)+\sinh^{2}r$, and the photon number variance
$\langle\Delta n^{2}\rangle = \frac{1}{2}(m^{2}+m+1)\sinh^{2}(2r)$,
which shows that the number uncertainty increases linearly with
the initial value $m$ of the photon number.
Nonclassical properties of the squeezed number states are
witnessed by extended regions of negativity of the Wigner function,
which are clearly visible in Fig. (\ref{DSFockPW}) (b), plotted
for the state $|0,1,15\rangle$. More generally, phase
space representations of displaced--squeezed number states and of
their superpositions have been extensively investigated in Refs.
\cite{DSFock4,DSFock5,DSFock6}. Higher-order squeezing and nonclassical
correlation properties, have been discussed in Ref. \cite{DSFock3}, and
it has been shown that further nonclassical effects arise
when displaced--squeezed number states are sent as inputs
in a Kerr medium \cite{DSFockKerr}. Finally, we wish to mention
a recent successful experiment in the production of displaced Fock
states of light \cite{DFockexpgen}; their synthesis has been obtained
by overlapping the pulsed optical single-photon number state with
coherent states at a beam splitter with high reflectivity.

\subsection{Displaced and squeezed Kerr states}

\begin{figure}[h]
\begin{center}
\includegraphics*[width=15cm]{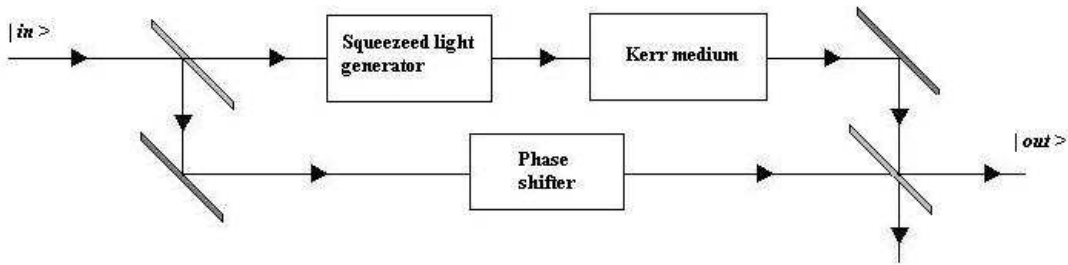}
\end{center}
\caption{Schematic diagram of a nonlinear Mach-Zehnder
interferometer for the generation of the displaced-Kerr-squeezed
state $|\psi\rangle_{DKS}$.  }
\label{displsqueezKerrgen}
\end{figure}

Another interesting method to realize nonclassical states of light
consists in associating displacement and squeezing with a
third-order nonlinear, unitary Kerr evolution of the form
$\exp\{-i\chi^{(3)} a^{\dag 2}a^{2}\}$. In Section \ref{section4},
we showed that a typical Kerr state $|\psi\rangle_{K}$ is
generated by applying the Kerr evolution operator to an initial
coherent state: $|\psi\rangle_{K}=\exp\{-i\chi^{(3)} a^{\dag
2}a^{2}\}|\alpha\rangle$. On the other hand, in the same Section,
we have seen that the Kerr interaction acts on the initial
coherent state by modifying the phases of the number states, but
leaves invariant the Poissonian statistics. To obtain
modifications of the photon statistics, one can apply further
interactions; for instance, the Kerr state $|\psi\rangle_{K}$ can
be displaced, yielding the so called displaced Kerr state
$|\psi\rangle_{DK}=D(\beta)|\psi\rangle_{K}$, which can be
generated by the action of a nonlinear Mach-Zehnder
interferometer. Considering this device, Kitagawa and Yamamoto
\cite{McZend2} showed that the photon number fluctuations $\langle
\Delta n^{2}\rangle$ in a displaced Kerr state are lowered to the
value $\langle n\rangle^{1/3}$, as compared to the value $\langle
n\rangle^{2/3}$ obtained in an amplitude-squeezed state. With some
algebra the displaced Kerr states can be expressed in the Fock
basis, $|\psi\rangle_{DK}=\sum_{n}c_{n}|n\rangle$, where the
coefficients $c_{n}$ are given, {\it e. g.} in Ref.
\cite{DSKerr1}. This expansion is handy when determining the
analytic expressions of the photon number distribution and of the
Husimi $Q$-function.
\begin{figure}[h]
\begin{center}
\includegraphics*[width=15cm]{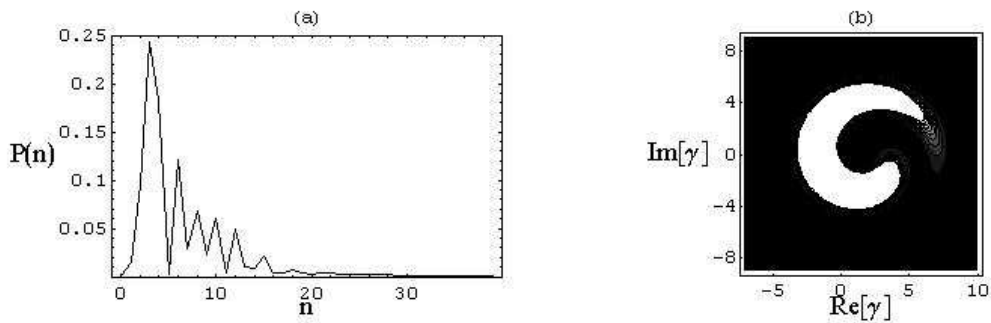}
\end{center}
\caption{(a) Photon number distribution in the state
$|\psi\rangle_{DK}=D(\beta)\exp\{-i\chi^{(3)} a^{\dag
2}a^{2}\}|\alpha\rangle$, plotted for $\alpha=3.5$, $\chi=0.25$,
and $\beta=2$. (b) Contour plot of $Q(\gamma)\equiv \pi^{-1}
|\langle\gamma|\psi\rangle_{DK}|^{2}$ for the same state.}
\label{DispKerr}
\end{figure}
In Fig. (\ref{DispKerr}) (a) and (b) we show the photon
number distribution $P(n)$ and the contour plot of $Q$,
respectively; we see that, for the super-Poissonian case of Fig.
(\ref{DispKerr}) (a), the contour plot of the $Q$-function tends
to become ring-shaped. It is worth mentioning
that the photon statistics of a squeezed Kerr state
$|\psi\rangle_{SK}=S(\varepsilon)|\psi\rangle_{K}$
is similar to that of the displaced Kerr states
$|\psi\rangle_{DK}$. A detailed discussion of the
statistical properties of the displaced and of
the squeezed Kerr states can be found in Refs.
\cite{DSKerr1,McZendsund,DSKerr2,DSKerr3}.\\
A further generalization of Kerr states
is achieved by suitably combining squeezing, Kerr effect, and
displacement. In Fig.
(\ref{displsqueezKerrgen}) we represent a generalized version of a
nonlinear Mach-Zehnder interferometer, obtained by adding to the
standard device a quadrature-squeezed light generator.
This configuration has been proposed to produce the
displaced-Kerr-squeezed state
$|\psi\rangle_{DKS}=D(\beta)\exp\{-i\chi^{(3)} a^{\dag
2}a^{2}\}|\alpha,\varepsilon\rangle$, where
$|\alpha,\varepsilon\rangle$ is a generic two-photon squeezed
state. The displaced-Kerr-squeezed state $|\psi\rangle_{DKS}$
allows a further reduction in the photon number uncertainty down
to the value $\langle n\rangle^{1/5}$ \cite{McZendsund}.

\subsection{Intermediate (binomial) states of the radiation field}

\label{BinomialStates}

\begin{figure}[h]
\begin{center}
\includegraphics[width=10cm]{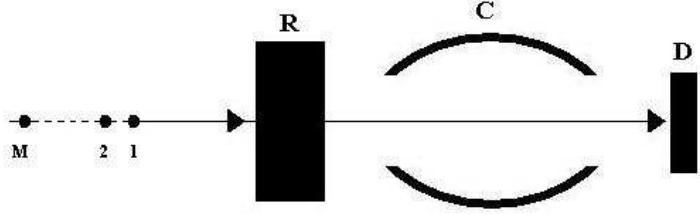}
\end{center}
\caption{Experimental setup for the generation of the reciprocal binomial states
(\ref{recbinst}). The $M$ atoms are prepared by a microwave field $R$;
they are injected in the cavity $C$; and, finally, they are counted in the ionization
detector $D$.}
\label{recbingen}
\end{figure}

The examples discussed so far can naturally be generalized to
conceive the engineering of states characterized by a specific
(\textit{a priori} assigned) probability distribution for the
number of photons. In previous Sections we have studied systems
associated with various statistical distributions, such as
coherent states (with Poissonian distribution), thermal states
(with Bose-Einstein distribution), squeezed states (with sub- or
super-Poissonian distribution), and Fock states. In 1985, Stoler
\textit{et al.} \cite{binomstoler} succeeded in introducing
single-mode quantum states which interpolate between a coherent
and a number state. They, in fact, introduced the binomial states,
characterized by a binomial photon number distribution. The
binomial states $|p,M\rangle_{B}$ can be written as a finite
combination of the first $M+1$ number states:
\be
|p,M\rangle_{B}
\, = \, \sum_{n=0}^{M}\left[\left(\begin{array}{c}
  M \\
  n \\
\end{array}\right)p^{n}(1-p)^{M-n}\right]^{1/2}|n\rangle
\; , \; \quad 0<p<1 \; ,
\label{binomst}
\ee
where
$\left(\begin{array}{c}
  M \\
  n \\
\end{array}\right)$ is the usual binomial coefficient. The state $|p,M\rangle_{B}$
reduces to the coherent state in the limit $p\rightarrow 0$,
$M\rightarrow\infty$, $pM=const$, while it realizes the number
state $|M\rangle$ for $p\rightarrow 1$. The properties of the
binomial states have been investigated in Refs.
\cite{binomstoler,binomlee,binomroversi}, and the methods for
their generation have been proposed in Refs.
\cite{vidiellaengine,binomstoler,binomlee,binomdattoli}. Binomial
states appear to be strongly nonclassical. In fact, they exhibit
intense second and fourth order squeezing (see definition
(\ref{highordsqueez})), whose maxima depend on $M$. Moreover, the
statistics of these states is intrinsically sub-Poissonian, as the
Mandel parameter $Q=-p$ \cite{binomroversi}, and their Wigner
function exhibits a negative region, more and more pronounced for
increasing $p$. Binomial states can be generated in a process of
emission of $M$ photons, each one emitted with the same
probability $p\;$; for instance, they can be produced from the
vibrational relaxation of an excited molecule \cite{binomstoler}.
A simple generalization  of the class of states (\ref{binomst}),
which is more flexible from the point of view of applications, can
be obtained in the form \cite{genBinost}
\bea
&&|p,M,q\rangle_{GB} \, =  \nonumber \\
&& \nonumber \\
&&\sum_{n=0}^{M}\left[\left(\begin{array}{c}
  M \\
  n \\
\end{array}\right)\frac{p}{1+Mq}
\left(\frac{p+nq}{1+Mq}\right)^{n-1}\left(1-\frac{p+nq}{1+Mq}\right)^{M-n}\right]^{1/2}|n\rangle
\; ,
\label{genbinomst}
\eea
where the additional parameter $q$
satisfies the constraint
$q\geq\max\left\{-\frac{p}{M}\,,-\frac{1-p}{M}\right\}$. The
parameter $q$ can be tuned to control the nonclassical behavior
and realize , for instance, a stronger sub-Poissonian statistics.
Comparing Eqs. (\ref{binomst}) and (\ref{genbinomst}), it is
evident that the emission events for different photons happen
with different probabilities, at variance with the case of the standard
binomial states. This implies that, in the process of generation
of the generalized binomial state (\ref{genbinomst}), the
probability of photon emission cannot be the same for the
different energy levels of an excited molecule \cite{genBinost}. \\ A
further typology of intermediate states is constituted by the
negative binomial states
\cite{negbinst,negbin2,negbin3,negbinagarw}. These states interpolate
between the coherent and the thermal states, and are
characterized by a \textit{negative binomial} photon number
distribution, which is, in a sense, the inverse of the binomial distribution,
as it can be guessed by noting the inverted order of the
factors in the binomial coefficient. In fact, the negative binomial
quantum states are expressed in the form \cite{negbinBarnet}
\be
|\alpha,M\rangle_{NB}\, = \, (1-|\alpha|^{2})^{(M+1)/2}\sum_{n=M}^{\infty}
\left(\begin{array}{c}
  n \\
  M \\
\end{array}\right)^{1/2}\alpha^{n-M}|n\rangle
\; ,
\label{negbinst}
\ee
where
$\alpha=|\alpha|e^{i\phi}$, $0<|\alpha|<1$, and $M$ is a
non-negative integer. A summary of the main properties of negative
binomial states, and a comparison of these states with the
binomial states and with other states of light, can be found in
Refs. (\cite{negbinagarw,negbinwang}). The states (\ref{negbinst})
can be prepared in optical multiphoton processes, in terms
of $M$-photon absorptions from a thermal beam of photons,
or by parametric amplification with
suitable initial conditions \cite{negbinagarw}. It is also possible
to conceive generation schemes of their superpositions
\cite{negbinguo,negbinliao}. \\ Finally, we wish to review
the class of the reciprocal binomial states $|\Phi,M\rangle_{RB}$, introduced by
Barnett and Pegg \cite{BarnetPegRecbin}. These states have been proposed
as fiducial reference for the experimental reconstruction
of the quantum optical phase probability distribution, in a scheme
that mixes the signal field and the reference state $|\Phi,M\rangle_{RB}$
by means of a beam splitter. The reciprocal binomial state is defined as
\be
|\phi,M\rangle_{RB} \, = \, \mathcal{N}^{-1}\sum_{n=0}^{M}\left(\begin{array}{c}
  M \\
  n \\
\end{array}\right)^{-1/2}e^{i n(\phi-\pi/2)}|n\rangle \; ,
\label{recbinst}
\ee
where $\mathcal{N}$ is a normalization factor. A scheme for the
generation of the states (\ref{recbinst}), that can be used as well
for the production of states (\ref{binomst}) and (\ref{negbinst}),
has been proposed in Ref. \cite{recbinstgen}; this proposal is related
to the one by Vogel \textit{et al.}
\cite{fockgen4}. The proposed experimental setup is shown in Fig.
(\ref{recbingen}); it is composed by $M$ two-level (Rydberg)
atoms, a Ramsey zone (R), a high-$Q$ cavity (C), and a field
ionization detector (D).
The atoms are prepared, by a microwave field $R$ (Ramsey zone),
in a superposition of the ground state $|g\rangle$ and of the excited state $|e\rangle
\; $:
 $\; c_{g}^{(n)}|g\rangle_{(n)}+c_{e}^{(n)}|e\rangle_{(n)}\; $ (with $n$ labelling the
$n$-th atom). The atoms are then injected one-by-one in the
cavity. The on-resonant interaction of each atom with the cavity
field is described by the Hamiltonian
$H_{onres}=\omega_{1}(|e\rangle\langle g|a+|g\rangle\langle
e|a^{\dag})$, where we have dropped the index $n$. When all the
$M$ atoms are detected in the ground state at the output of the
cavity, the state of the cavity field becomes
$$
\mathcal{N}^{-1}\sum_{n=0}^{M}\Lambda_{n}^{(M)}e^{-in
\pi/2}|n\rangle \; ,
$$
with the coefficients $\Lambda_{n}^{(M)}$
given by the recurrence formula
$$
\Lambda_{n}^{(M)}=(1-\delta_{n,0})\Lambda_{n-1}^{(M-1)}c_{e}^{(M)}\sin(\sqrt{n}\omega_{1}\tau_{M})+
(1-\delta_{n,M})\Lambda_{n}^{(M-1)}c_{g}^{(M)}\cos(\sqrt{n}\omega_{1}\tau_{M})\; .
$$
The procedure is concluded by sending an auxiliary atom in the
cavity with an off-resonant interaction of the form
$H_{offres}=\omega_{2}a^{\dag}a(|e\rangle\langle e|-|g\rangle\langle
g|)$. If the atom is prepared in the ground state, the
off-resonant interaction produces, in the atomic states, a
conditional phase shift $\phi$ controlled by the photon number in
the cavity field. The reciprocal binomial state
(\ref{recbinst}) is finally obtained by imposing a suitable form for the
coefficients $\Lambda_{n}^{(M)}$, with $\phi=\omega_{2}T$ ($T$
being the duration of interaction) \cite{recbinstgen}.\\
Besides the binomial states and their generalizations, several
other classes of intermediate states of light have been
constructed and investigated, such as logarithmic states
\cite{logarst} and multinomial states \cite{multinomst}. For a
complete bibliography on this subject we refer to the review
article \cite{revDodonov}.

\subsection{Photon-added, photon-subtracted, and vortex states}

\begin{figure}[h]
\begin{center}
\includegraphics*[width=15cm]{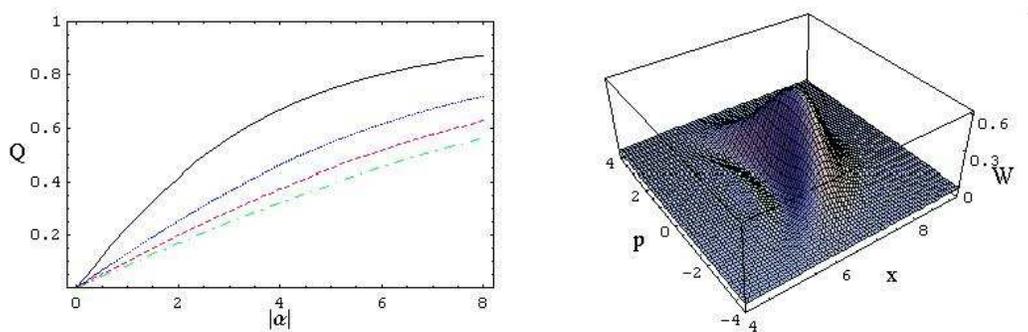}
\end{center}
\caption{(a) Plot of the $Q$ parameter of the photon-added coherent states
(\ref{phaddcohst}) as a function of the coherent amplitude $|\alpha|$,
for $k=5$ (full line), $k=15$ (dotted line), $k=25$ (dashed line),
and $k=35$ (dot-dashed line). (b) Plot of the Wigner function $W(x,p)$
(\ref{wigfunctphaddcohstates}) for $k=15$ and $\alpha=2$.}
\label{photonaddedcohst}
\end{figure}

\begin{figure}[h]
\begin{center}
\includegraphics*[width=7cm]{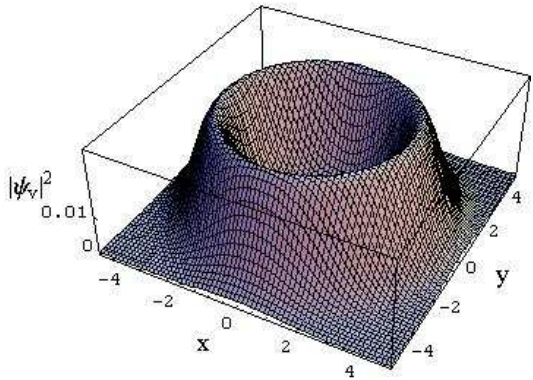}
\end{center}
\caption{Intensity distribution function for the vortex state
(\ref{wfvortexst}) with $\sigma=\sqrt{2}$ and $m=5$.}
\label{wfvortexplt}
\end{figure}

\begin{figure}[h]
\begin{center}
\includegraphics*[width=14cm]{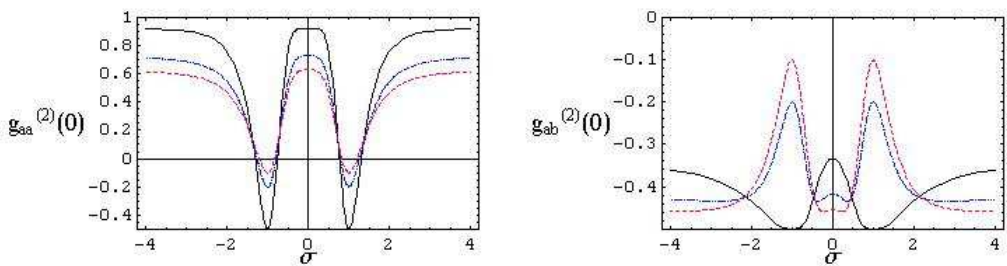}
\end{center}
\caption{(a): plot of $g^{(2)}_{aa,T}(0)$. (b): plot
of $g^{(2)}_{ab,T}(0)$. Both plots are dawn as functions
of $\sigma$, and with $m=2$ (full
line), $m=5$ (dotted line), and $m=10$ (dashed line).}
\label{vortexcorr}
\end{figure}

Nonclassical states with very interesting properties are the
so-called photon-added states $|\psi\,,k\rangle$, that are produced by repeated
applications of the photon creation operator on an arbitrary
quantum state
$|\psi\rangle$:
\be
|\psi\,,k\rangle \, = \, \mathcal{N}a^{\dag
k}|\psi\rangle \; ,
\label{phaddst}
\ee
where $\mathcal{N}$ is a
normalization factor. Agarwal and Tara introduced
a first type of states (\ref{phaddst}), the photon-added
coherent states, by choosing for $|\psi\rangle$ the coherent state
$|\alpha\rangle$ \cite{photaddagarwal}:
\be
|\alpha\,,k\rangle \, = \, \mathcal{N}(\alpha\,,k)a^{\dag k}|\alpha\rangle
\; , \label{phaddcohst}
\ee
where the normalization factor
$\mathcal{N}(\alpha\,,k)=\langle\alpha|a^{k}a^{\dag
k}|\alpha\rangle^{-1/2}=[L_{k}(-|\alpha|^{2})k!]^{-1/2}$, with
$L_{k}(x)$ the Laguerre polynomial of degree $k$. The photon-added
coherent states (\ref{phaddcohst}) exhibit squeezing,
and their photon number distribution is a shifted Poisson distribution
of the form
\be
P(n) \, = \, \mathcal{N}(\alpha\,,k)^{2}
\frac{n!|\alpha|^{2(n-k)}}{[(n-k)!]^{2}}e^{-|\alpha|^{2}} \; .
\label{Pndphadd}
\ee
In Fig. (\ref{photonaddedcohst}) (a) we have
plotted the Mandel $Q$ parameter as a function of $|\alpha|$ for
different choices of the number $k$ of added photons.
We see that the Mandel parameter remains always
below one, and decreases with increasing $k$, and the photon
number distribution (\ref{Pndphadd}) describes a growingly stronger
sub-Poissonian statistics \cite{photaddagarwal}.
A further signature of nonclassicality
is provided by the fact that the Wigner function for
these states
\be
W\left(\xi=\frac{x+i
p}{\sqrt{2}}\right) \, = \,
\frac{2(-1)^{k}L_{k}(|2\xi-\alpha|^{2})}{\pi
L_{k}(-|\alpha|^{2})}e^{-2|\xi-\alpha|^{2}} \; ,
\label{wigfunctphaddcohstates}
\ee
plotted in Fig. (\ref{photonaddedcohst}) (b),
takes negative values due to the weights of the Laguerre
polynomials.
Agarwal and Tara have proposed also a scheme for the production of
photon-added states, based on nonlinear processes in a cavity. The
interaction Hamiltonian describing the passage of two-level
excited atoms ($k$-photon medium) through a cavity can
be written in the form \be H_{I} \, = \, \eta
\sigma^{+}a^{k}+\eta^{*}\sigma^{-}a^{\dag k} \; , \ee where $a$ is
the cavity field mode, $\sigma^{+}=|e\rangle\langle g|$,
$\sigma^{-}=|g\rangle\langle e|$, and the notation $\sigma^{\pm}$
refers to the fact that such atomic raising and lowering operators
can be represented by Pauli matrices. If the initial state of the
field-atom system is the factorized state
$|\alpha\rangle|e\rangle$, then, for short (reduced,
dimensionless) times $|\eta| t \ll 1$ the evolved state can be
approximated by \be |\psi(t)\rangle \, \approx \,
|\alpha\rangle|e\rangle \, - \, i\eta^{*}ta^{\dag
k}|\alpha\rangle|g\rangle \; . \ee When the atom is detected in
the ground state $|g\rangle$, then the photon-added coherent state
will be generated in the process, apart from a normalization
factor. Alternatively, the state $|\alpha\,,k\rangle$ may also be
produced by methods based on state reduction and feedback
\cite{BjorkYamam}. \\
Dodonov \textit{et al.} \cite{phadddodonov} have studied
the dynamics of the states $|\alpha\,,k\rangle$, when the field
mode eigenfrequency $\omega$ is time-dependent, and the
Hamiltonian ruling the time evolution is \be H\, = \;
\frac{1}{2}[P^{2}+\omega^{2}(t)Q^{2}] \; , \quad\quad \omega(t) \;
= \; 1+2\gamma\cos 2t \; , \; \; |\gamma|\ll 1 \; . \ee The action
of this Hamiltonian on photon-added coherent states produces new
states that exhibit, under certain conditions, a larger degree of
squeezing with respect to two-photon squeezed states, and
a transition from the initial sub-Poissonian statistics to a super-Poissonian one. \\
Varying the choices of the reference state $|\psi\rangle$, many other
types of photon added states can be obtained, such as the photon-added
squeezed states \cite{photonaddedsqueezst}, the even/odd
photon-added states \cite{photonaddedevenoddst}, and the
photon-added thermal states \cite{photonaddedthst}. Symmetrically
to the photon-added states, one can as well define the class of the
photon-subtracted states. The latter can be obtained by simply replacing the
creation operator $a^{\dag}$ with the annihilation operator $a$ in
Eq. (\ref{phaddst}) \cite{phsubtst,phsubtractKim}. Proposals for the
generation of photon-added and photon-subtracted states, based on
conditional measurements, have been discussed in Refs.
\cite{phsubtst,phaddsubgenban}. \\
A two-mode generalization of the degenerate photon-added states
has been considered in Ref. \cite{phadd2modehong}.
Similar to these two-mode photon-added states are the two-mode vortex
states of the radiation field \cite{vortexst}, which are characterized by
a wave function, in the two-dimensional
configuration representation, of the form
\be
\psi_{v}^{(m)}(x,y) \, = \,
\frac{1}{\sqrt{m!\pi\sigma^{2m+2}}}(x-i
y)^{m}e^{-\frac{1}{2\sigma^{2}}(x^{2}+y^{2})} \; .
\label{wfvortexst}
\ee
The vortex structure of these states shows up
in the intensity distribution function
$|\psi_{v}^{(m)}(x,y)|^{2}$, plotted in Fig.
(\ref{wfvortexplt}).
 In terms of the field mode
operators $a$ and $b$, the vortex state can be recast in
the form
\bea
&&|\psi_{v}^{(m)}\rangle \, = \, \mathcal{N}_{v}(a^{\dag}-i
b^{\dag})^{m}e^{r(a^{\dag 2}-a^{2})} e^{r(b^{\dag
2}-b^{2})}|0,0\rangle \; , \nonumber \\
&& \nonumber \\
&&\mathcal{N}_{v} \, = \, \frac{2^{-m/2}(1+\xi)^{m}}{\sqrt{m!}\sigma^{m}}
\; , \quad \quad \xi \, = \, \frac{\sigma^{2}-1}{\sigma^{2}+1} \;,
\label{vortexop}
\eea
where $\sigma=e^{2r}$. The nonclassical
character of the state (\ref{vortexop}) emerges from
the behavior of the second-order correlation functions.
For different choices of $m$, we have plotted in
Fig. (\ref{vortexcorr}) (a) and
(b), respectively, the quantities
$g^{(2)}_{aa,T}(0)=g^{(2)}_{aa}(0)-1$ and
$g^{(2)}_{ab,T}(0)=g^{(2)}_{ab}(0)-1$,
where $g^{(2)}_{aa}(0)$ is the auto-correlation of mode $a$,
and $g^{(2)}_{ab}(0)$ is the cross-correlation between modes
$a$ and $b$. The graphic shows that
the field mode $a$ exhibits sub-Poissonian
statistics, and that modes $a$ and $b$ are anticorrelated.
A simple scheme to produce the vortex states, based on a
three-level $\Lambda$ system interacting with two polarized fields
on resonance, has been proposed in Ref. \cite{vortexst}.

\subsection{Higher-power coherent and squeezed states}

Among the many nonclassical generalizations of Glauber's
coherent states, the even and odd coherent states are of
particular relevance. They are defined as two particular
classes of eigenstates of $a^{2}$, respectively with
even and odd number of photons. All the remaining
eigenstates can then be obtained as arbitrary linear
combinations of even and odd coherent states.
These states were introduced by Dodonov, Malkin, and
Man'ko \cite{evenoddcohstat}, and can be written in
the following form:
\bea
|\alpha\rangle_{even}& \, = \,&(\cosh
|\alpha|^{2})^{-1/2}\sum_{n=0}^{\infty}\frac{\alpha^{2n}}{\sqrt{(2n)!}}|2n\rangle
\; , \\
&& \nonumber \\
|\alpha\rangle_{odd}& \, = \, &(\sinh
|\alpha|^{2})^{-1/2}\sum_{n=0}^{\infty}\frac{\alpha^{2n+1}}{\sqrt{(2n+1)!}}|2n+1\rangle
\; .
\eea
These two classes of coherent states have been later generalized
by introducing the $j$-th order coherent states, defined
as the eigenstates of $a^{j}$ \cite{jexquang,ajeigenst,ajNietoTruax}:
\be
a^{j}|\alpha;j,k\rangle \, = \, \alpha^{j}|\alpha;j,k\rangle \; ,
\label{ajcohst}
\ee
where, as will be clarified, the notation $j, k$
indicates that the states $|\alpha;j,k\rangle$ can be written as
superpositions of the Fock states $\{|jn+k\rangle\}$, with the
additional integer parameter $k$ restricted to be $0\leq k\leq
j-1$. We recall that a subclass of these states can be generated
at suitable times by a Kerr interaction (see Section
\ref{section4}), and that the even and odd coherent states are a
particular realization of the Barut-Girardello coherent
states, already introduced in this Section. \\
The solution of Eq. (\ref{ajcohst}) is given by the infinite
superposition
\be
|\alpha;j,k\rangle \, = \, S^{-1/2}(j,k,|\alpha|^{2})
\sum_{n=0}^{\infty}\frac{\alpha^{jn+k}}{\sqrt{(jn+k)!}}|jn+k\rangle
\; ,
\label{ajcohstsol}
\ee
where
\be
S(j,k,z) \, = \, \sum_{n=0}^{\infty}\frac{z^{jn+k}}{(jn+k)!}
\; .
\ee
To any fixed eigenvalue $\alpha$
in the equation (\ref{ajcohst}) are associated $j$ solutions for
the $j$-th order coherent states.
It can be seen that the standard coherent states coincide with
the unique solution $|\alpha;1,0\rangle$ for $j=1$ and $k=0$.
Moreover, the two solutions with $j=2 \, , k=0$ and $j=2 \, , k=1$,
$|\alpha;2,0\rangle$ and $|\alpha;2,1\rangle$, coincide, respectively,
with the even and odd coherent states.
Nieto and Truax \cite{ajNietoTruax} have computed the wave function solution of
the equation (\ref{ajcohst}). It can be expressed in terms
of the following generalized generating functions for the Hermite polynomials
\be
G(j,k,x,z) \, = \, \sum_{n=0}^{\infty}\frac{z^{jn+k}H_{jn+k}(x)}{(jn+k)!} \; ,
\label{gengenfunHpoly}
\ee
and, consequently, cast in the form \cite{ajNietoTruax}
\be
\langle
x|\alpha\,,jk\rangle \, = \,
\frac{e^{-\frac{1}{2}x^{2}}G(j,k,x,\alpha/\sqrt{2})}{\pi^{1/4}S^{1/2}(j,k,|\alpha|^{2})}
\; .
\ee
Among the main properties of the
states $|\alpha;j,k\rangle$, the following relations of
orthogonality can be easily verified
\bea
&&\langle\alpha;j,k|\alpha;j,k'\rangle \, =0 \, \; ,
\quad \quad k \neq
k' \, = \, 0,1,...,j-1 \; , \nonumber \\
&& \nonumber \\
&&\langle\alpha;j,k|\alpha;j',k'\rangle \, = \, \delta_{jn+k\, , \, j'n'+k'}
\; , \quad \quad j,j'\geq 3 \; , \nonumber \\
&& \nonumber \\
 && k \, = \, 0,1,...,j-1 \; , \quad k' \, = \, 0,1,...,j'-1 \; , \quad n,n' \, = \, 0,1,... \; ,
\eea
while the relation
\be
\frac{1}{\pi}\int
d^{2}\alpha\sum_{k=0}^{j-1}|\alpha;j,k\rangle \,
\langle\alpha;j,k|=1
\ee
gives a complete representation
with respect to $\alpha$. Moreover, for $\alpha\neq\alpha'$ the
states $|\alpha;j,k\rangle$ and $|\alpha';j,k\rangle$ are not
orthogonal; therefore, the generalized coherent states
$|\alpha;j,k\rangle$ form an overcomplete set with respect to $\alpha$.
In addition, any orthonormalized
eigenstate of $a^{j}$ can be recast in the form of a superposition
of $j$ coherent states with different phases:
\be
|\alpha;j,k\rangle \, = \, \frac{e^{\frac{1}{2}|\alpha|^{2}}}{j
S^{1/2}(j,k,|\alpha|^{2})}\sum_{l=0}^{j-1}e^{i\frac{2\pi}{j}k(j-l)}|\alpha
e^{i\frac{2\pi}{j}l}\rangle \; .
\ee
The states
$|\alpha;j,k\rangle$ do not enjoy second-order squeezing, because
they are not minimum uncertainty states of the canonically conjugated
quadrature operators. On the other hand, all the generalized
coherent states $|\alpha;j,k\rangle$ exhibit antibunching
effects, and satisfy the definition of $N$-th order squeezing in
the sense of Zhang \textit{et al.} \cite{zhangsqueez}. Given the
operators
\be
Z_{1}(N) \, = \, \frac{1}{2}(a^{\dag N}+a^{N}) \; , \quad \quad
Z_{2}(N) \, = \, \frac{i}{2}(a^{\dag N}-a^{N}) \; ,
\ee
these authors, at variance with the more common criterion (\ref{highordsqueez}),
define the $N$-th order squeezed states as those which
satisfy the relation
\be
\langle \Delta Z_{i}^{2}(N)\rangle \,
< \, \frac{1}{4}\langle [a^{N},a^{\dag N}]\rangle \; , \quad \quad i=1,2 \; .
\label{zhangsqueez}
\ee
J. Sun, J. Wang, and C. Wang have proved
that the generalized coherent states $|\alpha;j,k\rangle$ exhibit
$N$-th order squeezing, in the sense of Eq. (\ref{zhangsqueez}),
if $N=(m+\frac{1}{2})j$, $m=0,1,...$, and $j$ even.
On the contrary, no state $|\alpha;j,k\rangle$
with $j$ odd possesses higher-order squeezing \cite{ajeigenst}.
The same authors have shown that, both for odd and
for even degree $j$, all the states $|\alpha;j,k\rangle$ are
minimum uncertainty states for the pair of operators $Z_{1}(N)$
and $Z_{2}(N)$, with $N=mj$ ($m=1,2,...$).
Finally, we mention that in Ref. \cite{jextorma}, the one-mode $j$-th
order coherent states have been generalized to the multimode
instance through the following definition of the multimode higher
order coherent states $|\Psi\rangle_{kl...m}$:
\be
a^{k}b^{l}...c^{m}|\Psi\rangle_{kl...m} \,
= \, \alpha^{k}\beta^{l}...\gamma^{m}|\Psi\rangle_{kl...m} \; .
\ee
Analogous techniques can be exploited to define the $j$-th order
squeezed states (which must not be confused with the concept of
higher order squeezing of a state), as the solutions of the
eigenvalue equation \cite{ajNietoTruax}
\be
\left[\frac{1}{2}(1+\lambda) a^{j}+\frac{1}{2}(1-\lambda) a^{\dag
j}
\right]|\beta,\lambda; j, k\rangle \, = \,
\beta^{j}|\beta,\lambda,jk\rangle \; .
\label{jorderssNieTrua}
\ee
When
$\lambda\rightarrow 1$, the $j$-th order squeezed states reduce to
$j$-th order coherent states, and they reduce to the standard Glauber
coherent states for $j=\lambda=1$. However, they never reduce to
the standard two-photon squeezed states.
The second order ($j=2$) squeezed
states present interesting features
\cite{secondorderssMarian}. For instance, the variances of the
amplitude-squared operators $\langle \Delta Z_{i}^{2}\rangle$, and
the second order correlation function at the initial time
$g^{(2)}(0)$ for a second order squeezed state are functions of
the mean photon number $\langle a^{\dag}a\rangle$, and take the
forms:
\be
\langle \Delta Z_{1}^{2}\rangle \, = \, \lambda \left(\langle
a^{\dag}a\rangle+\frac{1}{2}\right) \; ,
\quad \quad \langle \Delta
Z_{2}^{2}\rangle \, = \, \frac{1}{\lambda} \left(\langle
a^{\dag}a\rangle+\frac{1}{2}\right) \; ,
\label{Marianeq1}
\ee
\be
g^{(2)}(0) \, = \, \frac{1}{\langle
a^{\dag}a\rangle^{2}}\left\{\frac{(\lambda-1)^{2}}{\lambda}\left(\langle
a^{\dag}a\rangle+\frac{1}{2}\right)+Re[\beta^{2}]^{2}+\frac{1}{\lambda^{2}}Im[\beta^{2}]^{2}\right\}
\; .
\label{Marianeq2}
\ee
From the two relations in Eq.
(\ref{Marianeq1}), it follows that the parameter $\lambda$ is a
squeezing factor, and can be expressed as the square root of the
ratio $\langle\Delta Z_{1}^{2}\rangle /\langle\Delta
Z_{2}^{2}\rangle$. \\ Finally, in Ref. \cite{xinwang} it has been
shown that the orthonormal eigenstates of an arbitrary power
$b^{j}$ of the linear combination $b=\mu a+\nu a^{\dag}$ (with
$\mu$, $\nu$ satisfying the Bogoliubov condition of canonicity)
can be constructed by applying the squeezing operator
$S(\varepsilon)$ to the $j$-th order coherent states
(\ref{ajcohstsol})
\bea
&&|\beta;j,k\rangle_{g} \, = \, S(\varepsilon)|\beta;j,k\rangle
\, = \, \{\langle\beta|P_{k}^{j}|\beta\rangle\}^{-1/2}S(\varepsilon)P_{k}^{j}|\beta\rangle
\; , \nonumber \\
&& \nonumber \\
&&k=0,1,...,j-1 \,, \label{bjcohst}
\eea
where $P_{k}^{j}$ is a
generalized projection operator defined by
\be
P_{k}^{j} \, = \, \sum_{n=0}^{\infty}|jn+k\rangle\langle jn+k| \; ,
k=0,1,...,j-1 \; .
\ee
For the particular choice $j=2$, the
operators $P_{k}^{j}$ $(k=0,1)$ and $S(\varepsilon)$ commute, and
the states (\ref{bjcohst}) become the so-called even and odd
two-photon coherent states, given by
\bea
&&|\beta;2,0\rangle_{g} \, = \, (\cosh
|\beta|^{2})^{-1/2}e^{\frac{|\beta|^{2}}{2}}P_{0}^{2}|\beta\rangle_{g}
\; , \nonumber \\
&& \nonumber \\
&&|\beta;2,1\rangle_{g} \, = \, (\sinh
|\beta|^{2})^{-1/2}e^{\frac{|\beta|^{2}}{2}}P_{1}^{2}|\beta\rangle_{g}
\; ,
\eea
where $|\beta\rangle_{g}$ is a the standard Yuen two-photon coherent state.
In Ref. \cite{xinwang}, it is shown that both even and odd
two-photon coherent states exhibit squeezing for suitable choices
of the parameters, and that a strong antibunching effect is
exhibited by the odd states, while this
antibunching effect, although still present, is sensibly weaker in
the case of even states.

\subsection{Cotangent and tangent states of the electromagnetic field}

The class of cotangent and tangent states of the electromagnetic
field have been introduced and discussed in Refs. \cite{cotangtang}.
These states can be generated
in a high-$Q$ micromaser cavity by the evolution of the
harmonic oscillator (cavity mode) coupled to a ``quantum current''
consisting of a beam of two-level atoms, whose initial state is
given by a superposition of the upper and lower atomic states.
The effective interaction is described, as usual, by a Jaynes-Cummings
Hamiltonian, with exact resonance between the field frequency and the
atomic transition frequency. Moreover, it is assumed that only one atom at a time
is present inside the resonator, and that the state of the atom is not measured
as it leaves the cavity. However, the passage of each atom induces a nonselective
measurement \cite{Daviestx}, described by a partial trace operation
on the density matrix of the total system over the atom variables.
Cotangent and tangent states are generated during the time evolution
as finite superpositions of the form:
\begin{equation}
|\Psi\rangle_{c,t} \, = \, \sum_{n=N}^{M} s_{n}|n\rangle \; .
\label{tangcotang}
\end{equation}
Here $N, \, M$ are the number indices associated to ``trapping states'',
and determined by the conditions
\begin{equation}
\kappa \sqrt{N} \tau \, = \, q \, \pi \qquad , \qquad \kappa \sqrt{M+1} \tau = \, p \, \pi \; ,
\label{trapstates}
\end{equation}
where, $q, \, p$ are integer numbers, $\kappa$ is the Jaynes-Cummings cavity-atom coupling,
and $\tau$ is the interaction time between a single atom and the cavity.
The coefficients $s_{n}$ in Eq. (\ref{tangcotang}) are given by
\begin{equation}
s_{n} \, = \, \mathcal{N}_c (- i)^n (\alpha/\beta)^n
\prod_{j=1}^{n} \cot (\kappa \sqrt{j} \tau /2) \; ,
\label{cotang}
\end{equation}
for the choice of $q$ even and $p$ odd in Eq. (\ref{trapstates})
(cotangent states), and by
\begin{equation}
s_{n} \, = \, \mathcal{N}_t (i)^n (\alpha/\beta)^n
\prod_{j=1}^{n} \tan (\kappa \sqrt{j} \tau /2) \; ,
\label{tang}
\end{equation}
for the choice of $q$ odd and $p$ even in Eq. (\ref{trapstates})
(tangent states), with $\mathcal{N}_c, \, \mathcal{N}_t$ normalization constants.
Cotangent and tangent states, under suitable conditions, can exhibit
sub-Poissonian photon statistics or, alternatively, they resemble
macroscopic superpositions \cite{cotangtang}; these nonclassical
properties are in principle remarkably robust under the
effects of cavity damping \cite{cotangtangrobust}.
Finally, in Ref. \cite{Wilkens2}
it has been shown that these states, under a wide range of conditions,
are highly squeezed, and that the phase distribution can exhibit
oscillations resulting from the formation of states that are
again reminiscent of macroscopic superpositions.

\subsection{Quantum state engineering}

In the previous Subsections we have reviewed several schemes
for the generation of nonclassical multiphoton states of light
beyond the standard two-photon squeezed states. However, all
these schemes are tied to particular contexts and specific
descriptions. It would be instead desirable to implement a
general procedure which allows, at least in principle, to
engineer nonclassical multiphoton states of generic form.
Several attempts in this direction
have resulted in two main approaches. One approach is based on
the time evolution generated by a generic, controlling
Hamiltonian which drives an initial state to the final target
state (pure state, unitary evolution). Another approach is realized in
two steps: first, the quantum system of interest is correlated (entangled)
with another auxiliary system; next, a measurement is performed
on the auxiliary system, reducing the state of the system of interest
to the desired target state. In the following we briefly review
some of the most relevant methods, thus providing a compact introduction
to quantum state engineering in the framework of multiphoton quantum
optics.

a) \hspace{2mm} Clearly, the simplest theoretical and experimental
ways to produce nonclassical states of light is based on the processes
of amplification in $\chi^{(2)}$ and $\chi^{(3)}$ media.
The signal and the idler beams of the (nonclassical) squeezed light,
generated in nondegenerate parametric down conversion,
exhibit strong space-time correlations.
A destructive measurement on the idler output may lead to the generation
and$/$or manipulation of nonclassical states of light at the signal output.
Many papers have been devoted to various applications of this strategy
\cite{YuenNearPhotNumb,TapsternegfeedbacsubPoiss,WatanabeYam,UedaContstatereduc,Ban2msqstred1msqst,SotoGeneralmeasur,LvovskycondmeasnegWig,PariscondmeasnegWig,FiurasekSubPoisscondgen};
in the following we review some of the most relevant results.
A feedback measurement scheme has been proposed by Yuen to produce
near-photon-number-eigenstate fields \cite{YuenNearPhotNumb}.
A quantum nondemolition measurement of the number of photons can be performed to generate
a number-phase squeezed state \cite{QNDKitag}.
In Ref. \cite{BjorkYamam} the outcome of a homodyne measurement of the idler wave
is used to manipulate the signal wave by means of feed forward, linear attenuators,
amplifiers, or phase modulators.
A detailed analysis of the effect on the signal output of a measurement on the
idler output, described by a POVM (positive operator-valued measure
or generalized projection) \cite{Daviestx}, has been performed by Watanabe and
Yamamoto \cite{WatanabeYam}.
In particular, these authors have studied the application of three
different state-reduction measurements (expressed in terms of projection operators)
leading to different output signal fields.
If the idler photon number is measured by a photon counter,
the signal wave is reduced to a number-phase squeezed state;
if the idler single-quadrature amplitude is measured by homodyne detection,
the signal wave is reduced to a quadrature-amplitude squeezed state;
if the idler two-quadrature amplitudes are measured by heterodyne detection,
the signal wave is reduced to a coherent state \cite{WatanabeYam}.
In Ref. \cite{UedaContstatereduc}, Ueda \textit{et al.} have developed a general theory
of continuous state reduction, by substituting the POVM with suitable
superoperators. More recently, arbitrary generalized measurements
have been considered in Ref. \cite{SotoGeneralmeasur}.
It has been shown \cite{PariscondmeasnegWig} that the action
on a twin beam of an avalanche photodetector,
described by a specific POVM, may lead to a highly nonclassical reduced state,
characterized by a Wigner function with negative regions.
The possibility of generating nonclassical states
with sub-Poissonian photon-number statistics has been investigated as well.
For instance, the experimental generation of sub-Poissonian light has been obtained
by using a negative feedback sent to the pump from the idler wave of a twin beam
\cite{TapsternegfeedbacsubPoiss}.
In Ref. \cite{FiurasekSubPoisscondgen}, a particular homodyne detection,
with randomized phases and in a defined range of the quadrature eigenvalues,
has been proposed to reduce an initial twin beam to a sub-Poissonian field. \\
We remark that these methods have recently led to the
successful experimental generation of several nonclassical states of light
\cite{ZavattaScience,phsubtractexp,ReschCondCoher,DeMartiniquantinjec,LamasScience,LauratSubPoisslight}.
All these eperiments combine parametric amplification and state-reduction techniques.

\begin{figure}[h]
\begin{center}
\includegraphics*[width=12cm]{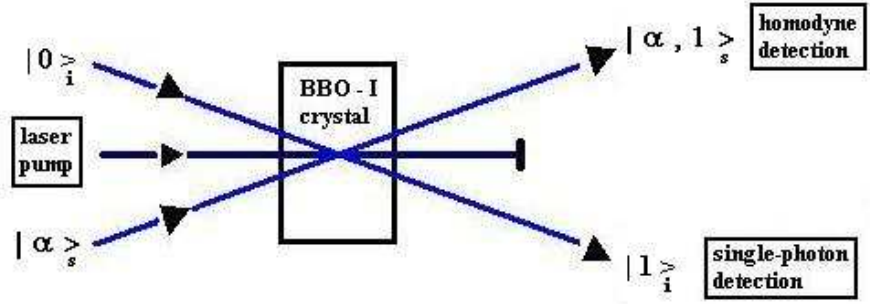}
\end{center}
\caption{Experimental scheme for the preparation of a single-photon-added coherent state.
It is based on a parametric down conversion process
in a type-I beta-barium borate (BBO) crystal;
initially, the signal field is in a coherent state (seed field),
while the idler field is in the vacuum state.
The generation of the one-photon-added coherent state is conditioned to the detection
of one photon in the output idler mode.
The output signal state is then characterized and reconstructed by standard
quantum tomographical techniques.}
\label{expZavatta}
\end{figure}

\begin{figure}[h]
\begin{center}
\includegraphics*[width=10cm]{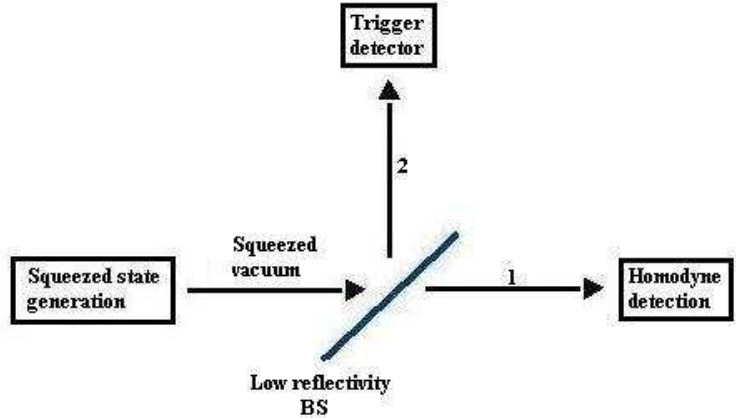}
\end{center}
\caption{Experimental scheme for the preparation of a one-photon-subtracted state.
The twin beam, obtained in the process of degenerate parametric amplification
in a crystal of potassium niobate (KNbO$_{3}$) is sent to a low reflectivity beam splitter.
The reflected wave is detected by a silicon avalanche photodiode, while the transmitted wave
is sent to a homodyne detection. This scheme provides a degaussification protocol
that maps pulses of squeezed light onto non-Gaussian states.}
\label{expGrangier}
\end{figure}

As relevant examples we describe in some detail the beautiful and encouraging
experimental observation of single-photon-added states
reported by Zavatta, Viciani, and Bellini \cite{ZavattaScience},
and of single-photon-subtracted states reported
by Wenger, Tualle-Brouri, and Grangier \cite{phsubtractexp}.
The experimental setup used by Zavatta \textit{et al.}
is schematically depicted in Fig. (\ref{expZavatta}); the production of the state
$|\alpha,\,1\rangle = (1+|\alpha|^{2})^{-1/2} a^{\dag}|\alpha\rangle$,
with $|\alpha\rangle$ denoting a coherent state, is obtained
exploiting the process of nondegenerate parametric amplification
of the initial state $|\alpha\rangle_{s} |0\rangle_{i}$, where the subscripts $s$ and $i$
denote the signal and the idler beam, respectively.
By keeping  the parametric gain $g$ sufficiently low,
the output state $|\psi_{out}\rangle$
can be written in the form
\begin{equation}
|\psi_{out}\rangle \, \propto \, (1+g a_{s}^{\dag}a_{i}^{\dag})|\alpha\rangle_{s} |0\rangle_{i}
\, = \, |\alpha\rangle_{s} |0\rangle_{i} + g a_{s}^{\dag}|\alpha\rangle_{s} |1\rangle_{i} \; .
\label{eqZavatta}
\end{equation}
From Eq. (\ref{eqZavatta}) it is clear that if the idler field is detected in the state
$|1\rangle_{i}$, then the signal output field is just the state $|\alpha,\,1\rangle$,
that is the single-photon-added state for $\alpha \neq 0$, and the one-photon number state
for $\alpha=0$.
The output single-photon-added coherent state
is then reconstructed by exploiting techniques of quantum tomography
(homodyne measurements). By gradually increasing the amplitude $\alpha$,
a spectacular transition occurs from the quantum one-photon number state
$|1\rangle_{s}$ at $\alpha=0$ to a classical (high intensity) coherent state
at sufficiently high values of $|\alpha|$.
The single-photon-added coherent state describes
the continuous and smooth intermediate regimes between these
two extremes. \\
The experimental realization of one-photon-subtracted states
has been recently reported by Wenger \textit{et al.} \cite{phsubtractexp}.
A simplified description of the experimental setup
is depicted in Fig. (\ref{expGrangier}).
A single-mode squeezed vacuum $|sv\rangle$, with the squeezing parameter fixed at $r=0.43$,
is sent to a beam splitter with low reflectivity $R \ll 1$;
the statistics of the transmitted signal wave is reconstructed
by using a balanced homodyne detector (with detection efficiency $\eta=0.75$),
while the small reflected fraction of the squeezed vacuum beam
is detected by an avalanche photodiode. The postselection then
provides non-Gaussian (nonclassical) statistics. The reconstruction of the state
is obtained numerically by considering the expansion of the squeezed vacuum state
in the Fock basis $\{|n\rangle\}$ up to $n=10$. However, for the sake of
simplicity and to illustrate the procedure, the authors report the calculations
by truncating at $n=4$:
\begin{equation}
|sv\rangle = \gamma_{0} |0\rangle + \gamma_{1} |2\rangle + \gamma_{2} |4\rangle \; ,
\label{GrangierTwB}
\end{equation}
with $\gamma_{0}=0.96,\,\gamma_{1}=0.27,\,\gamma_{2}=0.10$.
The passage through the beam splitter yields the output state $|sv\rangle_{BS}$:
\begin{eqnarray}
|sv\rangle_{BS} & = & (\gamma_{0} |0\rangle_{1}
+ T^{2}\gamma_{1} |2\rangle_{1} +
T^{4}\gamma_{2} |4\rangle_{1}) |0\rangle_{2}
\nonumber \\
& & \nonumber \\
& + & (\sqrt{2}RT \gamma_{1}|1\rangle_{1} + 2R T^{3}\gamma_{2}|3\rangle_{1})
|1\rangle_{2}+\mathcal{O}(2) \; ,
\label{GrangierBSout}
\end{eqnarray}
where $T$ and $R$ denote the transmittance and the reflectivity of the beam splitter, respectively;
$|\cdot\rangle_{1}$ and $|\cdot\rangle_{2}$ denote, respectively,
the states of the transmitted and reflected modes;
and $\mathcal{O}(2)$ represents higher-order terms containing Fock states $|m\rangle_{2}$ with $m>1$.
Conditional detection of the state $|1\rangle_{2}$ leads to the generation of the non-Gaussian state:
\begin{equation}
|sv\rangle_{cond} \propto \gamma_{1}|1\rangle_{1} + \sqrt{2} \gamma_{2} T^{2} |3\rangle_{1} \; .
\end{equation}
Having started from the truncated superposition that includes $|0\rangle$,  $|2\rangle$, and
$|4\rangle$, the above superposition is clearly the associated one-photon-subtracted state. \\
Using a postselection scheme similar to that of Ref. \cite{ZavattaScience},
Resch \textit{et al.} have experimentally engineered a superposition of the form
$\alpha |0\rangle + \beta |1\rangle$ \cite{ReschCondCoher}.
In Refs. \cite{DeMartiniquantinjec,LamasScience}, conditional parametric amplification
has been used to produce high-fidelity quantum clones of a single-photon input state.
Finally, the conditional preparation of a bright sub-Poissonian beam (from a twin beam)
has been reported in Ref. \cite{LauratSubPoisslight}, and theoretically analyzed
in Ref. \cite{LauratSubPoisslight2}.

b) \hspace{2mm} In Ref. \cite{fockgen4} Vogel, Akulin and Schleich
have developed the second approach, and provided a recipe
to construct, in principle, a generic quantum state of the
radiation field. Their scheme, already illustrated in Subsection \ref{BinomialStates},
Fig. (\ref{recbingen}), for the production of the reciprocal binomial states,
is based on the interaction of $N$ two-level atoms with a resonant mode in a cavity.
This interaction can be typically described by the
Jaynes-Cummings Hamiltonian \cite{BarnettRadmoretx}. Initially, the cavity
field is prepared in the vacuum state. One atom at a time is
injected in the cavity in a superposition of the excited state
$|e\rangle$ and of the ground state $|g\rangle$:
$|e\rangle+i\epsilon_{k}|g\rangle$, where $\epsilon_{k}$ denotes a
complex controlling parameter associated to the $k$-th atom. Let
us suppose that $k-1$ atoms have been injected in the cavity
without performing any measurement; then, the cavity field will
evolve towards a state of the general form
$|\psi^{(k-1)}\rangle=\sum_{n=0}^{k-1}\psi_{n}^{(k-1)}|n\rangle$,
where the coefficients $\psi_{n}^{(k-1)}$ must be suitably
determined recursively, as we will see later.
Let us suppose now that a measurement is performed on
the $k$-th atom at the output of the cavity. If this atom is
detected in the excited state, the whole procedure must be
repeated. If instead the $k$-th atom is detected in its ground
state, it can be easily proved that the new state of the
cavity field is
\bea
&&|\psi^{(k)}\rangle \, = \, \sum_{n=0}^{k}\psi_{n}^{(k)}|n\rangle \; , \\
&& \nonumber \\
&&\psi_{n}^{(k)} \, = \, \sin
(\lambda \tau_{k}\sqrt{n})\psi_{n-1}^{(k-1)}-\epsilon_{k}\cos(\lambda \tau_{k}\sqrt{n})\psi_{n}^{(k-1)}
\; ,
\eea
where $\lambda$ denotes the atom-field coupling constant, and
$\tau_{k}$ denotes the interaction time. Therefore, the passage of
each atom increases by one unit the number of Fock states appearing
in the finite superposition. If the
target state to be engineered is of the generic form
$|\psi_{tar}\rangle=\sum_{n=0}^{N}c_{n}|n\rangle$, where the $N+1$
coefficients $c_{n}$ have been fixed \textit{a priori}, one must
send in the cavity $N$ two-level atoms. Moreover, it can be shown
that the initial internal states of the $N$ atoms can be prepared
in such a way that the controlling parameters $\epsilon_{1},\ldots
,\epsilon_{N}$ assure the identification $\psi_{n}^{(N)}=c_{n}$,
$\forall\; n$. This completes
the procedure.

c) \hspace{2mm} A scheme
similar to that described above, exploiting Jaynes-Cummings models
in high-$Q$ cavities, has been proposed for the generation of
superpositions states, and in particular of macroscopic quantum
superpositions \cite{MQSjaycum1,MQSjaycum2,MQSjaycum3,MQSjaycum4}.
A nice proposal is due to
Meystre, Slosser and Wilkens \cite{MQSjaycum1}. These authors
consider, at very low temperature, a micromaser cavity
pumped by a stream of
polarized two-level atoms, and show that macroscopic
superpositions are generated in the cavity. Moreover, they show that the
onset of these superpositions can be interpreted in terms of a first order
phase transition; at variance with the usual bistable systems
involving incoherent mixture of the states localized at the two minima
of an effective potential, this transition
is characterized by coherent superpositions of such states. \\
In Refs. \cite{MQSjaycum3,MQSjaycum4} it has been introduced
a method for the
generation of mesoscopic (macroscopic) superpositions, that
is based on a two-photon resonant Jaynes-Cummings model, in which
a cascade of two atomic transitions of the kind
$|e\rangle\rightarrow|i\rangle\rightarrow|g\rangle$ is resonant
with twice the field frequency $\omega$: $\omega_{eg}=2\omega$.
Here and in the following $\omega_{eg}$, $\omega_{ig}$, will
denote the transition frequencies from the excited state
$|e\rangle$ and the intermediate state $|i\rangle$ of an atom to
its ground state $|g\rangle$, $\omega_{ei}$ will denote the transition frequency
from the excited state $|e\rangle$ to the intermediate state $|i\rangle$, and
$\Omega_{eg}$, $\Omega_{ig}$, $\Omega_{ei}$ will
denote the Rabi frequencies associated to the same pairs of states.
The intermediate transition frequences
$\omega_{ig}$ and $\omega_{ei}$ are
assumed strongly detuned from $\omega$ by
$\Delta/2=\omega-\omega_{ei}=\omega_{ig}-\omega$.
The dynamic evolution
in the two-photon resonant Jaynes-Cummings model can be obtained
in complete analogy to the ordinary one-photon model, by noting
that the intermediate state $|i\rangle$ can be eliminated. In fact,
by introducing the effective Rabi frequency \cite{harochemas}
$\Omega_{n}=[(\Omega_{ei}^{2}+2\Omega_{ig}^{2})
+n(\Omega_{ei}^{2}+\Omega_{ig}^{2})]/\Delta$,
corresponding to an $n$-photon Fock state between $|e\rangle$ and
$|g\rangle$, the state $|i\rangle$ remains unpopulated during
the atom-fied interaction time $t$ if the
condition $\Omega_{n}^{2} \, t/ |\Delta| \ll \pi$ is
satisfied \cite{Toor}. Let us consider the case
in which a single atom is initially in the excited state
$|e\rangle$, and the cavity field is in an arbitrary state; the
whole atom-field state is then of the factorized form
$|\Psi(0)\rangle \equiv |e\rangle |\Psi_{c}(0)\rangle
=|e\rangle\sum_{n=0}^{\infty}c_{0}(n)|n\rangle$.
After a time $t$, in which the atom-field state evolves from
$|\Psi(0)\rangle$ to $|\Psi(t)\rangle$, we can perform a
conditional measurement on the atom, obtaining as output the
excited state with a probability $P_{e}=|\langle
e|\Psi(t)\rangle|^{2}$. If state $|e\rangle$ is
effectively observed, the coefficients $c_{1}(n)$ of the new state
of the field in the cavity, $|\Psi_{c}(t)\rangle=\sum_{n=0}^{\infty}c_{1}(n)|n\rangle$,
can be computed in terms of the initial coefficients $\{c_{0}(n)\}$,
the effective Rabi frequencies, and the probability $P_{e}$, and
they read $c_{1}(n)=(P_{e})^{-1/2}c_{0}(n)\cos(\Omega_{n+2}t/2)$.
At this stage of the procedure any quantum state can be, in
principle, obtained. However, if an initial coherent state
$|\alpha\rangle$ is chosen (i.e. if the initial coefficients
$c_{0}(n)$ are the coefficients of the coherent superposition),
the state $|\Psi_{c}(t)\rangle$ describes a mesoscopic
superposition of coherent states. The method can be easily
generalized \cite{MQSjaycum4} to conditional excitation
measurements on $M$ atoms, leading to more complex superpositions,
and to a better control on the statistics of the states. \\
It is to be remarked that decoherence effects make very difficult to
implement the methods to create mesoscopic or
macroscopic superpositions of states of the radiation field.
However, exploiting high-$Q$ cavities, in 1996, Brune
\textit{et al.} have succeeded in
obtaining an experimental realization of a mesoscopic
superposition of an "atom $+$ measuring apparatus" (atom $+$
cavity field) of the form
$|\Psi_{cat}\rangle=2^{-1/2}(|e,\alpha e^{i\phi}\rangle+|g,\alpha
e^{-i\phi}\rangle)$, and in observing its progressive decoherence
\cite{mesoscopBruneexp}. Moreover, in 1997, Raimond \textit{et
al.} have prepared a Sch\"odinger cat made of few photons, and
studied the dynamics of its decoherence \cite{mesoscopBruneexp2}.
Both these experiments constitute a cornerstone in the
investigation of the quantum/classical boundary by quantum
optical methods. For a review on
this subject see Ref. \cite{RevHaroche}.

d) \hspace{2mm} An interesting proposal to control quantum states
of a cavity field by a pure unitary evolution is based on a
two-channel approach (both classical and quantum controlling
radiation fields) \cite{contreberly}. A two-level atom in the
cavity interacts with an external \textit{classical} field
$E_{ext}(t)$, and is coupled as well with the quantum cavity field
$a$, with coupling constant $g(t)$. In resonance conditions, the
controlling interaction Hamiltonian is given by \cite{contreberly}
\be
H_{I}(t) \, = \, [E_{ext}(t)+\lambda(t)a]\sigma^{+} \, + \, H.c. \; ,
\label{Heberly}
\ee
where $\sigma^{\pm}$ represent the standard Pauli matrix
notation for the atomic transitions. The controlling interaction
can force the initial state $|\Psi(0)\rangle=|0,g\rangle$ of the
system, i.e. the product of the vacuum cavity field state and of
the atomic ground state, to evolve during a time $\tilde{t}$
towards the general  (again factorized) final form
$|\Psi(\tilde{t})\rangle=\sum_{n=0}^{M}c_{n}|n,g\rangle$. The
procedure can be engineered as follows. The time interval
$[0,\tilde{t}]$ is divided in the subintervals
of equal lengths, $0 < \tau < 2 \tau \cdots
 < j \tau < (j+1) \tau \cdots < (2M-1) \tau < \tilde{t}\, , \;$ $\tau=\tilde{t}/2M$.
Further, the classical and the quantum channel are led to act
alternatively on the time subintervals, by assuming for the
functions $E_{ext}(t)$ and $\lambda(t)$ the following step-periodic
time modulation:
\bea &&\left\{\begin{array}{c}
  E_{ext}(t)=E_{j} \\
  \lambda(t)=0
\end{array} \right. \; , \quad \quad for \quad 2(j-1) \tau \, < \, t \, < \,
(2j-1)\tau \;  \\  &&\left\{\begin{array}{c}
  E_{ext}(t)=0 \\
  \lambda(t) \, = \, \lambda_{j}
\end{array} \right. \; , \quad \quad for \quad (2j-1) \tau \, < \, t \, < \, 2j\tau \; ,
\eea
where $E_{j}$ and $\lambda_{j}$ $(1\leq j\leq M)$ are complex
constants. The time evolution operator of the system is, thus,
given by a product of evolution operators associated with each
time interval, in the form \be
U(\tilde{t})=Q_{M}C_{M}Q_{M-1}C_{M-1}\cdots Q_{2}C_{2}Q_{1}C_{1}
\,, \ee where $Q_{j}$ and $C_{j}$ represent, respectively,
the quantum and the classical evolutions in the $j$-th interval,
and can be expressed in the form of $2\times 2$ matrices
\cite{contreberly}. The controlling values $\{E_{j}\}$ and
$\{\lambda_{j}\}$ of the external classical field and of the quantum
coupling can be finally determined by solving the equation of motion
of the time reversed evolution
\be
|0,g\rangle \, = \, U(-\tilde{t})|\Psi(\tilde{t})\rangle \, = \, C_{1}^{\dag}Q_{1}^{\dag}\cdots
C_{M}^{\dag}Q_{M}^{\dag}|\Psi(\tilde{t})\rangle \; ,
\ee
which, by connecting the desired final state to the given initial state,
completes the procedure. The possible realization of this model in a cavity
through a two-channel Raman interaction has been discussed as well in
Ref. \cite{contreberly}. A further scheme for the
engineering of a general field state in a cavity, which, at variance
with the previous case, is not empty but prepared in a coherent
state, has been considered in Ref. \cite{engineermous}.

e) \hspace{2mm} Arbitrary multiphoton states of a single-mode
electromagnetic field can be produced as well by exploiting discrete
superpositions of coherent states along a straight line or on a
circle in phase space \cite{stenginchsup,stenginchsup2}. To this
aim, let us consider, for instance, the discrete superposition of
$n+1$ coherent states symmetrically positioned on a circle with
radius $r$ \cite{stenginchsup}:
\be
|n,r\rangle \, = \, \mathcal{N}(r)\frac{\sqrt{n!}e^{r^{2}/2}}{(n+1)r^{n}}\sum_{k=0}^{n}e^{\frac{2\pi
i }{n+1}k}|re^{\frac{2\pi i }{n+1}k}\rangle \; ,
\ee
where the coherent amplitude $\alpha_{k}$ of the $k$th coherent state
entering the superposition, reads $\alpha_{k}=r\exp\{(2ik\pi)/(n+1)\}$,
and $\mathcal{N}(r)$ is a normalization constant. It can be shown that
$\lim_{r\rightarrow 0}|n,r\rangle=|n\rangle$ and that, in general,
the superposition state $|n,r\rangle$ can approximate the number
state $|n\rangle$ with good accuracy. Thus it is clear that a
general quantum state can be constructed by suitable
superpositions of coherent states, and an experimental scheme has been
proposed in Ref. \cite{stenginchsup2}.

f) \hspace{2mm} Another general method for preparing a quantum
state is based on the use of conditional output measurements on a
beam splitter \cite{Dakna1,Dakna2,Dakna3}. This method can be
applied both to cavity-field modes and to traveling-field modes.
For instance, a simple experimental setup for the generation of photon-added
states has been proposed in Ref. \cite{Dakna1}.
A signal mode, corresponding to a density matrix
$\rho^{(in)}=\sum_{\Phi}\tilde{p}_{\Phi}|\Phi\rangle\langle\Phi|$
which describes a general state, pure or mixed, with
$\sum_{\Phi}\tilde{p}_{\Phi}=1$ and $0\leq \tilde{p}_{\Phi}\leq
1$, and a reference mode, prepared in a number state $|n\rangle$,
are mixed at the two input ports of a beam splitter. It can be
shown that zero-photon conditional measurement at one output port
of the beam splitter can be used to generate, at the other output
port, photon added-states for a large class of possible initial
states of the signal mode: coherent states, squeezed states, and displaced
number states \cite{Dakna1}. In fact, denoting, as usual, by
$\eta$ the transmittance of the beam splitter, if no photons are
detected at one output port, the quantum state $\rho^{(out)}(n,0)$
of the mode at the other output port, which depends on the
reference mode $|n\rangle$, collapses, with a certain probability,
to the photon-added state
\be
\rho^{(out)}(n,0)\, = \, \sum_{\Phi}\tilde{p}_{\Phi}|\Psi_{n0}\rangle\langle\Psi_{n0}|
\; , \quad \quad |\Psi_{n0}\rangle \, = \, \mathcal{N}^{-1/2}a^{\dag
n}\eta^{a^{\dag}a}|\Phi\rangle \; ,
\ee
where $\mathcal{N}^{-1/2}$
is the normalization factor. The same scheme can be implemented to
produce photon-subtracted or photon-added Jacobi polynomial states
\cite{Dakna2}. At variance with the previous scheme, in this case
the measurement is conditioned to record $m$ photons at the output
channel, and the collapsed state $\rho^{(out)}(n,m)$ takes the form
\bea
&&\rho^{(out)}(n,m) \, = \, \sum_{\Phi}\tilde{p}_{\Phi}|\Psi_{nm}\rangle\langle\Psi_{nm}|
\; , \nonumber \\
&& \nonumber \\
&&|\Psi_{nm}\rangle \, = \, \mathcal{N}^{-1/2}\sum_{k=\mu}^{n}\frac{(-|R|^{2})^{k}}{(k-\nu)!}
\left(\begin{array}{c}
  n \\
  k
\end{array}\right) a^{k-\nu}a^{\dag k}\eta^{a^{\dag}a}|\Phi\rangle \; ,
\eea
where $R$ is the reflectance of the beam splitter:
$|R|^{2}=1-|\eta|^{2}$, $\nu=n-m$, and $\mu=\max \{0,\nu\}$.
Finally, a generalized procedure \cite{Dakna3}, based on this method,
has been devised to engineer an arbitrary quantum state, that can be
approximated to any desired degree
of accuracy by a finite superposition $|\Psi\rangle$ of number
states: $|\Psi\rangle=\sum_{n=0}^{N}\psi_{n}|n\rangle$. The state
$|\Psi\rangle$ can be written in the form
\be
|\Psi\rangle\, = \, \sum_{n=0}^{N} \frac{\psi_{n}}{\sqrt{n!}}a^{\dag
n}|0\rangle \, = \, (a^{\dag}-\beta_{N}^{*})(a^{\dag}-\beta_{N-1}^{*})\cdots
(a^{\dag}-\beta_{1}^{*})|0\rangle \; ,
\label{Daknasteng}
\ee
where
$\beta_{j}^{*}$ $(j=1,\ldots N)$ are the complex roots of the
characteristic polynomial
$\sum_{n=0}^{N}\frac{\psi_{n}}{\sqrt{n!}}(\beta^{*})^{n}=0$. Using
the relation $a^{\dag}-\beta=D(\beta)a^{\dag}D^{\dag}(\beta)$
(where $D(\beta)$ is the Glauber displacement operator), the state
(\ref{Daknasteng}) can be expressed in the form
\be
|\Psi\rangle \, = \, D(\beta_{N})a^{\dag}D^{\dag}(\beta_{N})D(\beta_{N-1})a^{\dag}D^{\dag}(\beta_{N-1})\cdots
D(\beta_{1})a^{\dag}D^{\dag}(\beta_{1})|0\rangle \; .
\label{Dakfinsteng}
\ee
Therefore, the truncated state
$|\Psi\rangle$ can be obtained from the vacuum by a sequence of
two-step procedures, each one constituted by a state displacement
and a single-photon adding. Such a sequence of operations can be realized,
for instance, according to the following proposed experimental
setup \cite{Dakna3}. The scheme is depicted in Fig. (\ref{engDak}), and
employs an array of $2N+1$ beam splitters and $N$ highly efficient
avalanche photodiodes.
\begin{figure}[h]
\begin{center}
\includegraphics[width=12cm]{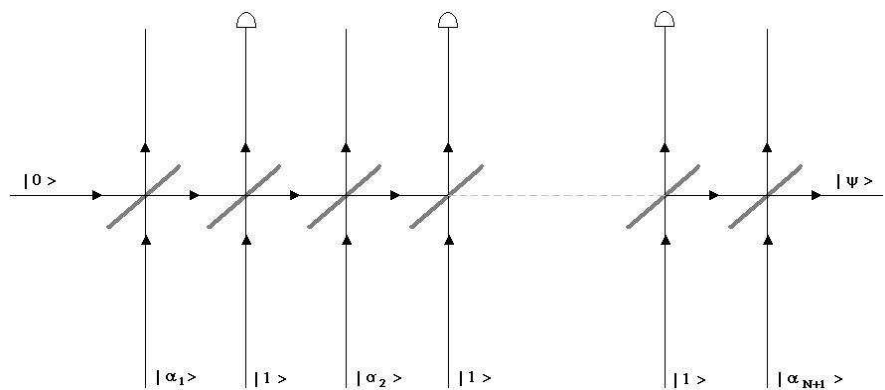}
\end{center}
\caption{Experimental setup for the engineering of arbitrary states of a traveling-field mode.
The array consists of $2N+1$ identical beam splitters all characterized by the same
transmittance $\eta$.
Coherent states $|\alpha_{i}\rangle$ and single-photon Fock states $|1\rangle$ in alternate sequence
enter one input port of the beam splitters. The desired output state $|\Psi\rangle$ is prepared
by a suitable choice of the coherent amplitudes $\alpha_{i}$.}
\label{engDak}
\end{figure}
Each two-step procedure corresponds to the action of a couple of
beam splitters, and is repeated $N$ times. If, for all the two-step
blocks, the photodiode detectors do not record photons, the
two-step procedure supplies each block with the operation
$a^{\dag}\eta^{a^{\dag}a}D(\alpha_{i})$ $(i=1,\ldots N)$, while
the last ($2N+1$)-th beam splitter realizes the final displacement
$D(\alpha_{N+1})$. Thus, with a certain probability,
the output state is:
\be |\Psi\rangle \, \sim \,
D(\alpha_{N+1})a^{\dag}\eta^{a^{\dag}a}D(\alpha_{N})a^{\dag}\eta^{a^{\dag}a}D(\alpha_{N-1})\cdots
a^{\dag}\eta^{a^{\dag}a}D(\alpha_{1})|0\rangle \; .
\label{Dakpsiout}
\ee
Finally, it can be shown that the states (\ref{Dakfinsteng})
and (\ref{Dakpsiout}) coincide for a suitable choice of the
experimental parameters $\alpha_{i}$ $(i=1,\ldots N+1)$
\cite{Dakna3}. This method has been proposed for the engineering of
the reciprocal binomial states and of the polynomial
states of the electromagnetic field \cite{ValverdeRB,AvelarPoly}.
Fiur\'{a}\v{s}ek \textit{et al.} have introduced an interesting
variation of the scheme of Ref. \cite{Dakna3}.
At variance with the first proposal, in Ref. \cite{FiurasekEnginee}
the engineering of an arbitrary quantum state is achieved by repeated
photon subtractions. This scheme is motivated by the fact that photon
subtractions are experimentally much more practicable than photon
additions. In their scheme the initial vacuum is replaced by a
squeezed vacuum, and photon subtractions are realized by conditioned
photodetections.

g) \hspace{2mm} Here we consider two other methods for the engineering
of nonclassical states by means of multiphoton processes in nonlinear
media. A first proposal is due to Leo\'{n}ski \cite{leonFDHS},
and it is based on a set of models leading to the generation of
quantum states that are very close to the coherent states of
finite dimensional Hilbert spaces. The $(s+1)$-dimensional
coherent states, finite dimensional approximations to the Glauber
coherent states, have been defined in Refs.
\cite{BuzekFDCS,MiranowFDCS}. They can be expressed in the
standard form
\be
|\alpha\rangle_{(s)} \, = \, \sum_{n=0}^{s}c_{n}^{(s)}|n\rangle \; ,
\quad \quad
\,_{(s)}\langle\alpha|\alpha\rangle_{(s)}\, = \, \sum_{n=0}^{s}|c_{n}^{(s)}|^{2}
\, = \, 1 \; ,
\label{FDCS}
\ee
where the coefficients $c_{n}^{(s)}$ have
been numerically computed by Bu\v{z}ek \textit{et al.}
\cite{BuzekFDCS}, and later analytically determined by Miranowicz
\textit{et al.} \cite{MiranowFDCS}. Leo\'{n}ski
has shown that, starting from an initial vacuum state, superpositions
very close to $|\alpha\rangle_{(s)}$ can be generated
by suitable interactions between an external field and a nonlinear
medium in a cavity\cite{leonFDHS}. \\
The second proposal has been presented in Ref.
\cite{vidiellaengine} by A. Vidiella-Barranco and J. A. Roversi,
and it is a generalization of the method introduced in Ref.
\cite{fockgen5}. The authors show that
an arbitrary pure state $|\psi\rangle$, expressed in the Fock
basis: $|\psi\rangle=\sum_{n=0}^{M}c_{n}|n\rangle$, can be obtained
as the results of the unitary evolution generated by the Hamiltonian
\be
H_{mph}^{M} \, = \, f_{0}(a^{\dag}a) \, + \,
\sum_{m=1}^{M}\frac{c_{m}}{\sqrt{m!}}[F(a^{\dag}a)a^{m} \, + \,
a^{\dag m}F(a^{\dag}a)] \; ,
\ee
where
\be
f_{0}(a^{\dag}a) \, = \, c_{0}[2(1-a^{\dag}a)F(a^{\dag}a)-1] \; , \quad \quad
F(a^{\dag}a) \, = \, \sum_{l=0}^{M}A_{l}\;(a^{\dag}a)^{l} \; .
\ee
The coefficients $A_{l}$ can be determined by the $M+1$ conditions
$A_{0}=1$ and $1+pA_{1}+p^{2}A_{2}+...+p^{M}A_{M}=0$, with
$p=1,...,M$. As an example, the state
$|\psi_{2}\rangle=c_{0}|0\rangle+c_{2}|2\rangle$ can be generated
by the Hamiltonian \cite{vidiellaengine}
\bea
&&H_{2ph}^{2} \,= \, c_{0}[1-5a^{\dag}a+4(a^{\dag}a)^{2}-(a^{\dag}a)^{3}]+
\frac{c_{2}}{\sqrt{2}}[F(a^{\dag}a)a^{2}+H.c.]
\label{Hvid2} \; , \\
&&F(a^{\dag}a) \, = \, 1-\frac{3}{2}a^{\dag}a+\frac{1}{2}(a^{\dag}a)^{2}
\; ,
\label{Hvidc2}
\eea
which, however, would require, to be
realized in nonlinear media, contributions from susceptibilities of
very high order.

We conclude this Subsection by briefly mentioning some
more recent proposals of quantum state engineering.
A scheme based on the use of conditional measurements
on entangled twin-beams to produce and manipulate nonclassical
states of light has been investigated in Ref. \cite{Parisengin}.
Another proposal has been presented in Ref. \cite{squeezengin}
in order to engineer squeezed cavity-field states
via the interaction of the cavity field with a driven three-level
atom. Finally, the high nonlinearities available in the
electromagnetically induced transparency regime offer further
possibilities to quantum state engineering. For example, in
Ref. \cite{Paternostro} the generation of entangled coherent states
via the cross-phase-modulation effect has been carefully analyzed.

\subsection{Nonclassicality of a state: criteria and measures}

\label{NonclassicalScriter}

Although a universal criterion for detecting nonclassicality of a
quantum state is a somewhat elusive concept, in this Subsection
we will discuss some interesting proposals aimed at qualifying
and quantifying nonclassicality, limiting our attention to
single-mode states of the radiation field. As we
have already seen, the concept of nonclassicality of a state of
the radiation field is generally associated with some
emerging physical property with no classical counterpart.
Typically, nonclassicality can be measured in a variety of
ways, e.g. by computing the degree of squeezing
\cite{naturewalls,HongMandelSqueez,zhangsqueez,HongNC}; by
observing the sub-Poissonian behavior of the statistics
\cite{MandelQpar,ShortMandel} or the presence of oscillations in
the photon number distribution \cite{SchillerOscillPND}; and by
investigating the non-existence or the negativity of the
phase-space quasi-probability distributions
\cite{Glauber3,negativeWig}. Besides these traditional and useful
tests, many other types of indicators, conditions, and
measures highlighting different nonclassical signatures, have been
introducedin the literature. Among the most interesting, we mention
the distance of a considered state to a set of reference states (e.g. the coherent states)
\cite{HilleryNC,DodonovNC,MarianNC,ZyczkowskiNC}; some peculiar behaviors of
quasi-probability distributions or characteristic functions
\cite{negativeWig,negativeWig2,LeeNC,LutkenhausNC,JanszkyNC,VogelNC,Lvovskystatmixture01};
the violation of inequalities involving the moments of the annihilation and creation operators
\cite{catsnonclasprop2,AgarwalNC,KlyshkoNC,ArvindNC,DArianoNC,VogelNC2,VogelNC3};
and, finally, the violation of inequalities involving the characteristic
functions \cite{VogelCharactNC,KorbiczNC}. \\
The definition of a nonclassical distance in the domain of quantum optics has been introduced
by Hillery \cite{HilleryNC}.
Such a distance $\delta_{H}(\rho,\rho_{cl})$
of a certain state $\rho$ from a set
of reference classical states $\rho_{cl}$ is defined as
\begin{equation}
\delta_{H}(\rho,\rho_{cl})=\inf_{\rho_{cl}}\parallel\rho-\rho_{cl}\parallel_{1} \,,
\label{Hillerydist}
\end{equation}
where $\rho_{cl}$ ranges over all classical density matrices and $\parallel \cdot\parallel_{1}$
denotes the (trace) norm of an operator:
\begin{equation}
\parallel A \parallel_{1}=Tr[|A|]=Tr[(A^{\dag}A)^{1/2}] \,,
\end{equation}
for a generic operator $A$.
Following the path opened by Hillery, other distances have been considered
to define computable indicators of nonclassicality.
For instance, the Hilbert-Schmidt distance $d_{HS}$
\begin{equation}
d_{HS}(\rho,\rho_{cl}) =
\parallel\rho-\rho_{cl}\parallel_{2}=\{Tr[(\rho-\rho_{cl})^{2}]\}^{1/2} \; ,
\end{equation}
has been used to define the following measure of nonclassicality \cite{DodonovNC}:
\begin{equation}
\delta_{HS}(\rho,\rho_{cl}) = \min_{\rho_{cl}} d_{HS}(\rho,\rho_{cl}) \; .
\end{equation}
The Bures distance $d_{B}(\rho,\rho_{cl})$ \cite{BuresDist} is defined as
\begin{equation}
d_{B}(\rho,\rho_{cl})=\left\{2-2
\sqrt{\mathcal{F}(\rho,\rho_{cl})}\right\}^{1/2} \; ,
\label{BuresDistance}
\end{equation}
where
\begin{equation}
\mathcal{F}(\rho,\rho_{cl})=\{Tr[(\sqrt{\rho}\,\rho_{cl}\,\sqrt{\rho})^{1/2}]\}^{2} \; .
\label{InfidelFidelity}
\end{equation}
The Bures distance has been exploited in Ref. \cite{MarianNC}
to introduce a measure of nonclassicality of the form
\begin{equation}
\delta_{B}^{2}(\rho,\rho_{cl}) =
\min_{\rho_{cl}} \frac{1}{2}d_{B}^{2}(\rho,\rho_{cl}) \; .
\label{BuresMarian}
\end{equation}
The quantity in Eq. (\ref{BuresMarian}) is obviously more difficult to compute
compared to measures based on the trace norm of operators.
We should finally mention that in phase space it is possible to define a
further distance-based measure of nonclassicality, the so-called Monge
distance between the Husimi distribution of the chosen state and that
of the reference coherent state \cite{ZyczkowskiNC}. \\
A very interesting approach to the detection, and even to the
measurement, of the nonclassicality of a state is based
on the behavior of the corresponding quasi-probability distributions or
characteristic functions in phase space.
A nonclassical depth in phase space can be defined by introducing
a proper distribution $R$ interpolating between the $P$- and
the $Q$- functions \cite{LeeNC}
\begin{equation}
R(\alpha,\tau)=
\frac{1}{\pi \tau}\int d^{2}w \exp\left\{-\frac{1}{\tau}|\alpha-w|^{2}
\right\}P(w) \; ,
\end{equation}
where $\tau$ is a continuous parameter
(for $\tau=0,\,1/2,\,1$ the distribution $R$ reduces
to the $P$-, $W$-, and $Q$- function, respectively).
The nonclassical depth $\tau_{ncl}$ is defined as the lower bound of the set
of values of $\tau$ that lead to a smoothing of the $P$-function of a
quantum state. It turns out that $\tau_{ncl}=0$ for a coherent state,
$\tau_{ncl}=(e^{2r}-1)/2e^{2r}$ for a squeezed state ($r$ being the
squeezing parameter), and $\tau_{ncl}=1$ for a number state \cite{LeeNC};
this finding gives support to the validity of $\tau_{ncl}$ as an
estimator of nonclassicality.
According to L\"{u}tkenhaus and Barnett \cite{LutkenhausNC},
a quantitative measure of nonclassicality can be associated
with the $p$ parameter of the $p$-ordered phase-space distributions $W(\alpha,\,p)$
(see Eq. (\ref{pOrderedDistributions})). In fact, two quasi-probability
distributions for the same state but different values of $p$ are related by
a convolution of the form
\begin{equation}
W(\alpha,\,p')=\int d^{2}\beta \; W(\beta,\,p'') \frac{2}{\pi(p''-p')}
\exp\left\{-\frac{2|\alpha-\beta|^{2}}{p''-p'}\right\} \; ,
\label{Lutkenhausconv}
\end{equation}
with $p''>p'$. Eq. (\ref{Lutkenhausconv}) can be viewed as the solution to
a sourceless diffusion equation with $p$, playing the role of backward time:
in fact $W(\alpha,\,p)$ is smoothed in the direction of decreasing $p$.
The critical value of $p$ for which the boundary of the
well-behaved quasi-probability distributions is reached,
can be assumed as a measure of nonclassicality \cite{LutkenhausNC}.
Probably the most intuitive indicator of nonclassicality for a
quantum state $|\psi\rangle$ is the one based on the volume of the
negative part of the Wigner function \cite{negativeWig,negativeWig2}:
\begin{equation}
\delta_{\psi} = \int\int dx_{\theta} \, dp_{\theta} \;
|W_{\psi}(x_{\theta},p_{\theta})| -1 \; ,
\end{equation}
where $W_{\psi}(x_{\theta},p_{\theta})$ is the Wigner function
corresponding to the state $|\psi\rangle$ in the quadrature
representation. A further phase space related criterion has been
introduced by Vogel, who has proposed to define that a quantum
state is nonclassical if the modulus of its characteristic
function exceeds that of the vacuum state at all points in the
space of definition \cite{VogelNC}. This definition has been used
to verify experimentally, via quantum tomographic reconstruction,
the nonclassical nature of statistical mixtures of the vacuum
state $|0\rangle$ and of the single-photon Fock state $|1\rangle$
\cite{Lvovskystatmixture01}. Unfortunately, it has been shown that
some nonclassical states violate this criterion \cite{Diosi}. \\
Another different typology of criteria is based on hierarchies of
inequalities involving the statistical moments of the annihilation
and creation operators $a$ and $a^{\dag}$. Agarwal and Tara have
introduced sufficient conditions for nonclassicality based on the
moments of the photon number operator \cite{catsnonclasprop2}. The
starting point is the diagonal coherent state expansion of the
density matrix $\rho$
\begin{equation}
\rho = \int d^{2} \alpha \, P(\alpha) |\alpha\rangle\langle\alpha| \; .
\end{equation}
Recalling the definition given in Section \ref{section2}, a state $\rho$ is said
to be "classical" if the $P$-function is pointwise nonnegative,
and nowhere more singular than a $\delta$-function
(that is a classical behavior of the $P$-function as a probability density).
Let us now consider a Hermitian operatorial function $\hat{F}_{n}(a^{\dag},a)$
of normally ordered creation and annihilation operators,
i.e. $\langle\alpha|\hat{F}_{n}(a^{\dag},a)|\alpha\rangle = F_{n}(\alpha^{*},\alpha)$.
The quantum mechanical expectation value of $\hat{F}_{n}$ in the state $\rho$ is
\begin{equation}
\langle \hat{F}_{n} \rangle = Tr[\rho \hat{F}_{n}] =
\int d^{2} \alpha \, P(\alpha)F_{n}(z^{*},z) \; .
\label{normorderF}
\end{equation}
The state $\rho$ is then defined to be nonclassical if
$\langle \hat{F}_{n}\rangle < 0$ ("quantum negativity")
for some $F_{n}(z^{*},z) \geq 0$.
Let us choose
\begin{equation}
\hat{F}_{n}(a^{\dag},a) =
\sum_{r,s=0}^{n-1} c_{r}^{*}c_{s} \, a^{\dag (r+s)}a^{(r+s)} \; ,
\end{equation}
with $c_{k}$ arbitrary constants, and let us define
$m^{(k)}=\langle a^{\dag k}a^{k}\rangle$. Then, if
$P(\alpha)$ behaves like a classical distribution,
the following quadratic form in the constants $\{c_{k}\}$:
\begin{equation}
F_{n}(\{c_{k}\}) = \sum_{r,s=0}^{n-1} c_{r}^{*}c_{s} \, m^{(r+s)} \; ,
\label{AgarTaraQuadform}
\end{equation}
should be positive, thus implying the positivity of the matrix
\begin{equation}
M_{n}=\left( \begin{array}{ccccc}
1 & m^{(1)} & m^{(2)} & \cdots & m^{(n-1)} \\
m^{(1)} & m^{(2)} & m^{(3)} & \cdots & m^{(n)} \\
m^{(2)} & m^{(3)} & m^{(4)} & \cdots & m^{(n+1)} \\
\vdots  & \vdots & \vdots & \ddots & \vdots \\
m^{(n-1)}  & m^{(n)} & m^{(n+1)} & \cdots & m^{(2n-2)} \\
\end{array}
\right) \; .
\end{equation}
It is easy to check that for $n=2$ the condition $M_{2} > 0$
reduces to the condition of positivity of the Mandel $Q$
parameter Eq. (\ref{Qmandel}). For $n>2$ we get conditions
on higher order moments, that are needed for
the full characterization of non-Gaussian states.
An example of such states is the photon-added thermal state,
described by the density operator
$\rho_{add-th}\propto a^{\dag m}e^{-\beta a^{\dag}a} a^{m}$ $(\beta >0)$,
that exhibits neither squeezing nor sub-Poissonian statistics in some ranges of
the inverse temperature $\beta$.
In order to test the nonclassicality of $\rho_{add-th}$, one can for instance
invoke the negativity of the matrix $M_{3}$. To this aim one can introduce
the following numerical form, bounded from below by $-1$ \cite{catsnonclasprop2}:
\begin{equation}
A_{3}=\frac{\det M_{3}}{\det W_{3} - \det M_{3} } \; ,
\end{equation}
where the matrix $W_{n}$ is constructed as $M_{n}$ with the
replacement $m^{(n)}\rightarrow w^{(n)}=\langle (a^{\dag}a)^{n}\rangle$.
The condition $A_{3} < 0$ determines the region of nonclassicality,
and it can be shown that for the state $\rho_{add-th}$ there exist ranges of $\beta$
such that $A_{3}<0$ even in the absence of squeezing and sub-Poissonian
statistics. The nonclassical nature of a superposition of coherent states
can be also tested in a similar way \cite{catsnonclasprop2}.
Different versions and generalizations of this criterion
have been given in Refs. \cite{AgarwalNC,KlyshkoNC}. \\
Among all the possible Hermitian operatorial functions $\hat{F}_{n}(a^{\dag},a)$
of normally ordered creation and annihilation operators, one can consider
the set of phase invariant (number conserving) operators, i.e. the ones such
that $[\hat{F}_{n},a^{\dag}a]=0$. Arvind \textit{et al.}
have introduced a phase-averaged $P$-function \cite{ArvindNC}
\begin{equation}
\mathcal{P}(I)=\int_{0}^{2\pi}
\frac{d\theta}{2\pi}P(I^{1/2}e^{i\theta}) \; .
\label{PhaseAveraged}
\end{equation}
From definition Eq. (\ref{PhaseAveraged}), the quantities
$\langle\hat{F}_{n}\rangle$ can be written as
\begin{equation}
\langle \hat{F}_{n}(a^{\dag},a) \rangle = \int_{0}^{\infty}dI \,
\mathcal{P}(I) \, F_{n}(I^{1/2},I^{1/2}) \; .
\label{Mukundaaverage}
\end{equation}
By requiring the nonnegativity of the expectations values (\ref{Mukundaaverage})
for a classical state, then the following finer definition of nonclassicality can be
stated \cite{ArvindNC}: a quantum state $\rho$ is weakly
nonclassical if $\mathcal{P}(I)\geq 0$, but $P(\alpha)\ngeq 0$;
and strongly nonclassical if also the quantity $\mathcal{P}(I)$
ceases to be a probability distribution, that is:
 $\mathcal{P}(I)\ngeq 0$, and $P(\alpha)\ngeq 0$.
D'Ariano \textit{et al.} have proposed an
experimental test of these conditions by exploiting homodyne
tomography \cite{DArianoNC}, taking into account also imperfect
quantum efficiency of the homodyne detection. Along the same line
followed in Refs. \cite{catsnonclasprop2,ArvindNC} very recent
criteria of nonclassicality, constructed as an infinite series of
inequalities, have been formulated in terms of normally ordered
moments of the annihilation
and creation operators \cite{VogelNC2,VogelNC3}. Let us briefly outline the
method to derive these conditions. Let $\hat{f}(a^{\dag},a)$ be an
operatorial function whose normally ordered form exists. The
occurrence of negative mean values of the form
\begin{equation}
\langle : \hat{f}^{\dag}\hat{f} : \rangle =
\int d^{2}\alpha \, |f(\alpha)|^{2} P(\alpha) \; < 0 \,
\label{ShchukincondNC}
\end{equation}
is a signature of nonclassicality.
The condition (\ref{ShchukincondNC}) also implies the violation of the Bochner theorem
for the existence of a classical characteristic function \cite{BochnerTheorem}
(a characteristic function $\chi(z,w)$, satisfying the condition $\chi(0,0)=1$,
is a classical characteristic function if and only if it is positive semidefinite).
If $\hat{f}$ is chosen as
\begin{equation}
\hat{f}(a^{\dag},a)=\sum_{k=0}^{K} \sum_{l=0}^{L} c_{kl} \, a^{\dag k}a^{l} \,,
\end{equation}
with arbitrary constants $c_{kl}$, then, for a classical state, the quadratic form
\begin{equation}
\langle : \hat{f}^{\dag} \hat{f} : \rangle =
\sum_{k,r=0}^{K} \sum_{l,s=0}^{L} c_{rs}^{*} c_{kl} \, \langle a^{\dag l+r}a^{k+s} \rangle \,,
\end{equation}
should be positive.
Equivalently, the condition for the classicality of a state is provided by the positivity
of a hierarchy of determinants $d_{N}$ of square matrices of dimension $N\times N$
(where $N=K+L$, and $K$, $L$ arbitrary) of the form:
\begin{equation}
d_{N}=\left| \begin{array}{ccccccc}
1 & \langle a \rangle & \langle a^{\dag} \rangle & \langle a^{2} \rangle & \langle a^{\dag}a
\rangle & \langle a^{\dag 2} \rangle & \cdots \\
\langle a^{\dag} \rangle & \langle a^{\dag}a \rangle & \langle a^{\dag 2} \rangle &
\langle a^{\dag}a^{2} \rangle & \langle a^{\dag 2}a \rangle & \langle a^{\dag 3} \rangle & \cdots \\
\langle a \rangle & \langle a^{2} \rangle & \langle a^{\dag}a \rangle & \langle a^{3}
\rangle & \langle a^{\dag}a^{2} \rangle & \langle a^{\dag 2}a \rangle & \cdots \\
\langle a^{\dag 2} \rangle & \langle a^{\dag 2}a \rangle & \langle a^{\dag 3} \rangle &
\langle a^{\dag 2}a^{2} \rangle & \langle a^{\dag 3}a \rangle & \langle a^{\dag 4} \rangle & \cdots \\
\langle a^{\dag}a \rangle & \langle a^{\dag}a^{2} \rangle & \langle a^{\dag 2}a \rangle &
\langle a^{\dag}a^{3} \rangle & \langle a^{\dag 2}a^{2} \rangle & \langle a^{\dag 3}a \rangle & \cdots \\
\langle a^{2} \rangle & \langle a^{3} \rangle & \langle a^{\dag}a^{2} \rangle & \langle a^{4}
\rangle & \langle a^{\dag}a^{3} \rangle & \langle a^{\dag 2}a^{2} \rangle & \cdots \\
\cdots & \cdots & \cdots & \cdots & \cdots & \cdots & \cdots \\
\end{array}
\right| \; .
\end{equation}
By this procedure one can prove that a state is nonclassical if and only if
at least one of the determinants satisfies \cite{VogelNC3}
\begin{equation}
d_{N}<0 \; ,
\end{equation}
with $N > 2$, because $d_{2}$ represents the incoherent part of the photon
number $\langle a^{\dag} a \rangle - \langle a^{\dag} \rangle
\langle a \rangle$ which is always nonnegative and thus cannot be used
as a quantifier of nonclassicality.
It is possible to provide other sufficient conditions for nonclassicality
by considering subdeterminants of $d_{N}$ obtained by deleting pairs of
lines and columns that cross at a diagonal element of the matrix.
Refs. \cite{VogelNC3,KorbiczNC} investigate the connection
between nonclassicality criteria, the quantumness of a state,
and the 17th Hilbert problem \cite{Reznick}. The latter
states that not every positive semidefinite
polynomial must be a sum-of-squares of other polynomials. \\
We finally wish to mention another very interesting approach
that relates the nonclassicality of a single-mode quantum state
to the amount of two-mode entanglement that can be generated by doubling
the original field mode through linear optical elements (beam splitters),
auxiliary classical states, and ideal photodetectors \cite{AsbothNC}.

\newpage

\section{Canonical multiphoton quantum optics}
\label{section6}

In Section \ref{section5} we have reviewed several interesting
approaches sharing a common goal: the generalization of two-photon
interactions to multiphoton processes associated to
higher order nonlinearities, for the engineering of
multiphoton nonclassical states of the electromagnetic
field. However relevant in various aspects, all these methods
fail in extending the elegant canonical formalism of two-photon
processes, that, via the linear Bogoliubov transformation (or the
equivalent unitary squeezing operator) determines completely
the physical and mathematical structure of two-photon quantum
optics, including the diagonalizable two-photon Hamiltonians,
the corresponding normal modes, and the exactly computable
two-photon squeezed states.
Inspired by its power and generality, recently a series of papers has
appeared, that generalizes the canonical formalism to multiphoton
quantum optics \cite{noi1,noi2,noi3,noi4,noi5}.
This generalized canonical formalism has been constructed both
for one mode \cite{noi1,noi3}, and for two modes \cite{noi2}
of the electromagnetic field.
For single-mode systems
the canonical structure has been determined by introducing
canonical transformations that depend on generic nonlinear
functions of the homodyne combinations of pairs of canonically
conjugated quadratures \cite{noi1}. This homodyne canonical
formalism defines the class of the single--mode, homodyne
multiphoton squeezed states (HOMPSS), which include the
single--mode, single--quadrature multiphoton squeezed states
as a particular case \cite{noi3,noi4}.
The nonlinear canonical
formalism has been extended to define two-mode
multiphoton nonclassical states \cite{noi2}.
This further generalization is achieved by introducing canonical
two-mode transformations that depend on nonlinear functions of heterodyne
variables \cite{heterodet1,heterodet2}.
The corresponding canonical structure, strictly related to the
entangled state representation \cite{entanrep,entanrep2,entanrep3},
defines highly nonclassical, two-mode entangled states,
the heterodyne multiphoton
squeezed states (HEMPSS).
A very nice feature of canonical multiphoton quantum optics is
that the canonical transformations depend in general on tunable
physical parameters. In the case of HOMPSS, the adjustable parameter
is the local-oscillator mixing angle. In the case of HEMPSS, there
may be several adjustable quantities, as will be discussed
in detail in the following. The existence of free parameters allows
to arbitrarily vary the statistics of the states and
to interpolate between different Hamiltonian models of multiphoton
processes. Finally, one can envisage
relatively simple and in principle feasible experimental schemes
for the production of multiphoton nonclassical states in nonlinear
media by realizing elementary interaction models based on the
nonlinear canonical structure \cite{noi2}.

\subsection{One-mode homodyne multiphoton squeezed states: definitions and statistical properties}

In this Subsection we describe the single-mode, nonlinear
canonical transformations \cite{noi1}, introduce the definition
of the HOMPSS, and discuss the main aspects of the
formalism. We generalize the one-mode, linear two-photon
Bogoliubov transformation by defining a
quasi-photon operator $b$ of the form
\begin{equation}
b \, = \, {\tilde\mu} a_{\theta} \, + \, {\tilde\nu} a_{\theta}^{\dagger}
\, + \, \gamma F(X_{\theta}) \; , \quad \quad
X_{\theta}\, = \, \frac{a_{\theta} + a_{\theta}^{\dagger}}{\sqrt{2}} \; ,
\label{boperg}
\end{equation}
where $\tilde{\mu}$, $\tilde{\nu}$, and $\gamma$
are generic complex parameters, $a_{\theta}=e^{-i\theta}a$ is the rotated
annihilation operator, and $F$ is an arbitrary hermitian nonlinear
function of the homodyne quadrature observable
$X_{\theta}= X\cos{\theta} + P\sin{\theta}$ (where $X$ and $P$
are the original quadrature observables). The transformation is canonical,
namely $[b,b^{\dag}]=1$, if the parameters satisfy
\begin{equation}
|\tilde{\mu}|^{2} - |\tilde{\nu}|^2 \, = \, 1 \; , \quad \quad  \quad
Re[{\tilde\mu} \gamma^{*} - {\tilde\nu}^{*} \gamma] \, = \, 0 \; .
\label{canonlin}
\end{equation}
The first condition in Eq. (\ref{canonlin}) correctly reproduces
the constraint corresponding to the linear
Bogoliubov transformation. Remarkably, the additional condition in
Eq. (\ref{canonlin}) does not depend on the specific choice of the
nonlinear function $F$, but only on its strength $\gamma$.
The first contraint is automatically satisfied with the
standard parametrizations ${\tilde\mu}\equiv
e^{i\theta}\mu=e^{i\theta} \cosh r$, ${\tilde\nu}\equiv
e^{-i\theta}\nu=e^{i(\phi-\theta)}\sinh r$.
Writing $\gamma=|\gamma|e^{i\delta}$,
the second constraint in Eq. (\ref{canonlin}) is then recast in
the two (equivalent) forms
\begin{eqnarray}
&&\cosh r
\cos(\delta-\theta) \, - \, \sinh r
\cos(\delta+\theta-\phi) \, = \, 0 \; ,
\label{canonlin1} \\
&& \nonumber \\
&&\tan\left(\delta-\frac{\phi}{2}\right)\tan\left(\theta-\frac{\phi}{2}\right)
\, = \, -e^{-2r} \; .
\label{canonlin2}
\end{eqnarray}
The form (\ref{canonlin1}) suggests immediately, by imposing the vanishing
of the trigonometric functions, one exact solution that preserves the
freedom on the choice of the squeezing parameter $r$:
\begin{equation}
\delta-\theta \, = \, \pm\frac{\pi}{2}
\;, \quad \quad \quad \delta+\theta-\phi \, = \, \pm\frac{\pi}{2}
\; .
\label{canconlimit}
\end{equation}
Alternatively, the second form (\ref{canonlin2}) allows
to determine, in general numerically, all the possible
solutions, either by constraining the squeezing parameter
$r$ alone, or only one of the phases. \\
We denote by $|\beta,\gamma\rangle_{b}$ the generalized coherent
states associated with the transformation (\ref{boperg}), defined
as the eigenstates of the transformed annihilation operator $b$:
\begin{equation}
b \, |\beta,\gamma\rangle_{b} \, =
\, \beta \, |\beta,\gamma\rangle_{b} \; ,
\label{eigeneq}
\end{equation}
where $\beta$ is a generic complex number $\beta \, = \,
|\beta|e^{i \varsigma}$.
The solutions of Eq.~(\ref{eigeneq}) define, in a sense
that will become clear in the following, the (one-mode)
{\it homodyne multiphoton squeezed states}, or HOMPSS in short.
In the rotated quadrature representation $\{|x_{\theta}\rangle\}$,
their wave function reads
\begin{eqnarray}
\psi_{\beta,\gamma}(x_{\theta})&& \, = \, \langle
x_{\theta}|\beta,\gamma\rangle_{b} \, = \nonumber \\
 && \nonumber \\
 &&\pi^{-1/4} \left( Re\left[\frac{{\tilde\mu}+{\tilde\nu}}{{\tilde\mu}
-{\tilde\nu}}\right]\right)^{1/4}\exp\left\{-\frac{1}{2}|\beta|^{2}
-\frac{1}{2}\frac{{\tilde\mu}^{*}-{\tilde\nu}^{*}}{{\tilde\mu}-{\tilde\nu}}\beta^{2}\right\}
\nonumber \\
&& \nonumber \\
&&\times\exp\left\{-\frac{1}{2}\frac{{\tilde\mu}+{\tilde\nu}}{{\tilde\mu}-{\tilde\nu}}
\, x_{\theta}^{2}+\frac{\sqrt{2}\beta}{{\tilde\mu}-{\tilde\nu}} \, x_{\theta}\right\}
\nonumber \\
&& \nonumber \\
&&\times\exp\left\{-\frac{\sqrt{2}\gamma}{{\tilde\mu}
-{\tilde\nu}}\int^{x_{\theta}}F(y)dy\right\} \; ,
\label{wavefunc}
\end{eqnarray}
where, due to the canonical conditions
(\ref{canonlin}), the coefficient which multiplies the integral in
the last exponential is purely imaginary, thus ensuring
normalizability. We thus see that the effect of the nonlinearity,
namely the presence of multiphoton interactions, amounts to the
appearance of a non Gaussian phase factor in the wave
function.\\
For the special choices
$\theta=0\,,\,\frac{\pi}{2}$ the HOMPSS reduce to single-mode,
single-quadrature multiphoton squeezed states \cite{noi3,noi4}.
Of course, for
$\gamma=0$ the expression (\ref{wavefunc}) reduces to the Gaussian
wave function of the two-photon squeezed states. The HOMPSS can be also obtained
from the vacuum state by applying on it a compound unitary transformation:
\begin{equation}
|\beta,\gamma \rangle_{b} \, = \, U_{\theta}(X_{\theta}) \,
D_{\theta}(\alpha_{\theta}) \, S_{\theta}(\zeta_{\theta}) \, |0
\rangle \; ,
\label{HOMPSSoper}
\end{equation}
where $D_{\theta}(\alpha_{\theta}) \, = \, \exp \left( \alpha_{\theta}
a^{\dag}_{\theta} - \alpha^{*}_{\theta} a_{\theta} \right)$ is the
standard Glauber displacement operator with $\alpha_{\theta} \, =
\, {\tilde{\mu}}^{*} \beta - \tilde{\nu} \beta^{*}$. Next,
$S_{\theta}(\zeta_{\theta}) \, = \, \exp
\left(-\frac{\zeta_{\theta}}{2}a^{\dag2}_{\theta} +
\frac{\zeta_{\theta}^{*}}{2}a^{2}_{\theta} \right)$ is the
squeezing operator with $\zeta_{\theta} \, = \, r
e^{i(\phi-2\theta)}$, and, finally,
$$
U_{\theta}(X_{\theta}) \, = \, \exp
\left[ -\frac{\sqrt{2}\gamma}{{\tilde\mu} -{\tilde\nu}}
\int^{X_{\theta}} dY F(Y) \right]
$$ is the unitary, linear operator
of ``nonlinear mixing'', which is responsible
for the non Gaussian part of the HOMPSS,
and is a consequence of the associated multiphoton nonlinear processes.
The expression (\ref{HOMPSSoper}) shows, once more, that the
HOMPSS are a generalization of the two-photon squeezed states
for nonvanishing nonlinear strength $\gamma$, and smoothly reduce
to them in the limit $\gamma \rightarrow 0$. The HOMPSS admit
resolution of the unity
\begin{equation}
\mathbb{I}_{b} \, = \, \frac{1}{\pi}\int
d^{2}\beta |\beta,\gamma\rangle_{b}\;_{b}\langle\beta,\gamma| \; ,
\label{complet}
\end{equation}
and are nonorthogonal
\begin{equation}
\,_{b}\langle\beta_{1},\gamma|\beta_{2},\gamma\rangle_{b} \, = \,
\exp\left\{-\frac{1}{2}|\beta_{1}|^{2}
-\frac{1}{2}|\beta_{2}|^{2}+\beta_{1}^{*}\beta_{2}\right\} \; .
\label{nonorthogonal}
\end{equation}
Therefore, they constitute an
overcomplete basis. \\
The nonlinear unitary transformation
(\ref{boperg}) can be inverted; in fact, by using conditions
(\ref{canonlin}),
\begin{eqnarray}
&&a_{\theta}
\, = \,
\tilde{\mu}^{*}b-\tilde{\nu}b^{\dag}-(\tilde{\mu}^{*}\gamma
-\tilde{\nu}\gamma^{*})F(X_{\theta})
\;,  \nonumber \\
&& \nonumber \\
&&X_{\theta} \, = \,
\frac{1}{\sqrt{2}}[(\tilde{\mu}^{*}-\tilde{\nu}^{*})b+(\tilde{\mu}-\tilde{\nu})b^{\dag}]
\; .
\label{homoinversfor}
\end{eqnarray}
These inversion formulae allow the exact
computation of all the physically important statistical
quantities, in particular, the expectation values of all the
operatorial functions of the form $\sum c_{nm}a^{\dag n}a^{m}$,
for any arbitrary choice of the coefficients $c_{nm}$. \\
For the uncertainties on the conjugate homodyne quadratures
$X_{\theta}$ and
$P_{\theta} = -i(a_{\theta} - a_{\theta}^{\dagger})/\sqrt{2}$
in a generic HOMPSS, one finds:
\begin{eqnarray}
\langle\Delta^{2}X_{\theta}\rangle && \, = \,
\frac{1}{2}|\tilde{\mu} -
\tilde{\nu}|^{2} \; , \label{homouncX} \\
&& \nonumber \\
\langle\Delta^{2}P_{\theta}\rangle && \, = \,
\frac{1}{2}|\tilde{\mu} + \tilde{\nu}|^{2} +
2Im^{2}[{\tilde{\mu}}^{*}\gamma - \tilde{\nu}\gamma^{*}] (\langle
F^{2}\rangle_{\beta} - \langle F
\rangle_{\beta}^{2}) \nonumber \\ && \nonumber \\
\quad &&-2Im[{\tilde{\mu}}^{*}\gamma - \tilde{\nu} \gamma^{*}]
Im[(\tilde{\mu} + \tilde{\nu}) \langle [F,b^{\dag}]
\rangle_{\beta}] \; , \label{homouncP}
\end{eqnarray}
where $\langle \cdot
\rangle_{\beta}$ denotes the expectation value in the HOMPSS
$|\beta ,\gamma \rangle_{b}$. From these expressions one can
compute the general form of the uncertainty product for a generic
choice of the nonlinear function $F$. However, at this point,
the explicit expression for a specific example can be illuminating.
As in Ref. \cite{noi1}, we consider the case
of the lowest possible nonlinearity
\begin{equation}
F(X_{\theta}) \, = \,
X_{\theta}^{2} \; .
\label{homoFX2}
\end{equation}
In this case, choosing the optimal solutions of the canonical
constraints (\ref{canconlimit}), and using the expressions
(\ref{homouncX}), (\ref{homouncP}), we obtain the following
explicit form of the uncertainty product
\begin{equation}
\langle\Delta^{2}X_{\theta}\rangle\langle\Delta^{2}P_{\theta}\rangle
 \, = \, \frac{1}{4} \, + \, \frac{1}{2}|\gamma|^{2}e^{-
4r}\{1+4|\beta|^{2}+4|\beta|^{2}\cos 2[\varsigma-\theta]\} \; ,
\label{homouncprod}
\end{equation}
where we recall that $\varsigma$ denotes
the phase of $\beta$. The r.h.s. in Eq. (\ref{homouncprod})
attains its minimum value for $\varsigma-\theta \, = \,
\pm\frac{\pi}{2}\;$:
\begin{equation}
\langle\Delta^{2}X_{\theta}\rangle\langle\Delta^{2}P_{\theta}\rangle
\, = \, \frac{1}{4} \, + \, \frac{1}{2}|\gamma|^{2}e^{- 4r} \; .
\end{equation}
Therefore, either by assuming a very weak strength $|\gamma|$
of the nonlinearity, and/or a sufficiently high squeezing $r$,
we see that the HOMPSS are to all practical purposes
minimum uncertainty states.

\begin{figure}[h]
\begin{center}
\includegraphics*[width=14cm]{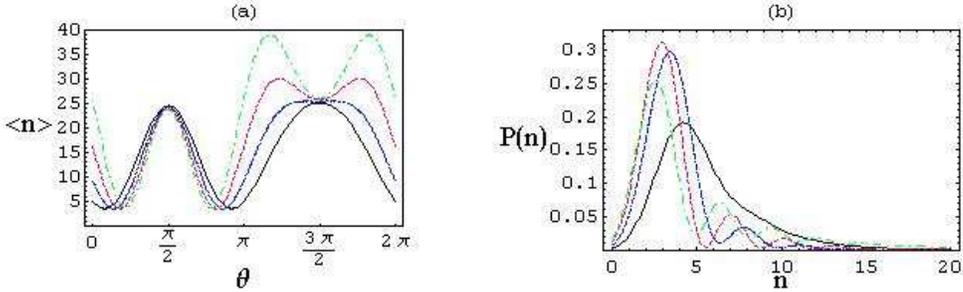}
\end{center}
\caption{(a) The mean photon number $\langle n \rangle$ as a
function of the homodyne angle $\theta$, for a HOMPSS with $r=0.5$, $\beta=3$, and
nonlinear strengths: $|\gamma| = 0$ (full line); $|\gamma| =
0.1$ (dotted line); $|\gamma| = 0.2$ (dashed line); and $|\gamma|
= 0.4$ (dot-dashed line). (b) The photon number distribution
$P(n)$ for a HOMPSS with $\theta=\frac{\pi}{6}$, and with the same
choices of (a) for $r$, $\beta$, and $|\gamma|$ (with the same
plot styles). In plots (a) and (b) the HOMPSS are associated to
the canonical conditions $\delta-\theta=-\frac{\pi}{2}$,
$\delta+\theta-\phi=-\frac{\pi}{2}$.} \label{homoNPn}
\end{figure}

\begin{figure}[h]
\begin{center}
\includegraphics*[width=14cm]{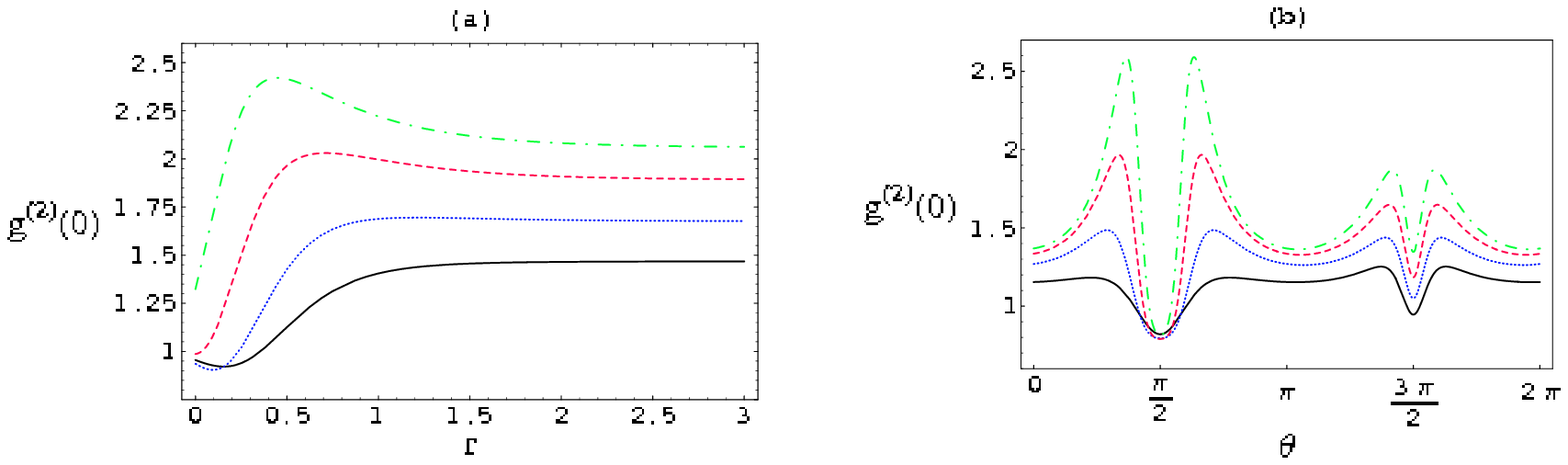}
\end{center}
\caption{(a) Plot of the second order correlation function
$g^{(2)}(0)$, as a function of $r$, for a HOMPSS with $\beta=3$,
$\theta=\frac{\pi}{3}$, and nonlinear strengths: $|\gamma| = 0.1$
(full line); $|\gamma| = 0.2$ (dotted line); $|\gamma| = 0.3$
(dashed line); and $|\gamma| = 0.4$ (dot-dashed line). (b) Plot of
$g^{(2)}(0)$, as function of $\theta$, for a HOMPSS with
$\beta=3$, $r=0.5$, and for the same values of $|\gamma|$ (with
the same plot styles). In plots (a) and (b) the HOMPSS are
associated to the canonical conditions
$\delta-\theta=-\frac{\pi}{2}$,
$\delta+\theta-\phi=\frac{\pi}{2}$.} \label{homog2}
\end{figure}

\begin{figure}[h]
\begin{center}
\includegraphics*[width=14cm]{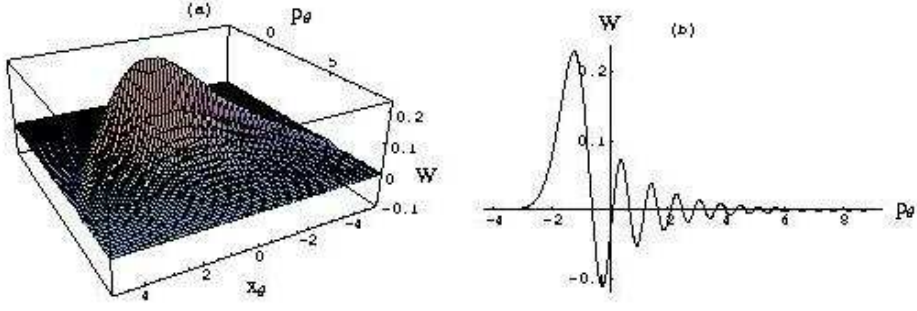}
\end{center}
\caption{(a) The Wigner function $W(x_{\theta},p_{\theta})$ for a
HOMPSS with $r=1.0$, $\beta=3$, $|\gamma| = 0.5$, and
$\theta=\frac{\pi}{2}$. (b) Section of the Wigner function
$W(x_{\theta}=x_{0},p_{\theta})$, with $x_{0}=0$. The canonical
conditions $\delta-\theta=-\frac{\pi}{2}$,
$\delta+\theta-\phi=\frac{\pi}{2}$ have been assumed.}
\label{homoW}
\end{figure}

In order to analyze the statistical properties of the HOMPSS, we
assume the particular solution (\ref{canconlimit}) for the
canonical conditions (\ref{canonlin1}), and adopt the minimal
quadratic form (\ref{homoFX2}) for the nonlinear function $F$.
The average photon number
$\langle
n\rangle \, = \,
\,_{b}\langle\beta,\gamma|n|\beta,\gamma\rangle_{b}$, in the original
mode variables $(n \, = \,
a^{\dag}a)$, can be easily computed by exploiting the inversion
formulae (\ref{homoinversfor}). In Fig. (\ref{homoNPn}) (a)
we show $\langle n \rangle$ as a function of the homodyne phase
$\theta$, with $r=0.5$, $\beta=3$, and for different values of
the nonlinear strength $|\gamma|$.
We see that the average photon number is strongly dependent on $|\gamma|$;
however, it is the homodyne phase $\theta$ that plays the main role in
determining the variation of $\langle n \rangle$. In Fig.
(\ref{homoNPn}) (b) we plot the photon number distribution
\begin{eqnarray}
P(n) \,  \equiv \, |\langle n|\beta,\gamma\rangle_{b}|^{2} \, = \,
&&\left|\int dx_{\theta}\langle n|x_{\theta}\rangle\langle
x_{\theta}|\beta,\gamma\rangle_{b}\right|^{2} \, = \nonumber \\
&& \nonumber \\
= \, && \frac{1}{2^{n}n!\pi^{1/2}}\left|\int dx_{\theta}
e^{-\frac{x_{\theta}^{2}}{2}}H_{n}(x_{\theta})\psi_{\beta,\gamma}(x_{\theta})\right|^{2}
\; ,
\end{eqnarray}
for $\theta = \frac{\pi}{6}$, where $H_{n}(x_{\theta})$
denotes the Hermite polynomial of degree $n$. The distribution
$P(n)$ shows an oscillatory behavior, which is more pronounced for
increasing $|\gamma|$. In Fig. (\ref{homog2}) (a) we plot the
second order correlation function  as a function of $r$, and in
Fig. (\ref{homog2}) (b) the second order correlation function as a
function of $\theta$, in both cases for different values of the
nonlinear strength $|\gamma|$.
Fig. (\ref{homog2}) (a) shows that the correlation function, at
fixed $\theta$, exhibits a typical behavior: as the degree of
squeezing $r$ is increased, it tends to a costant asymptote
whose numerical value increases with increasing $|\gamma|$.
Fig. (\ref{homog2}) (b) shows instead that the dependence
of the correlation function on the homodyne angle $\theta$
is so strong that, as the latter is varied, the nature
of the statistics can change, and one can thus select
super- or sub-Poissonian regimes. The strong and flexible
nonclassical character of the HOMPSS can be further verified
by determining their Wigner quasi-probability distribution (\ref{Wigxplambda})
for the classical phase space variables $x_{\theta}$ and $p_{\theta}$
corresponding to the canonically conjugate orthogonal quadrature
components $X_{\theta}$ and $P_{\theta}$. Figures
(\ref{homoW}) (a) and (b) show, respectively, a global projection
of $W(x_{\theta},p_{\theta})$, and an orthogonal section
$W(x_{\theta}=x_{0},p_{\theta})$ with $x_{0}=0$. The two plots show
that the Wigner function of canonical multiphoton squeezed states exhibits
nonclassical features, beyond a squeezed shape, that are
in general much stronger than those of the corresponding
canonical two-photon squeezed states, including interference
fringes and negative values.

\subsection{Homodyne multiphoton squeezed states:
diagonalizable Hamiltonians and unitary evolutions}
\label{subsection6.2}

In this Subsection we discuss the multiphoton Hamiltonians and the
multiphoton processes which can be associated to the one-mode
nonlinear canonical transformations that we have introduced.
We will then show how to determine the associated
exact, unitary time-evolutions.

The most elementary multiphoton interaction which can be associated with the
nonlinear transformation (\ref{boperg}) is defined by the
diagonal Hamiltonian
$$
H_{1}^{diag} \, = \, b^{\dag}b \; .
$$
The evolved state $|\psi(t)\rangle=e^{-itH_{1}^{diag}}|in\rangle$,
generated by the action of Hamiltonian $H_{1}^{diag}$
on a generic initial state $|in\rangle$, can be easily
determined by using the overcomplete set of states
$\{|\beta,\gamma\rangle_{b}\}$. In fact, one can insert in the
expression for $|\psi(t)\rangle$ the completeness relation
(\ref{complet}):
$|\psi(t)\rangle=e^{-itH_{1}^{diag}}\mathbb{I}_{b}|in\rangle$, and
exploit the eigenvalue equation (\ref{eigeneq}). For the choice
(\ref{homoFX2}), the Hamiltonian $H_{1}^{diag}$, expressed in terms
of the original mode operators $a$ and $a^{\dag}$, takes the form
\bea
H_{1G}^{4ph} \, = \, &&A_{0}+(A_{1}a^{\dag} + A_{2}a^{\dag2} +
A_{3}a^{\dag3} + A_{4}a^{\dag4} + H. c. ) \nonumber \\
&& \nonumber \\
&& + \, B_{0}a^{\dag}a + B_{1}a^{\dag2}a^{2} + (C a^{\dag2}a + D
a^{\dag3}a + H. c.) \; .
\label{homoH4ph}
\eea
We see that
$H_{1G}^{4ph}$ contains all the $n$-photon processes up to $n=4$.
That is, Eq. (\ref{homoH4ph}) contains the one-,
two-, three-, and four-photon down conversion terms $a^{\dag n}$
$(n=1,...,4)$, the Kerr term $a^{\dag 2}a^{2}$, and the terms
$a^{\dag 2}a $ and $a^{\dag 3}a $. The two last terms can be
interpreted as higher order one- and two-photon interactions
modulated by the intensity $a^{\dag}a$, or, as energy-assisted
one-photon and two-photon hopping terms. Obviously, it would
be important that the exactly diagonalizable
Hamiltonians constructed in terms of the transformed  mode $b$
describe realistically feasible multiphoton processes.
To this aim, it is crucial to have a sufficiently large
number of tunable parameters which
can model the multiphoton Hamiltonians. Therefore, one can
generalize the ``free'' Hamiltonian $H_{1}^{diag}$ by introducing the
``displaced-squeezed'' Hamiltonian
\begin{eqnarray}
H_{1}^{su11} & \, = \, & \frac{1}{2} \Omega
b^{\dag}b+\frac{1}{2}\eta b^{\dag 2}+\frac{1}{2}\eta^{*} b^{2}+\xi
b^{\dag}+\xi^{*}b+\frac{1}{4}\Omega \nonumber \\
&& \nonumber \\
&\, = \, & \Omega K_{0}+\eta K_{+}+\eta^{*}K_{-}+\xi b^{\dag}+\xi^{*}b \; ,
\label{homoHsu11h4}
\end{eqnarray}
where we have used the generators (\ref{su11degalgebra}) of the $SU(1,1)$
algebra. Here, $\Omega$ is a real, and $\eta$ and $\xi$ are complex
time-independent \emph{c}-numbers. The Hamiltonian $H_{1}^{su11}$
underlies a $SU(1,1)\oplus h(4)$ symmetry, and can be
reduced to a pure $SU(1,1)$ structure by means of the trivial
scale transformation
\begin{equation}
c \, = \, b+\Delta \; , \quad \quad
c^{\dag}\, = \, b^{\dag} +\Delta^{*} \; , \quad \quad
\Delta \, = \, 2\frac{\Omega\xi-2\xi^{*}\eta}{\Omega^{2}-4|\eta|^{2}}
\; .
\label{rescalingb}
\end{equation}
Of course, the HOMPSS are
eigenvectors of the displaced operator $c$,
with eigenvalue $\beta+\Delta$. The
Hamiltonian (\ref{homoHsu11h4}) can be written in the form
\begin{equation}
H_{1}^{su11} \, = \, \Omega {\tilde K}_{0}+\eta {\tilde K}_{+}+\eta^{*}{\tilde
K}_{-}+\Lambda \; ,
\label{homoHc}
\end{equation}
where the operators ${\tilde
K}_{0}$, ${\tilde K}_{+}$ and ${\tilde K}_{-}$ are defined in the
$c$ basis according to Eq. (\ref{su11degalgebra}), and
\begin{equation}
\Lambda \, = \, \frac{1}{2}\Omega|\Delta|^{2}+Re[\eta \Delta^{*
2}]-2Re[\xi \Delta^{*}] \; .
\end{equation}
The unitary evolution generated by Hamiltonian (\ref{homoHsu11h4}) is
again exactly computable, as we will show in the final part of
this Subsection. \\
Both the Hamiltonians $H_{1}^{diag}$ and
$H^{su11}$ are quadratic in $b$; then, for any finite $N$-sum choice
of the nonlinearity
\begin{equation}
F(X_{\theta}) \, = \, \sum_{j=1}^{N}c_{j}X_{\theta}^{j} \; , \quad \quad \quad
c_{j}\in\mathbb{R} \; ,
\label{homoF}
\end{equation}
the highest order nonlinear process entering in both Hamiltonians
will be of the form $a^{2N}$.
On the other hand, the Hamiltonian (\ref{homoHsu11h4})
offers a greater possibility of selecting the desired
multiphoton processes.
In fact, writing $H_{1}^{su11}$ in terms of the original modes
$a$ and $a^{\dag}$
\begin{eqnarray}
&&H_{1}^{su11} \, = \,
\left\{\frac{1}{2}\Omega(|\mu|^{2}+|\nu|^{2})+2Re[\eta\mu^{*}\nu^{*}]\right\}a^{\dag}a
+\frac{1}{4}\Omega+\frac{1}{2}\Omega|\nu|^{2}
\nonumber \\
&& \nonumber \\
&&+ Re[\eta\mu^{*}\nu^{*}]+\left\{(\mu\xi^{*}+\nu^{*}\xi)a +
\frac{1}{2}(\Omega\mu\nu^{*}+\eta^{*}\mu^{2}+\eta\nu^{*2})a^{2}\right.
\nonumber \\
&& \nonumber \\
&&+\left.\frac{1}{2}(\Omega\gamma^{*}\mu+\eta^{*}\gamma\mu+\eta\gamma^{*}\nu^{*})Fa
+\frac{1}{2}
(\Omega\gamma\nu^{*}+\eta^{*}\gamma\mu+\eta\gamma^{*}\nu^{*})a F
+H.c. \right\} \nonumber \\
&& \nonumber \\
&&+2Re[\xi\gamma^{*}]F+\left(\frac{1}{2}\Omega|\gamma|^{2}+Re[\eta\gamma^{*2}]\right)F^{2}
\; . \label{homoHImp}
\end{eqnarray}
Assuming for $F$ the general polynomial
form of generic degree $N$ Eq.
(\ref{homoF}), the expression (\ref{homoHImp}) will contain the
multiphoton terms:
\be
\sum_{n,m=0}^{2N}c_{n,m}a^{\dag n}a^{m} \,,
\label{Hmpschrod}
\ee
with $c_{n,m}$ time-independent $c$-numbers,
and $(n+m)\leq 2N$. The fast decrease of the strengths of higher-order
processes in nonlinear crystals suggests to consider
interaction Hamiltonians which contain at most up to third- or
fourth-order processes, as in Eq. (\ref{homoH4ph}).
This goal is achieved in the formalism of canonical multiphoton
quantum optics by considering the form (\ref{homoF}), with $N=3$,
for the nonlinearity $F$, and
imposing in Eq. (\ref{homoHImp}) the vanishing of the coefficient
of the term $F^{2}$:
\be
\frac{1}{2}\Omega|\gamma|^{2}+Re[\eta\gamma^{*2}] \, = \, 0 \; .
\label{condnullF2}
\ee
The remaining free parameters can be used to reduce
Hamiltonian $H_{1}^{su11}$ to even simpler forms.
Imposing the conditions $c_{2}=0$ and
$Re[\xi\gamma^{*}]=0$, respectively in Eqs. (\ref{homoF})
and (\ref{homoHImp}), the coefficients of the terms
$a^{\dag 3}$ and $a^{\dag 2}a$ vanish, $(A_{3}=C=0)$, and the
Hamiltonian $H_{1}^{su11}$ becomes, apart from harmonic or linear terms,
the reduced, interacting four-photon Hamiltonian
\begin{equation}
H^{4ph}_{1IR} \, = \, B_{1}a^{\dag2}a^{2}  +
\left(A_{2}a^{\dag2} + A_{4}a^{\dag4}+D a^{\dag3}a + H. c. \right)
\; .
\label{ham'4ph}
\end{equation}
Letting, instead, $c_{3}=0$ in Eq.
(\ref{homoF}), the consequent,
simultaneous vanishing of the coefficients of the powers
$a^{\dag 4}$, $a^{\dag 3}a$, $a^{\dag 2}a^{2}$ $(A_{4}=B_{1}=D=0)$
leads, apart from harmonic or linear terms, to the reduced,
interacting three-photon Hamiltonian:
\begin{equation}
H_{1IR}^{3ph} \, = \, A_{2}a^{\dag2} +
A_{3}a^{\dag3}+C a^{\dag 2}a + H. c. \; .
\label{ham3ph}
\end{equation}
The degenerate processes described by the Hamiltonians (\ref{ham'4ph})
and (\ref{ham3ph}) can be realized by suitably exploiting the
third- and the fourth-order susceptibilities $\chi^{(3)}$ and
$\chi^{(4)}$. Note that the expressions (\ref{ham'4ph}) and
(\ref{ham3ph}) still contain several tunable parameters, which can be
exploited to fit the experimental needs in realistic situations.

We now move to discuss the general forms and properties
of the time evolutions generated by the action of
Hamiltonian $H_{1}^{su11}$ on the vacuum of the original mode
operator $a$.
In the interaction
picture, the evolved state is
\begin{eqnarray}
|\Psi (t)\rangle& \,= \, & U_{1}^{su11}(t)|0\rangle \, = \,
\exp\{-it
H_{1}^{su11}\}|0\rangle  \nonumber \\
&& \nonumber \\
&\, = \, & \exp\{-it\Lambda\}\exp\{-it[\Omega{\tilde K}_{0}+\eta{\tilde
K}_{+}+\eta^{*}{\tilde K}_{-}]\}|0\rangle \; .
\label{homoTevol}
\end{eqnarray}
Applying the disentangling formula for the $SU(1,1)$ group
\cite{disentangWeiNor,disentangTruax,disentangDattoli}, the
evolution operator $U_{1}^{su11}(t)$ can be rewritten as
\begin{equation}
U_{1}^{su11}(t) \, = \, \exp\{-it\Lambda\}\exp\{\Gamma_{+}{\tilde
K}_{+}\}\exp\{(\ln\Gamma_{0}){\tilde
K}_{0}\}\exp\{\Gamma_{-}{\tilde K}_{-}\} \; ,
\label{disentform}
\end{equation}
where
\begin{eqnarray}
\Gamma_{0}& \, = \, &\left\{\cosh (\sigma
t)+\frac{i\Omega}{2\sigma}\sinh
(\sigma t)\right\}^{-2} \; , \nonumber \\
&& \nonumber \\
\Gamma_{+}& \, = \, &\frac{-2i\eta\sinh (\sigma t)}{2\sigma\cosh (\sigma
t)+i\Omega\sinh (\sigma t)} \; , \nonumber \\
&& \nonumber \\
\Gamma_{-}&\, = \, &\frac{-2i\eta^{*}\sinh (\sigma t)}{2\sigma\cosh
(\sigma t)+i\Omega\sinh (\sigma t)} \; , \nonumber \\
&& \nonumber \\
\sigma^{2}&\, = \, &\left(|\eta|^{2}-\frac{1}{4}\Omega^{2}\right) \; .
\label{Gamma1}
\end{eqnarray}
The condition (\ref{condnullF2}) implies
$\sigma^{2}=|\eta|^{2}\sin^{2}(\arg\eta-2\delta)$. Therefore,
$\sigma$ is real, and $1/|\sigma|$ defines a characteristic time
associated to the system. The disentangled form (\ref{disentform})
of $U_{1}^{su11}(t)$, and the two resolutions of the unity
$\mathbb{I}_{b}$ (given by Eq. (\ref{complet})), and
$\mathbb{I}_{x_{\theta}}=\int dx_{\theta}|x_{\theta}\rangle\langle
x_{\theta}|$, allow to express the state $|\Psi_{I}(t)\rangle$ as
\begin{eqnarray}
|\Psi (t)\rangle
& \, = \, & \mathbb{I}_{x_{\theta}}\mathbb{I}_{b}U_{1}^{su11}(t)
\mathbb{I}_{b}\mathbb{I}_{x_{\theta}}
|0\rangle \, = \nonumber \\
&& \nonumber \\
&\, = \, &\Gamma_{0}^{1/4}e^{-it\Lambda}\int\int
dx'_{\theta}dx''_{\theta}\;\frac{1}{\pi^{2}}\int\int
d^{2}\beta_{1}d^{2}\beta_{2}\; K(\beta_{1}^{*},\beta_{2};t)
\nonumber \\
&& \nonumber \\
&\, \times \, & |x'_{\theta}\rangle \,\langle
x'_{\theta}|\beta_{1},\gamma\rangle_{b}
\,_{b}\langle\beta_{1},\gamma|\beta_{2},
\gamma\rangle_{b}\,_{b}\langle\beta_{2},\gamma|x''_{\theta}\rangle
\langle x''_{\theta}|0\rangle \; ,
\label{homopsit}
\end{eqnarray}
with
\begin{eqnarray}
K(\beta_{1}^{*},\beta_{2};t)& \, = \, &
\exp\left\{\frac{\Gamma_{+}}{2}(\beta_{1}^{*}+\Delta^{*})^{2}\right.
\nonumber \\
&& \nonumber \\
& \, + \, &
\left.(\Gamma_{0}^{1/2}-1)(\beta_{1}^{*}+\Delta^{*})(\beta_{2}+\Delta)+
\frac{\Gamma_{-}}{2}(\beta_{2}+\Delta)^{2}\right\}\,.
\label{Ubetat}
\end{eqnarray}
Defining
\begin{eqnarray}
\mathcal{I}(x'_{\theta},x''_{\theta};t)&& \, = \,
\frac{1}{\pi^{2}}\int\int
d^{2}\beta_{1}d^{2}\beta_{2}\;K(\beta_{1}^{*},\beta_{2};t)\; \nonumber \\
&& \nonumber \\
&& \, \times \, \langle x'_{\theta}|\beta_{1},\gamma\rangle_{b}
\,_{b}\langle\beta_{1},\gamma|\beta_{2},\gamma\rangle_{b}
\,_{b}\langle\beta_{2},\gamma|x''_{\theta}\rangle \; ,
\label{homointegrals}
\end{eqnarray}
Eq. (\ref{homopsit}) reduces to
\begin{equation}
|\Psi(t)\rangle \, = \, \Gamma_{0}^{1/4}e^{-it\Lambda}\int\int
dx'_{\theta}dx''_{\theta}\;|x'_{\theta}\rangle \,
\mathcal{I}(x'_{\theta},x''_{\theta};t)\,\langle
x''_{\theta}|0\rangle \; .
\label{homopsit2}
\end{equation}
In Eqs.
(\ref{homopsit}) and (\ref{homointegrals}), $\langle
x_{\theta}|\beta,\gamma\rangle_{b}$ is the HOMPSS expressed
in the quadrature representation
(\ref{wavefunc}), and the scalar product $\,_{b}\langle
\beta_{1},\gamma|\beta_{2},\gamma\rangle_{b}$ is given by Eq.
(\ref{nonorthogonal}). Choosing different specific forms of $H_{1}^{su11}$, the
integrals in Eq. (\ref{homopsit}) can be computed exactly, leading
to an explicit expression for the evolved state
$\Psi(x_{\theta};t)=\langle x_{\theta}|\Psi(t)\rangle$.
Moreover, the $SU(1,1)$ invariance, and the invertibility of
the nonlinear Bogoliubov transformations (\ref{boperg}),
allow to study analytically the statistical properties of
the evolved states.

\subsection{Two-mode heterodyne multiphoton squeezed states: definitions and statistical properties}

\begin{figure}[h]
\begin{center}
\includegraphics*[width=14cm]{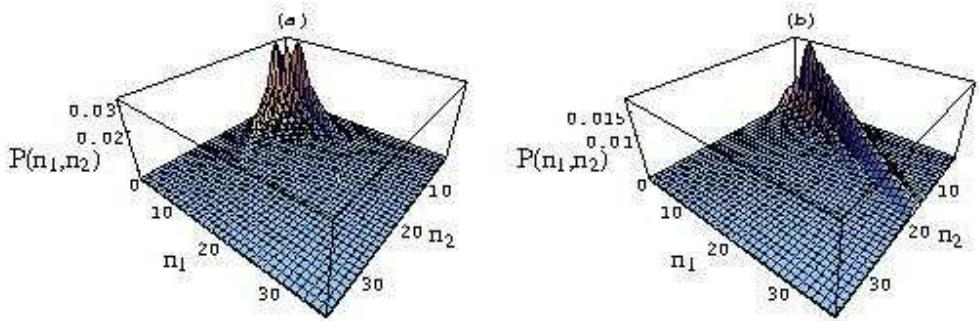}
\end{center}
\caption{(a) Joint probability $P(n_{1},n_{2})$ of two-mode,
two-photon squeezed states ($|\gamma|=0$),
for the symmetric choice $\alpha_{1}=\alpha_{2}=1$
of the coherent amplitudes, and with a squeezing parameter
$r=1.5$. (b) Joint probability $P(n_{1},n_{2})$ of HEMPSS
with lowest order nonlinearities $F(Z)=Z^{2}$ of equal strength
$|\gamma| = 0.3$. The canonical conditions
$\theta_{1}+\theta_{2}-\phi=0$, and
$\delta_{1}+\delta_{2}-\phi=\pi$ have been assumed, with assigned
values $\delta_1 = \pi/3$, and $\theta_1 = -\theta_2 = \pi/4$.
Same values of the squeezing parameter and of the coherent amplitudes
as in (a).}
\label{tmpnd}
\end{figure}

\begin{figure}[h]
\begin{center}
\includegraphics*[width=14cm]{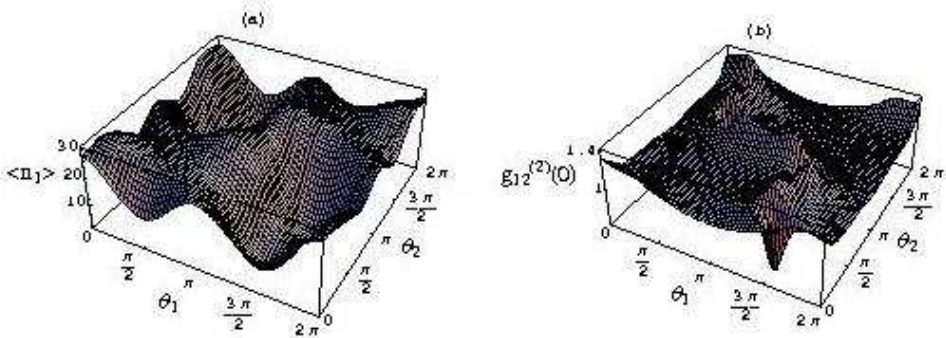}
\end{center}
\caption{(a) Average number of photons $\langle n_{1}\rangle$
in the original field mode $a_{1}$, as a function of the heterodyne
angles $\theta_{1}$, $\theta_{2}$, for a HEMPSS with $\beta_{1}=\beta_{2}=3$,
$r=0.5$, $|\gamma|=0.3$, and $\delta_{1}=\pi$. (b) Cross-correlation
$g^{(2)}(0)$ as a function of $\theta_{1}$ and
$\theta_{2}$, for the same values of the parameters as in (a).
In both cases, the imposed canonical constraints read
$\theta_{1}+\theta_{2}-\phi=0$ and
$\delta_{1}+\delta_{2}-\phi=\pi$.}
\label{heteroNg2}
\end{figure}

The multiphoton canonical formalism,
developed for a single mode of the electromagnetic
field, will be extended in this Subsection to bipartite systems of two
correlated modes \cite{noi2}. This generalization is realized by a
proper two-mode extension of the relation (\ref{boperg}),
obtained by promoting the argument of the nonlinear function $F$
to be the heterodyne variable
\begin{eqnarray}
&&Z \, = \, \frac{e^{-i\theta_{2}}a_{2}+
e^{i\theta_{1}}a^{\dag}_{1}}{\sqrt{2}}
\, \equiv \, \; \frac{a_{\theta_{2}}+
a^{\dag}_{\theta_{1}}}{\sqrt{2}} \; , \nonumber \\
&& \nonumber \\
&&a_{\theta_{i}} \, = \, e^{-i\theta_{i}}a_{i} \; , \; \quad \quad i \, = \, 1, 2 \; ,
\label{hetvar}
\end{eqnarray}
that can
be interpreted as the output photocurrent of an ideal heterodyne
detector \cite{noi2,heterodet1,heterodet2}. The
two-mode nonlinear canonical transformations read
\begin{eqnarray}
&& b_{1} \, = \, \mu' a_{\theta_{1}}
\, +  \, \nu'' a^{\dag}_{\theta_{2}} \, + \,
\gamma_{1} F\left(\frac{a_{\theta_{2}}+
a^{\dag}_{\theta_{1}}}{\sqrt{2}}\right) \; , \nonumber \\
&& \nonumber \\
&& b_{2} \, = \, \mu'' a_{\theta_{2}}
\, + \, \nu' a^{\dag}_ {\theta_{1}} \, + \,
\gamma_{2} F\left(\frac{a_{\theta_{1}}+
a^{\dag}_{\theta_{2}}}{\sqrt{2}}\right) \; ,
\label{heterob12}
\end{eqnarray}
with
\begin{equation}
\mu' \, = \, \mu e^{i\theta_{1}} \, , \quad \nu'' \, = \,\nu e^{-i\theta_{2}} \, ,
\quad \mu'' \, = \, \mu e^{i\theta_{2}} \, , \quad \nu' \, = \,\nu e^{-i\theta_{1}} \; ,
\end{equation}
where $\mu, \nu, \gamma_{1}, \gamma_{2}$ are complex parameters, and $\theta_{1}, \theta_{2}$
are heterodyne angles. For the transformations (\ref{heterob12}) to be canonical,
the nonlinear strenghts $\gamma_{j}\, , \; (j=1,2)$ must
share the same modulus: $\gamma_{j}=|\gamma|e^{i\delta_{j}}$.
The bosonic canonical commutation relations
$[b_{i} \, , \,b_{j}^{\dag}]=\delta_{ij}$ and $[b_{i}\, , \, b_{j}]=0$
impose the futher constraints
\begin{eqnarray}
&&|\mu|^{2} - |\nu|^{2} \, = \, 1 \; , \\
&& \nonumber \\
&&\mu \, \gamma_{2}^{*} \, e^{i\theta_{1}} - \nu \, \gamma_{2}^{*} \, e^{-i\theta_{2}}
+ \mu^{*} \, \gamma_{1} \, e^{-i\theta_{2}} - \nu^{*} \, \gamma_{1} \,e^{i\theta_{1}}
\, = \, 0 \; .
\label{hetcanco}
\end{eqnarray}
With the standard parametrization
$\mu=\cosh r$, $\nu=e^{i\phi}\sinh r$, the relation
(\ref{hetcanco}) can be recast in the form
\begin{equation}
\tanh r \, = \, \frac{\cos(\theta_{1}+\theta_{2}-\phi)+
\cos(\delta_{1}+\delta_{2}-\phi)}{1+\cos(\delta_{1}+\delta_{2}+\theta_{1}+\theta_{2}-2\phi)} \; .
\label{heterocanco}
\end{equation}
In analogy with the single-mode case, the conditions (\ref{heterocanco})
are independent of the
strength $|\gamma|$ and the specific form of the nonlinear
function $F$. There are many possible solutions to the constraints
(\ref{heterocanco}), most ones numerical. Here and in the following,
we will make use of the following six exact, analytical solutions:
\begin{equation}
\delta_{1}+\delta_{2}-\phi \, = \, 0,\, \pm\pi \; \quad  , \quad \;
\theta_{1}+\theta_{2}-\phi \, = \, \pm \pi , \, 0 \; .
\label{heterocancolim}
\end{equation}
We now move to define the (two-mode) heterodyne multiphoton
squeezed states (HEMPSS) associated with the
canonical transformations (\ref{heterob12}). To this aim, it is
convenient to exploit the so-called  {\it entangled-state representation}
\cite{entanrep,entanrep2,entanrep3}. This representation is
based on the pair of non-Hermitian operators $Z$ and
$P_{z}=i(e^{i\theta_{2}}a_{2}^{\dag}-
e^{-i\theta_{1}}a_{1})/\sqrt{2}$, that satisfy the commutation
relations $[Z,Z^{\dag}]=[P_{z},P_{z}^{\dag}]=0$, and
$[Z,P_{z}]=i$. The entangled-state
representation  of a generic, pure two-mode state $|\Phi\rangle$ is
the representation $\langle z|\Phi\rangle$ in the basis
$\{|z\rangle\}$ of the orthonormal eigenvectors of $Z$ and
$Z^{\dag}$
\begin{equation}
Z|z\rangle \, =\, z  \, |z\rangle \, , \; \quad \quad
Z^{\dag}|z\rangle \, = \, z^{*} \, |z\rangle \; .
\label{zstates}
\end{equation}
Here $z$ denotes an arbitrary
complex number: $z=z_{1}+iz_{2}$, and the
states $|z\rangle$ can be expressed in the form \cite{entanrep}
\begin{equation}
|z\rangle \, = \,
\exp\left[-|z|^{2}+\sqrt{2}z a_{\theta_{2}}^{\dag}+\sqrt{2}z^{*}
a_{\theta_{1}}^{\dag}-
a_{\theta_{1}}^{\dag}a_{\theta_{2}}^{\dag}\right]|00\rangle \,,
\label{zbasis}
\end{equation}
where $|00\rangle \equiv |0\rangle \otimes |0\rangle$
denotes the two-mode vacuum. Recalling the definitions of
the rotated homodyne quadratures $X_\theta$ and $P_\theta$,
the states (\ref{zbasis}) satisfy the eigenvalue equations
$(X_{\theta_{1}}+X_{\theta_{2}})|z\rangle =2z_{1}|z\rangle$ and
$(P_{\theta_{2}}-P_{\theta_{1}})|z\rangle =2z_{2}|z\rangle$.
Therefore, they are common eigenstates of the total
``coordinate'' and relative ``momentum'' for a generic (two-mode)
two-component quantum system, namely the celebrated
Einstein-Podolsky-Rosen (EPR) entangled states \cite{EPR}.
They satisfy the orthonormality and completeness relations
\begin{equation}
\langle z'|z\rangle \, = \, \pi \delta^{(2)}(z'-z) \; , \quad \quad
\frac{2}{\pi}\int d^{2}z |z\rangle\langle z| \, = \, 1 \; ,
\end{equation}
with $d^{2}z=dz\,dz^{*}$.
Thanks to the entangled-state representation, it is possible
to recast the canonical transformations (\ref{heterob12})
in terms of the operators $Z$, $Z^{\dag}$, $P_{z}$, and $P_{z}^{\dag}$,
and, as a consequence, to compute the eigenstates of the transformed
modes $b_1$ and $b_2$. In fact, the heterodyne multiphoton
squeezed states (HEMPSS) $|\Psi\rangle_{\beta}$,
are the common eigenstates of the mode
operators  $b_1$ and $b_2$, determined by
\begin{equation}
b_{i}|\Psi\rangle_{\beta} \, = \, \beta_{i}|\Psi\rangle_{\beta} \; , \;
\quad \quad i=1,2 \; .
\label{heteroeigeneq}
\end{equation}
In the entangled-state representation,
the solution of Eq. (\ref{heteroeigeneq}) reads
\begin{equation}
\Psi_{\beta}(z,z^{*}) \, \equiv \, \langle
z|\Psi\rangle_{\beta} \, = \, \mathcal{N} \exp\{ -A|z|^{2}+B_{1}z+B_{2}z^{*}-
\mathcal{F}(z,z^{*})\} \; .
\label{heterowf}
\end{equation}
In Eq.~(\ref{heterowf}), $\mathcal{N}$ is a normalization factor, the
function $\mathcal{F}(z,z^{*})$ is defined as
\begin{equation}
\mathcal{F}(z,z^{*}) \, = \, d_{1}\int^{z}d\xi
F(\xi)\, + \, d_{2}\int^{z^{*}}d\xi^{*} F(\xi^{*}) \; ,
\end{equation}
and the coefficients are
\begin{eqnarray}
&& A \, = \, \frac{\mu'+\nu''}{\mu'-\nu''} \, = \, \frac{\mu''+\nu'}{\mu''-\nu'}
 \; , \nonumber \\
&& \nonumber \\
&&d_{1} \, = \, \frac{\sqrt{2}|\gamma|e^{i\delta_{1}}}{\mu'-\nu''} \; ,
\; \quad \quad d_{2} \, = \, \frac{\sqrt{2}|\gamma|e^{i\delta_{2}}}{\mu''-\nu'}
\; , \nonumber \\
&& \nonumber \\
&&B_{1} \, = \, \frac{\sqrt{2}\beta_{1}}{\mu'-\nu''} \; ,
\; \quad \quad B_{2} \, = \, \frac{\sqrt{2}\beta_{2}}{\mu''-\nu'} \; .
\label{coeffhetss}
\end{eqnarray}
The canonical conditions
(\ref{heterocanco}) imply $Re[A]>0$, $Im[A]=0$ and
$Re[\mathcal{F}(z,z^{*})]=0$, thus ensuring the normalizability of
the wave function (\ref{heterowf}). The states (\ref{heterowf})
satisfy, moreover, the (over)completeness relation
\begin{equation}
\frac{1}{\pi^{2}}\int d^{2}\beta_{1}d^{2}\beta_{2}
|\Psi\rangle_{\mathbf{\beta}} \,_{\mathbf{\beta}}\langle\Psi|
\, = \, 1 \; .
\label{heterocomplet}
\end{equation}
From their expression in the entangled-state representation,
the HEMPSS can be recast in the heterodyne
quadrature representation, in terms of the two real homodyne
variables $X_{\theta_1}$ and $X_{\theta_2}$ \cite{enthetrepdarian}.
Of course, for $\gamma=0$ the
states (\ref{heterowf}) reduce to the standard two-mode squeezed
states, and for a single-mode they reduce to the HOMPSS. The
HEMPSS can be unitarily generated from the vacuum:
\begin{equation}
|\Psi\rangle_{\mathbf{\beta}} \, = \,
U(Z,Z^{\dag})D_{1}(\alpha_{1})
D_{2}(\alpha_{2})S_{12}(g)|00\rangle \; ,
\label{heterop}
\end{equation}
where $D_{i}(\alpha_{i}) \; i=1,2$ are the single-mode displacement
operators with $\alpha_{i}=\mu^{*}\beta_{i}-\nu\beta_{j}^{*}$
($i\neq j =1,2$), $S_{12}(g)$ is the two-mode squeezing
operator, and $g=\pm r$, with the choice of the sign depending
on the particular choice of the parameters.
The operator $U$ can be cast in the form
\begin{equation}
U(Z,Z^{\dag}) \, = \,
\exp\{-\mathcal{F}(Z,Z^{\dag})\} \; ,
\label{hetmixop}
\end{equation}
and the canonical
conditions assure that it is unitary. In the case of lowest
nontrivial nonlinearity, $F(Z)=Z^{2}$, the operator $U$ takes
the form
$$
U(Z,Z^{\dag})=e^{-\Delta^{*} Z^{3}+\Delta Z^{\dag 3}},
$$
with $\Delta$ denoting a complex number. As in the one-mode case,
the two-mode nonlinear canonical transformations can be inverted,
yielding the following expressions for the original mode variables
$a_{\theta_{1}}$ and $a_{\theta_{2}}$:
\begin{eqnarray}
&&a_{\theta_{1}} \, = \, \mu'^{*}b_{1}-\nu'b_{2}^{\dag}-|\gamma|
(\mu'^{*}e^{i\delta_{1}}-\nu'e^{-i\delta_{2}})F(Z)
\; , \nonumber \\
&& \nonumber \\
&&a_{\theta_{2}}\, = \, \mu''^{*}b_{2}-\nu''b_{1}^{\dag}
-|\gamma|(\mu''^{*}e^{i\delta_{2}}-\nu''e^{-i\delta_{1}})F(Z^{\dag})
\,, \nonumber \\
&& \nonumber \\
&&Z=\frac{1}{\sqrt{2}}[(\mu'-\nu'')b_{1}^{\dag}+(\mu''^{*}-\nu'^{*})b_{2}]
\,. \label{heteroinv}
\end{eqnarray}
Relations (\ref{heteroinv}) allow the
exact computation of the correlations
$$
\,_{\beta}\langle\Psi|(a_{\theta_{i_{1}}}^{\dag})^{
n_{i_{1}}}(a_{\theta_{i_{2}}}^{\dag})^{ n_{i_{2}}}\cdots
(a_{\theta_{i_{k}}}^{\dag})^{
n_{i_{k}}}(a_{\theta_{j_{1}}})^{n_{j_{1}}}(a_{\theta_{j_{2}}})^{n_{j_{2}}}\cdots
(a_{\theta_{j_{l}}})^{n_{j_{l}}} |\Psi\rangle_{\beta} \; ,
$$ and of
their combinations. The photon statistics of the HEMPSS, for the
lowest nonlinearity $F(Z)=Z^{2}$, has been analyzed in Ref.
\cite{noi2}, and compared to that of the two-mode,
two-photon squeezed states.
The two-mode photon number distribution $P(n_{1},n_{2})=|\langle
n_{1},n_{2}|\Psi\rangle_{\mathbf{\beta}}|^{2}$ can be computed by
exploiting the completeness relation and the expression
\begin{eqnarray}
\langle n_{1},n_{2}|z\rangle && \, = \, (-1)^{m}2^{(M-m)/2}
\exp\{i(n_{1}\theta_{1}+n_{2}\theta_{2})\}\exp\{-|z|^{2}\}
\nonumber \\
&& \nonumber \\
&&\times\left(\frac{m!}{M!}\right)^{1/2} z^{* n_{1}-m}z^{n_{2}-m}
L_{m}^{(M-m)}(2|z|^{2}) \; ,
\label{zn}
\end{eqnarray}
where $m=\min
(n_{1},n_{2})$, $M=\max (n_{1},n_{2})$ and $L_{n}^{(k)}(x)$
are the generalized Laguerre polynomials of indices $n,k$.
As in the one-mode instance, the additional parameters associated to the nonlinear
part can strongly modify the statistics compared to the linear case.
In Fig. (\ref{tmpnd}) (a) we report the joint probability
$P(n_{1},n_{2})$ in a two-mode, two-photon squeezed state
(HEMPSS with $\gamma=0$), for a symmetric choice of the coherent
amplitudes at a given squeezing. Viceversa, in
Fig. (\ref{tmpnd}) (b), for the same values of the coherent
amplitudes and of the squeezing parameter, we report the
joint probability  $P(n_{1},n_{2})$ of a HEMPSS with
nonvanishing $\gamma$ and lowest order nonlinearity $F(Z)=Z^{2}$,
with a set of phases $\{ \phi, \delta_{i}, \theta_{i} \}$
satisfying the canonical conditions (\ref{heterocancolim}).
Comparing the two cases, we see that the additional parameters
associated to the nonlinearity destroy the symmetry possessed
by the joint probability in the absence of the nonquadratic interaction.
In order to recover symmetric forms of the joint probability
$P(n_{1},n_{2})$ for HEMPSS, the various phases must be suitably
balanced, as illustrated in Ref. \cite{noi2}. In general, the
many adjustable parameters allow to vary the properties of the joint
photon number distribution. In a similar way, we can control and
sculpture the mean photon numbers $\langle n_1 \rangle$ and
$\langle n_2 \rangle$, and the second-order correlation functions,
in particular, the cross-correlation function
$g^{(2)}_{12}(0)=\langle
a_{1}^{\dag}a_{1}a_{2}^{\dag}a_{2}\rangle/\langle
a_{1}^{\dag}a_{1}\rangle\langle a_{2}^{\dag}a_{2}\rangle$. In
Fig. (\ref{heteroNg2}) (a) and (b) we have plotted the mean photon
number $\langle n_{1}\rangle$ of the first mode, and the
cross-correlation $g^{(2)}_{12}(0)$, as functions of the
two heterodyne phases $\theta_{1}$ and $\theta_{2}$, for fixed
values of the other parameters.
Once more, one can see that the tuning of the heterodyne
phases realizes wild variations of the statistical properties
of the HEMPSS.
In particular, Fig.
(\ref{heteroNg2}) (b) shows that both the strong suppression
($g^{(2)}_{12}(0)>0$) and the strong enhancement ($g^{(2)}_{12}(0)<0$)
of the anticorrelations between modes $a_1$ and $a_2$ can be
realized as $\theta_{1}$ and $\theta_{2}$ are varied.

\subsection{Heterodyne multiphoton squeezed states: diagonalizable Hamiltonians and unitary evolutions}

The most elementary Hamiltonian that can be associated to the two-mode
nonlinear transformations (\ref{heterob12}) is the diagonal
Hamiltonian of the form
\begin{equation}
H_{2}^{diag} \, = \, b_{1}^{\dag}b_{1} \, + \, b_{2}^{\dag}b_{2} \; .
\end{equation}
Inserting
Eqs.~(\ref{heterob12}) for the transformed variables in this
expression, yields multiphoton Hamiltonians written
in terms of the fundamental mode variables $a_{i}$
and $a_{i}^{\dag}$, each one characterized by a specific choice
of the nonlinear function $F$. In analogy with the fully degenerate
case, in this two-mode context one must look again for interactions
that may allow  particularly interesting and physically realistic
implementations.
To this aim, let us consider Hamiltonians associated to nonlinear
functions $F$ that are simple, positive powers of their argument:
$F(\zeta) = \zeta^n$. These Hamiltonians describe a combination of
nondegenerate and degenerate multiphoton interaction processes,
up to $2n$-th nonlinear order. As usual, we concentrate our attention on the simplest
choice of the lowest nonlinearity $F(\zeta) = \zeta^2$, describing up
to four-photon processes. In this case, the general four-photons
Hamiltonian of the form $H_{2}^{diag}$ becomes
\begin{eqnarray}
H_{2G}^{4ph}& \, = \, &A_{0}+B_{0}(a_{1}^{\dag}a_{1}+a_{2}^{\dag}a_{2})
+ C_{0}(a_{1}^{\dag 2}a_{1}^{2}+a_{2}^{\dag2}
a_{2}^{2}+2a_{1}^{\dag}a_{1}a_{2}^{\dag}a_{2}) \nonumber \\
&& \nonumber \\
&+&[D_{1}a_{1}^{\dag}a_{2}^{\dag}+D_{2}a_{1}^{\dag
2}a_{2}+D_{2}'a_{1}a_{2}^{\dag 2}+D_{3}a_{1}^{\dag
3}+D_{3}'a_{2}^{\dag 3}
+D_{4}a_{1}^{\dag 2}a_{2}^{\dag 2} \nonumber \\
&& \nonumber \\
&+&D_{5}(a_{1}^{\dag 2}a_{1}a_{2}^{\dag}+a_{1}^{\dag}a_{2}^{\dag
2}a_{2}) + H.c.] \; .
\label{heteroH}
\end{eqnarray}
This Hamiltonian contains, besides the free harmonic terms, two-mode down
conversion processes, degenerate and semi-degenerate three-photon
processes, all the Kerr processes, two-mode down conversion
processes modulated by the intensities $a_{i}^{\dag}a_{i}$
$\;(i=1,2)$, and semi-degenerate four-photon down conversion
processes. \\
As in the single-mode formalism, it is possible to reduce the
number of simultaneously present interactions by suitably tuning
the adjustable parameteres appearing in the Hamiltonian.
Again, in analogy to the single-mode case, one can even
increase the number of available free parameters by generalizing
the Hamiltonian $H_{2}^{diag}$ to the form:
\begin{equation}
H_{2}^{su11} \, = \, \sum_{i=1}^{2} \left[ \Omega_{i} b_{i}^{\dag}b_{i} \, + \, \xi_{i}
b_{i}^{\dag} \, + \, \xi_{i}^{*}b_{i}\right] \, + \, \eta
b_{1}^{\dag}b_{2}^{\dag} \, + \, \eta^{*} b_{1}b_{2} \; ,
\label{heteroHsu11}
\end{equation}
that once more underlies a $SU(1,1)$ symmetry.
The freedom in fixing the many adjustable parameters
should allow the reduction of such Hamiltonians to
the simplest possible forms describing
realistically implementable
two-mode multiphoton interactions. \\
Finally, by exploiting the invertible, two-mode nonlinear canonical
transformations (\ref{heterob12}), the $SU(1,1)$ invariance of the
Hamiltonian (\ref{heteroHsu11}), the eigenvalue equations
(\ref{heteroeigeneq}), and the completeness relation
(\ref{heterocomplet}), one can exactly compute the unitary
evolution from the initial vacuum state generated by $H_{2}^{su11}$.

\subsection{Experimental realizations: possible schemes and perspectives}

In this Subsection, we discuss the possible experimental
feasibility of multiphoton processes described by the
one-mode and two-mode Hamiltonians introduced above, and the
corresponding HOMPSS and HEMPSS . As shown
in Section \ref{section3}, the main difficulties in implementing
multiphoton processes in nonlinear media are related both to the
weakness of the nonlinear susceptibilities, and to the constraints
imposed by energy conservation and phase matching. In the same
Section we have seen that success in the fabrication of new
nonlinear materials, and the exploitation of highly nontrivial
effects as coherent population trapping and electromagnetically
induced transparency, have led to a sensible enhancement of the
third-order $\chi^{(3)}$ nonlinearity, and to the prospect of
enhancing the magnitude of higher-order susceptibilities by several
orders of magnitude. However, even putting aside this problem,
it becomes increasingly difficult to satisfy
the energy and phase matching constraints with increasing order
of the multiphoton process to be implemented. \\
In order to show the complexity of practical implementations, we
consider now the simple, but relevant, case of \textit{collinear}
multiphoton processes with nonlinear terms of the
general form $a_{1}^{\dag}a_{2}^{\dag}\cdots
a_{s}^{\dag}a_{s+1}\cdots a_{n}E^{\pm}_{1}\cdots E^{\pm}_{m}$,
where the $\{E^{\pm}_{i}\}$ are classical pumps. A collinear process is
defined by the condition that the fields associated to all the
modes share the same direction of propagation (i.e. the wave vectors
are parallel or anti-parallel). We denote by $\omega_{i}$ and
$k_{i}$ the frequency and the wave vector associated to the mode
$a_{i}$, and by $\Omega_{i}$ and $K_{i}$ the frequency and
momentum associated to a classical pump $E_{i}$. We implicitly
include in the frequencies $\Omega_{i}$ associated to a classical
pump a sign $\pm$ corresponding to a positive or negative
frequency choice of the pump itself; moreover, we include in each
momentum $K_{i}$ associated to a classical pump a sign which, in a
collinear process, discriminates between backward and forward
propagation. Energy conservation and phase matching conditions, in
this collinear case, read
\begin{eqnarray}
&&\sum_{i=1}^{s}\omega_{i} \, - \, \sum_{i=s+1}^{n}\omega_{i}
\, = \, \sum_{i=1}^{m} \Omega_{i} \; , \label{phasmatchsyst1} \\
&& \nonumber \\
&&\sum_{i=1}^{s}k_{i} \, - \, \sum_{i=s+1}^{n}k_{i} \, = \, \sum_{i=1}^{m} K_{i}
\; ,
\label{phasmatchsyst2}
\end{eqnarray}
Note that the momenta $k_{i}$ and
$K_{i}$, apart the sign, coincide, respectively, with
$n(\omega_{i})\omega_{i}$ and $n(\Omega_{i})\Omega_{i}$,
where $n(\cdot)$ denotes the
frequency-dependent refractive index associated to a given
frequency. Therefore, the second equation is a nonlinear equation,
with the nonlinearity due to phenomenological quantities such
as the refractive indices. Therefore, the solution of the problem
defined by the set of coupled equations
(\ref{phasmatchsyst1}) and (\ref{phasmatchsyst2}) is highly
nontrivial. To try to solve it, one must choose a
suitable number of classical pumps, and a convenient medium
that allows birefringence (i.e. ordinary and extraordinary
refractive indices) or quasi-phase matching (periodic behavior of
the refractive index). However, in arbitrary cases, the equations
(\ref{phasmatchsyst1}) and (\ref{phasmatchsyst2}) may not
have physically meaningful solutions, and must thus be analyzed
case by case in concrete, specific instances. Here then we
discuss the possible experimental realization of the effective
interaction described by the specific Hamiltonian
(\ref{ham3ph}), associated to the single-mode canonical structure. \\
We consider the collinear processes generated by the lowest possible
nonlinearity in a centrosymmetric crystal, namely the one associated
to the third-order susceptibility $\chi^{(3)}$, because, in such
a material $\chi^{(2n)}=0$ for all $n$. The crystal is illuminated
by two highly intense laser pumps, $E_{1}$, of
frequency $\omega$ and ordinary polarization, and
$E_{2}$, of frequency $3\omega$ and extraordinary
polarization. To achieve the phase matching condition, in the
following we refer to the discussion of Section
\ref{section3}, in particular to Eq. (\ref{phasmatchbiref})
and to the definition of positive and negative uniaxial crystals.
We consider a negative uniaxial crystal, and we exploit the
birefringence to compensate the dispersion effect of the
frequency-dependent refractive indices $n_\omega$. In particular,
a suitable choice of the phase matching angle $\theta$ between the
propagation vector and the optic axis can lead to the equality
$n^{ord}_\omega =n^{ext}_{3\omega}(\theta)$, needed to realize the
three-photon down conversion processes that we wish to engineer. The
combined actions of the two classical pumps on the crystal generate,
besides the free linear and harmonic terms, the multiphoton processes
described in the following, where $\kappa_{lm}$ will denote the coefficient of the
generic interaction term $a^{\dag l}a^{m}$, responsible for
the creation of $l$ photons and, correspondingly, the simultaneous
annihilation of $m$ photons. As
previously remarked, all these processes will be
generated only by the third-order $\chi^{(3)}$ susceptibility. \\
The interaction of the pump at frequency $3\omega$ with the
nonlinear crystal generates the three-photon down conversion
process $\kappa_{30}E_{2}a^{\dag 3}+H.c. \,$, with
$\kappa_{30}\propto\chi^{(3)}(3\omega;-\omega,-\omega,-\omega)\,$.
Moreover, the first pump at frequency $\omega$ generates the
additional four-wave process $\kappa_{21}E_{1}a^{\dag 2}a +H.c.
\,$, with
$\kappa_{21}\propto\chi^{(3)}(\omega;-\omega,-\omega,\omega)\,$.
The Kerr term $a^{\dag 2}a^{2}$ is obviously generated as well.
However, the coupling of this term does not
contain contributions due to the intense classical pumps;
therefore, it can be neglected in comparison to the other,
much stronger interactions.
Finally, the third-order nonlinearity generates also the
two-photon down conversion process $
[\kappa_{20}^{'}E_{1}^{2}+\kappa_{20}^{''}E_{1}^{*}E_{2}] a^{\dag
2} + H. c.\,$, where $\kappa_{20}^{'}$ and $\kappa_{20}^{''}$ are
proportional to $\kappa_{21}$ and $\kappa_{30}$, respectively. The
contributions associated to the couplings $\kappa_{21}$ and
$\kappa_{20}^{'}$ are due to the self-phase-modulation mechanism,
and are automatically phase-matched, while the contribution
associated to the coupling $\kappa_{20}^{''}$ is generated by the
simultaneous interaction of the two external pumps with the
medium. The total effective interaction Hamiltonian is then finally
the experimentally feasible form of the desired reduced
three-photon Hamiltonian (\ref{ham3ph}), and, in more detail,
it reads
\begin{equation}
H_{exp}^{3ph} \, = \,
[\kappa_{20}^{'}E_{1}^{2}+\kappa_{20}^{''}E_{1}^{*}E_{2}]
a^{\dag 2} \, + \, \kappa_{21} E_{1}a^{\dag 2}a \, + \,
\kappa_{30}E_{2}a^{\dag 3} \, + \, H.c. \quad .
\label{Hexp3ph}
\end{equation}
In the interaction picture, the coefficients appearing
in Eq. (\ref{Hexp3ph}) are time-independent, and thus
Hamiltonian $H_{exp}^{3ph}$ describes the same
processes as does Hamiltonian (\ref{ham3ph}). The residual freedom
(allowed by the canonical conditions) in
fixing the parameters of the theoretical Hamiltonian (\ref{ham3ph}),
and the possibility of selecting the desired intensities of the classical pumps in
the ``experimental'' Hamiltonian (\ref{Hexp3ph}), are then sufficient
to assure that the two expressions coincide. As a consequence,
at least a particular form of the general single-mode
Hamiltonian $H_{1}^{su11}$ of canonical multiphoton quantum
optics can be experimentally realized. This fact implies
that there exist, at least in principle,
physically implementable multiphoton processes that can be
associated to nonquadratic Hamiltonian evolutions,
exactly determined with the canonical methods described in Subsection
\ref{subsection6.2}, which generate nonclassical, non Gaussian, single-mode
multiphoton squeezed states of the electromagnetic field.

\newpage

\section{Application of multiphoton quantum processes and states to
quantum communication and information}

\label{section7}

\subsection{Introduction and general overview}

In the last two decades, systems of quantum optics are
progressively emerging as an important test ground for
the study and the experimental realization of quantum
computation protocols and quantum information processes.
In this Section, after a very brief resume of some
essential concepts of quantum information theory,
we discuss the role and possible applications
of nonlinear interactions and
multiphoton states in this rapidly growing
discipline. Clearly, in this short summary, we
will only sketch some guidelines into this very extended and
complex field of research. The main issues to be discussed
here will be the use of multiphoton quantum states as carriers of
information, and the manipulation of the latter by the methods of
quantum optics.\\
Mainstream research directions in the theory of quantum communication
and information include, among others, quantum cryptography, teleportation, quantum
error correction, approximate quantum state cloning, entanglement
swapping, and, of course, quantum computation
\cite{QIbook,QIbook1,QIbook2,QIbook3,QIrev1,QIrev2,QIrev3,QIrev4,QIrev5,QIrev6,QIrev7}.
The growing interest in quantum information and quantum computation arises from
the fact that the basic aspects of quantum mechanics, the superposition
principle and the Heisenberg uncertainty principle, lead to the existence
of non factorizable (nonseparable) dynamics and states,
and their associated peculiar quantum correlations.
In fact, the nonseparable states of quantum mechanics are endowed with
a structure of correlations, that by now goes under the commonplace
name of {\it entanglement}, allowing, in principle, computational and
informational tasks that are believed to be impossible in the realm
of classical informatics. This peculiar structure of
the correlations in nonseparable quantum states, this entanglement,
can thus be regarded as true {\it physical resources} for the realization
of logical gates, computational protocols, information processes.
From the specific point of view of quantum optics, quantum computation
and information are becoming very close benchmarks, together with
nanotechnology, because, due the fast and constant trend towards
the reduction in the energy content of optical signals and the
miniaturization of optical components, the consequences of
the underlying quantum behaviours begin to emerge in ways that
are progressively more pronounced. In particular, as one gets
closer and closer to the quantum regime, the direct consequences
of the uncertainty principle begin to matter, and the omnipresent
fluctuations of the electromagnetic vacuum constitute a source of
noise that affects the information carried by optical signals from
the initial generation during all its time evolution, and imposes
limitations on the fidelity of the transmission. Fortunately, the
progresses in nonlinear optics and cavity quantum electrodynamics
allow at present several strategies for the control and reduction
of vacuum fluctuations and quantum noise in computational and
informational processes. This is why quantum optical implementations
may turn out to be of paramount importance in the future of quantum
information: as already seen, the most common and
efficient methods for the reduction of quantum noise are based on
squeezed light generation via parametric down conversion, quantum
nondemolition measurements, and confinement of electromagnetic
fields in cavities. For further discussions and bibliographical sources
on noise limitations in optical communication systems, the interested
reader may profit of Ref. \cite{QIrev8}. \\
We have already mentioned that quantum information science is rooted
in two key concepts, respectively the quantum superposition principle
and quantum entanglement.
Quantum state superpositions allow an enrichment of the
information content of the signal field, as they can assume
different values at the same time, while the information, through
the entanglement effects, becomes a nonlocal property of the whole
system. Finally, the requirements of scalability and resilience to decoherence,
necessary to achieve efficient quantum computation and secure quantum communication,
impose the need for large sets of macroscopic superpositions sharing
degrees of multipartite entanglement as large as possible.
Consequently, in order to implement efficiently quantum information
protocols in a quantum optical setting, it would be of primary
importance to devise methods for the engineering of
{\it entangled} multiphoton superposition states, possibly robust against
dissipation and environmental decoherence.
Thus motivated, in the following we will review and discuss the theory
and the present status of the experimental realizations of these states,
and assess their relevance and use in view of efficient implementations
of quantum information protocols.

\subsection{Qualifying and quantifying entanglement}

The entanglement of quantum states of bipartite and multipartite systems
is one of the fundamental resources in quantum information. Therefore,
in this Subsection we will give a self-contained review on the most relevant
measures devised to quantify  entanglement in a quantum state,
and on the most significant criteria of inseparability. \\
For bipartite pure quantum states there is a unique measure of entanglement,
named entropy of entanglement, defined by the von Neumann entropy of any
of the reduced states \cite{CiracEntanglement}.
Let $\rho_{12} \equiv \rho_{P}$ be a pure state of two parties,
its entropy of entanglement $E(\rho_{P})$ is given by
\begin{equation}
E(\rho_{P})=S(\rho_{1})=S(\rho_{2}) \,,
\label{vonNeumannE}
\end{equation}
where $S(\sigma)=-Tr[\sigma \log_{2}\sigma]$, and
$\rho_{i}=Tr_{j\neq i}[\rho_{P}]$ $(i,j=1,2)$.
For a system composed of two $N$-level subsystems,
the quantity $E$ ranges from zero, corresponding to a product state,
to $\log_{2}N$, corresponding to a maximally entangled state.
The minimum set of essential properties that should be shared by any
{\it bona fide} measure of entanglement are
the following: it must be positive semi-definite, achieving zero only
for separable states; it must be additive; and, finally, it
must be conserved under local unitary operations and
must be non-increasing
on average under local operations and classical communication (LOCC).
The entropy of entanglement of pure quantum states satisfies all of
the above properties.
Under realistic conditions, only partially entangled mixed states
can be generated, rather than fully entangled pure states.
Unfortunately, in the case of mixed states the von Neumann
entropy of the reduced states fails to distinguish classical and
quantum correlations and is therefore no longer a good measure
of quantum entanglement.
Another entropy-based measure for a state $\rho_{12}$ is the von
Neumann mutual information \cite{CiracEntanglement}:
\begin{equation}
I_{N}(\rho_{1};\rho_{2};\rho_{12}) = S(\rho_{1})+S(\rho_{2})-S(\rho_{12}) \; .
\label{mutualinformation}
\end{equation}
However, the mutual information $I_{N}$ can increase under local nonunitary operations,
therefore it cannot be a good measure of entanglement.\\
Although the problem of quantifying the amount of entanglement in mixed states is still open,
several measures have been proposed in the last years
\cite{BennettEnt,Vedral1Ent,Vedral2Ent,Vedral3Ent,VidalWerner}.
For a mixed state $\rho_{12} \equiv \rho_{M}$, Bennett \textit{et al.} have introduced the so called
{\it entanglement of formation} $E_{F}(\rho_{M})$, defined as the infimum of the entropy
of entanglement of the state $\rho_{M}$ computed over all its possible (in general, infinite)
pure state decompositions \cite{BennettEnt}.\\
The same authors have defined another important measure of mixed-state entanglement,
the {\it entanglement of distillation} $E_{D}(\rho_{M})$, defined as
the number of bipartite maximally entangled states (Bell pairs)
that can be distilled (purified) from the state $\rho_{M}$ \cite{BennettEnt}.
We recall that the purification or distillation protocol is the procedure of concentrating
maximally entangled pure states from partially entangled mixed states by means of local operations
and classical communication \cite{BennettDistillation,DuanDistillation,PanDistillation}.
Nevertheless, and remarkably, it is not always possible to distill
the entanglement of any inseparable states; in fact there exist states,
the so called bound entangled states, whose entanglement
is not distillable \cite{HorodeckiBoundEnt,Horodecki2BoundEnt}.\\
The entanglement of formation and the entanglement of distillation,
although conceptually well defined, are extremely hard to compute
due to the infinite dimension of the minimization procedure they
involve. It is thus remarkable that the entanglement of formation
has been computed exactly in the case of an arbitrary mixed state
of two qubits \cite{WoottersEnt}, and for an arbitrary mixed
symmetric two-mode Gaussian state of continuous
variable systems \cite{EntFormsymmGaussst}.\\
Vedral \textit{et al.} have introduced
a further measure of entanglement, the relative entropy,
which is a suitable generalization of the mutual information
Eq. (\ref{mutualinformation}) and is defined as \cite{Vedral1Ent}:
\begin{equation}
E_{re}(\rho_{12})= \min_{\sigma\in \mathcal{D}} S(\rho_{12}\parallel \sigma) \,,
\label{EntVedral}
\end{equation}
where $\rho_{12}$ is the state whose entanglement one wishes to compute;
$S(\rho_{12}\parallel \sigma)=Tr[\rho_{12}(\ln\rho_{12}-\ln\sigma)]$
is an entropic measure of distance between the two density matrices
$\rho_{12}$ and $\sigma$; and $\mathcal{D}$ is the set of the density
matrices corresponding to all disentangled states. Extensions
of the relative entropy to quantify the entanglement of states of systems
composed by more than two subsystems are also possible
\cite{Vedral1Ent}.
From an operational point of view,
the entanglement measured by Eq. (\ref{EntVedral}) can be interpreted as
the quantity that determines ``the least number of measurements needed
to distinguish a given state from a disentangled state'' \cite{Vedral2Ent}.
The relative entropy of entanglement $E_{re}$
interpolates between the entanglement
of formation $E_{F}$ and the entanglement of distillation $E_{D}$,
according to hierarchy
$E_{D} \leq E_{re} \leq E_{F}$ \cite{HorodeckiLimitEnt}. \\
An effectively computable measure of entanglement, the negativity,
has been introduced independently by several researchers
\cite{VidalWerner,EisertNegativity,SanperaNegativity,PlenioNegativity,LeeNegativity,HorodeckiNegativity}.
This measure is based on the Peres-Horodecki criterion of positivity
under partial transposition (PPT criterion) \cite{PeresPPT,HorodeckiPPT}.
Let us recall that the partial transposition corresponds to
a partial time reversal transformation or
to a partial mirror reflection in phase space.
By definition, a quantum state of a bipartite system is separable,
and we will write it as $\rho_{12}^{(sep)}$, if and only if it can
be expressed in the form \cite{WernerSeparableState}
\begin{equation}
\rho_{12}^{(sep)} = \sum_{i} p_{i} \, \rho_{i1}\otimes\rho_{i2} \,, \qquad
p_{i} \geq 0 \,, \qquad \sum_{i}p_{i}=1 \; ,
\label{WernerSepar}
\end{equation}
that is, as a convex combination of product states of the two subsystems,
where $\rho_{i1}$ and $\rho_{i2}$ are, respectively, the normalized states
of the first and of the second party.
The operation of partial transposition $T_{j}\rho$
($j=1,2$) on a generic quantum state $\rho$ of a bipartite
system with respect to one party is symmetric under the
exchange of the two parties. Then, choosing, say,
party $j=2$ this operation is defined as
follows: $\rho \rightarrow \rho^{T_{2}}$ generates the
partially transposed matrix $\rho^{T_{2}}$, whose elements, expressed in
an orthonormal basis $|i_{1},j_{2}\rangle$, read:
$$
\langle i_{1},j_{2}| \rho^{T_{2}} |k_{1},l_{2} \rangle =
\langle i_{1},l_{2}| \rho |k_{1},j_{2} \rangle \; .
$$
Let us denote by $\tilde{\rho}$ the partially transposed state
$\rho^{T_{2}}$. Then, if the original state is separable,
partial transposition must yield again a separable state, i.e.
$$
\tilde{\rho}_{12}^{(sep)} \, \equiv \,
\left( \rho_{12}^{(sep)} \right)^{T_{2}} \, = \,
\sum_{i} p_{i} \, \rho_{i1} \otimes \rho_{i2}^{T_{2}} \; .
$$
Hence, for a generic bipartite state $\rho_{12}$ a necessary
condition of separability is that the partially transposed matrix
$\tilde{\rho}_{12}$ be a {\it bona fide} density matrix, i.e.
that it be positive semidefinite. This is equivalent to require
the nonnegativity of all the eigenvalues of $\tilde{\rho}_{12}$
\cite{PeresPPT}. Viceversa, the existence of negative eigenvalues
of $\tilde{\rho}_{12}$ is a sufficient condition for entanglement.
By exploiting this relevant result, a natural measure of entanglement
for the state $\rho_{12}$ can be defined as the sum of the moduli
of the negative eigenvalues of $\tilde{\rho}_{12}$. This quantity
defines the {\it negativity}, and is denoted by $\mathcal{N}(\rho_{12})$
\cite{VidalWerner}:
\begin{equation}
\mathcal{N}(\rho_{12}) \, = \,
\frac{\parallel \tilde{\rho}_{12} \parallel_{1}-1}{2} \; .
\label{negativityEnt}
\end{equation}
An alternative definition is the {\it logarithmic negativity}
$E_{\mathcal{N}}(\rho_{12})$
\cite{VidalWerner}:
\begin{equation}
E_{\mathcal{N}}(\rho_{12})= \log_{2}\parallel \tilde{\rho}_{12} \parallel_{1} \; .
\label{lognegativityEnt}
\end{equation}
The quantities (\ref{negativityEnt}) and (\ref{lognegativityEnt})
are monotone under LOCC and can be easily computed in general, at least numerically. \\
Let us consider the important instance of two-mode Gaussian states $\rho_{G}$ of
continuous variable (CV) systems, for which the PPT criterion is a necessary
and sufficient condition for separability, as shown by Simon \cite{InsepSimon}.
These states define the set of two-mode states
with Gaussian characteristic functions and quasi-probability distributions,
such as the two-mode squeezed thermal states.
By defining the vector of quadrature operators
$R=(X_{1},\,P_{1},\,X_{2},\,P_{2})$,
these states are completely characterized by
the first statistical moments $\langle R_{i} \rangle$
and by the covariance matrix $\mathbf{\sigma}$, defined as
$\sigma_{ij} = \frac{1}{2}\langle R_{i}R_{j}+R_{j}R_{i}\rangle
-\langle R_{i}\rangle \langle R_{j}\rangle$,
or in the convenient block-matrix form
\begin{equation}
\mathbf{\sigma} \,= \, \left(
\begin{array}{cc}
\mathbf{A} & \mathbf{C} \\
\mathbf{C}^{T} & \mathbf{B} \
\end{array}
\right) \; ,
\label{Gausscovmatrix}
\end{equation}
where $\mathbf{A}$, $\mathbf{B}$, and $\mathbf{C}$ are $2\times 2$ submatrices.\\
Let us recall that, due to Williamson theorem \cite{WilliamsonTheor},
the covariance matrix of an $n$-mode Gaussian state
can always be written as $\mathbf{\sigma}=S^{T}\mathbf{\nu}S$,
where $S\in Sp(2n,\mathbb{R})$ is a symplectic transformation, and
$\mathbf{\nu}=diag (\nu_{1},\nu_{1},\ldots,\nu_{n},\nu_{n})$ is the covariance matrix
of a thermal state written in terms of the symplectic spectrum of
$\mathbf{\sigma}$ whose meaning will be clarified below \cite{SimpletticiIndiani}.
Here, the elements of the (real) symplectic group $Sp(2n,\mathbb{R})$ are
the linear transformations $S$ in phase space with $\det S =1$ that preserve the
symplectic form $\Omega$, i.e. such that $S^{T}\Omega S = \Omega$, where
\begin{equation}
\Omega = \bigoplus_{1}^{n} \mathbf{J} \; , \qquad
\mathbf{J} = \left( \begin{array}{cc}
0 & 1 \\
-1 & 0
\end{array} \right) \; ,
\label{symplecticmatrix}
\end{equation}
and $\mathbf{J}$ is the symplectic matrix. By the Stone-von Neumann theorem,
symplectic transformations on quasi-probability distributions in phase space
correspond to unitary transformations on state vectors in Hilbert space.
The quantities $\nu_i$ that constitute the symplectic spectrum are
the eigenvalues of the matrix $|i\Omega \sigma |$
\cite{SimpletticiIndiani,NostroSimplettico1}.
In order to compute the negativity (\ref{negativityEnt}) and
the logarithmic negativity (\ref{lognegativityEnt}) for two-mode
Gaussian states, one has to compute the two symplectic
eigenvalues $\tilde{\nu}_{\mp}$ (with $\tilde{\nu}_{-}<\tilde{\nu}_{+}$)
of the matrix $|i\Omega \tilde{\sigma} |$, where $\tilde{\sigma}$ is
the covariance matrix of the partially transposed state $\tilde{\rho}_{G}$.
Given the two $Sp(2,\mathbb{R}) \oplus Sp(2,\mathbb{R})$ symplectic
invariants $\det \mathbf{\sigma}$ and
$\tilde{\Delta}(\sigma)=\det\mathbf{A}+\det\mathbf{B}-2\det\mathbf{C}$,
the two symplectic eigenvalues $\tilde{\nu}_{\mp}$ read
\cite{NostroSimplettico1,NostroSimplettico2}:
\begin{equation}
\tilde{\nu}_{\mp}=
\sqrt{\frac{\tilde{\Delta}(\sigma)\mp(\tilde{\Delta}(\sigma)^{2}-4\det\mathbf{\sigma})^{1/2}}{2}}
\; .
\label{PTsimplecteigenv}
\end{equation}
Finally, it is easy to show that for any two-mode Gaussian state $\rho_{G}$
the negativity is a simple decreasing function of $\tilde{\nu}_{-}$
\cite{NostroSimplettico2,NostroSimplettico3}:
\begin{eqnarray}
&&\parallel \tilde{\rho}_{G} \parallel_{\,_{1}} \, = \, \frac{1}{\tilde{\nu}_{-}} \; ,
\\ && \nonumber \\
&&\mathcal{N}(\rho_{G}) = \max \left[ 0,\, \frac{1-\tilde{\nu}_{-}}{2\tilde{\nu}_{-}} \right] \;,
\\ && \nonumber \\
&&E_{\mathcal{N}}(\rho_{G}) = \max [0,\, -\log_{2} \tilde{\nu}_{-}] \; .
\end{eqnarray}
As already mentioned before, for two-mode symmetric Gaussian states
(i.e. for two-mode Gaussian states with $\det\mathbf{A}=\det\mathbf{B}$)
the entanglement of formation $E_{F}(\rho_{G})$
can be computed as well \cite{EntFormsymmGaussst}, and coincides in this
case with the logarithmic negativity but for irrelevant scale factors.
It is likely that this equivalence breaks down in more general instances
of mixed Gaussian states. For instance, it has been recently shown that
a rigorous bound for the true entanglement of formation, the so-called
Gaussian entanglement of formation (i.e. the entanglement of formation
computed only on pure Gaussian state convex decompositions) \cite{GEOF}
does not coincide in general with the logarithmic negativity for particular
classes of mixed nonsymmetric two-mode Gaussian states \cite{Ordering}.
Moreover, for these states the two measures induce different orderings,
in the sense that, taken a reference state, another state will be more
or less entangled than the reference state depending whether one chooses
the Gaussian entanglement of formation or the logarithmic negativity to
quantify entanglement \cite{Ordering}. Generalizations are
straightforward for particular classes of fully symmetric and bisymmetric
multimode Gaussian states. In these instances the negativities are again
amenable to a complete analytic computation, and the entanglement quantified
by them possesses interesting properties of scaling and unitary
localization \cite{Scaling,Unitarily}.
Finally, in order to analyze entanglement in a multipartite system,
the different bipartite splittings of the whole system can be considered
\cite{MultipartiteEntBipSplit}. This procedure can be used to define
multipartite negativities, as shown in Ref. \cite{VidalWerner}.

Besides facing the hard task of quantifying the exact amount of entanglement
in a given quantum state, it is in principle important, and often interesting
for practical purposes, to establish its inseparability properties.
To this end, it is necessary to look for reliable (in)separability
criteria, and several ones have been proposed for bipartite states
\cite{PeresPPT,HorodeckiPPT,InsepSimon,InsepReid,InsepTan,InsepDGCZ,InsepGKLC,InsepGiovannetti,InsepRaymer,InsepToth,InsepManko,InsepBiswas,InsepHZ,InsepShchukin,InsepSerafini}.
Such criteria often take the form of
inequalities and usually provide sufficient conditions
for inseparability. However, in the case of two-mode Gaussian states
and of multimode Gaussian states of $1 \times N$ bipartitions,
it is possible to establish criteria
that are necessary and sufficient for inseparability.
As already mentioned, the continuous-variable version of
the PPT criterion is a necessary and sufficient condition
for separability of two-mode Gaussian states, as first proved
by Simon \cite{InsepSimon}, and of multimode Gaussian states of $1 \times N$
bipartitions, as shown by Werner and Wolf \cite{WernerWolf}.
A further criterion, due to Duan \textit{et al.}
\cite{InsepDGCZ}, based on suitably scaled Einstein-Podolski-Rosen (EPR) observables,
turns out to be necessary of sufficient for the separability of two-mode Gaussian states.\\
On the other hand, as seen in Sections \ref{section4} and \ref{section5},
the most recent theoretical and experimental efforts are concerned with the engineering of highly
nonclassical, non-Gaussian states of the radiation field, in order to enhance either
the entanglement \cite{KimBuzekKnight,Kazzigawa}, or other useful properties as
the quantum robustness against decoherence \cite{DodonovDeSouza}.
Furthermore, even if at present Gaussian states continue to play a central role
in quantum optics and quantum information, it has been shown rigorously
that some fundamental physical properties, such as the entanglement and the
distillable secret key rate, are minimized by Gaussian states \cite{ExtremalGaussian}.
In a sense, this result is not surprising. Gaussian states are the most
semiclassical among quantum states, and entanglement is a genuine quantum
property that essentially originates from the superposition principle.
The inseparability criteria that have been introduced so far provide only
sufficient conditions on the inseparability of non-Gaussian states.
Therefore, they can fail in detecting quantum correlations
for some sets of genuinely entangled states \cite{InsepHZ}.\\
Before reviewing some of the most significant inseparability criteria,
we describe a sufficiently general approach to their construction
for bipartite systems of continuous variables.
A natural starting point to search for a necessary condition for separability
(sufficient condition for entanglement) is the general definition
by Werner of separable states, Eq. (\ref{WernerSepar}).
One then proceeds to select a suitable operatorial function $f$ of the
annihilation and creation operators associated to the two subsystems,
considers its mean values in a generic, separable state written in the
Werner form Eq. (\ref{WernerSepar}), and derives an inequality constraint.
Let us consider first the two EPR-like operators of the form
\begin{equation}
U=|a| X_{1} + \frac{1}{a} X_{2} \,, \quad \quad V=|a| P_{1} -
\frac{1}{a} P_{2} \; ,
\label{EPRlikeop}
\end{equation}
where $X_{j}$ and $P_{j}$ $(j=1,2)$ are the position and
momentum operators corresponding to the mode $j$, and $a$ is an
arbitrary real number. Applying the Cauchy-Schwartz inequality to the
total variance
$\langle (\Delta U)^{2} \rangle_{\rho} + \langle (\Delta V)^{2} \rangle_{\rho}$,
it is easy to show that for any separable quantum state $\rho$
the following inequality must hold
\begin{equation}
\Delta_{EPR}(a) = \langle (\Delta U)^{2}\rangle_{\rho} +
\langle (\Delta V)^{2}\rangle_{\rho} - a^{2} - \frac{1}{a^{2}}
\geq 0 \; ,
\label{totalvariance}
\end{equation}
for any choice of the constant $a$ \cite{InsepDGCZ}.
As an immediate consequence, if at least a
value of $a$ exists such that the inequality (\ref{totalvariance}) is
violated, i.e. $\Delta_{EPR}(a) < 0$, then $\rho$ is an
entangled state. This sufficient condition for inseparability is
precisely the criterion formulated by Duan \textit{et al.}
\cite{InsepDGCZ}. The criterion is optimized by minimizing
$\Delta_{EPR}(a)$ with respect to $a$, while
keeping the other variables fixed.\\
As a second example, let us now consider the criterion
due to Simon to obtain a sufficient condition for
inseparability in terms of the second order statistical
moments of the position and momentum operators \cite{InsepSimon}.
Simon defines the operatorial function
$$
\hat{f} \, = \, c_{1} X_{1}+c_{2} P_{1} + c_{3} X_{2}+c_{4} P_{2} \; ,
$$
and observes that the Heisenberg uncertainty principle is
equivalent to the statement
$$
\langle \hat{f}^{\dag}\hat{f} \rangle_{\rho} \, = \, Tr[\rho \hat{f}^{\dag}\hat{f}]
\, \geq \, 0 \; .
$$
Finally, by applying the PPT requirement (a separable density operator $\rho$
must transform into a non-negative, bona fide density operator $\tilde{\rho}$
under partial transposition), Simon obtains the following necessary condition
for separability:
\begin{equation}
Tr[\tilde{\rho} \hat{f}^{\dag}\hat{f}] \, \geq \, 0 \; .
\label{SepcondSimon}
\end{equation}
This condition, in terms of the matrices defined
in Eq. (\ref{Gausscovmatrix}), reads
\begin{eqnarray}
&&\det \mathbf{A} \det \mathbf{B} +\left( \frac{1}{4}-|\det \mathbf{C}|
\right)^{2}- Tr[\mathbf{AJCJBJC^{T}J}] \nonumber \\ &&
\nonumber \\
&&-\frac{1}{4}(\det \mathbf{A} +\det \mathbf{B}) \, \geq \, 0 \; ,
\label{Simonmatrixinequal}
\end{eqnarray}
where $\mathbf{J}$ is the $2\times 2$ symplectic matrix defined
in Eq. (\ref{symplecticmatrix}).
Let us recall again that one can easily show that
for all two-mode Gaussian states the Duan and the Simon criteria are
necessary and sufficient for separability, and thus coincide. Moreover,
the Simon criterion given by Eq. (\ref{SepcondSimon}) or
(\ref{Simonmatrixinequal}) reduces to a simple inequality
for the lowest symplectic eigenvalue: $\tilde{\nu}_{-} \geq 1$;
or, equivalently, for the covariance matrix:
$\tilde{\Delta}(\mathbf{\sigma}) \leq \det\mathbf{\sigma} +1$
(with $\mathbf{\sigma}$ given by Eq. (\ref{Gausscovmatrix})). \\
Other interesting criteria include the so-called ``product'' condition
introduced by Giovannetti \textit{et al.} \cite{InsepGiovannetti}.
Through a method similar to that of Ref. \cite{InsepDGCZ},
thes authors have found a more general inequality condition,
containing as a special case Eq. (\ref{totalvariance}).
Man'ko \textit{et al.} have proposed a generalization of the PPT
criterion based on partial transposition by introducing a {\it scaled partial
transposition} \cite{InsepManko}. The associated criterion of
positivity under scaled partial transposition (PSPT)
might be particularly suitable for the detection of multipartite entanglement.\\
The EPR- and PPT-type criteria discussed above, i.e. the EPR and PPT criteria
suitably adapted to the case of Gaussian states,
naturally involve only second order statistical moments of the quadratures.
Moving to analyze non Gaussian states, criteria limited to second
order moments do not detect entanglement efficiently, and inclusion
of higher order moments is needed.
Inspired by this motivation, Agarwal and Biswas
have proposed new criteria based on inequalities involving higher order
correlations \cite{InsepBiswas}. In particular, these authors have
combined the method of partial transposition and the uncertainty relations for the
Schwinger realizations of the $SU(2)$ and $SU(1,1)$ algebras on generic states.
In a two-mode system, the realization of the $SU(1,1)$ algebra,
provided by the operators
$K_{x}=2^{-1}(a_{1}^{\dag}a_{2}^{\dag}+a_{1}a_{2})$,
$K_{y}=(2i)^{-1}(a_{1}^{\dag}a_{2}^{\dag}-a_{1}a_{2})$,
$K_{z}=2^{-1}(a_{1}^{\dag}a_{1}+a_{2}^{\dag}a_{2}+1)$;
leads to an uncertainty relation that reads:
$$
\langle (\Delta K_{x})^{2} \rangle_{\rho} \langle (\Delta K_{y})^{2}
\rangle_{\rho} \, \geq \, \frac{1}{4}|\langle K_{z} \rangle_{\rho}|^{2} \; .
$$
The PPT prescription under partial transposition, e.g.
$\Leftrightarrow$ $a_{2}\leftrightarrow a_{2}^{\dag}$,
introduced in the above inequality yields a necessary condition
of separability in terms of fourth order statistical moments.
In Ref. \cite{InsepBiswas} this approach has been successfully tested
on the two-mode non-Gaussian state described by the wave
function
$$
\psi (x_{1},x_{2}) \, = \, (2/\pi)^{1/2} (\gamma_{1} x_{1} +
\gamma_{2} x_{2}) \exp\{-(x_{1}^{2}+x_{2}^{2})/2\} \; ,
$$
with $|\gamma_{1}|^{2}+|\gamma_{2}|^{2}=1$. This state is
always entangled for all values of $\gamma_{1}$ and $\gamma_{2}$,
as can be seen as well by direct application of PPT to the associated
density matrix. However, according to the criteria based on second
order moments, like the ones by Duan {\it et al} and by Giovannetti
{\it et al}, this state appears to be always separable. The inequality
based on fourth order moments is instead sufficient to detect
inseparability over the whole range of values of
$\gamma_{1}$ and $\gamma_{2}$. Clearly, as already suggested by
Agarwal and Biswas, it is desirable to develop criteria based
on hierarchies of statistical moments of arbitrary order.
Progress along this program has been achieved first
by Hillery and Zubairy \cite{InsepHZ}, and later, in much greater
generality by Shchukin and Vogel \cite{InsepShchukin}.
In Ref. \cite{InsepHZ}, Hillery and Zubairy have
considered moments of arbitrary order of
the form $|\langle a_{1}^{k} a_{2}^{h} \rangle_{\rho}|$ or of the
form $|\langle a_{1}^{k} a_{2}^{\dag h} \rangle_{\rho}|$.
Then, starting for instance from the first form, one can prove
that a necessary condition for the
separability of a generic two-mode quantum state
is the validity of all the infinite inequalities
\begin{equation}
\Delta_{kh} \, = \, \langle a_{1}^{\dag k}a_{1}^{k} \rangle \langle
a_{2}^{\dag h}a_{2}^{h} \rangle - |\langle
a_{1}^{k}a_{2}^{h}\rangle|^{2} \geq 0 \; ,
\label{HZinequality}
\end{equation}
where $k$ and $h$ are two arbitrary positive integers. Similar relations
cab be obtained by starting from the second form. Obviously,
the violation of any one of these infinite inequalities is then
a sufficient condition for inseparability. It is important to
notice that the moments involved in the
set of inequalities derived by Hillery and Zubairy
do not coincide with the moments involved in the inequalities
derived by Agarwal and Biswas when considering the fourth order.
However a further generalization
containing all the cases previously discussed has been recently introduced
by Shchukin and Vogel using an elegant and unifying approach
\cite{InsepShchukin}. These authors succeed to express the PPT criterion
for two-mode states of continuous variable systems through an infinite
hierarchy of inequalities involving all moments of arbitrary order.
They consider Eq. (\ref{SepcondSimon}), with $\hat{f}$ given by
\begin{equation}
\hat{f} \, = \, \sum_{n,m,k,l=0}^{\infty} c_{nmkl} \,
a_{1}^{\dag n}a_{1}^{m}a_{2}^{\dag k}a_{2}^{l} \; ,
\end{equation}
and prove that
the PPT separability condition is equivalent to imposing
the nonnegativity of an infinite hierarchy of suitably
chosen determinants \cite{InsepShchukin}
\begin{equation}
D_{N} \, = \, \left| \begin{array}{cccccc}
1 & \langle a_{1} \rangle & \langle a_{1}^{\dag} \rangle & \langle a_{2}^{\dag}
\rangle & \langle a_{2} \rangle  & \cdots \\
\langle a_{1}^{\dag} \rangle & \langle a_{1}^{\dag}a_{1} \rangle & \langle a_{1}^{\dag 2}
\rangle & \langle a_{1}^{\dag}a_{2}^{\dag} \rangle & \langle a_{1}^{\dag}a_{2} \rangle & \cdots \\
\langle a_{1} \rangle & \langle a_{1}^{2} \rangle & \langle a_{1}a_{1}^{\dag} \rangle &
\langle a_{1}a_{2}^{\dag} \rangle & \langle a_{1}a_{2} \rangle & \cdots \\
\langle a_{2} \rangle & \langle a_{1}a_{2} \rangle & \langle a_{1}^{\dag}a_{2} \rangle &
\langle a_{2}^{\dag}a_{2} \rangle & \langle a_{1}^{2} \rangle & \cdots \\
\langle a_{2}^{\dag} \rangle & \langle a_{1}a_{2}^{\dag} \rangle & \langle a_{1}^{\dag}a_{2}^{\dag}
\rangle & \langle a_{2}^{\dag 2} \rangle & \langle a_{2}a_{2}^{\dag} \rangle & \cdots \\
\cdots & \cdots & \cdots & \cdots & \cdots & \cdots \\
\end{array}
\right| \; ,
\end{equation}
i.e. $D_{N} \geq 0$ $\forall N$. Viceversa, if there exists at least one
negative determinant, i.e. if
$$
\exists N \; : \qquad D_{N} \, < \, 0 \; ,
$$
this is a sufficient condition for inseparability of the state
\cite{InsepShchukin}.

\subsection{Engineering and applications of multiphoton entangled states:
theoretical proposals and experimental realizations}

\begin{figure}[h]
\begin{center}
\includegraphics*[width=13.5cm]{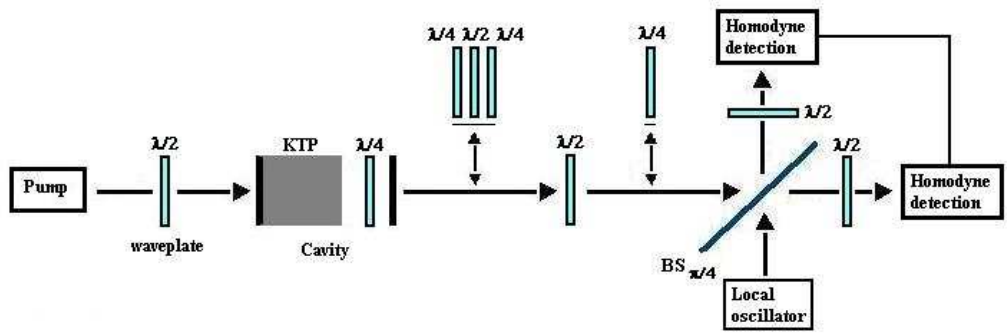}
\end{center}
\caption{Experimental setup for the production of two-mode squeezed states,
and for the characterization and manipulation of their entanglement.
A continuous-wave frequency-doubled $Nd:YAG$ laser pumps
a type-II OPO ($KTP$ crystal) with a (quarter)
$\lambda/4$ waveplate inserted inside a cavity.
By suitably inserting the quarter and half waveplates, the covariance matrix of the
two-mode Gaussian state can be reconstructed, and the entanglement can be optimized.}
\label{LauratExpScheme}
\end{figure}

\begin{figure}[h]
\begin{center}
\includegraphics*[width=12cm]{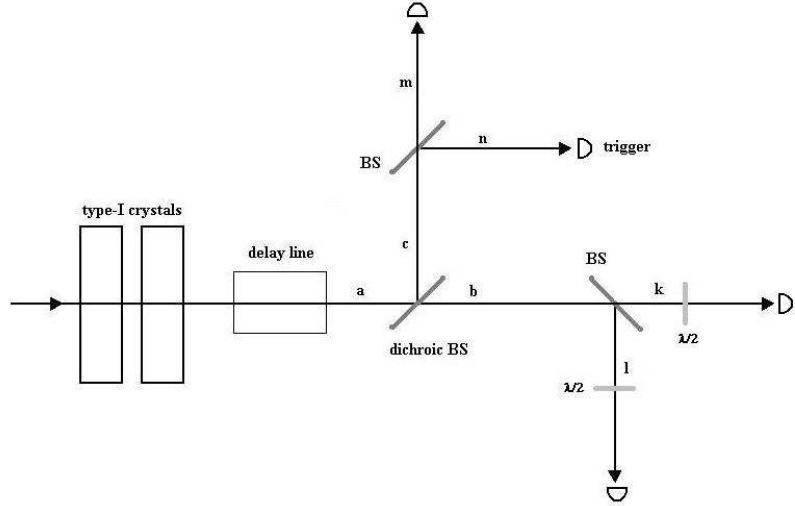}
\end{center}
\caption{Scheme for the generation and the detection of the
four-photon entangled state $|\Phi^{(4)}\rangle$.}
\label{4phstLi}
\end{figure}

A very nice feature of quantum optics is that it is flexible and
versatile enough to provide the necessary tools for quantum
information based either on discrete or on continuous variables.
Discrete-variable quantum information deals with elaborating,
storing, manipulating, and transmitting information encoded in
quantum states with a discrete spectrum of eigenvalues, and thus
belonging to finite-dimensional Hilbert spaces. In the simplest
instance of two-dimensional Hilbert spaces, the elementary units
of information are then two-state (two-level) systems, universally
known as qubits in the quantum-informatic jargon. All the possible
superposition states of a two-level system can be expressed in terms
of linear combinations of two-dimensional orthonormal states defining
the standard computational basis. In a quantum optical setting, these
basis states can be realized as the two orthonormal eigenstates of
the photon polarization observable, respectively, the horizontally
$(H)$ and vertically $(V)$ polarized one-photon states
\begin{equation}
|H\rangle \, = \, \left(\begin{array}{c}
  1 \\
  0 \\
\end{array}\right) \; , \quad \quad \quad |V\rangle\, = \,
\left(\begin{array}{c}
  0 \\
  1 \\
\end{array}\right) \; .
\label{qubitpol}
\end{equation}
On the other hand, to the electromagnetic field are associated
other physical objects, such as the phase and
amplitude quadratures ("position" and "momentum"), that
are continuous variables, i.e. generalized
observables with a continuous spectrum of eigenvalues.
Quantum information based on continuous variables is then
characterized by the fact that the information is carried
by quantum states belonging to infinite-dimensional Hilbert
spaces. Thus, quantum optics enables in principle to realize
the encoding of quantum information either with polarization
qubits or with continuous quadratures. In fact, it allows
much more. Namely, all intermediate possibilities for encoding
information can be realized by exploiting superpositions of a finite
number of macroscopically distinguishable states
\cite{logqubit,logqubit2}, so that, in a quantum optical framework,
a generic superposition state of a qubit can be constructed, e. g.,
as the superposition of two distinguishable continuous variables states,
for example the two coherent states of $\pi$-opposite phase
$|\alpha\rangle$ and $|-\alpha\rangle$.
Then, the generic continuous-variable qubit state in this case reads
\begin{equation}
|\psi\rangle \, = \, \xi_{1}|\alpha\rangle+\xi_{2}|-\alpha\rangle \; ,
\label{logqubit}
\end{equation}
where the complex coefficients $\xi_{i}$
satisfy the normalization condition
$|\xi_{1}|^{2}+|\xi_{2}|^{2}+2Re[\xi_{1}^{*}\xi_{2}\langle\alpha|-\alpha\rangle]=1$.
For large values of $|\alpha|^{2}$ the two states $|\pm \alpha \rangle$ will
be approximately orthogonal, and thus they will realistically approximate
the vectors of the computational basis.\\
We will now turn to discuss both
discrete-variable and continuous-variable schemes of
quantum computation and information. We begin by
introducing the general concept of quantum computer, probably the
most fascinating object of theoretical investigation in the field
of quantum information. In principle, it is expected that
a quantum computer will allow to
solve problems exponentially faster than any classical computer
\cite{Benioff1,Benioff2,Feynman1,Feynman2,Feynman3,Shorcomp}.
A simple prescription for the construction of
such a machine was given by Deutsch \cite{Deutsch}: it is a set of
$n$ qubits on which unitary operations or quantum logic gates can
be applied. It is well known \cite{QIbook} that a unitary transformation
(one-bit gate) and a two-bit conditional quantum phase gate are
sufficient to build a universal quantum computer. The one-bit gate
is the so-called universal quantum gate which by its repeated use
can generate all unitary transformations of the qubits, and it reads
\begin{equation}
U_{\theta,\phi} \, = \, \left(\begin{array}{cc}
\cos\theta & -ie^{-i\phi}\sin\theta \\
-ie^{i\phi}\sin\theta & \cos\theta \\
\end{array}\right) \; .
\end{equation}
The transformation for a two-bit phase gate is instead given by
\be
Q_{\phi}^{(2)}|\xi,\eta\rangle \, \equiv \,
e^{i\phi\delta_{\xi,V}\delta_{\eta,V}}|\xi,\eta\rangle
\; , \label{2bitgate}
\ee
where $|\xi,\eta\rangle=|\xi\rangle|\eta\rangle$ and $\xi,\eta=H,V$. It
is to be noted that the set of the linear unitary transformations
is not sufficient to implement the universal quantum computation;
this implies that nonlinear transformations are needed too. From
this point of view, several schemes have been proposed to realize
qubit-based quantum computation \cite{DVcomp1,DVcomp2,DVcomp3}. In
particular, Knill, Laflamme, and Milburn have demonstrated that,
in principle, robust and efficient quantum computing is possible
by combining linear optical elements with photodetectors inducing
nonlinear transformations \cite{DVcomp2}. On the other hand, resonantly enhanced
nonlinearities can offer some advantages in the implementation of
quantum computing protocols. Let us consider some examples
\cite{EITkerr3,chi5ena} which refer to recent proposals for the realization
of phase gates. The transformation (\ref{2bitgate})
can be realized by exploiting an enhanced $\chi^{(3)}$ cross-Kerr
interaction \cite{EITkerr3}, whose corresponding unitary operator
reads $Q_{\phi}^{(2)}=e^{i\phi
a_{V}^{\dag}a_{V}b_{V}^{\dag}b_{V}}$. The interaction
$H_{5}=\kappa^{(5)}
a_{V}^{\dag}a_{V}b_{V}^{\dag}b_{V}c_{V}^{\dag}c_{V}$, realizable
in an enhanced $\chi^{(5)}$ medium \cite{chi5ena}, can be used for
the implementation of Grover's quantum search algorithm
\cite{Grover}. A weak cross-Kerr nonlinearity has also been used
to construct a CNOT gate based on a quantum nondemolition
measurement scheme \cite{CNOTqnd}. \\
All the examples that we have just cited, refer to schemes involving
multiphoton processes in nonlinear media for the realization of
information protocols with discrete variables. However, as shown by Lloyd
and Braunstein \cite{QIbook4,QIrev9,CVcomp1}, a universal quantum computer can
be in principle realized over continuous variables as well, although limited
to the classes of transformations that are polynomial in
those variables. Considering as continuous
variables the quadrature amplitudes $\{X,P\}$ of the
electromagnetic field, these transformations might be implemented
by means of simple optical devices such as beam splitters, phase shifters, squeezers,
and Kerr-effect fibers \cite{CVcomp1}. Let us consider the application of a
sequence of Hamiltonians $H_{i}$, $H_{j}$, $-H_{i}$, $-H_{j}$,
each for a time $\delta t\rightarrow 0$; in this limit the
following relation holds
\be
e^{iH_{i}\delta t}e^{iH_{j}\delta
t}e^{-iH_{i}\delta t}e^{-iH_{j}\delta
t} \, = \, e^{(H_{i}H_{j}-H_{j}H_{i})\delta t^{2}}
\, + \, \mathcal{O}(\delta
t^{3}) \; .
\label{CVxprel}
\ee
Relation (\ref{CVxprel}) means that
the application of the sequence of Hamiltonians is equivalent to
the application of the Hamiltonian $i[H_{i},H_{j}]$ for a time
$\delta t^{2}$. In general, starting from a set of Hamiltonians
$\{\pm H_{i}\}$, any Hamiltonian of the form $\pm i[H_{i},H_{j}]$,
$\pm [H_{i},[H_{j},H_{k}]]$, etc. can be constructed
\cite{CVcomp1}. That is, one can construct the Hamiltonians in the
algebra generated from the original set of operators by repeated
commutations. If one
chooses as original set the operators $X$, $P$,
$H_{0}=X^{2}+P^{2}$, and $S=(XP+PX)/2$, corresponding to
translations, phase shifts, and squeezers, one can construct any
Hamiltonian quadratic in $X$ and $P$. Therefore one- and two-photon
passive and active optical transformations are enough to simulate
any quadratic Hamiltonian associated to second-order (two-photon)
processes. In order to obtain Hamiltonians that are arbitrary-order
self-adjoint polynomials of $X$ and $P$, it is instead necessary
the exploitation of nonlinear, nonquadratic processes like the Kerr
interaction $H^{2}=(X^{2}+P^{2})^{2}$.
Alternatively, as for the discrete variable case, the photon
number measurement can be used to induce a nonlinear
transformation \cite{CVcomp2}. For instance, by exploiting squeezing,
phase-space displacement, and photon counting, it is possible to
generate the so-called "cubic-phase state" \cite{CVcomp2} defined
as
\be
|\gamma\rangle \, = \, \int dx e^{i\gamma x^{3}}|x\rangle \; .
\label{cubicphst}
\ee
One can easily see that these cubic phase states coincide with
the homodyne multiphoton squeezed states (HOMPSS) \cite{noi1}
discussed in Section \ref{section6}, in the case of lowest
quadratic nonlinearity $F(x)=x^{2}$, and can thus be realized
by purely Hamiltonian, unitary evolutions, albeit nonlinear.\\
The realization of quantum gates can be also implemented by
means of teleportation protocols. Quantum teleportation has been
originally proposed by Bennett \textit{et al.} \cite{Bennetelep}
for systems of discrete variables, and successively extended by
Vaidman \cite{Vaidmtelep} and by Braunstein and Kimble
\cite{Brauntelep} to continuous variables. It consists in the
transmission of unknown quantum states between distant users
exploiting shared entangled states (resources), and with the
assistance of classical communication. Quantum teleportation
has been experimentally demonstrated by Bouwmeester \textit{et al.} \cite{QTelepexp}
and Boschi \textit{et al.} \cite{Boschi} for the discrete variable case, and by
Furusawa \textit{et al.} \cite{QTelepexp2,Furunocloning} in the case of continuous
variables. These are important results {\it per se} and from
the point of view of quantum computing as well, because it has been
shown that exploiting quantum teleportation reduces the necessary resources
for universal quantum computation \cite{QcompTelep}.
Actually, single-qubit operations, Bell-basis measurements and
entangled states as the three-photon Greenberger-Horne-Zeilinger
(GHZ) states \cite{GHZstates} result to be sufficient for the
realization of a universal quantum computer \cite{QcompTelep}.
However, the fundamental tools for the effective realization of
teleportation remain bipartite entangled states like the so-called
Bell states and devices for Bell-state measurements.
Let us first consider the discrete variable instance.
The four two-photon maximally polarization-entangled Bell states,
\be
|\phi_{\pm}\rangle \, = \, \frac{1}{\sqrt{2}}(|VV\rangle\pm|HH\rangle) \; ,
\qquad
|\psi_{\pm}\rangle \, = \, \frac{1}{\sqrt{2}}(|VH\rangle\pm|HV\rangle) \; ,
\label{4Bellst}
\ee
can be experimentally generated by means of
type-II parametric down conversion, as reported by Kwiat
\textit{et al.} \cite{PDCbellgenexp}. The notation, assumed in Eq.
(\ref{4Bellst}) and in the following relations for the discrete
variables polarization states, refers to the general definition
of the $N$-mode, $N$-photon state $|p_{1} \, p_2 \, \ldots \, p_{N}\rangle\equiv
\bigotimes_{k=1}^{N}|p_{k}\rangle$, in which each mode $k$ is populated by exactly
one photon having polarization $p_{k}=H,V$.
The possibility to produce such states in nonlinear photonic crystals
has been investigated as well \cite{phcrysBellgen}.
Very recently, the generation of Bell
states has been obtained experimentally by Li \textit{et al.}
\cite{fiberBellgenexp} in an optical fiber in the $1550\; nm$ band
of telecommunications. As a complete Bell measurement is
very difficult to achieve by means of linear optics
\cite{Bellmeaslin,Bellmeaslin2}, schemes for the
discrimination of the Bell states, based on nonlinear effects as
resonant atomic interactions \cite{Bellmeasatom} and Kerr
interactions \cite{BellmeasKerr1,BellmeasKerr2}, have been proposed.\\
On the other hand, moving to a continuous variable setting,
the Bell measurement can be easily performed by homodyne detection,
but it is difficult to generate states with large quadrature entanglement.
The continuous variable state $|\Psi_{EPR-B}\rangle$ containing the necessary
correlations for teleportation can be obtained by the
Einstein-Podolsky-Rosen (EPR) state \cite{EPR} (already introduced, in
a different form, in Section \ref{section6}), defined as the
simultaneous eigenstate of the two commuting
observables, total momentum and relative position, of a two-mode radiation
field. Let us consider the nondegenerate parametric
amplifier $H_{PA}=ir(a_{1}^{\dag}a_{2}^{\dag}-a_{1}a_{2})$; the
solution of the Heisenberg equations for the field modes yield the
squeezed joint quadratures
\be
X_{1}(t)-X_{2}(t)=(X_{1}-X_{2})e^{-rt} \; , \quad
P_{1}(t)+P_{2}(t)=(P_{1}+P_{2})e^{-rt} \; .
\ee
In the limit $rt\rightarrow\infty$ in which
squeezing becomes infinite, or, equivalently,
the waiting time becomes very large,
these operators commute
and admit the common eigenstate
\be
|\Psi_{EPR-B}\rangle \; = \; \frac{1}{\sqrt{2\pi}}\int
dx|x\rangle_{1}|x\rangle_{2} \; .
\ee
These two limits are unrealistic for concrete applications.
In a recent paper, van Enk has proposed a scheme that would allow
in principle to produce,
{\it in arbitrarily short time}, an arbitrarily large amount of entanglement
in a bipartite system \cite{vanEnk}. The scheme is based on the following simple
two-step interaction: in the first step one induces
the unitary evolution of an initial coherent state in a Kerr medium,
and subsequently, in the second step, the output state generated in the first
step is feeded to a $50-50$ beam splitter.
The final state $|\Psi_{ent} (\tau)\rangle$ is entangled and reads
\begin{eqnarray}
&& |\Psi_{ent} (\tau)\rangle \, = \, U_{K,BS} |\alpha\rangle_{a} |0\rangle_{b} \, = \nonumber \\
&& \nonumber \\
&& \exp \left\{ \frac{\pi}{4}(a^{\dag}b-b^{\dag}a)\right\} \,
\exp\{-i\tau a^{\dag 2}a^{2}\} |\alpha\rangle_{a} |0\rangle_{b} \; ,
\label{vanEnkevol}
\end{eqnarray}
where $\tau$ is the dimensionless time. Fixing $\tau \equiv \tau_{m}=\pi/m$,
the final state $|\Psi_{ent} (\tau_{m})\rangle$
is a superposition of $m$ two-mode coherent states of the form
\begin{eqnarray}
&& |\Psi_{ent} (\tau_{m})\rangle \, = \nonumber \\
&& \nonumber \\
&& \left\{
\begin{array}{cc}
\sum_{q=0}^{m-1}f_{q}^{(o)} |\alpha e^{-\frac{2\pi i q}{m}}/\sqrt{2}\rangle
|\alpha e^{-\frac{2\pi i q}{m}}/\sqrt{2}\rangle & \qquad m \quad odd \; , \\
& \\
\sum_{q=0}^{m-1}f_{q}^{(e)} |\alpha e^{\frac{i\pi}{m}} e^{-\frac{2\pi i q}{m}}/\sqrt{2}\rangle
|\alpha e^{\frac{i\pi}{m}} e^{-\frac{2\pi i q}{m}}/\sqrt{2}\rangle & \qquad m \quad even \; ,
\end{array}
\right.
\label{vanEnkstates}
\end{eqnarray}
where the coefficients $f_{q}^{(o)}$ and $f_{q}^{(e)}$ coincide with
those of Eq. (\ref{gattiagarwal}). The two-mode states (\ref{vanEnkstates}) are
entangled with entanglement $E(\tau_{m}) \simeq \log_{2}m$ ebits;
thus the entanglement increases with increasing $m$ or,
equivalently, with decreasing interaction time,
leading to an arbitrarily large entanglement in arbitrarily short time.
Note that, as mentioned in Section \ref{section4}, a similar scheme, and the states
(\ref{vanEnkstates}), have been proposed
to produce macroscopic superposition states in a short interaction
time in the presence of small Kerr nonlinearity \cite{superpossmallkerr}.\\
Continuous variable entanglement has
been experimentally observed in an optical fiber
\cite{CVentfibexp} and in the interaction between a linearly
polarized coherent field and cold cesium atoms in a cavity
\cite{CVentcesexp}. \\
An exhaustive investigation of continuous variable entanglement
of two-mode Gaussian states has been performed
by Laurat \textit{et al.} \cite{Lauratetc}, who,
after reviewing the general framework for the characterization and quantification
of entanglement, have experimentally produced two-mode squeezed states
through parametric amplification in a type-II OPO, and have showed how to
manipulate such states to maximize the entanglement.
The experimental setup realized in Ref. \cite{Lauratetc}
is similar to that of the seminal work
by Ou \textit{et al.} \cite{OuEntExp};
a simplified scheme of the setup is depicted in Fig. (\ref{LauratExpScheme}).
The two-mode squeezed state is produced by a frequency-degenerate type-II OPO
below threshold; the insertion of a $\lambda/4$ birefringent plate
inside the optical cavity allows to produce symmetric Gaussian states,
although not in standard form
(for the definition of a standard form of a covariance matrix
see e.g. Refs. \cite{InsepDGCZ,InsepSimon}).
Laurat \textit{et al.} show how to theoretically compute and experimentally determine
the elements of the matrix (\ref{Gausscovmatrix})
through a simultaneous double homodyne detection,
and how to optimize the entanglement by using a phase
shifter \cite{Lauratetc}. An improvement of the above experiment has been
recently proposed by D'Auria \textit{et al.} \cite{SolimenoPeris}.
The improvement amounts to exploiting
only one homodyne detection, with the consequent
repeated measurements of single-mode quadratures. \\
In order to implement quantum teleportation by
means of linear optics and squeezed light, multipartite
entanglement has been exploited by van Loock and Braunstein
\cite{multipartentCV}. They have shown that a $N$-partite,
multiphoton entangled state can be generated from $N$ squeezed field modes that are
then combined in a suitable way by $N-1$ beam splitters, and that quantum
teleportation between two of the $N$ parties is possible with the assistance
of the other $N-2$. This quantum teleportation network has been recently demonstrated
experimentally for $N=3$ modes \cite{demonetwork}. Moreover,
it has been shown that the success of a $N$-user continuous variable
teleportation network, quantified by the maximal achievable fidelity between
the input and the teleported state, is qualitatively and
quantitatively {\it equivalent} to the presence of genuine multipartite quadrature
entanglement in the shared resource states \cite{telepoppate}. More generally,
multipartite entanglement, either of continuous or of discrete
variables, represents a powerful resource to enlarge the field of
applications and possible implementations of quantum information
protocols, and its study presents very interesting aspects for the
understanding of the foundations of quantum theory \cite{QIbook3}. \\
Let us now consider some examples of multipartite
entangled states and methods for their generation in the quantum
optical domain. The most representative multipartite polarization-entangled
(discrete variable) states are the $N$-photon GHZ states \cite{GHZstates}
\be
|{\rm GHZ}_{N}\rangle \, = \, \frac{1}{\sqrt{2}}\left(
|\underbrace{VV\ldots V}_{N}\rangle \, + \,
|\underbrace{HH\ldots H}_{N}\rangle \right) \; ,
\label{GHZNst}
\ee
and the $N$-photon $W$ states \cite{Dur}
\be
|W_{N}\rangle \, = \, \frac{1}{\sqrt{N}}\left(|\underbrace{H\ldots
HV}_{N} \rangle \, + \, |\underbrace{H\ldots HVH}_{N} \rangle \, + \, \ldots
\, + \, |\underbrace{VH\ldots H}_{N} \rangle \right) \; .
\label{WNst}
\ee
The GHZ and $W$ states are inequivalent as they
cannot be converted into each other under stochastic local
operations and classical communications, and therefore differ
greatly in their entanglement properties \cite{Dur,Verstr}.
The GHZ state possesses maximal $N$-partite entanglement,
i.e. it violates Bell inequalities maximally; but its entanglement
is fragile, because if one or more parties are lost due, for instance,
to decoherence, then the entanglement is destroyed.
The $W$ state instead possesses less $N$-partite entanglement but it is more
robust against the loss of one or more parties, in the sense that
each pair of qubits in the reduced state has maximal possible
bipartite entanglement. Several successful experiments have been
reported on the generation of multiphoton entangled states,
in particular the three-, four-, and five-photon GHZ states, and
the three-photon $W$ state
\cite{GHZ3exp,GHZ4exp,GHZ4exp2,GHZ5exp,W3exp}. Furthermore, many
schemes have been proposed for the generation of the four-photon $W$
state and of other four-photon entangled states
\cite{W4,W42,W43,4phentst,4phentst2,4phentst3}. We want also to
mention the recent observation \cite{4phstexpeibl} of the
four-photon entangled state  \bea
|\Psi^{(4)}\rangle=&&\frac{1}{\sqrt{3}}[|HHVV\rangle +|VVHH\rangle
 \nonumber \\
 && \nonumber \\
 &&-\frac{1}{2}
(|HVHV\rangle+|VHVH\rangle\pm|HVVH\rangle\pm|VHHV\rangle)] \,,
\eea which can be expressed as a superposition of a four-photon
GHZ state and of a product of two Bell states. The state
$|\Psi^{(4)}\rangle$ can be generated directly by a second order
parametric down conversion process \cite{4phentst}. In fact, let
us consider the multiple emission events in type II parametric
down conversion during a single pump pulse, that is
\be
\exp \{-i\kappa(a_{V}^{\dag}b_{H}^{\dag}+a_{H}^{\dag}b_{V}^{\dag})\} |0\rangle
\; .
\ee
The second order term of the exponential expansion
corresponds to four-photon effects, and is proportional to
\bea
&&(a_{V}^{\dag}b_{H}^{\dag}+a_{H}^{\dag}b_{V}^{\dag})^{2}|0\rangle=
\nonumber \\
&& \nonumber \\
&&|2H_{a},2V_{b}\rangle+|2V_{a},2H_{b}\rangle+|1H_{a},1V_{a},1H_{b},1V_{b}\rangle
\,.
\eea
If this state enters two nonpolarizing beam splitters,
the output state $|\Psi^{(4)}\rangle$ can be obtained
\cite{4phentst}. Because it is relatively easy to generate, and shows
interesting correlation properties, this four-photon state has been proposed
as a candidate for the implementation of several communication schemes
\cite{4phstexpeibl}. A similar four-photon polarization-entangled state,
belonging to the class of so-called cluster states \cite{raussendorf}, that reads
\be
|\Phi_{cluster}\rangle \, = \, \frac{1}{2}\left(|HHHH\rangle  + |HHVV\rangle +
|VVHH\rangle - |VVVV\rangle \right) \; ,
\ee
has been recently realized experimentally, and exploited for the
demonstration of the Grover algorithm \cite{Grover}
on a $4$-photon quantum computer \cite{naturecluster}.
Other kinds of four-photon entangled states
can be obtained from type I down conversion in a cascaded
two-crystal geometry \cite{4phentst2,4phentst3}, see Fig.
(\ref{cascade2nonlin}). In the case of collinear nondegenerate
down conversion \cite{4phentst3}, the Hamiltonians for the the two
crystals write
\be
H_{1DC} \, = \, \eta_{1}a_{sH}^{\dag}a_{iH}^{\dag} \, + \, H.c.
\,, \quad \quad H_{2DC}=\eta_{2}a_{sV}^{\dag}a_{iV}^{\dag} \, + \, H.c. \; ,
\ee
where $a_{sq}$ and $a_{iq}$ $(q=H,V)$ represent the signal and the
idler modes, and $\eta_{i}$ are the coupling constants depending
on the strength of the nonlinearity and on the intensity of the pump
pulse. After a delay line, the information about the origin
(crystal) of the down converted fields is lost, and the state of the field
reads
\bea
|\Phi(t)\rangle \, & = & \, \exp\{-(iH_{2DC}t + iH_{1DC}t)\}|0\rangle \, \nonumber \\
&& \nonumber \\
& \approx & \,
[1-iH_{DC}t+(-iH_{DC}t)^{2}]|0\rangle \; ,
\label{4phstatLi}
\eea
with $H_{DC}=H_{1DC}+H_{2DC}$.
The four-photon contribution to the state (\ref{4phstatLi}) is
\be
\frac{1}{2}[\kappa_{1}^{2} a_{sH}^{\dag 2}a_{iH}^{\dag
2}+\kappa_{2}^{2} a_{sV}^{\dag 2}a_{iV}^{\dag
2}+2\kappa_{1}\kappa_{2}
a_{sH}^{\dag}a_{sV}^{\dag}a_{iH}^{\dag}a_{iV}^{\dag}]|0\rangle \,,
\ee
with $\kappa_{i}=\eta_{i}t$. In order to separate the signal
and idler fields, a dichroic beam splitter can be used; it directs
the signal beam to a mode which we denote by $b_{sq}$ $(q=H,V)$,
and the idler beam to a mode which we denote by $c_{iq}$. Each of
the modes $b_{sq}$ and $c_{iq}$ are directed to $50:50$
nonpolarizing beam splitters, leading to the transformations
$b_{sq}=\frac{1}{\sqrt{2}}(k_{sq}+l_{sq})$ and
$c_{iq}=\frac{1}{\sqrt{2}}(m_{iq}+n_{iq})$, where
$\{k_{sq}, m_{iq}\}$, and $\{l_{sq},n_{iq}\}$ are
the pairs of transmitted and reflected modes, respectively.
Successively, the modes $k_{sq}$ and
$l_{sq}$ pass through a half wave plate, which rotates their
polarization by $\pi/2$, leading to the exchange $H\rightarrow V$
and $V\rightarrow H$. The experimental setup is schematically
depicted in Fig. (\ref{4phstLi}). The resulting four-photon state
can be finally written down in the form
\bea
|\Phi^{(4)}\rangle=&&
\frac{1}{2}\left[\kappa_{1}^{2}k_{sV}^{\dag}l_{sV}^{\dag}m_{iH}^{\dag}n_{iH}^{\dag}+
\kappa_{2}^{2}k_{sH}^{\dag}l_{sH}^{\dag}m_{iV}^{\dag}n_{iV}^{\dag}
\right. \nonumber \\
&& \nonumber \\
&&\left.
+\frac{1}{2}\kappa_{1}\kappa_{2}(k_{sV}^{\dag}l_{sH}^{\dag}
+k_{sH}^{\dag}l_{sV}^{\dag})(m_{iH}^{\dag}n_{iV}^{\dag}+m_{iV}^{\dag}n_{iH}^{\dag})\right]|0\rangle
\; .
\eea
If the photon $l_{sq}$ is measured in the computational basis
$\{|H\rangle,|V\rangle\}$ with the result $|V\rangle$, then the
entangled state is projected onto the state
\be
\frac{1}{2}\left[\kappa_{1}^{2}k_{sV}^{\dag}m_{iH}^{\dag}n_{iH}^{\dag}
+\frac{1}{2}\kappa_{1}\kappa_{2}k_{sH}^{\dag}(m_{iH}^{\dag}n_{iV}^{\dag}
+ m_{iV}^{\dag}n_{iH}^{\dag})\right]|0\rangle
\; ,
\ee
which, after normalization, and by
setting $\kappa_{2}=2\kappa_{1}$, reduces to a three-photon $W$
state. All the multiphoton multipartite entangled states discussed
above have been proposed for various applications in quantum
information. In particular, the GHZ- and $W$-class states have been
proposed for quantum teleportation with high nonclassical fidelity
\cite{GHZWQtelep}, dense coding
\cite{GHZWdenscod}, quantum secret sharing \cite{GHZQsecshar}, and
quantum key distribution \cite{GHZQKD}. \\ Multipartite
entanglement has been experimentally demonstrated also for
continuous variables \cite{CVmultientexp,CVmultientexp2}. In fact,
tripartite entangled states have been generated by combining three
independent squeezed vacuum states \cite{CVmultientexp}, or by
manipulating an EPR state with linear optical devices \cite{CVmultientexp2}.
In the context of nonlinear optics, a multistep interaction
model has been proposed as
source of three-photon entangled states of the form \cite{3phKeller}
\be
|\Psi_{3ph}\rangle \, = \, \sum_{k_{1},k_{2},k_{3}}F(k_{1},k_{2},k_{3})
a_{1}^{\dag}(k_{1})a_{2}^{\dag}(k_{2})a_{3}^{\dag}(k_{3})|000\rangle
\; ,
\label{3phentst}
\ee
where $F$ is the three-photon spectral
function. The state (\ref{3phentst}) can be generated through two
independent down conversion processes and a subsequent up
conversion process (sum-frequency generation) inside the same
crystal illuminated by a classical pump. In the down conversion
processes two pairs of photons are created; the up conversion
process, responsible for the generation of a third photon, is
originated by the annihilation of a couple of photons each created
by a different down conversion process. The three-photon entangled
state $|\Psi_{3ph}\rangle$ can be derived from third order
perturbation under conditions of energy conservation and phase
matching \cite{3phKeller}. Another possible experimental scheme
for the generation of a three-photon entangled state using
multiphoton processes in nonlinear optical media
is based on the four-wave mixing Hamiltonian (\ref{H14wmx}),
in which one treats the pump mode as a classical pump,
describing a particular form of
nondegenerate three-photon down conversion process. In Ref.
\cite{3DCHnilo} the experimental feasibility and the phase
matching conditions for this process are discussed, and some
quantitative estimates are presented in the particular case of
a calcite crystal. \\
Several other proposals based on the exploitation of cascaded nonlinearities
have been put forward for the realization of multimode entangled states
\cite{FerraroBondani1,FerraroBondani2,PfisterCV,GuoCV}. Hamiltonians of the form
Eq. (\ref{smitherHI}) have been considered for the generation of
continuous-variable entangled states. In particular, the
realization of multiphoton processes described by
Hamiltonian (\ref{smitherHI}), involving a $\beta-BaB_{2}O_{4}$
nonlinear crystal, has been suggested and is supported by some
preliminary experimental results \cite{FerraroBondani1,FerraroBondani2}. With
respect to previous realizations
\cite{simUpDwnCexp}, in the above mentioned schemes
noncollinear phase matching conditions are
satisfied through suitable choices of frequencies and propagation
angles. The Hamiltonian (\ref{smitherHI}) admits the following
constant of the motion:
\be
\Delta=N_{1}(t)-N_{2}(t)-N_{i}(t) \; ,
\ee
where $N_{j}(t)=\langle a_{j}^{\dag}(t)a_{j}\rangle$. For the
initial three-mode vacuum state $|0,0,0\rangle$, one has $\Delta=0$ and
$N_{1}(t)=N_{2}(t)+N_{i}(t)$ at all times $t$. By exploiting the
conservation law, the evolved state $|\psi^{2step}(t)\rangle$ can
be expressed in the form \cite{FerraroBondani1}
\bea
&&|\psi^{2step}(t)\rangle=e^{-it H_{I}^{2step}}|0,0,0\rangle = \frac{1}{\sqrt{1+N_{1}}}
\nonumber \\
&& \nonumber \\
&& \times\sum_{m,n}
\left(\frac{N_{2}}{1+N_{1}}\right)^{m/2}\left(\frac{N_{i}}{1+N_{1}}\right)^{n/2}
\left[\frac{(m+n)!}{m!n!}\right]^{1/2}|n+m,m,n\rangle \; .
\label{entangferraro}
\eea
The three-mode entangled state
(\ref{entangferraro}) can be used to achieve optimal telecloning
of coherent states \cite{FerraroBondani1}. Another cascaded scheme
has been investigated in Ref. \cite{PfisterCV}, based on the
following symmetrized three-mode interaction Hamiltonian
\be
H_{3CV} \, = \, ir(a_{1}^{\dag}a_{2}^{\dag}+a_{2}^{\dag}a_{3}^{\dag}+a_{3}^{\dag}a_{1}^{\dag})+H.c.
\; .
\ee
The Heisenberg equations yield the following relations for
the quadrature operators
\bea
&&X_{1}(t)-X_{2}(t)=(X_{1}-X_{2})e^{-rt} \,, \nonumber \\
&& \nonumber \\
&&X_{1}(t)-X_{3}(t)=(X_{1}-X_{3})e^{-rt} \,, \nonumber \\
&& \nonumber \\
&&P_{1}(t)+P_{2}(t)+P_{3}(t)=(P_{1}+P_{2}+P_{3})e^{-2rt} \,, \eea
whose common eigenstate, in the limit of infinite squeezing
$rt\rightarrow\infty$, is the continuous variable GHZ state
\be
|\Psi_{GHZ_{3}}\rangle \, = \, \frac{1}{(2\pi)^{3/2}}\int
|x\rangle_{1}|x\rangle_{2}|x\rangle_{3} \; .
\ee
The infinite squeezing limit is of course unrealistic. However,
at finite squeezing, the continuous variable infinitely entangled
three-mode GHZ state can be very well approximated by pure symmetric
three-mode Gaussian states with sufficiently high, but finite,
degree of squeezing \cite{contanglamemucho}. These states enjoy
several nice properties, as, in particular, the fact of possessing
at the same time maximal genuine tripartite entanglement and maximal
bipartite entanglement in all possible two-mode reduced states if one
of the original three modes is lost. For this reason, at variance with
their three-qubit counterparts, these continuous variable states allow
a {\it promiscuous} entanglement sharing, and have been therefore
named continuous-variable GHZ/$W$ states \cite{contanglamemucho}.\\
The procedure illustrated above for the production of entangled
states of three modes can be easily extended to entangle
an arbitrary number of modes \cite{PfisterCV}.
In this general case the entangling Hamiltonian becomes
\be
H_{NCV} \, = \, ir\sum_{i=1}^{N}\sum_{j>1}^{N}a_{i}^{\dag}a_{j}^{\dag} \, + \, H.c.
\; ,
\ee
and the joint operators are given by
\bea
&&\sum_{i=1}^{N}P_{i}(t) \, = \, e^{-(N+1)rt}\sum_{i=1}^{N}P_{i} \; , \nonumber \\
&& \nonumber \\
&&X_{i}(t)-X_{j}(t) \, = \, e^{-rt}(X_{i}-X_{j}) \; , \qquad i\neq j \; .
\eea
The time evolution generated by this Hamiltonian is able to realize for
asymptotically large times or asymptotically large squeezing the continuous
variable $N$-mode entangled GHZ state.
The effective feasibility of the scheme based on Hamiltonian
$H_{NCV}$ has been investigated for $N=3,4$, by resorting to quasi-phase matching
techniques, and the simultaneous concurrence of different nonlinearities
has been experimentally observed in periodically poled $RbTiOAsO_{4}$ \cite{PfisterCV}.

\subsection{Quantum repeaters and quantum memory of light}

\begin{figure}[h]
\begin{center}
\includegraphics*[width=13.5cm]{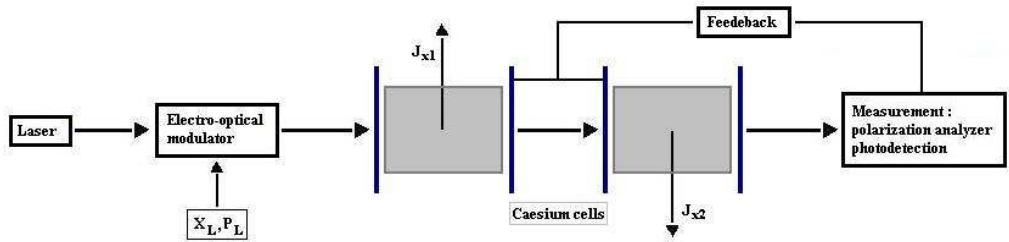}
\end{center}
\caption{Experimental setup for the realization of a quantum memory of light.
The electro-optic modulator is used to generate an input field with arbitrary
displacements $X_{L}$ and $P_{L}$.
The input field is encoded in the $y$ component
of a $x-y$ polarization-entangled state.
The two samples of cesium vapor are placed in paraffin-coated glass cells;
the cells, placed inside magnetic shields, constitute the storage units.
The bias magnetic field $\mathbf{H}$ along the $x$ axis
allows for encoding the memory at the Larmor frequency $\Omega$.
The canonical quadratures are detected by a polarization state analyzer
and by lock-in detection of the $\Omega$-component of the photocurrent.}
\label{ExpQuantMem}
\end{figure}

The main scope of quantum communication is the transmission of
quantum states between distant sites, a task that is particularly
suited for systems of continuous variables. For instance, it
can be in principle easily accomplished by resorting to photonic channels,
the main source of troubles in a realistic application being that
the error probability scales with the length $l$ of the channel.
For instance, in an optical fiber the probability for absorption
and depolarization of a photon grows exponentially with the length of the fiber.
The fidelity $\mathcal{F}$ of the transmitted state decays exponentially with $l$,
i.e. $\mathcal{F}\propto e^{-\gamma l}$,
with $\gamma$ being the characteristic decoherence rate;
in the same way, decoherence deteriorates entanglement \cite{Serafozzi}.
Hence, long-distance communication appears to be unrealistic, or at
least extremely difficult to achieve.
The standard purification schemes
\cite{BennettDistillation,DeutschDistillation,GisinDistillation}
are not sufficient to circumvent this problem, because they require
a threshold value of the fidelity $\mathcal{F}_{min}$ to operate, and
the latter is not achievable for increasing $l$.\\
However, an interesting strategy to solve this probleme has been
recently proposed \cite{longdistcomm}, based on the idea of
quantum repeaters \cite{Qrepeater}. The strategy works as follows.
Given a long channel, connecting two end sites $A$ and $B$, one can
divide it into $N$ segments, with $N$ such that the length
$l/N$ of each segment allows to support, between the nodes $A$ and $C_{1}$,
$C_{1}$ and $C_{2}$, $\ldots$, $C_{N-1}$ and $B$,
EPR pairs with sufficiently high initial fidelity, satisfying
$\mathcal{F}_{min}<\mathcal{F}\leq \mathcal{F}_{max}$.
Here, $\mathcal{F}_{max}$ denotes the maximum attainable fidelity,
and the end points of the intermediate segments
$C_{i}$ $(i=1,\ldots,N-1)$ are named auxiliary nodes.
After successful purification, all the EPR pairs between $A$ and $B$
are connected by performing Bell measurements at each node $C_{i}$,
and the result of this measurement is classically communicated to the next node.
At the end, a shared maximally entangled state between $A$ and $B$ has been obtained.
This process is called entanglement swapping \cite{EntSwap}.
A quantum repeater is thus based on a {\it nested purification protocol},
combining entanglement swapping and purification \cite{Qrepeater}.
It can be shown that the resources necessary for the efficient
functioning of the protocol grow only polynomially with $N$. This is
a crucial property for the succesfull implementation of the scheme.\\
In Ref. \cite{QuantRepLinearOpt}, an interesting proposal has been
put forward for the realization of a quantum repeater by combining
linear optical operations and a ``double-photon gun''.
The ``double-photon gun'' is defined as a source of a single
polarization-entangled photon pair at a time, on demand, and
with very high output fidelity. This is a theoretical concept,
however proposals for possible experimental realizations do
exist. For instance, an example of this source of entanglement
has been propesed by Benson \textit{et al.} \cite{SemicondQuantDot}.
It is based on a semiconductor quantum dot in which the electron-hole
recombination leads to the production of a single entangled photon pair,
with the advantage of vanishing probability of creating multiple pairs.\\
In general, the implementation of a quantum repeater, and of
other quantum information devices, such as linear optical
elements for quantum computation, requires suitable
{\it quantum memories}.
A quantum memory is a device that realizes the task of storing
an unknown quantum state of light with a fidelity
higher than that of the classical recording. Considering
coherent states as the states to be stored, the classical
memorization cannot overcome 50 per cent fidelity.
Examples of classical memories of light have been
experimentally realized exploiting the electromagnetically
induced transparency \cite{atomemoryclass1,atomemoryclass2}.
Quantum memories should allow a memorization of coherent states
of light with nonclassical fidelity, i.e. above the 50 per cent
classical limit. Several designes of quantum memories have been
proposed in Refs. \cite{photonstore1,photonstore2,photonstore3,photonstore4}.
Experimentally, important progress has been achieved in Refs.
\cite{atomemory1,atomemory2}, while the definitive experimental
demonstration of quantum memory of light has been realized
by storing coherent states of light onto quantum states of
atomic ensembles with 70 per cent fidelity  \cite{atomemory3}.\\
In the experiments of Refs. \cite{atomemory1,atomemory2},
the quantum correlations of photon pairs were generated in the
collective emission from an atomic ensemble with a controlled time
delay. These experiments have accomplished the task of controlled
entanglement between atoms and light, but are not yet examples
of quantum memory devices obeying the following two fundamental
requirements: 1) the state of light sent by a third party must
be unknown to the memory (atomic) party; 2) the state of light
is transferred in a quantum state (atomic state) of the memory
device with nonclassical fidelity. From this point of view, the
results reported Ref. \cite{atomemory3} are of particular relevance.
In this experiment, satisfying the two above criteria, a coherent state
of light is stored in the superposition of magnetic sublevels of the
ground state of an ensemble of cesium atoms.
The experimental setup is schematically depicted in Fig. (\ref{ExpQuantMem}).
The objective of the experiment is the storage of the values
of the two quadrature operators $X_{L}$ and $P_{L}$
characterizing an, in principle, arbitrary input quantum state
(a coherent state in the concrete experimental realization).
The $y$-polarized input quantum field $a(t)$ is generated
by an electro-optic modulator, and, before crossing an
array of paraffin-coated cells, is mixed, on a polarizing beam splitter,
with a strong entangling $x$-polarized pulse with photon flux $n(t)$.
The two samples of cesium atomic vapors are placed
in a bias magnetic field $\mathbf{H}$ oriented along the $x$ axis;
moreover, the atomic ensemble is prepared in states that
are approximatively coherent spin states, with the $x$
projection $J_{x 1}=-J_{x 2}=J_{x}=F \, N_{at}$,
where $F$ is the collective magnetic moment, and
$N_{at}$ is the number of atoms.
The array of the two cells allows to introduce the canonical variables
for the two ensembles $X_{A}=(J_{y1}-J_{y2})/\sqrt{2 J_{x}}$,
and $P_{A}=(J_{z1}+J_{z2})/\sqrt{2 J_{x}}$.
The light is transmitted through the atomic samples.
By considering the evolution of the Stokes operators, it can be shown that
in the presence of $\mathbf{H}$, the memory couples to the $\Omega$-sidebands
of light:
$X_{L}=T^{-1/2} \int_{0}^{T} dt \, (a^{\dag}(t)+a(t)) \cos(\Omega t)$ and
$P_{L}= i T^{-1/2} \int_{0}^{T} dt \, (a^{\dag}(t)-a(t)) \cos(\Omega t)$,
with $a(t)$ normalized to the photon flux, and $T$ being the pulse duration.
$X_{L}$ and $P_{L}$ are detected by a polarization state analyzer
and by lock-in detection of the $\Omega$-component of the photocurrent.
The polarization measurement of the transmitted light is followed
by the feedback onto atoms by means of a radio-frequency magnetic pulse
conditioned on the measurement result.
At the end, the desired mapping of the input quadratures
onto the atom-memory canonical variables, i.e.
 $X_{L}^{in}\rightarrow -P_{A}^{mem}$,
$P_{L}^{in}\rightarrow X_{A}^{mem}$, is realized;
thus, the quantum storage is achieved \cite{atomemory3}.
Let us note that this result has been obtained utilizing coherent states of light
as input states; however, since any arbitrary quantum state can be written in terms of
a superposition of coherent states, the approach has in principle a completely general
validity. Further proposals for the improvement of the above protocol have been put
forward in Refs. \cite{AtomemoryFiurasek1,AtomemoryFiurasek2}.

\newpage

\section{Conclusions and outlook}

\label{section8}

Quantum optics plays a key role in several branches of modern physics, involving
conceptual foundations and applicative fallouts as well, ranging from fundamental
quantum mechanics to lasers and astronomical observation.
To understand the vast and complex structure of quantum optics, it
is sufficient to list the phenomena that heavily involve
multiphoton processes: among them, ionization processes,
spontaneous emission, photo-association/dissociation of molecules,
quantum interference, quantum tunneling, quantum diffusion,
quantum dissipation, shaping of potentials, heating of plasmas,
and parametric processes in nonlinear media.\\
In this review, we have chosen to concentrate our attention
on those multiphoton processes that can be exploited to engineer
nonclassical and/or entangled states of light.
We are motivated by the fact that nonclassical states
of light can be considered a fundamental resource in a wide
range of applications, from quantum interferometry to
more recent and very promising fields of research such as
quantum information and quantum communication.
Besides the Fock states, the prototypes of nonclassical states
of light are the celebrated two-photon squeezed states, that can be
generated, for instance, by two-photon down-conversion processes.
Therefore, we have mainly considered the multiphoton
parametric processes occurring in nonlinear media, that can constitute
a useful device for producing new multiphoton quantum states with
appealing nonclassical properties that can include and extend
the ones exhibited by the two-photon squeezed states. \\
In the first part of this review we have discussed as well the
difficulties in implementing processes associated to susceptibilities
of higher nonlinear order, and the possible remarkable enhancements
of these quantities that could be achieved by exploiting the
recent successes in the engineering of new composite and
layered materials.
Besides the parametric processes, we have discussed other
methods for the generation of multiphoton quantum states,
based on the combination either of linear optics or of cavity fields
with quantum nondemolition (QND) measurements.
Furthermore, we have tried to give a sufficiently broad and fairly
complete review of the existing theoretical methods proposed
for the definition of different types and classes of multiphoton
nonclassical states, and we have briefly discussed the production
of multiphoton entangled states and their applications in the
rapidly growing research area of quantum computation and information.\\
The reader should be aware of the fact that, due to limitations in
length and scopes, we have often discussed effective Hamiltonian models
and realization schemes for multiphoton states under the assumption
of devices with ideal efficiency and/or in the absence of losses and dissipation.
However, during the time evolution of a real quantum system,
the unavoidable interaction with the environment will in general
induce loss of phase coherence between the constituent
states of the system itself: this phenomenon is known as
phase damping, or decoherence \cite{RevZurek,RevSchlosshauer}.
In quantum optics, the influence of amplitude damping, dissipation,
and quantum noise on the statistical properties of a quantum
open system can be investigated by using different approaches.
A first family includes phenomenological master equations
for the density matrix and classical Fokker-Planck equations
for the corresponding quasi-probability distributions, or
Langevin equations. including stochastic forces
\cite{Louiselltx,WallsMilburntx,txtPerina,GardinerZollertx}.
Microscopic approaches are also widely used, based on tracing
out environmental degrees of freedom in suitable system $+$ bath
models \cite{Revleggett}, and, finally, many of these different
descriptions find often an harmonic merging in the axiomatic theory
based on the mathematical structure of completely positive dynamical
maps and semigroups, and the Lindblad-Skraus superoperators
\cite{Daviestx,Gorini,Lindblad,AlickiLendi}. It is important to keep in
mind a clear distinction between decoherence and amplitude damping,
especially because the former may often take place on much shorter
time scales than the latter, and moreover can be realized by interactions
with minimal environments.
Methods and schemes to produce multiphoton nonclassical
and entangled states via nonlinear interactions in media and cavities
are then threatened by the disruptive effects of decoherence and dissipation.
In recent years many strategies have been devised
to protect quantum systems from losses and dissipation.
These strategies, based on quantum control implemented by
active feedback mechanisms \cite{quantfeed}, or by passive protections
like decoherence-free subspaces \cite{zanardirasettidecfree},
open realistic perspectives, in reasonable times, at least for
partial solutions of the problems related to
dissipation and decoherence.\\
In conclusion, we would like to comment on some possible future directions
in the physics of multiphoton processes and in the applications of
multiphoton states. A first direction will surely concern the theoretical
modelling and the experimental realization of larger, better, and more
stable (against noise and decoherence) quantum superposition states of
the radiation field. Multiphoton coherent and squeezed states of high
nonlinear order would be required for such a task, and therefore the
attempts to obtain strongly enhanced nonlinear susceptibilities of
higher order should play an increasingly important role in this field
of research. Another important directions will be along the route to
the understanding of the nature of quantum correlations between many
parties, and a better understanding of the entanglement properties
of multimode multiphoton nonclassical states together with their
experimental realizations would be very important tools in the study
of multipartite entanglement and, more generally, of fundamental
quantum theory. Finally, besides these important conceptual and
foundational aspects, realization of larger and better multiphoton
nonclassical states should allow further, and partly unexplored,
chances and perspectives to quantum optical realizations of quantum
information and communication processes.

\newpage

{\bf Aknowledgments}

\vspace{0.6cm}

We acknowledge financial support from MIUR under project ex 60\%, INFN,
and Coherentia CNR-INFM.

\end{document}